{}

\documentclass[10pt,a4paper]{article}
\newif\ifpublic
\newif\ifanswers
\newif\iflectures
\publictrue
\answerstrue
\lecturestrue
\lecturesfalse

\newif\ifextra\extrafalse

\ifpublic\else\usepackage{showkeys}\fi
\usepackage[bookmarks=true,hyperfigures=true,hidelinks,colorlinks=true,linkcolor=black,citecolor=black,urlcolor=black,bookmarksnumbered]{hyperref}
\usepackage[nosort]{cite}
\usepackage[parsep]{collref}
\usepackage[english]{babel}
\usepackage[T1]{fontenc}
\usepackage[latin1]{inputenc}
\usepackage[en-GB]{datetime2}
\usepackage{amsfonts,amsbsy,dsfont,euscript,mathrsfs,fixmath}
\usepackage{amsmath,amssymb}
\usepackage{scalerel}
\usepackage{enumerate}
\usepackage{multirow,booktabs}
\usepackage[x11names]{xcolor}
\usepackage{graphicx}
\usepackage{tikz,pgfplots}
\usepackage[tikz]{mdframed}
\usetikzlibrary{arrows,calc,chains,decorations.pathmorphing,decorations.pathreplacing,fadings,fit,matrix,positioning,shapes,shapes.geometric,shapes.symbols,trees}
\pgfplotsset{compat=1.9}

\def\showkeysrefformat#1{{\normalfont\tiny\ttfamily#1}}
\makeatletter
\def\SK@@ref#1>#2\SK@{{\@inlabelfalse\leavevmode\vbox to\z@{\vss\SK@refcolor\rlap{\vrule\raise .75em \hbox{\showkeysrefformat{#2}}}}}}
\makeatother

\usepackage[a4paper,text={160mm,247mm},centering]{geometry}
\allowdisplaybreaks[3]
\numberwithin{equation}{section}
\def\[{\begin{equation}\begin{aligned}}
\def\]{\end{aligned}\end{equation}}

\usepackage[font=small,labelfont=bf,width=0.85\textwidth]{caption}

\expandafter\def\expandafter\bfseries\expandafter{\bfseries\ifmmode\else\boldmath\fi}
\expandafter\def\expandafter\mdseries\expandafter{\mdseries\ifmmode\else\unboldmath\fi}
\expandafter\def\expandafter\normalfont\expandafter{\normalfont\ifmmode\else\unboldmath\fi}

\RequirePackage{verbatim}

\makeatletter
\newwrite\bibinl@out
\newenvironment{bibtex}[1][\jobname]{%
\immediate\openout\bibinl@out #1.bib%
\immediate\write\bibinl@out{\@percentchar generated from `\jobname' starting line \the\inputlineno^^J}%
\def\verbatim@processline{\immediate\write\bibinl@out{\the\verbatim@line}}%
\@bsphack\let\do\@makeother\dospecials\catcode`\^^M\active\verbatim@start%
}
{\immediate\closeout\bibinl@out\@esphack}
\makeatother

\let\barefrac=\frac
\renewcommand{\frac}[2]{\mathinner{\barefrac{#1}{#2}}}

\let\baresqrt=\sqrt
\makeatletter
\renewcommand{\sqrt}{\@ifnextchar[\@sqrt@space@a\@sqrt@space@b}
\def\@sqrt@space@a[#1]#2{\mathinner{\mathchoice{\mkern-3mu}{\mkern-3mu}{}{}\baresqrt[#1]{#2}}}
\def\@sqrt@space@b#1{\mathinner{\mathchoice{\mkern-3mu}{\mkern-3mu}{}{}\baresqrt{#1}}}
\makeatother

\makeatletter
\let\per@dot@old=\.
\def\.{\ifmmode\def\per@dot@sel{\mkern3mu}\else\def\per@dot@sel{\per@dot@old}\fi\per@dot@sel}
\makeatother

\let\barefootnote=\footnote
\renewcommand{\footnote}[1]{\barefootnote{#1\vspace{3pt}}}

\newcommand{\vfrac}[2]{\ifmmode\mathinner{\textstyle^{#1}\!/\!_{#2}}\else$^{#1}\!/\!_{#2}$\fi}

\DeclareMathOperator{\tr}{tr}
\DeclareMathOperator{\Tr}{Tr}

\DeclareMathOperator{\im}{im}

\newcommand{\Order}{O}

\newcommand{\Real}{\mathds{R}}
\newcommand{\Complex}{\mathds{C}}
\newcommand{\Integer}{\mathds{Z}}

\newcommand{\ind}[1]{{\scriptscriptstyle{#1}}}
\newcommand{\indrm}[1]{{\mathrm{\scriptscriptstyle{#1}}}}

\makeatletter
\newcommand*\bigcdot{\mathpalette\bigcdot@{.5}}
\newcommand*\bigcdot@[2]{\mathbin{\vcenter{\hbox{\scalebox{#2}{$\m@th#1\bullet$}}}}}
\makeatother

\newcommand{\alg}[1]{\mathfrak{#1}}
\newcommand{\grp}[1]{\mathrm{#1}}
\DeclareMathOperator{\rank}{rank}
\DeclareMathOperator{\Lie}{Lie}
\DeclareMathOperator{\ad}{ad}
\DeclareMathOperator{\Ad}{Ad}

\newcommand{\Lax}{\mathcal{L}}
\newcommand{\Mon}{\mathcal{M}}

\def\lang{\big\langle\!\big\langle}
\def\rang{\big\rangle\!\big\rangle}

\DeclareMathOperator*{\bigdotplus}{\scalerel*{\dotplus}{\bigoplus}}

\newcommand{\geom}[1]{\mathrm{#1}}

\newcommand{\AdS}{\geom{AdS}}

\newcommand{\Sp}{\geom{S}}
\newcommand{\Ball}{\geom{B}}

\newcommand{\To}{\geom{T}}
\newcommand{\CP}{\Complex\mathbf{P}}

\newcommand{\PexpL}{\mathrm{P}\hspace{-3pt}\stackrel{\longleftarrow}{\exp}}

\newcommand{\Act}{\mathcal{S}}

\DeclareFontEncoding{LS1}{}{}
\DeclareFontSubstitution{LS1}{stix}{m}{n}
\DeclareSymbolFont{stixsymbols}{LS1}{stixscr}{m}{n}
\SetSymbolFont{stixsymbols}{bold}{LS1}{stixscr}{b}{n}
\DeclareMathSymbol{\kay}{\mathalpha}{stixsymbols}{"6B}
\DeclareMathSymbol{\hay}{\mathalpha}{stixsymbols}{"68}
\DeclareMathAlphabet{\mathdsl}{U}{bbm}{m}{sl}
\newcommand{\gdsl}{\mathdsl{g}}

\providecommand{\href}[2]{#2}

\makeatletter
\def\mr@ignsp#1 {\ifx\:#1\@empty\else #1\expandafter\mr@ignsp\fi}
\newcommand{\multiref}[1]{\begingroup%
\xdef\mr@no@sparg{\expandafter\mr@ignsp#1 \: }%
\def\mr@comma{}\def\mr@dash{-}%
\@for\mr@refs:=\mr@no@sparg\do{%
\ifx\mr@refs\mr@dash\def\mr@comma{}--\else%
\mr@comma\def\mr@comma{,}\ref{\mr@refs}\fi}%
\endgroup}
\renewcommand{\eqref}[1]{(\multiref{#1})}
\makeatother

\makeatletter
\newcommand{\namedref}[2]{\hyperref[#2]{#1~\ref*{#2}}}
\newcommand{\secref}{\@ifstar{\namedref{Section}}{\namedref{sec.}}}
\newcommand{\appref}{\@ifstar{\namedref{Appendix}}{\namedref{app.}}}
\newcommand{\tabref}{\@ifstar{\namedref{Table}}{\namedref{tab.}}}
\newcommand{\figref}{\@ifstar{\namedref{Figure}}{\namedref{fig.}}}
\newcommand{\footref}{\@ifstar{\namedref{Footnote}}{\namedref{foot.}}}
\makeatother

\let\oldbib=\thebibliography
\def\thebibliography{\phantomsection\addcontentsline{toc}{section}{\refname}\oldbib}

\providecommand{\hypersetup}[1]{}
\providecommand{\texorpdfstring}[2]{#1}
\hypersetup{plainpages=false}
\hypersetup{pdfpagemode=UseNone}
\hypersetup{bookmarksnumbered=true}
\hypersetup{pdfstartview=FitH}
\hypersetup{colorlinks=false}
\hypersetup{citebordercolor={1 1 1}}
\hypersetup{urlbordercolor={1 1 1}}
\hypersetup{linkbordercolor={1 1 1}}

\makeatletter
\let\@keywords\@empty
\let\@subject\@empty
\providecommand{\keywords}[1]{\gdef\@keywords{#1}}
\providecommand{\subject}[1]{\gdef\@subject{#1}}
\def\thetitle{\@title}
\def\theauthor{\@author}
\def\thesubject{\@subject}
\def\thedate{\@date}
\def\thekeywords{\@keywords}
\makeatother
\AtBeginDocument{
\hypersetup{pdftitle={\thetitle}}
\hypersetup{pdfauthor={\theauthor}}
\hypersetup{pdfsubject={\thesubject}}
\hypersetup{pdfkeywords={\thekeywords}}}
\DTMlangsetup[en-GB]{ord=omit}
\DTMsettimestyle{iso}

\newif\ifshownote
\shownotetrue

\ifpublic\shownotefalse\fi

\ifshownote

\ifpdf\else\RequirePackage[active]{srcltx}\fi
\RequirePackage{xcolor}
\RequirePackage{ulem}

\newcommand{\remark}[2][]{{\normalfont\sffamily\hspace{1ex}
\def\emph{\textsl}\def\textbullet{$\bullet$}
\def\tmparga{#1}
\def\tmpargb{BH}\ifx\tmparga\tmpargb\color[rgb]{0.7,0,0}\fi
\def\tmpargb{}\ifx\tmparga\tmpargb\color{red}\fi
\def\tmpargb{}\ifx\tmparga\tmpargb\else \textbf{#1:}\fi
#2\hspace{1ex}}}

\else
\newcommand{\remark}[2][]{\ignorespaces}

\fi

\newcounter{question}
\stepcounter{question}
\mdfdefinestyle{questionstyle}{outerlinewidth = 0.2 , roundcorner = 2pt , leftmargin = 0pt , rightmargin = 0pt , backgroundcolor = SlateGray4!15 , linecolor = SlateGray4!25 , innertopmargin = 10pt , innerbottommargin = 10pt , everyline = true , splittopskip = 18pt , splitbottomskip = 10pt}
\newcommand{\question}[1]{
\begin{mdframed}[style=questionstyle]
\small
\textbf{Exercise \thequestion} \quad #1
\normalsize
\end{mdframed}
\stepcounter{question}}

\ifanswers
\newcounter{answer}
\stepcounter{answer}
\mdfdefinestyle{answerstyle}{outerlinewidth = 0.2 , roundcorner = 2pt , leftmargin = 0pt , rightmargin = 0pt , backgroundcolor = DarkSlateGray4!15 , linecolor = DarkSlateGray4!25 , innertopmargin = 10pt , innerbottommargin = 10pt , everyline = true , splittopskip = 18pt , splitbottomskip = 10pt}
\newcommand{\answer}[1]{
\begin{mdframed}[style=answerstyle]
\small
\textbf{Solution \theanswer} \quad #1
\normalsize
\end{mdframed}
\stepcounter{answer}}
\else
\newcommand{\answer}[1]{\ignorespaces}
\fi

\title{Integrable Deformations of Sigma Models}
\author{Ben Hoare}

\begin{document}

\pdfbookmark[1]{Title Page}{title}
\thispagestyle{empty}

\iflectures
\vspace*{2cm}
\begin{center}
\begingroup\Large\bfseries\thetitle
\par\endgroup
\vspace{1cm}
\begingroup Please send comments and corrections to\par \theauthor\par at \href{mailto:ben.hoare@durham.ac.uk}{\texttt{ben.hoare@durham.ac.uk}} \par\endgroup
\vspace{1cm}
\else
\vspace*{2cm}
\begin{center}
\begingroup\Large\bfseries\thetitle\par\endgroup
\vspace{1cm}

\begingroup\theauthor\par\endgroup
\vspace{1cm}

\textit{Department of Mathematical Sciences \\ Durham University \\ Durham DH1 3LE, UK}
\vspace{5mm}

\begingroup\ttfamily\small
ben.hoare@durham.ac.uk\par
\endgroup
\vspace{5mm}
\fi

\begin{abstract}\noindent
In \iflectures these lecture notes \else this pedagogical review \fi we introduce systematic approaches to deforming integrable 2-dimensional sigma models.
We use the integrable principal chiral model and the conformal Wess-Zumino-Witten model as our starting points and explore their Yang-Baxter and current-current deformations.
There is an intricate web of relations between these models based on underlying algebraic structures and worldsheet dualities, which is highlighted throughout.
We finish with a discussion of the generalisation to other symmetric integrable models, including some original results related to $\Integer_T$ cosets and their deformations, and the application to string theory.
\iflectures These notes were \else This review is based on notes \fi written for lectures delivered at the school ``Integrability, Dualities and Deformations,'' which ran from 23 to 27 August 2021 in Santiago de Compostela and virtually.
\end{abstract}

\small

\vspace{0.5cm}

\iflectures

\hfill Durham, August 2021
\\
\hfill last updated \today

\fi

\end{center}

\vspace{0.5cm}

\tableofcontents

\normalsize

\newpage

\section{Introduction}\label{sec:introduction}

The goal of \iflectures these lecture notes \else this pedagogical review \fi is to introduce systematic approaches to constructing integrable deformations of 2-d sigma models.
In order to keep the presentation pedagogical, we will use the integrable principal chiral model, the principal chiral model plus Wess-Zumino term and the conformal Wess-Zumino-Witten model as our starting points.
After introducing these models in \secref{sec:pcmwz}, we will start our exploration by considering different integrable deformations of the $\grp{SU}(2)$ models in \secref{sec:su2pcm}.
These include the trigonometric, elliptic and Yang-Baxter deformations.

In \secref{sec:ybpcm} we will investigate the generalisation of the Yang-Baxter deformation for general group $\grp{G}$.
There are two classes of such models, the homogeneous and inhomogeneous deformations, which correspond to twisting and q-deforming the symmetry algebra respectively.
We will see the close relationship between homogeneous deformations and worldsheet dualities, and study inhomogeneous deformations based on the well-known Drinfel'd-Jimbo solution to the modified classical Yang-Baxter equation.
Investigating the underlying algebraic structure of Yang-Baxter deformations will lead us to the notion of Poisson-Lie T-duality in \secref{sec:dd} and current-current deformations of the WZW model, which we will explore in more detail in \secref{sec:wzwmodel}.
Finally, in \secref{sec:generalisations} we finish with an overview of the literature including references for the material presented in \iflectures these notes, \else this review, \fi discuss the generalisations to other symmetric integrable models, including some original results related to $\Integer_T$ cosets and their deformations, and highlight applications to string theory.

\bigskip

Before we begin, let us take a moment to discuss some of the reasons why it is of interest to study integrable deformations of sigma models.
Due to their large hidden symmetry, 2-d integrable models are amongst a small class of interacting field theories that can potentially be solved exactly both classically and at a quantum level.
A large toolkit of methods and techniques has been developed that can be applied to these models.
This motivates their classification, which, while a formidable task, has seen much progress in recent years.
Constructing integrable deformations of known integrable sigma models is a systematic approach that has been particularly fruitful, and allows us to investigate different corners of the space of integrable models.

On the other hand, deformations can also provide new understanding of the structure of known integrable models.
For example, applying standard quantum field theory techniques to many 2-d sigma models is complicated by the fact that the perturbative excitations are massless.
In the principal chiral model these massless excitations are the Goldstone bosons associated to symmetry breaking.
In the quantum theory the global symmetry is restored and the elementary degrees of freedom are actually massive.
By deforming the theory, additional structure, such as new fixed points, can be introduced.
This allows alternative first-principle approaches to be employed that may be more amenable to quantization and verifying conjectures following from integrability.

While 2-d integrable sigma models are interesting in their own right, much of the recent motivation to study them has come from applications to string theory.
Indeed, worldsheet string theories are described by 2-d sigma models.
There are a number of examples of superstrings supported by Ramond-Ramond fluxes whose worldsheet theories are integrable.
This includes the $\AdS_5 \times \Sp^5$ and $\AdS_4 \times \CP^3$ superstrings amongst others.
While it is difficult to apply traditional conformal field theory methods to these models, the toolkit of integrability methods can be used, leading to a proposal for the exact spectrum of string energies in the free string limit.
Finding new examples of solvable string theories is always interesting, and studying integrable deformations is one approach to achieving this.
A more detailed discussion of some of these applications together with the relevant references can be found in \secref{sec:generalisations}.

\section{The principal chiral model and the Wess-Zumino term}\label{sec:pcmwz}

\subsection{The principal chiral model}

The principal chiral model (PCM) is the prototypical example of a 2-d integrable sigma model.
The field content is simply a field $g(t,x)$ valued in the Lie group $\grp{G}$, which plays the role of the target space.
For convenience we will take this Lie group to be simple throughout \iflectures these lecture notes, \else this review, \fi and in any examples it will be $\grp{SU}(2)$ or one of its non-compact counterparts.
Since the Lie group $\grp{G}$ is simple, there is a unique invariant bilinear form, up to normalisation, on its Lie algebra $\Lie(\grp{G}) = \alg{g}$, which we denote by $\tr$.
\unskip\footnote{\label{foot:index}Often it is convenient to take the group-valued field to be in a matrix representation of $\grp{G}$.
In order to avoid introducing ambiguities related to the choice of representation, we define $\tr$ in the adjoint representation to be
\begin{equation*}
\tr\big(XY\big) = \frac{1}{2h^\vee} \Tr\big(\ad_X\ad_Y\big) \qquad X,Y \in \alg{g} ~,
\end{equation*}
where $h^\vee$ is the dual Coxeter number of $\alg{g}$.
In a general representation this means that
\begin{equation*}
\tr\big(XY\big) = \frac{1}{2\chi_\ind{R}} \Tr\big(XY\big) ~, \qquad X,Y \in \alg{g} ~,
\end{equation*}
where $\chi_\ind{R}$ is the index of the representation and $\Tr$ is the usual matrix trace.
It follows that $\tr$ is independent of the choice of representation.}
This is the Killing form.
We also introduce the left- and right-invariant Maurer-Cartan forms
\begin{equation}
j = g^{-1} d g \in \alg{g} ~, \qquad k = - dg g^{-1} \in \alg{g} ~,
\end{equation}
which are the distinguished differential 1-forms on the Lie group manifold $\grp{G}$ valued in the Lie algebra $\alg{g}$, and their pull-backs to the 2-d worldsheet
\unskip\footnote{In \iflectures these notes \else this review \fi we typically refer to the 2-d space-time as the worldsheet.}
\begin{equation}\label{eq:lrmc}
j_\mu = g^{-1} \partial_\mu g \in \alg{g} ~, \qquad k_\mu = - \partial_\mu gg^{-1} \in \alg{g} ~,
\end{equation}
also valued in the Lie algebra $\alg{g}$.
Here the index $\mu = t,x$ runs over the 2-d worldsheet coordinates.
The flatness of the Maurer-Cartan forms then implies that
\begin{equation}\label{eq:mcflat}
\partial_\mu j_\nu - \partial_\nu j_\mu + [j_\mu,j_\nu] = 0 ~, \qquad
\partial_\mu k_\nu - \partial_\nu k_\mu + [k_\mu,k_\nu] = 0 ~.
\end{equation}

The action of the PCM is
\begin{equation}\label{eq:pcmaction}
\Act_\indrm{PCM} = \frac{\hay}{2} \int d^2x \, \tr\big(g^{-1}\partial_\mu g g^{-1}\partial^\mu g\big)
= \frac{\hay}{2} \int d^2x \, \tr\big(j_\mu j^\mu\big)
= \frac{\hay}{2} \int d^2x \, \tr\big(k_\mu k^\mu\big) ~.
\end{equation}
We assume that we are on a flat Minkowski worldsheet with metric $\eta_{\mu\nu}$, $-\eta_{tt} = \eta_{xx} = 1$ and $\eta_{tx} = \eta_{xt} = 0$.
The parameter $\hay$ is the inverse coupling constant of the model and is proportional to radius squared of the target space.
In the context of worldsheet string theories it plays the role of the string tension.
The overall sign of the action~\eqref{eq:pcmaction} ensures that for a compact Lie group $\grp{G}$, which has a negative-definite Killing form, the kinetic terms are positive and the target-space metric has Euclidean signature.
For a non-compact Lie group, the kinetic terms have mixed signs.
In such examples it may be of interest to reverse the overall sign of the action~\eqref{eq:pcmaction}, for example to give a target-space metric with Lorentzian signature.
This will be the case for models with $\grp{G} = \grp{SL}(2,\Real)$.

The PCM has a $\grp{G}_\ind{L} \times \grp{G}_\ind{R}$ global symmetry, which acts on the group-valued field as
\begin{equation}
g \to g_\ind{L} g g_\ind{R} ~, \qquad g_\ind{L},g_\ind{R} \in \grp{G} ~.
\end{equation}
The equations of motion of the PCM can be written as the conservation of either the right-acting or left-acting symmetry
\begin{equation}\label{eq:eomcons}
\partial^\mu j_\mu = 0 ~, \qquad \partial^\mu k_\mu = 0 ~.
\end{equation}
\iflectures

\question{
Consider the infinitesimal variation of the group-valued field $g \to g e^{\varepsilon} \sim g(1+\varepsilon)$, where $\varepsilon \in \alg{g}$, and compute the equations of motion of the PCM in terms of $j_\mu$.
Show that they can be written as the conservation equation $\partial_\mu j^\mu = 0$.
}

\answer{We consider the infinitesimal variation of the group-valued field $g \to g e^{\varepsilon} \sim g(1+\varepsilon)$, where $\varepsilon \in \alg{g}$.
Therefore, $\delta g = g \varepsilon$ and $\delta g^{-1} = -g^{-1}\delta g g^{-1} = -\varepsilon g^{-1}$.
Varying $j_\mu = g^{-1}\partial_\mu g$ gives
\begin{equation*}\begin{split}
\delta j_\mu = \delta(g^{-1}\partial_\mu g) = -\varepsilon g^{-1} \partial_\mu g + g^{-1} \partial_\mu (g\varepsilon)
= \partial_\mu \varepsilon + [g^{-1}\partial_\mu g, \varepsilon] = \partial_\mu\varepsilon + [j_\mu,\varepsilon] ~.
\end{split}\end{equation*}
Now varying the action~\eqref{eq:pcmaction} we find
\begin{equation*}\begin{split}
\delta \Act_\indrm{PCM} & = \hay \int d^2x \, \tr\big((\partial_\mu \varepsilon + [j_\mu, \varepsilon]) j^\mu\big)
\\ & = - \hay \int d^2x \, \tr\big(\varepsilon (\partial_\mu j^\mu + [j_\mu,j^\mu])\big) ~,
\end{split}\end{equation*}
where we have used integration by parts and the invariance of the bilinear form.
Given that $[j_\mu,j^\mu] = 0$, we immediately see that for the variation to vanish for arbitrary $\varepsilon \in \alg{g}$, the equation of motion is the conservation equation
\begin{equation*}
\partial_\mu j^\mu = 0 ~.
\end{equation*}
This demonstrates explicitly that the left-invariant $j_\mu$ is the conserved Noether current for the right-acting symmetry.
}
\else
To see this explicitly let us consider the infinitesimal variation of the group-valued field $g \to g e^{\varepsilon} \sim g(1+\varepsilon)$, where $\varepsilon \in \alg{g}$.
Therefore, $\delta g = g \varepsilon$ and $\delta g^{-1} = -g^{-1}\delta g g^{-1} = -\varepsilon g^{-1}$.
Varying $j_\mu = g^{-1}\partial_\mu g$ gives
\begin{equation}\begin{split}
\delta j_\mu = \delta(g^{-1}\partial_\mu g) = -\varepsilon g^{-1} \partial_\mu g + g^{-1} \partial_\mu (g\varepsilon)
= \partial_\mu \varepsilon + [g^{-1}\partial_\mu g, \varepsilon] = \partial_\mu\varepsilon + [j_\mu,\varepsilon] ~.
\end{split}\end{equation}
Now varying the action~\eqref{eq:pcmaction} we find
\begin{equation}\begin{split}
\delta \Act_\indrm{PCM} & = \hay \int d^2x \, \tr\big((\partial_\mu \varepsilon + [j_\mu, \varepsilon]) j^\mu\big)
= - \hay \int d^2x \, \tr\big(\varepsilon (\partial_\mu j^\mu + [j_\mu,j^\mu])\big)
= - \hay \int d^2x \, \tr\big(\varepsilon \partial_\mu j^\mu\big) ~,
\end{split}\end{equation}
where we have used integration by parts, the invariance of the bilinear form and that $[j_\mu,j^\mu] = 0$.
We immediately see that requiring the variation to vanish for arbitrary $\varepsilon \in \alg{g}$ implies that the equation of motion is the conservation equation $\partial_\mu j^\mu = 0$.
This also demonstrates that the left-invariant $j_\mu$ is the conserved Noether current for the right-acting symmetry.
A similar computation shows that the right-invariant $k_\mu$ is the conserved Noether current for the left-acting symmetry.
\fi

In general, we will work with light-cone coordinates on the worldsheet, which we define as
\begin{equation}
x^\pm = \frac{1}{2}(t\pm x) ~, \qquad \partial_\pm = \partial_t \pm \partial_x ~.
\end{equation}
In terms of light-cone coordinates the action of the PCM~\eqref{eq:pcmaction} becomes
\begin{equation}\label{eq:pcmact}
\Act_\indrm{PCM} = - \frac{\hay}{2} \int d^2x \, \tr\big(j_+ j_-\big) ~.
\end{equation}
The equations of motion~\eqref{eq:eomcons} are given by
\begin{equation}\label{eq:eompcm}
\partial_+ j_- + \partial_- j_+ = 0 ~,
\end{equation}
while the flatness condition~\eqref{eq:mcflat} is
\begin{equation}\label{eq:flatness}
\partial_+ j_- - \partial_- j_+ + [j_+,j_-] = 0 ~.
\end{equation}

The PCM can be written in the canonical form of a 2-d sigma model
\unskip\footnote{Conventionally, this action comes with an additional factor of $\frac{1}{2\pi\alpha'}$ where $\alpha'$ plays the role of the loop-counting parameter in the quantum theory.
In the context of string theory this factor plays the role of the string tension.
Given that we will mostly focus on classical sigma models, we set $\alpha' = \frac{1}{2\pi}$ for convenience.}
\begin{equation}\begin{split}\label{eq:gensigmamodel}
\Act_\indrm{SM} & = - \frac{1}{2} \int d^2x \, (G_\ind{MN}(\phi) \partial_\mu \phi^\ind{M} \partial^\mu \phi^\ind{N} + \epsilon^{\mu\nu} B_\ind{MN}(\phi) \partial_\mu \phi^\ind{M} \partial_\nu \phi^\ind{N})
\\ & = \frac{1}{2} \int d^2x \, (G_\ind{MN}(\phi) + B_\ind{MN}(\phi)) \partial_+ \phi^\ind{M} \partial_- \phi^\ind{N} ~.
\end{split}\end{equation}
Here $\phi^\ind{M}$ are local coordinates on the target space, $G = G_\ind{MN}(\phi) d\phi^\ind{M} d\phi^\ind{N}$ is the metric on the target space and the B-field $B = \frac12 B_\ind{MN}(\phi) d\phi^\ind{M} \wedge d\phi^\ind{N}$ is a target-space 2-form.
$\epsilon^{\mu\nu}$ is the antisymmetric tensor on the worldsheet with $\epsilon^{01} = - \epsilon^{10} = 1$.
Up to boundary terms, the action~\eqref{eq:gensigmamodel} is invariant under gauge transformations
\begin{equation}\label{eq:gauge}
B \to B + d \Lambda ~, \qquad
B_\ind{MN} \to B_\ind{MN} + \partial_\ind{M} \Lambda_\ind{N} - \partial_\ind{N} \Lambda_\ind{M} ~.
\end{equation}
We can therefore define the H-flux, a gauge-invariant closed 3-form,
\begin{equation}
H = dB ~, \qquad H_\ind{MNP} = \partial_\ind{M} B_\ind{NP} + \partial_\ind{N} B_\ind{PM} + \partial_\ind{P} B_\ind{MN} ~.
\end{equation}
If the 2-form $B$ is globally well-defined then $H$ is exact as well as closed.

To write the PCM in the form~\eqref{eq:gensigmamodel}, we introduce coordinates $\phi^\ind{M}$ on the Lie group manifold $\grp{G}$, that is $g(t,x) = g(\phi^\ind{M}(t,x))$, and generators $T_a$ of the Lie algebra $\alg{g}$.
We then expand $j_\mu$ in terms of left-invariant frame fields
\begin{equation}
j_\mu = L^a_\ind{M}(\phi)  \partial_\mu \phi^\ind{M} T_a ~.
\end{equation}
Substituting into the action~\eqref{eq:pcmaction} we see that it takes the form~\eqref{eq:gensigmamodel} with
\begin{equation}
G_\ind{MN}(\phi) = - L^a_\ind{M}(\phi) L^b_\ind{N}(\phi) \tr\big(T_a T_b\big) ~, \qquad B_\ind{MN}(\phi) = 0 ~.
\end{equation}
Similarly, we can introduce right-invariant frame fields
\begin{equation}
k_\mu = R^a_\ind{M}(\phi)  \partial_\mu \phi^\ind{M} T_a ~,
\end{equation}
in terms of which $G_\ind{MN}(\phi)$ and $B_\ind{MN}(\phi)$ are given by
\begin{equation}
G_\ind{MN}(\phi) = - R^a_\ind{M}(\phi) R^b_\ind{N}(\phi) \tr\big(T_a T_b\big) ~, \qquad B_\ind{MN}(\phi) = 0 ~.
\end{equation}
Therefore, we indeed find a 2-d sigma model whose target-space metric is the bi-invariant metric on the Lie group $\grp{G}$ and with vanishing B-field.

The PCM is an integrable sigma model.
To show this we start by constructing a Lax connection.
The Lax connection depends on a spectral parameter $z \in \Complex$ and its flatness for arbitrary $z$ should be equivalent to the equations of motion of the model.
One approach to constructing a Lax connection, which will prove useful later, is to observe that the current $j_\pm$ is both conserved and flat on-shell.
When we have such a current we can always write down the following Lax connection
\begin{equation}
\Lax_\pm(z) = \frac{j_\pm}{1\mp z} ~.
\end{equation}
Indeed, computing the curvature we find
\begin{equation}\begin{split}
\partial_+ \Lax_- - \partial_- \Lax_+ + [\Lax_+,\Lax_-] & = \frac{1}{1-z^2}\big((1-z)\partial_+j_- - (1+z) \partial_- j_+ + [j_+,j_-])
\\ & = \frac{1}{1-z^2}\big(\partial_+ j_- - \partial_- j_+ + [j_+,j_-] - z(\partial_+ j_- + \partial_- j_+)\big) ~.
\end{split}\end{equation}
We immediately see that the vanishing of the curvature for all $z$ implies that $j_\pm$ is both conserved and flat.
Conversely, if $j_\pm$ is both conserved and flat then the Lax connection is flat.
Note that we can also construct a Lax connection starting from $k_\pm$, which is also conserved and flat.
The two connections are equivalent in the sense that they are related by a formal gauge transformation, which leaves the flatness condition
\begin{equation}
\partial_+ \Lax_- - \partial_- \Lax_+ + [\Lax_+,\Lax_-]  = 0 ~,
\end{equation}
invariant, and a redefinition of the spectral parameter.
These formal gauge transformations do not correspond to a gauge symmetry of the model.
\iflectures

\question{Show that the two Lax connections
\begin{equation*}
\Lax_\pm(z) = \frac{j_\pm}{1\mp z} ~, \qquad
\tilde{\Lax}_\pm(z) = \frac{k_\pm}{1\mp z} ~,
\end{equation*}
are related by a formal gauge transformation and a redefinition of the spectral parameter.
}

\answer{
We have
\begin{equation*}
\Lax_\pm(z) = \frac{g^{-1}\partial_\pm g}{1\mp z} ~,
\qquad \tilde{\Lax}_\pm(z) = - \frac{\partial_\pm g g^{-1}}{1\mp z} ~.
\end{equation*}
It is then relatively straightforward to see that
\begin{equation*}\begin{split}
g \Lax_\pm (z) g^{-1} - \partial_\pm g g^{-1} & = \frac{\partial_\pm g g^{-1}}{1\mp z} - \partial_\pm g g^{-1}
\\ & = - \frac{\partial_\pm g g^{-1}}{1\mp z^{-1}} = \tilde{\Lax}_\pm(z^{-1}) ~,
\end{split}\end{equation*}
as required.
}
\else
Explicitly, if we define
\begin{equation}
\Lax_\pm(z) = \frac{g^{-1}\partial_\pm g}{1\mp z} ~,
\qquad \tilde{\Lax}_\pm(z) = - \frac{\partial_\pm g g^{-1}}{1\mp z} ~,
\end{equation}
then we have that
\begin{equation}\begin{split}
g \Lax_\pm (z) g^{-1} - \partial_\pm g g^{-1} & = \frac{\partial_\pm g g^{-1}}{1\mp z} - \partial_\pm g g^{-1}
= - \frac{\partial_\pm g g^{-1}}{1\mp z^{-1}} = \tilde{\Lax}_\pm(z^{-1}) ~,
\end{split}\end{equation}
as claimed.
\fi

Once we have a Lax connection, we can follow the usual procedure for constructing the monodromy matrix and conserved charges.
The monodromy matrix is given by
\begin{equation}
\Mon(t;z) = \PexpL\big(- \int_{-\infty}^{\infty} dx \, \Lax_x(t,x;z)\big) = \PexpL\big(- \int_{-\infty}^{\infty} dx \, \frac{j_x + z j_t}{1-z^2}\big) ~,
\end{equation}
where $\Lax_x(z) = \frac{1}{2}(\Lax_+-\Lax_-)$ is the spatial component of the Lax connection and $\PexpL$ denotes the path-ordered exponential with greater $x$ to the left.
The flatness of the Lax connection then implies that the monodromy matrix satisfies
\begin{equation}
\partial_t \Mon(t;z) = - \Lax_t(z)\big|_{x \to \infty} \Mon(t;z) + \Mon(t;z) \Lax_t(z)\big|_{x \to -\infty} ~,
\end{equation}
hence is conserved if we assume suitable decaying boundary conditions at spatial infinity.
Expanding the monodromy matrix in powers of $\frac{1}{z}$ gives
\begin{equation}
\Mon(t;z) = 1 + \frac{1}{z} \int_{-\infty}^{\infty} dx \, j_t(t,x) + \frac{1}{z^2} \big(\int_{-\infty}^{\infty} dx \, j_x(t,x) + \int_{-\infty}^{\infty} dx \int_{-\infty}^{x} dx' j_t(t,x) j_t(t,x') \big) + \dots ~.
\end{equation}
At the first non-trivial order we find the Noether charges associated to the right-acting global symmetry.
At higher orders in $\frac{1}{z}$ we find non-local conserved charges, which, together with the Noether charges, generate a classical Yangian algebra, an infinite-dimensional algebra that underlies the integrability of the model.
The non-locality of the higher conserved charges means that this is often referred to as a hidden symmetry.
It is also possible to extract an infinite number of independent local conserved charges in involution from the monodromy matrix following a procedure known as abelianisation.
This involves transforming to a diagonal gauge and expanding around the poles of the Lax connection at $z = \pm 1$.

\def\ob{(}
\def\cb{)}
The conserved charges found from the monodromy matrix have vanishing spin in the classical theory, where the spin is the charge under the $\grp{SO}^+(1,1)$ Lorentz symmetry of the 2-d sigma model.
It turns out that the PCM also has an infinite number of independent higher-spin local conserved charges in involution.
Let us take $\grp{G}$ to be a classical simple Lie group in its defining matrix representation.
The construction of the higher-spin local conserved charges is then based on the existence of the following conserved currents
\unskip\footnote{Here we work in the defining matrix representation as it is a convenient way to introduce the symmetric invariant tensors $d_{a_1\dots a_m} = \Tr(T_{\ob a_1}\dots T_{a_m\cb})$.
For $\alg{su}(N)$ these are non-vanishing for all integer $m \geq 2$, while for $\alg{so}(N)$ and $\alg{sp}(N)$ they are only non-vanishing when $m$ is even.}
\begin{equation}
\mathcal{J}_{m+} = \Tr(j_+^m) ~, \qquad
\mathcal{J}_{m-} = \Tr(j_-^m) ~.
\end{equation}
For $m=2$ these are proportional to the non-vanishing components of the energy-momentum tensor
\begin{equation}
T_{++} \propto \Tr(j_+j_+) ~, \qquad T_{--} \propto \Tr(j_-j_-) ~.
\end{equation}
The energy-momentum tensor is symmetric and, since the PCM is classically invariant under conformal transformations, its trace vanishes, $T_{+-} = T_{-+} = 0$.

Combining the equations of motion~\eqref{eq:eompcm} and the flatness condition~\eqref{eq:flatness} we have that
\begin{equation}
\partial_+ j_- = - \frac{1}{2} [j_+,j_-] ~, \qquad \partial_- j_+ = - \frac{1}{2}[j_-,j_+] ~.
\end{equation}
Therefore,
\begin{equation}
\partial_-\mathcal{J}_{m+} = m \Tr(j_+^{m-1} \partial_- j_+) = -\frac{m}{2} \Tr(j_+^{m-1}[j_-,j_+])
= \frac{m}{2} \Tr([j_+^{m-1},j_+] j_-) = 0 ~,
\end{equation}
where we have used the cyclicity and invariance of the trace.
Similarly, we can show that $\partial_+ \mathcal{J}_{-m} = 0$.
We can then define the following local conserved charges
\begin{equation}\label{eq:conscharge}
Q_{\pm s} = \int_{-\infty}^{\infty} dx \, \mathcal{J}_{\pm m} ~,
\end{equation}
which are labelled by their spin $s = m-1$.
For $m=2$ these charges are integrals of components of the energy-momentum tensor, hence they are proportional to the light-cone momenta $P_\pm = E \pm P$, where $E$ is the energy and $P$ is the spatial momentum.
We normalise such that the light-cone momenta have spin equal to $\pm 1$.
From the conserved charges~\eqref{eq:conscharge} it is possible to extract an infinite subset that are independent and in involution, thereby justifying the claim that the PCM is indeed integrable.
The existence of these conserved charges is again tied to the existence of a Lax connection, and expressing the former in terms of the latter provides a natural way to generalise the above construction to a larger class of integrable sigma models.

\subsection{The Wess-Zumino term}

Thus far, we have considered the PCM, which has $\grp{G}_\ind{L} \times \grp{G}_\ind{R}$ global symmetry.
This is a parity-invariant model and has vanishing B-field.
Introducing a parity-odd term that manifestly preserves the global symmetry is subtle, but it can be achieved by considering the Wess-Zumino (WZ) term.
To define the WZ term we introduce a fiducial third dimension, such that the 2-d worldsheet is the boundary of the resulting 3-d bulk space.
Extending the group-valued field $g(t,x)$ away from the boundary to the interior of the 3-d bulk space,
\unskip\footnote{Seeing as it is not central to our discussion, here we only give a brief outline of the argument explaining why this is possible.
We first consider the Euclidean theory and compactify the worldsheet to $\Sp^2$, such that the 3-d bulk space is the 3-d ball $\Ball^3$.
Given that the second homotopy group of a simple Lie group $\grp{G}$ is trivial, $\pi_2(\grp{G}) = 0$, we can always extend a map from $\Sp^2 \to \grp{G}$ to a map from $\Ball^3 \to \grp{G}$.
Subject to the group-valued field satisfying suitable boundary conditions, we may then use the stereographic projection and analytic continuation to return to a flat Minkowski worldsheet.}
we can write down the following WZ term
\begin{equation}\label{eq:WZterm}
\Act_\indrm{WZ} = \frac{\kay}{6} \int d^3 x\, \epsilon^{ijk} \tr\big(j_i[j_j,j_k]\big) ~,
\end{equation}
where the index $i = 0,1,2$ runs over the 3-d bulk space coordinates.
By construction this term is manifestly invariant under the $\grp{G}_\ind{L} \times \grp{G}_\ind{R}$ global symmetry.

The WZ term~\eqref{eq:WZterm} is defined in 3 dimensions.
However, we are interested in 2-d integrable sigma models.
Therefore, for this term to make sense it should be independent of how we extend the group-valued field to the 3-d bulk space.
To show this we start by considering infinitesimal deformations $g \to g e^{\varepsilon} \sim g(1+\varepsilon)$ where $\varepsilon$ vanishes on the 2-d boundary.
Under such a deformation we have
\begin{equation}
j_i \to e^{-\varepsilon} j_i e^{\varepsilon} + e^{-\varepsilon} \partial_i e^{\varepsilon} \sim
j_i + \partial_i \varepsilon + [j_i,\varepsilon] ~,
\end{equation}
hence
\begin{equation}
\delta j_i = \partial_i \varepsilon + [j_i,\varepsilon] ~.
\end{equation}
Therefore,
\begin{equation}\begin{split}
\delta \Act_\indrm{WZ}
& = \frac{\kay}{2} \int d^3 x \, \epsilon^{ijk} \tr\big((\partial_i \varepsilon + [j_i,\varepsilon]) [j_j,j_k]\big)
\\
& = - \frac{\kay}{2} \int d^3 x \, \epsilon^{ijk} \tr\big(\varepsilon (\partial_i [j_j,j_k] + [j_i,[j_j,j_k])\big)
\\
& = \frac{\kay}{2} \int d^3x  \, \epsilon^{ijk} \tr\big(\varepsilon \partial_i (\partial_j j_k - \partial_j j_k ) \big)
= 0 ~,
\end{split}\end{equation}
where we have used integration by parts, the invariance of the bilinear form, the flatness of $j_i$, that partial derivatives commute and that $\epsilon^{ijk} [j_i,[j_j,j_k]] = 0$ by the Jacobi identity.
Note that when we integrate by parts we drop the boundary term since we take the variation $\varepsilon$ to vanish on the 2-d boundary.
It therefore follows that continuous deformations of the group-valued field do not change the WZ term.

This is sufficient for our purposes given that we are primarily be interested in the equations of motion of the classical theory, which follow from the variational principle.
Let us note, however, that there may be extensions of the group-valued field to the 3-d bulk space that are not related by continuous deformations.
These are characterized by the third homotopy group, $\pi_3(\grp{G})$, which for a compact simple Lie group is equal to $\Integer$.
As a consequence, the WZ term is strictly speaking a multi-valued functional, which could lead to an ill-defined quantum theory.
If the coupling $\kay$ is quantized as
\begin{equation}
k = 4\pi\kay \in \Integer ~,
\end{equation}
then the contribution of the potential ambiguity to the path integral is proportional to $e^{2i\pi n}$ with $n\in \Integer$, hence the path integral is well-defined.
The integer $k$ is known as the level of the WZ term.

Note that the WZ term is proportional to the integral over the pull-back of the H-flux $H \propto \tr\big(j \wedge j \wedge j\big)$.
This 3-form is closed, $dH = 0$, but not exact.
That is we cannot write $H = d B$ where $B$ is globally well-defined and $\grp{G}_\ind{L} \times \grp{G}_\ind{R}$ invariant.
Nevertheless, from the discussion above we see that the WZ term defines a consistent local 2-d sigma model coupling.

\subsection{The PCM plus WZ term and the WZW model}

Having introduced the WZ term, we are now in a position to introduce the PCM plus WZ term (PCWZM), whose action is given by
\begin{equation}\label{eq:pcmwzterm}
\Act_\indrm{PCWZM} = - \frac{\hay}{2} \int d^2x \, \tr\big(j_+ j_-\big) + \frac{\kay}{6} \int d^3 x\, \epsilon^{ijk} \tr\big(j_i[j_j,j_k]\big) ~.
\end{equation}
\iflectures
The equations of motion of this model are given by
\begin{equation}\label{eq:eompcmwz}
(1+\frac{\kay}{\hay}) \partial_+ j_- + (1-\frac{\kay}{\hay}) \partial_- j_+ = 0 ~.
\end{equation}

\question{Show that the equations of motion of the PCWZM are given by
\begin{equation*}
(1+\frac{\kay}{\hay}) \partial_+ j_- + (1-\frac{\kay}{\hay}) \partial_- j_+ = 0 ~.
\end{equation*}
}

\answer{We start by recalling that under the infinitesimal variation $g \to g e^{\varepsilon} \sim g(1+\varepsilon)$ we have
\begin{equation*}
\delta j_i = \partial_i\varepsilon + [j_i,\varepsilon] ~.
\end{equation*}
Therefore, varying the action~\eqref{eq:pcmwzterm} we find
\begin{equation}\begin{split}
\delta \Act_\indrm{PCWZM}
& = - \frac{\hay}{2} \int d^2x \, \tr\big((\partial_+ \varepsilon + [j_+,\varepsilon]) j_- + j_+ (\partial_-\varepsilon + [j_-, \varepsilon])\big)
\\ & \qquad
+ \frac{\kay}{2} \int d^3 x\, \epsilon^{ijk} \tr\big( (\partial_i \varepsilon + [j_i,\varepsilon]) [j_j,j_k]\big)
\\ & = \frac{\hay}{2} \int d^2x \, \tr\big(\varepsilon(\partial_+j_- + [j_+,j_-] + \partial_- j_+ + [j_-,j_+]) \big)
\\ & \qquad
+ \frac{\kay}{2} \int d^3 x\, \epsilon^{ijk} \partial_i \tr\big( \varepsilon [j_j,j_k]\big)
- \frac{\kay}{2} \int d^3 x\, \epsilon^{ijk} \tr\big(\varepsilon (\partial_i [j_j,j_k] + [j_i,[j_j,j_k]) \big)
\\ & = \frac{\hay}{2} \int d^2x \, \tr\big(\varepsilon(\partial_+j_- + \partial_- j_+ ) \big)
- \frac{\kay}{2} \int d^2 x\, \tr\big( \varepsilon [j_+,j_-]\big)
\\ & = \frac{\hay}{2} \int d^2x \, \tr\big(\varepsilon(\partial_+j_- + \partial_- j_+ ) \big)
+ \frac{\kay}{2} \int d^2 x\, \tr\big(\varepsilon( \partial_+ j_- - \partial_- j_+)\big)
\\ & = \frac{\hay}{2} \int d^2x \, \tr\big(\varepsilon((1+\frac{\kay}{\hay}) \partial_+j_- + (1-\frac{\kay}{\hay})\partial_- j_+ ) \big) ~.
\end{split}\end{equation}
Requiring that the variation vanishes for arbitrary $\varepsilon \in \alg{g}$, we indeed find the equation of motion
\begin{equation*}
(1+\frac{\kay}{\hay}) \partial_+ j_- + (1-\frac{\kay}{\hay}) \partial_- j_+ = 0 ~.
\end{equation*}
Note that while we have dropped boundary terms when integrating by parts on the 2-d worldsheet, crucially we keep them when integrating by parts in the 3-d bulk space.
To write the boundary term as an integral over the 2-d worldsheet we have used Stokes' theorem
\begin{equation*}
\int d^3 x \, \epsilon^{ijk} \partial_i B_{jk} = \int d^2x \, \epsilon^{\mu\nu} B_{\mu\nu} ~,
\end{equation*}
and that $\epsilon^{+-} = - \epsilon^{-+} = -\frac{1}{2}$.
}
\else
Recalling that under the infinitesimal variation $g \to g e^{\varepsilon} \sim g(1+\varepsilon)$ we have $\delta j_i = \partial_i\varepsilon + [j_i,\varepsilon]$, we can vary the action of the PCWZM~\eqref{eq:pcmwzterm} to find
\begin{equation}\begin{split}
\delta \Act_\indrm{PCWZM}
& = - \frac{\hay}{2} \int d^2x \, \tr\big((\partial_+ \varepsilon + [j_+,\varepsilon]) j_- + j_+ (\partial_-\varepsilon + [j_-, \varepsilon])\big)
\\ & \qquad
+ \frac{\kay}{2} \int d^3 x\, \epsilon^{ijk} \tr\big( (\partial_i \varepsilon + [j_i,\varepsilon]) [j_j,j_k]\big)
\\ & = \frac{\hay}{2} \int d^2x \, \tr\big(\varepsilon(\partial_+j_- + [j_+,j_-] + \partial_- j_+ + [j_-,j_+]) \big)
\\ & \qquad
+ \frac{\kay}{2} \int d^3 x\, \epsilon^{ijk} \partial_i \tr\big( \varepsilon [j_j,j_k]\big)
- \frac{\kay}{2} \int d^3 x\, \epsilon^{ijk} \tr\big(\varepsilon (\partial_i [j_j,j_k] + [j_i,[j_j,j_k]) \big)
\\ & = \frac{\hay}{2} \int d^2x \, \tr\big(\varepsilon(\partial_+j_- + \partial_- j_+ ) \big)
- \frac{\kay}{2} \int d^2 x\, \tr\big( \varepsilon [j_+,j_-]\big)
\\ & = \frac{\hay}{2} \int d^2x \, \tr\big(\varepsilon(\partial_+j_- + \partial_- j_+ ) \big)
+ \frac{\kay}{2} \int d^2 x\, \tr\big(\varepsilon( \partial_+ j_- - \partial_- j_+)\big)
\\ & = \frac{\hay}{2} \int d^2x \, \tr\big(\varepsilon((1+\frac{\kay}{\hay}) \partial_+j_- + (1-\frac{\kay}{\hay})\partial_- j_+ ) \big) ~.
\end{split}\end{equation}
Therefore, requiring that the variation vanishes for arbitrary $\varepsilon \in \alg{g}$, we find the equation of motion
\begin{equation}\label{eq:eompcmwz}
(1+\frac{\kay}{\hay}) \partial_+ j_- + (1-\frac{\kay}{\hay}) \partial_- j_+ = 0 ~.
\end{equation}
Note that while we have dropped boundary terms when integrating by parts on the 2-d worldsheet, we crucially keep them when integrating by parts in the 3-d bulk space.
To write the boundary term as an integral over the 2-d worldsheet we use Stokes' theorem
\begin{equation}
\int d^3 x \, \epsilon^{ijk} \partial_i B_{jk} = \int d^2x \, \epsilon^{\mu\nu} B_{\mu\nu} ~,
\end{equation}
and that $\epsilon^{+-} = - \epsilon^{-+} = -\frac{1}{2}$.
\fi

While $j_\pm$ is still flat, it is no longer conserved.
In the presence of the WZ term the conserved current is modified to
\begin{equation}
J_\pm = \frac{1}{\xi}(1 \mp \frac{\kay}{\hay})j_\pm ~,
\end{equation}
where $\xi$ is a free constant.
We would now like to ask if there is a choice of $\xi$ such that $J_\pm$ is also flat on-shell.
We have
\begin{equation}\begin{split}
\partial_+ J_- - \partial_- J_+ + [J_+,J_-] & =
\frac{1}{\xi}(1+\frac{\kay}{\hay}) \partial_+ j_- - \frac{1}{\xi}(1-\frac{\kay}{\hay}) \partial_- j_+
+ \frac{1}{\xi^2}(1-\frac{\kay^2}{\hay^2}) [j_+,j_-]
\\ & = \big(\frac{1}{\xi}(1+\frac{\kay}{\hay}) - \frac{1}{\xi^2}(1-\frac{\kay^2}{\hay^2})\big) \partial_+ j_-
+ \big(-\frac{1}{\xi}(1-\frac{\kay}{\hay}) + \frac{1}{\xi^2}(1-\frac{\kay^2}{\hay^2})\big) \partial_- j_+ ~.
\end{split}\end{equation}
If we set $\xi = 1$ then
\begin{equation}
\partial_+ J_- - \partial_- J_+ + [J_+,J_-] = \frac{\kay}{\hay}\big((1+\frac{\kay}{\hay})\partial_+ j_-
+ (1-\frac{\kay}{\hay})\partial_- j_+\big) ~,
\end{equation}
which indeed vanishes on the equations of motion~\eqref{eq:eompcmwz}.

Therefore, we find that the current
\begin{equation}
J_\pm = (1\mp \frac{\kay}{\hay}) j_\pm ~,
\end{equation}
is both conserved and flat on-shell and it follows that we can write down the following Lax connection of the PCWZM
\begin{equation}
\Lax_\pm = \frac{J_\pm}{1\mp z} ~.
\end{equation}

Let us finish by very briefly commenting on the special point $\hay = \kay$.
The first thing to notice is that the equations of motion simplify to
\begin{equation}
\partial_+ j_- = 0 ~,
\end{equation}
which can be easily solved in closed form
\begin{equation}
g(x,t) = g_-(x^-) g_+(x^+) ~.
\end{equation}
That is left- and right-moving waves pass through each other without interference.
This is indicating that something special happens at this point.
Indeed, the one-loop beta function for the coupling vanishes for $\hay = \kay$.
It can then be argued that this is an exact result and that the PCWZM term at $\hay = \kay$ defines a conformal field theory known as the Wess-Zumino-Witten (WZW) model.

\subsection{The \texorpdfstring{$\grp{SU}(2)$}{SU(2)} PCM and PCWZM}

To gain a better understanding of the PCM and PCWZM let us consider the case $\grp{G} = \grp{SU}(2)$.
We start by explicitly constructing the target-space metric.
To do so we introduce the familiar matrix representation of $\alg{su}(2)$
\begin{equation}\label{eq:su2gen}
T_1 = \begin{pmatrix} 0 & i \\ i & 0\end{pmatrix} ~, \qquad
T_2 = \begin{pmatrix} 0 & 1 \\ -1 & 0\end{pmatrix} ~, \qquad
T_3 = \begin{pmatrix} i & 0 \\ 0 & -i \end{pmatrix} ~.
\end{equation}
In this representation $\tr = \Tr$, where $\Tr$ is the standard matrix trace.
\unskip\footnote{This follows from the definition of the invariant bilinear form in \footref{foot:index} since the index of the fundamental representation of $\alg{su}(N)$ is $\chi = \frac12$.}
Therefore, we have
\begin{equation}
\tr\big(T_aT_b\big) = -2\delta_{ab} ~, \qquad [T_a,T_b] = -2\epsilon_{ab}{}^c T_c ~,
\end{equation}
where the index $a=1,2,3$ is lowered and raised with the Kronecker delta $\delta_{ab}$ and its inverse $\delta^{ab}$, and $\epsilon_{123} = 1$ is completely antisymmetric.

If we now parametrise the group-valued field as
\begin{equation}\label{eq:gparam}
g = \exp\big(\frac{\varphi+\phi}{2} T_3\big) \exp\big(\theta T_1\big) \exp\big(\frac{\varphi - \phi}{2} T_3\big) ~,
\end{equation}
and substitute into the PCM action~\eqref{eq:pcmact} we find
\begin{equation}
\Act_\indrm{SU(2)-PCM} = \hay \int d^2x \, \big(\partial_+ \theta \partial_-\theta + \cos^2\theta \partial_+\varphi\partial_-\varphi + \sin^2\theta \partial_+ \phi \partial_-\phi\big) ~.
\end{equation}
We can then read off the explicit form of target-space metric in local coordinates
\begin{equation}
G = 2\hay (d\theta^2 + \cos^2\theta d\varphi^2 + \sin^2\theta d\phi^2 ) ~.
\end{equation}
As expected, we recover the bi-invariant (round) metric on $\grp{SU}(2) \cong \Sp^3$ with radius squared $2\hay$.
\iflectures

\question{$\Sp^3$ with unit radius can be defined as the locus of all points $(X_1,X_2,X_3,X_4) \in \Real^4$ satisfying
\begin{equation*}
X_1^2 + X_2^2 + X_3^2 + X_4^2 = 1 ~.
\end{equation*}
Use this definition to recover the bi-invariant (round) metric on $\Sp^3$
\begin{equation*}
G = d\theta^2 + \cos^2\theta d\varphi^2 + \sin^2\theta d\phi^2 ~.
\end{equation*}
}

\answer{If we solve the constraint equation
\begin{equation*}
X_1^2 + X_2^2 + X_3^2 + X_4^2 = 1 ~,
\end{equation*}
by setting
\begin{equation*}
X_1 + i X_2 = \cos\theta e^{i\varphi} ~, \qquad
X_3 + i X_4 = \sin\theta e^{i\phi} ~,
\end{equation*}
then the induced metric on $\Sp^3$ is given by
\begin{equation*}\begin{split}
G & = dX_1^2 + dX_2^2 + dX_3^2 + dX_4^2
\\ & = (-\sin\theta d\theta + i \cos\theta d\varphi) e^{i\varphi}(-\sin\theta d\theta - i \cos\theta d\varphi) e^{-i\varphi}
+ (\cos\theta d\theta + i \sin\theta d\phi) e^{i\phi}(\cos\theta d\theta - i \sin\theta d\phi) e^{-i\phi}
\\ & = \sin^2\theta d\theta^2 + \cos^2 \theta d\varphi^2 + \cos^2 \theta d\theta^2 + \sin^2\theta d\phi^2
\\ & = d\theta^2 + \cos^2\theta d\varphi^2 + \sin^2\theta d\phi^2 ~.
\end{split}\end{equation*}
}
\else
\unskip\footnote{$\Sp^3$ with unit radius can be defined as the locus of all points $(X_1,X_2,X_3,X_4) \in \Real^4$ satisfying $X_1^2 + X_2^2 + X_3^2 + X_4^2 = 1$.
Solving this equation by setting
\begin{equation*}
X_1 + i X_2 = \cos\theta e^{i\varphi} ~, \qquad
X_3 + i X_4 = \sin\theta e^{i\phi} ~,
\end{equation*}
the induced metric on $\Sp^3$ is given by
\begin{equation*}\begin{split}
G & = dX_1^2 + dX_2^2 + dX_3^2 + dX_4^2
\\ & = (-\sin\theta d\theta + i \cos\theta d\varphi) e^{i\varphi}(-\sin\theta d\theta - i \cos\theta d\varphi) e^{-i\varphi}
+ (\cos\theta d\theta + i \sin\theta d\phi) e^{i\phi}(\cos\theta d\theta - i \sin\theta d\phi) e^{-i\phi}
\\ & = \sin^2\theta d\theta^2 + \cos^2 \theta d\varphi^2 + \cos^2 \theta d\theta^2 + \sin^2\theta d\phi^2
\\ & = d\theta^2 + \cos^2\theta d\varphi^2 + \sin^2\theta d\phi^2 ~.
\end{split}\end{equation*}}
\fi

To construct the PCWZM let us compute the WZ term for $\grp{G} = \grp{SU}(2)$.
Extending the group-valued field~\eqref{eq:gparam} to the 3-d bulk space and substituting into the WZ term~\eqref{eq:WZterm}, we find
\begin{equation}\begin{split}
\Act_\indrm{SU(2)-WZ} & = - \kay \int d^3 x\, \epsilon^{ijk} \partial_i\cos 2\theta \partial_j\varphi\partial_k\phi
= - \kay \int d^3x \, \epsilon^{ijk} \partial_i (\cos 2\theta \partial_j\varphi\partial_k\phi) ~.
\end{split}\end{equation}
Here we see explicitly that when we write the action in terms of local coordinates the integrand of the WZ term becomes a total derivative.
Therefore, we can use Stokes' theorem to write
\begin{equation}\begin{split}
\Act_\indrm{SU(2)-WZ} & = - \kay \int d^2x \, \epsilon^{\mu\nu} \cos 2\theta \partial_\mu\varphi\partial_\nu \phi
= \frac{\kay}{2} \int d^2x \, \cos 2\theta (\partial_+\varphi \partial_-\phi - \partial_- \varphi \partial_+ \phi) ~.
\end{split}\end{equation}
From this expression we can read off the target-space B-field
\begin{equation}
B = \kay \cos 2\theta d\varphi \wedge d\phi ~.
\end{equation}
As expected, this B-field is not globally well-defined or invariant under the $\grp{SU}(2)_\ind{L} \times \grp{SU}(2)_\ind{R}$ global symmetry.
On the other hand, the target space H-flux
\begin{equation}
H = - 2\kay \sin 2\theta d\theta\wedge d\varphi \wedge d\phi ~,
\end{equation}
is globally well-defined and invariant under the global symmetry.
One way to see this is to simply note that the 3-form $H$ is proportional to the volume form of $\Sp^3$.
\iflectures

\question{Under infinitesimal global symmetry transformations the local coordinates $\varphi$, $\phi$ and $\theta$ transform as
\begin{equation*}\begin{split}
\delta\varphi & = \chi_\ind{L}^3 + \chi_\ind{R}^3 + \tan\theta ( \chi_\ind{L}^1 \sin(\varphi+\phi)+ \chi_\ind{R}^1 \sin(\varphi-\phi) + \chi_\ind{L}^2 \cos(\varphi+\phi)- \chi_\ind{R}^2\cos(\varphi-\phi) ) ~,
\\
\delta\phi & = \chi_\ind{L}^3 - \chi_\ind{R}^3 - \cot\theta (\chi_\ind{L}^1 \sin(\varphi+\phi) - \chi_\ind{R}^1 \sin(\varphi-\phi) + \chi_\ind{L}^2\cos(\varphi+\phi) + \chi_\ind{R}^2\cos(\varphi-\phi) ) ~,
\\
\delta\theta & = \chi_\ind{L}^1 \cos(\varphi+\phi) + \chi_\ind{R}^1 \cos(\varphi-\phi) - \chi_\ind{L}^2 \sin(\varphi+\phi) + \chi_\ind{R}^2 \sin(\varphi-\phi) ~,
\end{split}\end{equation*}
where $\chi_\ind{L}^a$ and $\chi_\ind{R}^a$ are infinitesimal parameters.
Show that the metric and H-flux of the $\grp{SU}(2)$ PCWZM are invariant under these transformations, while the B-field is invariant only up to a gauge transformation.
}

\answer{
Let us first define
\begin{equation*}
e_\varphi = \cos\theta d\varphi ~, \qquad
e_\phi = \sin\theta d\phi ~, \qquad
e_\theta = d\theta ~.
\end{equation*}
and consider their variations
\begin{equation*}\begin{split}
\delta e_\varphi & = - \sin \theta \delta\theta + \cos\theta d \delta \varphi
\\ & = e_\phi (\chi_\ind{L}^1 \cos(\varphi+\phi) - \chi_\ind{R}^1 \cos(\varphi-\phi) - \chi_\ind{L}^2 \sin(\varphi+\phi) - \chi_\ind{R}^2 \sin(\varphi-\phi))
\\ & \qquad + e_\theta \sec \theta (\chi_\ind{L}^1 \sin(\varphi+\phi) + \chi_\ind{R}^1 \sin(\varphi-\phi) + \chi_\ind{L}^2 \cos(\varphi+\phi)-  \chi_\ind{R}^2\cos(\varphi-\phi)) ~,
\\
\delta e_\phi & = \cos \theta \delta\theta + \sin\theta d \delta \phi
\\ & = - e_\varphi (\chi_\ind{L}^1 \cos(\varphi+\phi) - \chi_\ind{R}^1 \cos(\varphi-\phi) - \chi_\ind{L}^2 \sin(\varphi+\phi) - \chi_\ind{R}^2 \sin(\varphi-\phi))
\\ & \qquad + e_\theta \csc \theta (\chi_\ind{L}^1 \sin(\varphi+\phi) - \chi_\ind{R}^1 \sin(\varphi-\phi) + \chi_\ind{L}^2 \cos(\varphi+\phi)+ \chi_\ind{R}^2\cos(\varphi-\phi) ) ~,
\\
\delta e_\theta & = d \delta \theta
\\ & = - e_\varphi \sec\theta (\chi_\ind{L}^1 \sin(\varphi+\phi) + \chi_\ind{R}^1 \sin(\varphi-\phi)+ \chi_\ind{L}^2 \cos(\varphi+\phi)- \chi_\ind{R}^2\cos(\varphi-\phi) )
\\ & \qquad  - e_\phi \csc \theta (\chi_\ind{L}^1 \sin(\varphi+\phi) - \chi_\ind{R}^1 \sin(\varphi-\phi) + \chi_\ind{L}^2 \cos(\varphi+\phi)+ \chi_\ind{R}^2\cos(\varphi-\phi) ) ~.
\end{split}\end{equation*}
It is then straightforward to see that, since
\begin{equation*}
G = 2\hay (e_\varphi^2 + e_\phi^2 + e_\theta^2) ~, \qquad
H = -4\kay e_\theta \wedge e_\varphi \wedge e_\phi ~,
\end{equation*}
we have
\begin{equation*}
\delta G = 4 \hay(e_\varphi \delta e_\varphi + e_\phi \delta e_\phi + e_\theta\delta e_\theta) = 0 ~,
\end{equation*}
and
\begin{equation*}
\delta H = -4 \kay ( \delta e_\theta \wedge e_\varphi \wedge e_\phi + e_\theta \wedge \delta e_\varphi \wedge e_\phi + e_\theta \wedge e_\varphi \wedge \delta e_\phi ) = 0 ~,
\end{equation*}
that is the metric and H-flux are indeed invariant.
The variation of the B-field is
\begin{equation*}\begin{split}
\delta B & = 2\kay (\delta (\cot 2\theta) e_\varphi \wedge e_\phi
+ \cot 2\theta (\delta e_\varphi \wedge e_\phi + e_\varphi \wedge \delta e_\phi))
\\ & = 2\kay (-2\csc^22\theta (\chi_\ind{L}^1 \cos(\varphi+\phi) + \chi_\ind{R}^1 \cos(\varphi-\phi) - \chi_\ind{L}^2 \sin(\varphi+\phi) + \chi_\ind{R}^2 \sin(\varphi-\phi))e_\varphi \wedge e_\phi
\\ & \qquad\quad + \cot 2\theta \sec\theta (\chi_\ind{L}^1 \sin(\varphi+\phi) + \chi_\ind{R}^1 \sin(\varphi-\phi) + \chi_\ind{L}^2 \cos(\varphi+\phi)- \chi_\ind{R}^2\cos(\varphi-\phi) ) e_\theta \wedge e_\phi
\\ & \qquad\quad + \cot 2\theta \csc\theta (\chi_\ind{L}^1 \sin(\varphi+\phi) - \chi_\ind{R}^1 \sin(\varphi-\phi)+ \chi_\ind{L}^2 \cos(\varphi+\phi)+ \chi_\ind{R}^2\cos(\varphi-\phi) ) e_\varphi \wedge e_\theta)
\\ & = 2\kay (-\csc2\theta (\chi_\ind{L}^1 \cos(\varphi+\phi) + \chi_\ind{R}^1 \cos(\varphi-\phi) - \chi_\ind{L}^2 \sin(\varphi+\phi) + \chi_\ind{R}^2 \sin(\varphi-\phi))d\varphi \wedge d\phi
\\ & \qquad\quad + \cot 2\theta \tan\theta (\chi_\ind{L}^1 \sin(\varphi+\phi) + \chi_\ind{R}^1 \sin(\varphi-\phi) + \chi_\ind{L}^2 \cos(\varphi+\phi)- \chi_\ind{R}^2\cos(\varphi-\phi) ) d\theta \wedge d\phi
\\ & \qquad\quad + \cot 2\theta \cot\theta (\chi_\ind{L}^1 \sin(\varphi+\phi) - \chi_\ind{R}^1 \sin(\varphi-\phi) + \chi_\ind{L}^2 \cos(\varphi+\phi)+ \chi_\ind{R}^2\cos(\varphi-\phi) ) d\varphi \wedge d\theta)
\\ & = \kay (-\csc2\theta (\chi_\ind{L}^1 \cos\varphi_+ + \chi_\ind{R}^1 \cos\varphi_- - \chi_\ind{L}^2 \sin\varphi_+ + \chi_\ind{R}^2 \sin\varphi_-) d\varphi_- \wedge d\varphi_+
\\ & \qquad \ + \cot 2\theta \tan\theta (\chi_\ind{L}^1 \sin\varphi_+ + \chi_\ind{R}^1 \sin\varphi_- + \chi_\ind{L}^2 \cos\varphi_+ - \chi_\ind{R}^2\cos\varphi_- ) d\theta \wedge (d\varphi_+ - d \varphi_-)
\\ & \qquad \ - \cot 2\theta \cot\theta (\chi_\ind{L}^1 \sin\varphi_+ - \chi_\ind{R}^1 \sin\varphi_- + \chi_\ind{L}^2 \cos\varphi_+ + \chi_\ind{R}^2\cos\varphi_-) d\theta \wedge (d\varphi_+ + d\varphi_-))
\\ & = \kay (\chi_\ind{L}^1 ( \csc 2\theta \cos \varphi_+ d\varphi_+ \wedge d\varphi_-
- 2 \cot 2\theta \csc 2\theta \sin\varphi_+ d\theta \wedge d\varphi_-
- 2 \cot^2 2\theta \sin\varphi_+ d\theta\wedge d\varphi_+)
\\ & \quad \, -  \chi_\ind{R}^1 ( \csc 2\theta \cos \varphi_- d\varphi_- \wedge d\varphi_+
- 2 \cot 2\theta \csc 2\theta \sin\varphi_- d \theta \wedge d\varphi_+
- 2 \cot^2 2\theta \sin\varphi_- d\theta \wedge d\varphi_-)
\\ & \quad \, - \chi_\ind{L}^2 (\csc 2\theta \sin \varphi_+ d\varphi_+ \wedge d\varphi_-
+ 2 \cot 2\theta \csc 2\theta \cos\varphi_+ d\theta \wedge d\varphi_-
+ 2 \cot^2 2\theta \cos\varphi_+ d\theta \wedge d\varphi_+)
\\ & \quad \, - \chi_\ind{R}^2 (\csc 2\theta \sin \varphi_- d\varphi_- \wedge d\varphi_+
+ 2 \cot 2\theta \csc 2\theta \cos\varphi_- d\theta \wedge d\varphi_+
+ 2 \cot^2 2\theta \cos\varphi_- d\theta \wedge d\varphi_-))
\\ & = \kay (\chi_\ind{L}^1 d ( \csc 2\theta \sin \varphi_+ d\varphi_-
- 2 \cot^2 2\theta \cos\varphi_+ d\theta)
-  \chi_\ind{R}^1 d ( \csc 2\theta \sin \varphi_- d\varphi_+
- 2 \cot^2 2\theta \cos\varphi_- d\theta )
\\ & \quad \, + \chi_\ind{L}^2 d (\csc 2\theta \cos \varphi_+ d\varphi_-
+ 2 \cot^2 2\theta \sin\varphi_+ d\theta)
+ \chi_\ind{R}^2 d (\csc 2\theta \cos \varphi_- d\varphi_+
+ 2 \cot^2 2\theta \sin\varphi_- d\theta)) ~,
\end{split}\end{equation*}
where we have introduced $\varphi_\pm = \varphi \pm \phi$.
Therefore, we explicitly see that $\delta B = d\Lambda$, that is the B-field is only invariant up to a gauge transformation.
}
\else
To show it explicitly, we observe that under infinitesimal global symmetry transformations the local coordinates $\varphi$, $\phi$ and $\theta$ transform as
\begin{equation}\begin{split}
\delta\varphi & = \chi_\ind{L}^3 + \chi_\ind{R}^3 + \tan\theta ( \chi_\ind{L}^1 \sin\varphi_++ \chi_\ind{R}^1 \sin\varphi_- + \chi_\ind{L}^2 \cos\varphi_+- \chi_\ind{R}^2\cos\varphi_- ) ~,
\\
\delta\phi & = \chi_\ind{L}^3 - \chi_\ind{R}^3 - \cot\theta (\chi_\ind{L}^1 \sin\varphi_+ - \chi_\ind{R}^1 \sin\varphi_- + \chi_\ind{L}^2\cos\varphi_+ + \chi_\ind{R}^2\cos\varphi_- ) ~,
\\
\delta\theta & = \chi_\ind{L}^1 \cos\varphi_+ + \chi_\ind{R}^1 \cos\varphi_- - \chi_\ind{L}^2 \sin\varphi_+ + \chi_\ind{R}^2 \sin\varphi_- ~,
\end{split}\end{equation}
where $\chi_\ind{L}^a$ and $\chi_\ind{R}^a$ are infinitesimal parameters and we have introduced $\varphi_\pm = \varphi \pm \phi$.
Defining the components of the vielbein
\begin{equation}
e_\varphi = \cos\theta d\varphi ~, \qquad
e_\phi = \sin\theta d\phi ~, \qquad
e_\theta = d\theta ~,
\end{equation}
their variations are given by
\begin{equation}\begin{split}
\delta e_\varphi & = - \sin \theta \delta\theta + \cos\theta d \delta \varphi
= e_\phi (\chi_\ind{L}^1 \cos\varphi_+ - \chi_\ind{R}^1 \cos\varphi_- - \chi_\ind{L}^2 \sin\varphi_+ - \chi_\ind{R}^2 \sin\varphi_-)
\\ & \hphantom{= - \sin \theta \delta\theta + \cos\theta d \delta \varphi} \qquad + e_\theta \sec \theta (\chi_\ind{L}^1 \sin\varphi_+ + \chi_\ind{R}^1 \sin\varphi_- + \chi_\ind{L}^2 \cos\varphi_+-  \chi_\ind{R}^2\cos\varphi_-) ~,
\\
\delta e_\phi & = \cos \theta \delta\theta + \sin\theta d \delta \phi
= - e_\varphi (\chi_\ind{L}^1 \cos\varphi_+ - \chi_\ind{R}^1 \cos\varphi_- - \chi_\ind{L}^2 \sin\varphi_+ - \chi_\ind{R}^2 \sin\varphi_-)
\\ & \hphantom{= \cos \theta \delta\theta + \sin\theta d \delta \phi} \qquad + e_\theta \csc \theta (\chi_\ind{L}^1 \sin\varphi_+ - \chi_\ind{R}^1 \sin\varphi_- + \chi_\ind{L}^2 \cos\varphi_++ \chi_\ind{R}^2\cos\varphi_- ) ~,
\\
\delta e_\theta & = d \delta \theta
= - e_\varphi \sec\theta (\chi_\ind{L}^1 \sin\varphi_+ + \chi_\ind{R}^1 \sin\varphi_-+ \chi_\ind{L}^2 \cos\varphi_+- \chi_\ind{R}^2\cos\varphi_- )
\\ & \hphantom{= d \delta \theta} \qquad  - e_\phi \csc \theta (\chi_\ind{L}^1 \sin\varphi_+ - \chi_\ind{R}^1 \sin\varphi_- + \chi_\ind{L}^2 \cos\varphi_++ \chi_\ind{R}^2\cos\varphi_- ) ~.
\end{split}\end{equation}
It is then straightforward to see that, since
\begin{equation}
G = 2\hay (e_\varphi^2 + e_\phi^2 + e_\theta^2) ~, \qquad
H = -4\kay e_\theta \wedge e_\varphi \wedge e_\phi ~,
\end{equation}
we have
\begin{equation}
\delta G = 4 \hay(e_\varphi \delta e_\varphi + e_\phi \delta e_\phi + e_\theta\delta e_\theta) = 0 ~,
\end{equation}
and
\begin{equation}
\delta H = -4 \kay ( \delta e_\theta \wedge e_\varphi \wedge e_\phi + e_\theta \wedge \delta e_\varphi \wedge e_\phi + e_\theta \wedge e_\varphi \wedge \delta e_\phi) = 0 ~,
\end{equation}
that is the metric and H-flux are indeed invariant.
Similarly, it is also possible to explicitly show that $\delta B = d\Lambda$, that is the B-field is only invariant up to a gauge transformation, with
\begin{equation}\begin{split}
\Lambda
& = \kay (\chi_\ind{L}^1 ( \csc 2\theta \sin \varphi_+ d\varphi_-
- 2 \cot^2 2\theta \cos\varphi_+ d\theta)
-  \chi_\ind{R}^1 ( \csc 2\theta \sin \varphi_- d\varphi_+
- 2 \cot^2 2\theta \cos\varphi_- d\theta )
\\ & \quad \, + \chi_\ind{L}^2 (\csc 2\theta \cos \varphi_+ d\varphi_-
+ 2 \cot^2 2\theta \sin\varphi_+ d\theta)
+ \chi_\ind{R}^2 (\csc 2\theta \cos \varphi_- d\varphi_+
+ 2 \cot^2 2\theta \sin\varphi_- d\theta)) ~,
\end{split}\end{equation}
where $\varphi_\pm = \varphi \pm \phi$.
\fi

\section{Integrable deformations of the \texorpdfstring{$\grp{SU}(2)$}{SU(2)} PCM and PCWZM}\label{sec:su2pcm}

Now that we have introduced the PCM, the PCWZM and the WZW model, we turn to the main aim of \iflectures these lecture notes, \else this review, \fi which is to explore their integrable deformations.
Before we investigate integrable deformations for general Lie group $\grp{G}$, let us first see what we can learn by studying the example of $\grp{G} = \grp{SU}(2)$.
We begin by writing the right-invariant Maurer-Cartan form $k_\pm$~\eqref{eq:lrmc} as
\begin{equation}
k_\pm = k_\pm^a T_a ~,
\end{equation}
where $T_a$, $a=1,2,3$, are the generators~\eqref{eq:su2gen} of $\alg{su}(2)$.
In terms of $k_\pm^a$ the PCM action for $\grp{G} = \grp{SU}(2)$ can be written as
\begin{equation}
\Act_\indrm{SU(2)-PCM} = \hay \int d^2 x \, \sum_a k_+^a k_-^a ~.
\end{equation}
We can deform this model by introducing coefficients $\alpha_a$ breaking the left-acting $\grp{SU}(2)$ global symmetry
\begin{equation}\label{eq:actdefcomp}
\Act_{\alpha_a} = \hay \int d^2 x \, \sum_a \alpha_a^{-1} k_+^a k_-^a ~.
\end{equation}
Setting $\alpha_a = 1$ we recover the action of the $\grp{SU}(2)$ PCM.
Note that only two of the three parameters $\alpha_a$ are genuine deformation parameters since an overall factor can be absorbed into $\hay$.
Rather than working in terms of components, it is often easier to work with linear operators that map from the algebra to itself.
To this end, we introduce the constant invertible linear operator $\mathcal{A}_{\vec{\alpha}} : \alg{su}(2) \to \alg{su}(2)$ defined such that
\begin{equation}
\mathcal{A}_{\vec{\alpha}} T_a = \alpha_a^{-1} T_a ~.
\end{equation}
This map is symmetric with respect to the invariant bilinear form, that is
\begin{equation}\label{eq:asym}
\tr\big(X\mathcal{A}_{\vec{\alpha}}Y\big) = \tr\big((\mathcal{A}_{\vec{\alpha}}X) Y\big) ~, \qquad X,Y\in\alg{su}(2)~.
\end{equation}
The deformed action~\eqref{eq:actdefcomp} can then be written as
\begin{equation}\label{eq:actdef}
\Act_{\vec{\alpha}} = -\frac{\hay}{2} \int d^2 x \, \tr\big(k_+ \mathcal{A}_{\vec{\alpha}} k_-\big) ~.
\end{equation}
Under the infinitesimal variation $g \to e^{-\varepsilon} g \sim (1-\varepsilon)g$ we have
\begin{equation}
\delta k_\pm = \partial_\pm \varepsilon + [k_\pm, \varepsilon] ~.
\end{equation}
Varying the deformed action~\eqref{eq:actdef} leads to the following equations of motion
\begin{equation}
\partial_+k_- + \partial_- k_+ + \mathcal{A}_{\vec{\alpha}}^{-1}[k_+,\mathcal{A}_{\vec{\alpha}}k_-] + \mathcal{A}_{\vec{\alpha}}^{-1}[k_-,\mathcal{A}_{\vec{\alpha}}k_+] = 0 ~.
\end{equation}
Combining these with the flatness condition for $k_\pm$~\eqref{eq:mcflat} we find
\begin{equation}\begin{aligned}\label{eq:eomsplit}
\partial_+ k_- & = -\frac{1}{2}\big([k_+,k_-] + \mathcal{A}_{\vec{\alpha}}^{-1}[k_+,\mathcal{A}_{\vec{\alpha}}k_-] - \mathcal{A}_{\vec{\alpha}}^{-1}[\mathcal{A}_{\vec{\alpha}}k_+,k_-]\big) ~,
\\
\partial_- k_+ & = -\frac{1}{2}\big([k_-,k_+] + \mathcal{A}_{\vec{\alpha}}^{-1}[k_-,\mathcal{A}_{\vec{\alpha}}k_+] - \mathcal{A}_{\vec{\alpha}}^{-1}[\mathcal{A}_{\vec{\alpha}}k_-,k_+]\big) ~.
\end{aligned}\end{equation}

\subsection{The trigonometric and elliptic deformations of the \texorpdfstring{$\grp{SU}(2)$}{SU(2)} PCM}

First let us study the case $\alpha_1 = \alpha_2 = 1$ and $\alpha_3 = \alpha$, for which we write $\Act_{\vec{\alpha}} = \Act_\alpha$ and $\mathcal{A}_{\vec{\alpha}} = \mathcal{A}_\alpha$.
This deformation does not completely break the left-acting $\grp{SU}(2)$ global symmetry.
It preserves a $\grp{U}(1)$ subgroup, which acts as
\begin{equation}
g \to e^{\chi_\ind{L} T_3} g ~.
\end{equation}
\iflectures
This is a consequence of the identity
\begin{equation}
\mathcal{A}_\alpha (e^{\chi_\ind{L} T_3} X e^{-\chi_\ind{L} T_3}) = e^{\chi_\ind{L} T_3} (\mathcal{A}_\alpha X) e^{-\chi_\ind{L} T_3} ~, \qquad X\in\alg{su}(2) ~,
\end{equation}
which states that the adjoint action of the $\grp{U}(1)$ subgroup commutes with the action of $\mathcal{A}_\alpha$.
Therefore, the global symmetry of the deformed model is $\grp{U}(1)_\ind{L} \times \grp{SU}(2)_\ind{R}$.
Parametrising $g$ as in~\eqref{eq:gparam} and substituting into the action of the deformed model~\eqref{eq:actdef}, we can read off the explicit form of the deformed target-space metric in local coordinates
\begin{equation}\label{eq:su2def}
G = 2\hay (d\theta^2 +\cos^2\theta(\sin^2\theta + \alpha\cos^2\theta) d\varphi^2 + \sin^2\theta(\cos^2\theta + \alpha\sin^2\theta) d\varphi^2 + 2  (1-\alpha) \sin^2\theta\cos^2\theta d\varphi d\phi) ~.
\end{equation}

\question{Show that
\begin{equation*}
\mathcal{A}_\alpha (e^{\chi_\ind{L} T_3} X e^{-\chi_\ind{L} T_3}) = e^{\chi_\ind{L} T_3} (\mathcal{A}_\alpha X) e^{-\chi_\ind{L} T_3} ~, \qquad X\in\alg{su}(2) ~.
\end{equation*}
}

\answer{Given that both the adjoint action and $\mathcal{A}_\alpha$ are linear operators on $\alg{su}(2)$, it is sufficient to check the identity for each of the generators $T_a$.
Noting that
\begin{equation*}
e^{\chi_\ind{L} T_3} T_1 e^{-\chi_\ind{L} T_3} = \cos 2\chi_\ind{L} T_1 - \sin 2\chi_\ind{L} T_2 ~, \qquad
e^{\chi_\ind{L} T_3} T_2 e^{-\chi_\ind{L} T_3} = \cos 2\chi_\ind{L} T_2 + \sin 2\chi_\ind{L} T_1 ~,
\end{equation*}
that is the adjoint action of the $\grp{U}(1)$ subgroup rotates $T_1$ and $T_2$,
and using that $\mathcal{A}_\alpha T_1 = T_1$ and $\mathcal{A}_\alpha T_2 = T_2$, it immediately follows that the identity holds for $T_1$ and $T_2$.
For $T_3$ we have
\begin{equation*}\begin{split}
\mathcal{A}_\alpha (e^{\chi_\ind{L} T_3} T_3 e^{-\chi_\ind{L} T_3}) =
\mathcal{A}_\alpha T_3  = \alpha^{-1} T_3 ~,
\end{split}\end{equation*}
and
\begin{equation*}
e^{\chi_\ind{L} T_3} (\mathcal{A}_\alpha T_3) e^{-\chi_\ind{L} T_3} =
\alpha^{-1} e^{\chi_\ind{L} T_3} T_3 e^{-\chi_\ind{L} T_3}  = \alpha^{-1} T_3 ~.
\end{equation*}
}
\else
This is a consequence of the identity
\begin{equation}
\mathcal{A}_\alpha (e^{\chi_\ind{L} T_3} X e^{-\chi_\ind{L} T_3}) = e^{\chi_\ind{L} T_3} (\mathcal{A}_\alpha X) e^{-\chi_\ind{L} T_3} ~, \qquad X\in\alg{su}(2) ~,
\end{equation}
which states that the adjoint action of the $\grp{U}(1)$ subgroup commutes with the action of $\mathcal{A}_\alpha$.
\unskip\footnote{Given that both the adjoint action and $\mathcal{A}_\alpha$ are linear operators on $\alg{su}(2)$, it is sufficient to check this identity for each of the generators $T_a$.
Noting that
\begin{equation*}
e^{\chi_\ind{L} T_3} T_1 e^{-\chi_\ind{L} T_3} = \cos 2\chi_\ind{L} T_1 - \sin 2\chi_\ind{L} T_2 ~, \qquad
e^{\chi_\ind{L} T_3} T_2 e^{-\chi_\ind{L} T_3} = \cos 2\chi_\ind{L} T_2 + \sin 2\chi_\ind{L} T_1 ~,
\end{equation*}
that is the adjoint action of the $\grp{U}(1)$ subgroup rotates $T_1$ and $T_2$,
and using that $\mathcal{A}_\alpha T_1 = T_1$ and $\mathcal{A}_\alpha T_2 = T_2$, it immediately follows that the identity holds for $T_1$ and $T_2$.
For $T_3$ we have $\mathcal{A}_\alpha (e^{\chi_\ind{L} T_3} T_3 e^{-\chi_\ind{L} T_3}) = \mathcal{A}_\alpha T_3  = \alpha^{-1} T_3$
and
$e^{\chi_\ind{L} T_3} (\mathcal{A}_\alpha T_3) e^{-\chi_\ind{L} T_3} = \alpha^{-1} e^{\chi_\ind{L} T_3} T_3 e^{-\chi_\ind{L} T_3}  = \alpha^{-1} T_3$.}
Therefore, the global symmetry of the deformed model is $\grp{U}(1)_\ind{L} \times \grp{SU}(2)_\ind{R}$.
Parametrising $g$ as in~\eqref{eq:gparam} and substituting into the action of the deformed model~\eqref{eq:actdef}, we can read off the explicit form of the deformed target-space metric in local coordinates
\begin{equation}\label{eq:su2def}
G = 2\hay (d\theta^2 +\cos^2\theta(\sin^2\theta + \alpha\cos^2\theta) d\varphi^2 + \sin^2\theta(\cos^2\theta + \alpha\sin^2\theta) d\varphi^2 + 2  (1-\alpha) \sin^2\theta\cos^2\theta d\varphi d\phi) ~.
\end{equation}
\fi

Recalling the Lax connection of the undeformed model
\begin{equation}\label{eq:undeflax}
\tilde{\Lax}_\pm(z) = \frac{k_\pm}{1\mp z} ~,
\end{equation}
and given the unbroken $\grp{U}(1)$ symmetry, it is natural to consider the following ansatz for the Lax connection of the deformed model
\begin{equation}\label{eq:triglax}
\tilde{\Lax}_{\alpha\pm}(z) = \mathcal{F}_\pm(z,\alpha) k_\pm ~,
\end{equation}
where the linear operators $\mathcal{F}_\pm(z,\alpha)$ act as
\begin{equation}\label{eq:trigf}
\mathcal{F}_\pm(z,\alpha) T_1 = f_\pm(z,\alpha) T_1 ~, \qquad
\mathcal{F}_\pm(z,\alpha) T_2 = f_\pm(z,\alpha) T_2 ~, \qquad
\mathcal{F}_\pm(z,\alpha) T_3 = \tilde{f}_\pm(z,\alpha) T_3 ~.
\end{equation}
Computing the curvature of $\tilde{\Lax}_{\alpha\pm}(z)$ and eliminating the derivatives of $k_\pm$ using eq.~\eqref{eq:eomsplit} we find
\begin{equation}\begin{split}
& \partial_+ \tilde{\Lax}_{\alpha-} - \partial_- \tilde{\Lax}_{\alpha+} + [\tilde{\Lax}_{\alpha+},\tilde{\Lax}_{\alpha-}]
\\ & =
-\frac{1}{2}(\mathcal{F}_+ + \mathcal{F}_-) [k_+,k_-]
+\frac{1}{2}(\mathcal{F}_+ - \mathcal{F}_-) (\mathcal{A}_\alpha^{-1}[k_+,\mathcal{A}_\alpha k_-] - \mathcal{A}_\alpha^{-1}[\mathcal{A}_\alpha k_+,k_-])
+[\mathcal{F}_+k_+,\mathcal{F}_-k_-] ~.
\end{split}\end{equation}
Demanding that this expression vanishes leads to following three equations for the four functions $f_\pm(z,\alpha)$ and $\tilde{f}_\pm(z,\alpha)$
\begin{equation}\begin{split}
f_- - f_+ - 2\alpha f_-(1-\tilde{f}_+)= 0 ~, \qquad
f_- - f_+ + 2\alpha f_+(1-\tilde{f}_-)= 0 ~, \qquad
\tilde{f}_- + \tilde{f}_+ - 2 f_+ f_- = 0 ~.
\end{split}\end{equation}
The freedom in solving this system of equations allows us to introduce the spectral parameter $z$.
One way to do this is to set
\begin{equation}\label{eq:trigsol}
f_\pm(z,\alpha) = \frac{\sinh\nu}{\sinh(\nu(1\mp z))} ~, \qquad
\tilde{f}_\pm(z,\alpha) = \frac{\tanh\nu}{\tanh(\nu(1\mp z))} ~, \qquad
\alpha = \cosh^2\nu ~.
\end{equation}
This solution covers the regime $\alpha > 1$.
The regime $\alpha < 1$ can be covered by analytically continuing $\nu \to i \nu$.
The form of the deformed Lax connection explains why this model is often referred to as the trigonometric deformation of the $\grp{SU}(2)$ PCM.
The Lax connection \eqref{eq:undeflax} of the undeformed model is recovered in the rational limit, which is given by $\nu \to 0$.
\iflectures

\question{For the general deformed model~\eqref{eq:actdef} consider the following ansatz for the Lax connection
\begin{equation*}
\tilde{\Lax}_{\vec{\alpha}\pm}(z) = \mathcal{F}_\pm(z,\vec{\alpha}) k_\pm ~,
\end{equation*}
where the linear operators $\mathcal{F}_\pm(z,\vec{\alpha})$ act as
\begin{equation*}
\mathcal{F}_\pm(z,\vec{\alpha}) T_1 = f_{1\pm}(z,\vec{\alpha}) T_1 ~, \qquad
\mathcal{F}_\pm(z,\vec{\alpha}) T_2 = f_{2\pm}(z,\vec{\alpha}) T_2 ~, \qquad
\mathcal{F}_\pm(z,\vec{\alpha}) T_3 = f_{3\pm}(z,\vec{\alpha}) T_3 ~.
\end{equation*}
Compute the curvature of $\tilde{\Lax}_{\vec{\alpha}\pm}(z)$ and eliminate the derivatives of $k_\pm$ using~\eqref{eq:eomsplit}.
Determine the set of equations for the six functions $f_{a\pm}(z,\vec{\alpha})$ that should be satisfied for this to vanish.
Show that these equations are solved by
\begin{equation*}
f_{1\pm} = \frac{\operatorname{sc}(\nu,k^2)}{\operatorname{sc}(\nu(1\mp z),k^2)} ~, \qquad
f_{2\pm} = \frac{\operatorname{sd}(\nu,k^2)}{\operatorname{sd}(\nu(1\mp z),k^2)} ~, \qquad
f_{3\pm} = \frac{\operatorname{sn}(\nu,k^2)}{\operatorname{sn}(\nu(1\mp z),k^2)} ~,
\end{equation*}
with
\begin{equation*}
\operatorname{cn}^2(\nu,k^2) = \frac{\alpha_1}{\alpha_3} ~, \qquad
\operatorname{dn}^2(\nu,k^2) = \frac{\alpha_2}{\alpha_3} ~, \qquad
\operatorname{cd}^2(\nu,k^2) = \frac{\alpha_1}{\alpha_2} ~, \qquad
k^2 = \frac{\alpha_3 - \alpha_2}{\alpha_3 - \alpha_1} ~.
\end{equation*}
Here $k$ is the elliptic modulus, $\operatorname{sn}(x,k^2)$, $\operatorname{cn}(x,k^2)$ and $\operatorname{dn}(x,k^2)$ are the usual Jacobi elliptic functions and
\begin{equation*}
\operatorname{sc}(x,k^2) = \frac{\operatorname{sn}(x,k^2)}{\operatorname{cn}(x,k^2)} ~, \qquad
\operatorname{sd}(x,k^2) = \frac{\operatorname{sn}(x,k^2)}{\operatorname{dn}(x,k^2)} ~, \qquad
\operatorname{cd}(x,k^2) = \frac{\operatorname{cn}(x,k^2)}{\operatorname{dn}(x,k^2)} ~.
\end{equation*}
This model is often referred to as the elliptic deformation of the $\grp{SU}(2)$ PCM with the trigonometric limit ($\alpha_1 = \alpha_2$) given by $k \to 1$.
}

\answer{Computing the curvature of the Lax connection $\tilde{\Lax}_{\vec{\alpha}\pm}(z)$ and eliminating the derivatives of $k_\pm$ using~\eqref{eq:eomsplit} we find
\begin{equation*}\begin{split}
& \partial_+ \tilde{\Lax}_{\vec{\alpha}-} - \partial_- \tilde{\Lax}_{\vec{\alpha}+} + [\tilde{\Lax}_{\vec{\alpha}+},\tilde{\Lax}_{\vec{\alpha}-}]
\\ & =
-\frac{1}{2}(\mathcal{F}_+ + \mathcal{F}_-) [k_+,k_-]
+\frac{1}{2}(\mathcal{F}_+ - \mathcal{F}_-) (\mathcal{A}_{\vec{\alpha}}^{-1}[k_+,\mathcal{A}_{\vec{\alpha}} k_-] - \mathcal{A}_{\vec{\alpha}}^{-1}[\mathcal{A}_{\vec{\alpha}} k_+,k_-])
+[\mathcal{F}_+k_+,\mathcal{F}_-k_-] ~.
\end{split}\end{equation*}
Demanding that this expression vanishes leads to the following six equations for the six functions $f_{a\pm}(z,\vec{\alpha})$
\begin{equation*}\begin{split}
& f_{1-} + f_{1+} + (f_{1-} - f_{1+}) \big(\frac{\alpha_1}{\alpha_2} - \frac{\alpha_1}{\alpha_3}\big) - 2 f_{2-}f_{3+} = 0 ~,
\\
& f_{1-} + f_{1+} - (f_{1-} - f_{1+}) \big(\frac{\alpha_1}{\alpha_2} - \frac{\alpha_1}{\alpha_3}\big) - 2 f_{3-}f_{2+} = 0 ~,
\\
& f_{2-} + f_{2+} + (f_{2-} - f_{2+}) \big(\frac{\alpha_2}{\alpha_3} - \frac{\alpha_2}{\alpha_1}\big) - 2 f_{3-}f_{1+} = 0 ~,
\\
& f_{2-} + f_{2+} - (f_{2-} - f_{2+}) \big(\frac{\alpha_2}{\alpha_3} - \frac{\alpha_2}{\alpha_1}\big) - 2 f_{1-}f_{3+} = 0 ~,
\\
& f_{3-} + f_{3+} + (f_{3-} - f_{3+}) \big(\frac{\alpha_3}{\alpha_1} - \frac{\alpha_3}{\alpha_2}\big) - 2 f_{1-}f_{2+} = 0 ~,
\\
& f_{3-} + f_{3+} - (f_{3-} - f_{3+}) \big(\frac{\alpha_3}{\alpha_1} - \frac{\alpha_3}{\alpha_2}\big) - 2 f_{2-}f_{1+} = 0 ~.
\end{split}\end{equation*}
Note that these equations are not all independent and they can be solved introducing a spectral parameter $z$.
To show that these equations are solved by
\begin{equation*}
f_{1\pm} = \frac{\operatorname{sc}(\nu,k^2)}{\operatorname{sc}(\nu(1\mp z),k^2)} ~, \qquad
f_{2\pm} = \frac{\operatorname{sd}(\nu,k^2)}{\operatorname{sd}(\nu(1\mp z),k^2)} ~, \qquad
f_{3\pm} = \frac{\operatorname{sn}(\nu,k^2)}{\operatorname{sn}(\nu(1\mp z),k^2)} ~,
\end{equation*}
with
\begin{equation*}
\operatorname{cn}^2(\nu,k^2) = \frac{\alpha_1}{\alpha_3} ~, \qquad
\operatorname{dn}^2(\nu,k^2) = \frac{\alpha_2}{\alpha_3} ~, \qquad
\operatorname{cd}^2(\nu,k^2) = \frac{\alpha_1}{\alpha_2} ~, \qquad
k^2 = \frac{\alpha_3 - \alpha_2}{\alpha_3 - \alpha_1} ~.
\end{equation*}
we substitute in for $f_{a\pm}(z,\vec{\alpha})$ and $\alpha_a$, and use the identities
\begin{equation*}\begin{split}
\operatorname{sn}(x+y,k^2) & = \frac{\operatorname{sn}(x,k^2)\operatorname{cn}(y,k^2)\operatorname{dn}(y,k^2) + \operatorname{sn}(y,k^2)\operatorname{cn}(x,k^2)\operatorname{dn}(x,k^2)}{1-k^2 \operatorname{sn}^2(x,k^2) \operatorname{sn}^2(y,k^2)} ~,
\\
\operatorname{cn}(x+y,k^2) & = \frac{\operatorname{cn}(x,k^2)\operatorname{cn}(y,k^2) - \operatorname{sn}(x,k^2)\operatorname{sn}(y,k^2)\operatorname{dn}(x,k^2)\operatorname{dn}(y,k^2)}{1-k^2 \operatorname{sn}^2(x,k^2) \operatorname{sn}^2(y,k^2)} ~,
\\
\operatorname{dn}(x+y,k^2) & = \frac{\operatorname{dn}(x,k^2)\operatorname{dn}(y,k^2) - k^2\operatorname{sn}(x,k^2)\operatorname{sn}(y,k^2)\operatorname{cn}(x,k^2)\operatorname{cn}(y,k^2)}{1-k^2 \operatorname{sn}^2(x,k^2) \operatorname{sn}^2(y,k^2)} ~,
\end{split}\end{equation*}
and
\begin{equation*}
\operatorname{cn}^2(x,k^2) = 1 - \operatorname{sn}^2(x,k^2) ~, \qquad
\operatorname{dn}^2(x,k^2) = 1 - k^2\operatorname{sn}^2(x,k^2) ~.
\end{equation*}
}
\else

Now turning to the general deformed model~\eqref{eq:actdef}, we consider the following ansatz for the Lax connection
\begin{equation}
\tilde{\Lax}_{\vec{\alpha}\pm}(z) = \mathcal{F}_\pm(z,\vec{\alpha}) k_\pm ~,
\end{equation}
where the linear operators $\mathcal{F}_\pm(z,\vec{\alpha})$ act as
\begin{equation}
\mathcal{F}_\pm(z,\vec{\alpha}) T_1 = f_{1\pm}(z,\vec{\alpha}) T_1 ~, \qquad
\mathcal{F}_\pm(z,\vec{\alpha}) T_2 = f_{2\pm}(z,\vec{\alpha}) T_2 ~, \qquad
\mathcal{F}_\pm(z,\vec{\alpha}) T_3 = f_{3\pm}(z,\vec{\alpha}) T_3 ~.
\end{equation}
Computing the curvature of $\tilde{\Lax}_{\vec{\alpha}\pm}(z)$ and eliminating the derivatives of $k_\pm$ using~\eqref{eq:eomsplit} we find
\begin{equation}\begin{split}
& \partial_+ \tilde{\Lax}_{\vec{\alpha}-} - \partial_- \tilde{\Lax}_{\vec{\alpha}+} + [\tilde{\Lax}_{\vec{\alpha}+},\tilde{\Lax}_{\vec{\alpha}-}]
\\ & =
-\frac{1}{2}(\mathcal{F}_+ + \mathcal{F}_-) [k_+,k_-]
+\frac{1}{2}(\mathcal{F}_+ - \mathcal{F}_-) (\mathcal{A}_{\vec{\alpha}}^{-1}[k_+,\mathcal{A}_{\vec{\alpha}} k_-] - \mathcal{A}_{\vec{\alpha}}^{-1}[\mathcal{A}_{\vec{\alpha}} k_+,k_-])
+[\mathcal{F}_+k_+,\mathcal{F}_-k_-] ~.
\end{split}\end{equation}
Demanding that this expression vanishes leads to the following six equations for the functions $f_{a\pm}(z,\vec{\alpha})$
\begin{equation}\begin{split}
& f_{1-} + f_{1+} \pm (f_{1-} - f_{1+}) \big(\frac{\alpha_1}{\alpha_2} - \frac{\alpha_1}{\alpha_3}\big) - 2 f_{2\mp}f_{3\pm} = 0 ~,
\\
& f_{2-} + f_{2+} \pm (f_{2-} - f_{2+}) \big(\frac{\alpha_2}{\alpha_3} - \frac{\alpha_2}{\alpha_1}\big) - 2 f_{3\mp}f_{1\pm} = 0 ~,
\\
& f_{3-} + f_{3+} \pm (f_{3-} - f_{3+}) \big(\frac{\alpha_3}{\alpha_1} - \frac{\alpha_3}{\alpha_2}\big) - 2 f_{1\mp}f_{2\pm} = 0 ~.
\end{split}\end{equation}
These equations are not all independent and they can be solved introducing a spectral parameter $z$.
Indeed, they are solved by
\unskip\footnote{
To show this we use the identities
\begin{equation*}\begin{split}
\operatorname{sn}(x+y,k^2) & = \frac{\operatorname{sn}(x,k^2)\operatorname{cn}(y,k^2)\operatorname{dn}(y,k^2) + \operatorname{sn}(y,k^2)\operatorname{cn}(x,k^2)\operatorname{dn}(x,k^2)}{1-k^2 \operatorname{sn}^2(x,k^2) \operatorname{sn}^2(y,k^2)} ~,
\\
\operatorname{cn}(x+y,k^2) & = \frac{\operatorname{cn}(x,k^2)\operatorname{cn}(y,k^2) - \operatorname{sn}(x,k^2)\operatorname{sn}(y,k^2)\operatorname{dn}(x,k^2)\operatorname{dn}(y,k^2)}{1-k^2 \operatorname{sn}^2(x,k^2) \operatorname{sn}^2(y,k^2)} ~,
\\
\operatorname{dn}(x+y,k^2) & = \frac{\operatorname{dn}(x,k^2)\operatorname{dn}(y,k^2) - k^2\operatorname{sn}(x,k^2)\operatorname{sn}(y,k^2)\operatorname{cn}(x,k^2)\operatorname{cn}(y,k^2)}{1-k^2 \operatorname{sn}^2(x,k^2) \operatorname{sn}^2(y,k^2)} ~,
\end{split}\end{equation*}
and
\begin{equation*}
\operatorname{cn}^2(x,k^2) = 1 - \operatorname{sn}^2(x,k^2) ~, \qquad
\operatorname{dn}^2(x,k^2) = 1 - k^2\operatorname{sn}^2(x,k^2) ~.
\end{equation*}}
\begin{equation}
f_{1\pm} = \frac{\operatorname{sc}(\nu,k^2)}{\operatorname{sc}(\nu(1\mp z),k^2)} ~, \qquad
f_{2\pm} = \frac{\operatorname{sd}(\nu,k^2)}{\operatorname{sd}(\nu(1\mp z),k^2)} ~, \qquad
f_{3\pm} = \frac{\operatorname{sn}(\nu,k^2)}{\operatorname{sn}(\nu(1\mp z),k^2)} ~,
\end{equation}
with
\begin{equation}
\operatorname{cn}^2(\nu,k^2) = \frac{\alpha_1}{\alpha_3} ~, \qquad
\operatorname{dn}^2(\nu,k^2) = \frac{\alpha_2}{\alpha_3} ~, \qquad
\operatorname{cd}^2(\nu,k^2) = \frac{\alpha_1}{\alpha_2} ~, \qquad
k^2 = \frac{\alpha_3 - \alpha_2}{\alpha_3 - \alpha_1} ~.
\end{equation}
Here $k$ is the elliptic modulus, $\operatorname{sn}(x,k^2)$, $\operatorname{cn}(x,k^2)$ and $\operatorname{dn}(x,k^2)$ are the usual Jacobi elliptic functions and
\begin{equation}
\operatorname{sc}(x,k^2) = \frac{\operatorname{sn}(x,k^2)}{\operatorname{cn}(x,k^2)} ~, \qquad
\operatorname{sd}(x,k^2) = \frac{\operatorname{sn}(x,k^2)}{\operatorname{dn}(x,k^2)} ~, \qquad
\operatorname{cd}(x,k^2) = \frac{\operatorname{cn}(x,k^2)}{\operatorname{dn}(x,k^2)} ~.
\end{equation}
This model is often referred to as the elliptic deformation of the $\grp{SU}(2)$ PCM with the trigonometric limit ($\alpha_1 = \alpha_2$) given by $k \to 1$.
\fi

\subsection{The Yang-Baxter deformation of the \texorpdfstring{$\grp{SU}(2)$}{SU(2)} PCM}

The approach used above to construct Lax connections of the trigonometric and elliptic deformations highlights the appealing hierarchy of these models.
However, it turns out to be non-trivial to generalise this to general Lie group $\grp{G}$.
Given this, we now explore an alternative method that can be used to construct a Lax connection of the one-parameter deformation $\grp{SU}(2)$ PCM.
This is based on the approach introduced in~\secref{sec:pcmwz}, which is to construct a current that is both conserved and flat on-shell.

Using the invariance of the bilinear form, we start by writing the deformed action~\eqref{eq:actdef} with $\alpha_1 = \alpha_2 = 1$ and $\alpha_3 = \alpha$ as
\begin{equation}\label{eq:defa}
\Act_\alpha = - \frac{\hay}{2}\int d^2x \, \tr\big(j_+ \mathcal{A}_{\alpha g} j_- \big) ~,
\end{equation}
where
\begin{equation}
\mathcal{A}_{\alpha g} = \Ad_g^{-1} \mathcal{A}_\alpha \Ad_g ~.
\end{equation}
As usual, the adjoint action is given by $\Ad_g X = g X g^{-1}$ and it defines a linear operator on $\alg{su}(2)$ satisfying $\tr\big(X\Ad_g Y\big) = \tr\big((\Ad_g^{-1} X) Y\big)$.
Since $\mathcal{A}_\alpha$ is symmetric~\eqref{eq:asym}, it follows that $\mathcal{A}_{\alpha g}$ is also
\begin{equation}\label{eq:agsym}
\tr\big(X\mathcal{A}_{\alpha g}Y\big) = \tr\big((\mathcal{A}_{\alpha g}X) Y\big) ~, \qquad X,Y\in\alg{su}(2)~.
\end{equation}
It will also be convenient to introduce the index $\bar{a} = 1,2$ such that
\begin{equation}
\mathcal{A}_\alpha T_{\bar{a}} = T_{\bar{a}} ~, \qquad \mathcal{A}_\alpha T_3 = \alpha^{-1} T_3 ~,
\end{equation}
and
\begin{equation}
[T_{\bar{a}}, T_{\bar{b}}] = -2 \epsilon_{\bar{a}\bar{b}} T_3 ~, \qquad
[T_{\bar{a}}, T_3] = 2 \epsilon_{\bar{a}}{}^{\bar{b}} T_{\bar{b}} ~,
\end{equation}
where $\epsilon_{12} = -\epsilon_{21} = 1$, $\epsilon_{11} = \epsilon_{22} = 0$ and we lower and raise the index $\bar{a}$ with the Kronecker delta $\delta_{\bar{a}\bar{b}}$ and its inverse $\delta^{\bar{a}\bar{b}}$.

Given that the model is invariant under an $\grp{SU}(2)$ global symmetry acting as $g \to g g_\ind{R}$, there is an associated conserved current.
\iflectures
To compute this conserved current we consider the infinitesimal variation $g \to g e^\varepsilon \sim g(1+\varepsilon)$, under which
\begin{equation}\label{eq:variations}
\delta j_\pm = \partial_\pm \varepsilon + [j_\pm,\varepsilon] ~,
\qquad
\delta \mathcal{O}_g = [\mathcal{O}_g,\ad_\varepsilon] ~,
\end{equation}
where $\mathcal{O}_g$ is defined in terms of a constant linear operator $\mathcal{O}:\alg{su}(2)\to\alg{su}(2)$
\begin{equation}
\mathcal{O}_g = \Ad_g^{-1} \mathcal{O} \Ad_g ~,
\end{equation}
and, as usual, $\ad_\varepsilon X = [\varepsilon,X]$.
The resulting equations of motion indeed take the form of a conservation equation
\begin{equation}
\partial_+ J_{\mathcal{A}-} + \partial_- J_{\mathcal{A}+} = 0 ~, \qquad J_{\mathcal{A}\pm} = \frac{1}{\xi} \mathcal{A}_{\alpha g} j_\pm ~,
\end{equation}
where we have used that $\mathcal{A}_{\alpha g}$ is symmetric~\eqref{eq:agsym} and $\xi$ is again a free constant.

\question{For a general Lie group $\grp{G}$, show that under the infinitesimal variation $g\to g e^{\varepsilon} \sim g(1+\varepsilon)$
\begin{equation*}
\delta \mathcal{O}_g = [\mathcal{O}_g,\ad_\varepsilon] ~,
\end{equation*}
and that
\begin{equation*}
\partial_\pm \mathcal{O}_g = [\mathcal{O}_g,\ad_{j_\pm}] ~,
\end{equation*}
where $\mathcal{O}:\alg{g} \to \alg{g}$ is a constant linear operator.
}

\answer{Let us consider an arbitrary variation $g \to g + \delta g$ of the adjoint action of $g$ on $X \in \alg{g}$
\begin{equation*}
\delta \Ad_g X = \delta(g X g^{-1}) = \delta g X g^{-1} - g X g^{-1} \delta g g^{-1}
= \Ad_g [g^{-1}\delta g, X] = \Ad_g \ad_{g^{-1}\delta g} X ~,
\end{equation*}
where we do not vary $X$, that is we treat it as a constant.
We can then write $\delta \Ad_g = \Ad_g \ad_{g^{-1}\delta g}$.
From this it also follows that $\delta \Ad_g^{-1} = - \Ad_g^{-1} \Ad_g \ad_{g^{-1}\delta g} \Ad_g^{-1} = - \ad_{g^{-1}\delta g}\Ad_g^{-1}$.
Therefore, we find that
\begin{equation*}
\delta \mathcal{O}_g = \delta (\Ad_g^{-1} \mathcal{O} \Ad_g) = - \ad_{g^{-1}\delta g}\Ad_g^{-1} \mathcal{O} \Ad_g + \Ad_g^{-1} \mathcal{O} \Ad_g \ad_{g^{-1}\delta g} = [\mathcal{O}_g,\ad_{g^{-1}\delta g}] ~.
\end{equation*}

For the infinitesimal variation $g \to g e^{\varepsilon} \sim g (1+\varepsilon)$ we see that $g^{-1} \delta g = \varepsilon$, hence we indeed have that
\begin{equation*}
\delta \mathcal{O}_g = [\mathcal{O}_g,\ad_\varepsilon] ~.
\end{equation*}
On the other hand, setting $\delta = \partial_\pm$ gives
\begin{equation*}
\partial_\pm \mathcal{O}_g = [\mathcal{O}_g,\ad_{j_\pm}] ~.
\end{equation*}

}
\else
To compute this conserved current we consider the infinitesimal variation $g \to g e^\varepsilon \sim g(1+\varepsilon)$, under which
\unskip\footnote{For general Lie group $\grp{G}$ and constant linear operator $\mathcal{O}:\alg{g} \to \alg{g}$ we can consider an arbitrary variation $g \to g + \delta g$ of the adjoint action of $g$ on $X \in \alg{g}$
\begin{equation*}
\delta \Ad_g X = \delta(g X g^{-1}) = \delta g X g^{-1} - g X g^{-1} \delta g g^{-1}
= \Ad_g [g^{-1}\delta g, X] = \Ad_g \ad_{g^{-1}\delta g} X ~,
\end{equation*}
where we do not vary $X$, that is we treat it as a constant.
Therefore, we can write $\delta \Ad_g = \Ad_g \ad_{g^{-1}\delta g}$.
From this it also follows that $\delta \Ad_g^{-1} = - \Ad_g^{-1} \Ad_g \ad_{g^{-1}\delta g} \Ad_g^{-1} = - \ad_{g^{-1}\delta g}\Ad_g^{-1}$.
Therefore, we have that
\begin{equation*}
\delta \mathcal{O}_g = \delta (\Ad_g^{-1} \mathcal{O} \Ad_g) = - \ad_{g^{-1}\delta g}\Ad_g^{-1} \mathcal{O} \Ad_g + \Ad_g^{-1} \mathcal{O} \Ad_g \ad_{g^{-1}\delta g} = [\mathcal{O}_g,\ad_{g^{-1}\delta g}] ~.
\end{equation*}
For the infinitesimal variation $g \to g e^{\varepsilon} \sim g (1+\varepsilon)$ we see that $g^{-1} \delta g = \varepsilon$, hence $\delta \mathcal{O}_g = [\mathcal{O}_g,\ad_\varepsilon]$.
On the other hand, setting $\delta = \partial_\pm$ gives $\partial_\pm \mathcal{O}_g = [\mathcal{O}_g,\ad_{j_\pm}]$.}
\begin{equation}\label{eq:variations}
\delta j_\pm = \partial_\pm \varepsilon + [j_\pm,\varepsilon] ~,
\qquad
\delta \mathcal{O}_g = [\mathcal{O}_g,\ad_\varepsilon] ~,
\end{equation}
where $\mathcal{O}_g$ is defined in terms of a constant linear operator $\mathcal{O}:\alg{su}(2)\to\alg{su}(2)$
\begin{equation}
\mathcal{O}_g = \Ad_g^{-1} \mathcal{O} \Ad_g ~,
\end{equation}
and, as usual, $\ad_\varepsilon X = [\varepsilon,X]$.
The resulting equations of motion indeed take the form of a conservation equation
\begin{equation}
\partial_+ J_{\mathcal{A}-} + \partial_- J_{\mathcal{A}+} = 0 ~, \qquad J_{\mathcal{A}\pm} = \frac{1}{\xi} \mathcal{A}_{\alpha g} j_\pm ~,
\end{equation}
where we have used that $\mathcal{A}_{\alpha g}$ is symmetric~\eqref{eq:agsym} and $\xi$ is again a free constant.

\fi

Using that
\begin{equation}
\partial_\pm \mathcal{O}_g = [\mathcal{O}_g,\ad_{j_\pm}] ~,
\end{equation}
for a constant linear operator $\mathcal{O}:\alg{su}(2) \to \alg{su}(2)$, and the flatness of $j_\pm$~\eqref{eq:flatness}, we can compute
\begin{equation}\begin{split}
& \partial_+ J_{\mathcal{A}-} - \partial_- J_{\mathcal{A}+} + [J_{\mathcal{A}-},J_{\mathcal{A}+}]
\\ & \qquad = \frac{1}{\xi} \mathcal{A}_{\alpha g} [j_+,j_-] - \frac{1}{\xi} [j_+,\mathcal{A}_{\alpha g}j_-] - \frac{1}{\xi}[\mathcal{A}_{\alpha g} j_+,j_-] + \frac{1}{\xi^2} [\mathcal{A}_{\alpha g}j_+, \mathcal{A}_{\alpha g} j_-] ~.
\end{split}\end{equation}
We would now like to ask if there is a choice of $\xi$ such that $J_{\mathcal{A}\pm}$ is also flat on-shell.
This is equivalent to demanding
\unskip\footnote{Note that, as we vary $g$ over $\grp{SU}(2)$, the pull-backs of the Maurer-Cartan forms $j_\pm$ and $k_\pm$ cover the whole of $\alg{su}(2)$.}
\begin{equation}
\frac{1}{\xi} \mathcal{A}_{\alpha } [X,Y] - \frac{1}{\xi} [X,\mathcal{A}_{\alpha }Y] - \frac{1}{\xi}[\mathcal{A}_{\alpha} X,Y] + \frac{1}{\xi^2} [\mathcal{A}_{\alpha}X, \mathcal{A}_{\alpha} Y] = 0~, \qquad X,Y \in \alg{su}(2)~.
\end{equation}
This relation is antisymmetric upon interchanging $X$ and $Y$, therefore we can simply check whether it holds for the two cases: (i) $X=T_{\bar{a}}$, $Y=T_{\bar{b}}$ and (ii)  $X=T_{\bar{a}}$, $Y=T_3$, which leads to the following equations for $\xi$
\begin{equation}
\frac{1}{\alpha\xi} - \frac{2}{\xi} + \frac{1}{\xi^2} = 0 ~, \qquad
- \frac{1}{\alpha\xi} \big(1-\frac{1}{\xi}\big) = 0 ~.
\end{equation}
These two equations clearly have no solution for $\xi$ if $\alpha \neq 1$.
It therefore appears that we have reached an impasse.
There is no choice of $\xi$ for which $J_{\mathcal{A}\pm}$ is both conserved and flat on-shell.

To circumvent this, we observe that there is a second linear operator on $\alg{su}(2)$ that commutes with the adjoint action of the $\grp{U}(1)$ subgroup.
\iflectures
This is given by
\begin{equation}\label{eq:firstr}
\mathcal{R} T_{\bar{a}} = \epsilon_{\bar{a}}{}^{\bar{b}} T_{\bar{b}} ~, \qquad \mathcal{R} T_3 = 0 ~.
\end{equation}
This map is antisymmetric with respect to the invariant bilinear form, that is
\begin{equation}\label{eq:rsym}
\tr\big(X\mathcal{R}Y\big) = - \tr\big((\mathcal{R}X) Y\big) ~, \qquad X,Y\in\alg{su}(2)~.
\end{equation}

\question{Show that $\mathcal{R}$ commutes with the adjoint action of the $\grp{U}(1)$ subgroup, that is
\begin{equation*}
\mathcal{R} (e^{\chi_\ind{L} T_3} X e^{-\chi_\ind{L} T_3}) = e^{\chi_\ind{L} T_3} (\mathcal{R} X) e^{-\chi_\ind{L} T_3} ~, \qquad X\in\alg{su}(2) ~.
\end{equation*}
}

\answer{Recalling that
\begin{equation*}
e^{\chi_\ind{L} T_3} T_{\bar{a}} e^{-\chi_\ind{L} T_3} = \cos 2\chi_\ind{L} T_{\bar{a}} - \sin 2\chi_\ind{L} \epsilon_{\bar{a}}{}^{\bar{b}} T_{\bar{b}} ~, \qquad
e^{\chi_\ind{L} T_3} T_{3} e^{-\chi_\ind{L} T_3} = T_3 ~,
\end{equation*}
we have
\begin{equation*}\begin{split}
& \mathcal{R}(e^{\chi_\ind{L} T_3} T_{\bar{a}} e^{-\chi_\ind{L} T_3}) = \cos 2\chi_\ind{L} \epsilon_{\bar{a}}{}^{\bar{b}} T_{\bar{b}} - \sin 2\chi_\ind{L} \epsilon_{\bar{a}}{}^{\bar{b}} \epsilon_{\bar{b}}{}^{\bar{c}} T_{\bar{c}} ~,
\\
& e^{\chi_\ind{L} T_3} (\mathcal{R}T_{\bar{a}}) e^{-\chi_\ind{L} T_3} = \epsilon_{\bar{a}}{}^{\bar{b}} e^{\chi_\ind{L} T_3} T_{\bar{b}} e^{-\chi_\ind{L} T_3}
=  \cos 2\chi_\ind{L} \epsilon_{\bar{a}}{}^{\bar{b}} T_{\bar{b}} - \sin 2\chi_\ind{L} \epsilon_{\bar{a}}{}^{\bar{b}} \epsilon_{\bar{b}}{}^{\bar{c}} T_{\bar{c}} ~,
\end{split}\end{equation*}
and
\begin{equation*}\begin{split}
\mathcal{R}(e^{\chi_\ind{L} T_3} T_{3} e^{-\chi_\ind{L} T_3}) = \mathcal{R} T_3 = 0 ~, \qquad
e^{\chi_\ind{L} T_3} (\mathcal{R}T_3) e^{-\chi_\ind{L} T_3} = 0 ~,
\end{split}\end{equation*}
thereby proving that $\mathcal{R}$ commutes with the adjoint action of the $\grp{U}(1)$ subgroup.
}
\else
This is given by
\unskip\footnote{Recalling that $e^{\chi_\ind{L} T_3} T_{\bar{a}} e^{-\chi_\ind{L} T_3} = \cos 2\chi_\ind{L} T_{\bar{a}} - \sin 2\chi_\ind{L} \epsilon_{\bar{a}}{}^{\bar{b}} T_{\bar{b}}$ and $e^{\chi_\ind{L} T_3} T_{3} e^{-\chi_\ind{L} T_3} = T_3$ we have
\begin{equation*}\begin{gathered}
\mathcal{R}(e^{\chi_\ind{L} T_3} T_{\bar{a}} e^{-\chi_\ind{L} T_3}) = \cos 2\chi_\ind{L} \epsilon_{\bar{a}}{}^{\bar{b}} T_{\bar{b}} - \sin 2\chi_\ind{L} \epsilon_{\bar{a}}{}^{\bar{b}} \epsilon_{\bar{b}}{}^{\bar{c}} T_{\bar{c}}
= \epsilon_{\bar{a}}{}^{\bar{b}} e^{\chi_\ind{L} T_3} T_{\bar{b}} e^{-\chi_\ind{L} T_3}
= e^{\chi_\ind{L} T_3} (\mathcal{R}T_{\bar{a}}) e^{-\chi_\ind{L} T_3} ~,
\\
\mathcal{R}(e^{\chi_\ind{L} T_3} T_{3} e^{-\chi_\ind{L} T_3}) = \mathcal{R} T_3 = 0 = e^{\chi_\ind{L} T_3} (\mathcal{R}T_3) e^{-\chi_\ind{L} T_3} ~.
\end{gathered}\end{equation*}
Therefore $\mathcal{R} (e^{\chi_\ind{L} T_3} X e^{-\chi_\ind{L} T_3}) = e^{\chi_\ind{L} T_3} (\mathcal{R} X) e^{-\chi_\ind{L} T_3}$, $X\in\alg{su}(2)$, and we see that $\mathcal{R}$ indeed commutes with the adjoint action of the $\grp{U}(1)$ subgroup.}
\begin{equation}\label{eq:firstr}
\mathcal{R} T_{\bar{a}} = \epsilon_{\bar{a}}{}^{\bar{b}} T_{\bar{b}} ~, \qquad \mathcal{R} T_3 = 0 ~.
\end{equation}
The map $\mathcal{R}$ is antisymmetric with respect to the invariant bilinear form, that is
\begin{equation}\label{eq:rsym}
\tr\big(X\mathcal{R}Y\big) = - \tr\big((\mathcal{R}X) Y\big) ~, \qquad X,Y\in\alg{su}(2)~.
\end{equation}
\fi
Defining
\begin{equation}
J_{\mathcal{R}\pm} = \mp \mathcal{R}_g j_\pm ~, \qquad \mathcal{R}_g = \Ad_g^{-1} \mathcal{R} \Ad_g ~,
\end{equation}
we can compute
\begin{equation}\begin{split}
\partial_+ J_{\mathcal{R}-} + \partial_- J_{\mathcal{R}+} & = \mathcal{R}_g \partial_+ j_- + [\mathcal{R}_g,\ad_{j_+}] j_- - \mathcal{R}_g \partial_- j_+ - [\mathcal{R}_g,\ad_{j_-}] j_+
\\ & = \mathcal{R}_g [j_+,j_-] - [j_+,\mathcal{R}_g j_-] - [\mathcal{R}_g j_+,j_-] = 0 ~,
\end{split}\end{equation}
where we have used the flatness of $j_\pm$~\eqref{eq:flatness} and the identity
\unskip\footnote{Since this identity is antisymmetric upon interchanging $X$ and $Y$, to prove it we can simply check it for the two cases: (i) $X=T_{\bar{a}}$, $Y=T_{\bar{b}}$ and (ii)  $X=T_{\bar{a}}$, $Y=T_3$, as follows
\begin{equation*}\begin{split}
& \mathcal{R} [T_{\bar{a}},T_{\bar{b}}] - [T_{\bar{a}},\mathcal{R}T_{\bar{b}}] - [\mathcal{R}T_{\bar{a}},T_{\bar{b}}]
= -2 \epsilon_{\bar{a}\bar{b}} \mathcal{R} T_3 - \epsilon_{\bar{b}}{}^{\bar{c}} [T_{\bar{a}},T_{\bar{c}}] - \epsilon_{\bar{a}}{}^{\bar{c}} [T_{\bar{c}},T_{\bar{b}}]
= 2 \epsilon_{\bar{b}}{}^{\bar{c}} \epsilon_{\bar{a}\bar{c}} T_3 + 2 \epsilon_{\bar{a}}{}^{\bar{c}} \epsilon_{\bar{c}\bar{b}} T_3 = 0 ~,
\\
&\mathcal{R} [T_{\bar{a}},T_3] - [T_{\bar{a}},\mathcal{R}T_3] - [\mathcal{R}T_{\bar{a}},T_3]
= 2 \epsilon_{\bar{a}}{}^{\bar{b}} \mathcal{R} T_{\bar{b}} - \epsilon_{\bar{a}}{}^{\bar{b}} [T_{\bar{b}}, T_3]
= 2\epsilon_{\bar{a}}{}^{\bar{b}} \epsilon_{\bar{b}}{}^{\bar{c}} T_{\bar{c}} - 2 \epsilon_{\bar{a}}{}^{\bar{b}} \epsilon_{\bar{b}}{}^{\bar{c}} T_{\bar{c}}  = 0 ~.
\end{split}\end{equation*}}
\begin{equation}\label{eq:closedr}
\mathcal{R} [X,Y] - [X,\mathcal{R}Y] - [\mathcal{R}X,Y] = 0  ~, \qquad X,Y\in\alg{su}(2)~.
\end{equation}
Therefore, we see that $J_{\mathcal{R}\pm}$ is conserved without using the equations of motion.
In particular, the Hodge dual of the 1-form $J_{\mathcal{R}}$ is closed.
Moreover, it is also exact since it can be written in the form $J_{\mathcal{R}\pm} = \pm \partial_\pm (\frac12\Ad_g^{-1} T_3)$.
Therefore, its conservation can be found via the variational principle from the following boundary term
\unskip\footnote{To see that this is a boundary term we start by recalling the identity~\eqref{eq:closedr}
\begin{equation*}
\mathcal{R} [X,Y] - [X,\mathcal{R}Y] - [\mathcal{R}X,Y] = 0 ~, \qquad X,Y\in\alg{su}(2)~.
\end{equation*}
This identity implies that $\rho\big(X,Y\big) = \tr\big(X\mathcal{R}Y\big)$ is a closed 2-cocycle in the Lie algebra cohomology of $\alg{su}(2)$ since
\begin{equation*}\begin{split}
(d\rho)\big(X,Y,Z\big)
&= \rho\big([X,Y],Z\big) + \rho\big([Y,Z],X\big) + \rho\big([Z,X],Y\big)
\\ & = \tr\big([X,Y]\mathcal{R}Z + [Y,Z]\mathcal{R}X + [Z,X]\mathcal{R}Y\big)
\\ & = - \tr\big(Z(\mathcal{R} [X,Y] - [X,\mathcal{R}Y] - [\mathcal{R}X,Y] )\big) = 0 ~,
\qquad X,Y,Z \in \alg{su}(2) ~.
\end{split}\end{equation*}
Moreover, it turns out that $\rho$ is also exact, that is
\begin{equation*}
\rho\big(X,Y\big) = (d\tilde{\rho})\big(X,Y\big) = \tilde\rho\big([X,Y]\big) ~, \qquad X,Y \in \alg{su}(2) ~,
\end{equation*}
where
\begin{equation*}
\tilde\rho\big(X\big) = \frac12\tr\big(XT_3\big) ~, \qquad X\in\alg{su}(2)~.
\end{equation*}
It then follows that
\begin{equation*}
\tr\big(j_+\mathcal{R}_gj_-\big) = \tr\big(k_+\mathcal{R}k_-\big) = \rho\big(k_+,k_-\big) = \tilde\rho\big([k_+,k_-]\big) = \tilde\rho(-\partial_+k_- + \partial_-k_+\big) = \tilde\rho\big(2\epsilon^{\mu\nu} \partial_\mu k_\nu\big) = \epsilon^{\mu\nu} \partial_\mu \tilde\rho\big(2k_\nu\big) ~,
\end{equation*}
where we have used the flatness condition for $k_\pm$~\eqref{eq:mcflat}.
Therefore, we see explicitly that $\Act_{\mathcal{R}}$ is indeed a boundary term.}
\begin{equation}
\Act_{\mathcal{R}} = - \frac{\hay}{2} \int d^2 x \, \tr\big(j_+ \mathcal{R}_g j_-\big) ~,
\end{equation}
the integrand of which is a total derivative, hence the target-space B-field is closed and exact.

We can use the existence of this off-shell current to construct a more general conserved current
\begin{equation}
J_\pm = J_{\mathcal{A}\pm} + \frac{\eta}{\xi} J_{\mathcal{R}\pm} = \frac{1}{\xi} (\mathcal{A}_{\alpha g} \mp \eta\mathcal{R}_g) j_\pm ~,
\end{equation}
and ask if we can now choose $\xi$ and $\eta$ such that $J_\pm$ is also flat on-shell.
To this end, we compute $\partial_+ J_- - \partial_- J_+ + [J_+,J_-]$
and use the conservation of $J_{\mathcal{A}\pm}$ on-shell to replace $\partial_+ j_- + \partial_- j_+$ by $\mathcal{A}_{\alpha g}^{-1}\big([j_+,\mathcal{A}_{\alpha g}j_-] -[\mathcal{A}_{\alpha g}j_+,j_-]\big)$.
Demanding that $J_\pm$ is flat implies
\begin{equation}\begin{split}
& \frac{1}{\xi}\Big( \mathcal{A}_{\alpha} [X,Y]
+ \eta \mathcal{R}\mathcal{A}_{\alpha}^{-1}\big([X,\mathcal{A}_{\alpha}Y] -[\mathcal{A}_{\alpha}X,Y]\big)
- [X,(\mathcal{A}_\alpha + \eta \mathcal{R}) Y]
\\ & \hspace{75pt}
- [(\mathcal{A}_\alpha - \eta \mathcal{R}) X,Y]
+ \frac{1}{\xi} [(\mathcal{A}_\alpha - \eta \mathcal{R}) X,(\mathcal{A}_\alpha + \eta \mathcal{R}) Y]\Big)
= 0 ~, \qquad X,Y\in\alg{su}(2)~.
\end{split}\end{equation}
Substituting in the four cases:
(i) $X=T_{\bar{a}}$, $Y=T_{\bar{b}}$; (ii)  $X=T_{\bar{a}}$, $Y=T_3$, (iii) $X=T_3$, $Y = T_{\bar{a}}$ and (iv) $X=T_3$, $Y=T_3$, we find the following equations for $\xi$ and $\eta$
\begin{equation}\label{eq:eqstosol}
\frac{\eta(1-\xi)}{\xi^2} = 0 ~, \qquad
\frac{\eta(1-\xi)}{\alpha\xi^2} = 0 ~, \qquad
\frac{1-\xi}{\alpha\xi^2} = 0 ~, \qquad
\frac{1}{\alpha\xi} - \frac{2}{\xi} + \frac{1}{\xi^2} = \frac{\eta^2}{\xi^2} ~.
\end{equation}
These equations can be solved for general $\alpha$ by setting
\begin{equation}
\xi = 1 ~, \qquad \eta = \sqrt{\frac{1-\alpha}{\alpha}} ~.
\end{equation}
A second solution is given by taking $\eta$ to be the negative square root.

It therefore follows that if we modify the deformed action~\eqref{eq:defa} by a boundary term such that
\unskip\footnote{We have also rescaled $\hay \to \alpha \hay$ in order to match with the conventions used in the following sections.}
\begin{equation}\label{eq:ybsu2}
\Act_\indrm{YB} = - \frac{\hay}{2} \int d^2 x \, \tr \big(j_+ (\alpha\mathcal{A}_{\alpha g} + \sqrt{\alpha(1-\alpha)} \mathcal{R}_g ) j_- \big) ~,
\end{equation}
then the conserved current associated to the right-acting $\grp{SU}(2)$ global symmetry
\begin{equation}
J_\pm = (\mathcal{A}_{\alpha g} \mp \sqrt{\frac{1-\alpha}{\alpha}} \mathcal{R}_g ) j_\pm ~,
\end{equation}
is flat on-shell.
Therefore, we can write down the following Lax connection of this model
\begin{equation}
\Lax_\pm(z) = \frac{J_\pm}{1\mp z} ~.
\end{equation}
For reasons that will become clear in \secref{sec:ybpcm} this model is known as the Yang-Baxter (YB) deformation of the $\grp{SU}(2)$ PCM.
\iflectures
Let us conclude by noting that, in contrast to the trigonometric deformation, the action and Lax connection of the YB deformation are only real for $\alpha \leq 1$.

\question{Show that
\begin{equation*}
\alpha\mathcal{A}_\alpha + \sqrt{\alpha(1-\alpha)} \mathcal{R} = \frac{1}{1-\eta \mathcal{R}} ~,
\end{equation*}
where
\begin{equation*}
\eta =  \sqrt{\frac{1-\alpha}{\alpha}} ~.
\end{equation*}
Note that the range $\alpha \in [1,0)$ corresponds to $\eta \in [0,\infty)$.
Moreover, by allowing $\eta$ to take negative values we can simultaneously cover the second solution to eq.~\eqref{eq:eqstosol}.
}

\answer{Noting that
\begin{equation*}
\mathcal{R}^2 = \frac{1}{1-\alpha}(\alpha\mathcal{A}_\alpha-1) ~, \qquad \mathcal{R}^3 = -\mathcal{R} ~,
\qquad \mathcal{R} \mathcal{A}_\alpha = \mathcal{A}_\alpha \mathcal{R} = \mathcal{R} ~,
\end{equation*}
we observe that we can write
\begin{equation*}
\frac{1}{1-\eta \mathcal{R}} = \gamma_1 \mathcal{A}_\alpha + \gamma_2 \mathcal{R} + \gamma_3 ~,
\end{equation*}
where the parameters $\gamma_1$, $\gamma_2$ and $\gamma_3$ are to be determined.
Acting on both sides with $1 - \eta \mathcal{R}$ we find
\begin{equation*}\begin{split}
1 & = \gamma_1 \mathcal{A}_\alpha - \gamma_1 \eta \mathcal{R}\mathcal{A}_\alpha + \gamma_2 \mathcal{R} - \gamma_2 \eta \mathcal{R}^2 + \gamma_3 - \eta \gamma_3\mathcal{R}
\\ & = \gamma_1 \mathcal{A}_\alpha - \gamma_1 \eta \mathcal{R} + \gamma_2 \mathcal{R} - \frac{\gamma_2 \eta}{1-\alpha}(\alpha\mathcal{A}_\alpha - 1) + \gamma_3 - \eta \gamma_3\mathcal{R} ~.
\end{split}\end{equation*}
Equating the coefficients of the operators $1$, $\mathcal{R}$ and $\mathcal{A}_\alpha$ and using the given relation between $\eta$ and $\alpha$ implies that
\begin{equation*}
1= \frac{\gamma_2(1+\eta^2)}{\eta} + \gamma_3 ~,
\qquad
0 = -\gamma_1\eta+\gamma_2-\eta \gamma_3 ~,
\qquad
0 = \gamma_1- \frac{\gamma_2}{\eta} ~.
\end{equation*}
These are solved by
\begin{equation*}
\gamma_1 = \frac{1}{1+\eta^2} =  \alpha ~, \qquad \gamma_2 = \frac{\eta}{1+\eta^2} = \sqrt{\alpha(1-\alpha)} ~, \qquad \gamma_3 = 0 ~,
\end{equation*}
giving the required result.
}

\question{Consider the following deformation of the PCWZM for $\grp{G} = \grp{SU}(2)$
\begin{equation*}
\Act_\indrm{YB-PCWZM} = - \frac{\hay}{2} \int d^2x \, \tr\big(j_+ ( \alpha\mathcal{A}_{\alpha g} + \alpha \eta \mathcal{R}_g) j_-\big) + \frac{\kay}{6} \int d^3 x\, \epsilon^{ijk} \tr\big(j_i[j_j,j_k]\big) ~.
\end{equation*}
Show that the equations of motion are given by
\begin{equation*}
\partial_+ J_- + \partial_- J_+ = 0 ~, \qquad
J_\pm = \frac{1}{\xi}(\mathcal{A}_{\alpha g} \mp \eta \mathcal{R}_g \mp \frac{\kay}{\alpha\hay}) j_\pm ~.
\end{equation*}
Using that $J_{\mathcal{R}\pm} = \mp \mathcal{R}_g j_\pm$ is conserved off-shell for $\grp{G} = \grp{SU}(2)$, determine the values of $\xi$ and $\eta$ that ensure that the conserved current $J_\pm$ is also flat on-shell.
This model is known as the YB deformation of the $\grp{SU}(2)$ PCWZM.
}

\answer{Given that the model is invariant under a $\grp{SU}(2)$ global symmetry acting as $g \to g g_\ind{R}$, there is an associated conserved current.
To compute this conserved current we consider the infinitesimal variation $g \to g e^{\varepsilon} \sim g(1+\varepsilon)$.
Using the various results that we have already derived, it is straightforward to see that the resulting equations of motion indeed take the claimed form
\begin{equation*}
\partial_+ J_- + \partial_- J_+ = 0 ~, \qquad
J_\pm = \frac{1}{\xi}(\mathcal{A}_{\alpha g} \mp \eta \mathcal{R}_g \mp \frac{\kay}{\alpha\hay}) j_\pm ~.
\end{equation*}
Since $J_{\mathcal{R}\pm} = \mp \mathcal{R}_g j_\pm$ is conserved off-shell, it follows that $J_{\mathcal{A}\pm} = J_\pm - \frac{\eta}{\xi} J_{\mathcal{R}\pm} = \frac{1}{\xi}\big(\mathcal{A}_{\alpha g} \mp \frac{\kay}{\alpha\hay}\big) j_\pm$ is also conserved on-shell.
Now computing $\partial_+ J_- - \partial_- J_+ + [J_+,J_-]$, and using the conservation of $J_{\mathcal{A}\pm}$ on-shell and the flatness of $j_\pm$ to replace $\partial_+ j_- + \partial_- j_+$ by $\mathcal{A}_{\alpha g}^{-1}([j_+,\mathcal{A}_{\alpha g}j_-] -[\mathcal{A}_{\alpha g}j_+,j_-] + \frac{\kay}{\alpha\hay}[j_+,j_-])$, we find that $J_\pm$ is flat if
\begin{equation*}\begin{split}
& \frac{1}{\xi}\Big( \mathcal{A}_{\alpha} [X,Y]
+ (\eta\mathcal{R} + \frac{\kay}{\alpha\hay} )\mathcal{A}_{\alpha}^{-1}\big([X,\mathcal{A}_{\alpha}Y] -[\mathcal{A}_{\alpha}X,Y] + \frac{\kay}{\alpha\hay}[X,Y]\big)
-  [X,(\mathcal{A}_\alpha + \eta \mathcal{R}) Y]
\\ & \hspace{60pt}
-  [(\mathcal{A}_\alpha - \eta \mathcal{R}) X,Y]
+ \frac{1}{\xi} [(\mathcal{A}_\alpha - \eta \mathcal{R} - \frac{\kay}{\alpha\hay}) X,(\mathcal{A}_\alpha + \eta \mathcal{R}+\frac{\kay}{\alpha\hay}) Y]\Big)
= 0 ~, \qquad X,Y\in\alg{su}(2)~.
\end{split}\end{equation*}
Substituting in the four cases:
(i) $X=T_{\bar{a}}$, $Y=T_{\bar{b}}$; (ii)  $X=T_{\bar{a}}$, $Y=T_3$, (iii) $X=T_3$, $Y = T_{\bar{a}}$ and (iv) $X=T_3$, $Y=T_3$, the resulting set of equations can be solved for $\xi$ and $\eta$ by setting
\begin{equation*}
\xi = 1 ~, \qquad \eta = \sqrt{\frac{1-\alpha}{\alpha}\big(1 - \frac{\kay^2}{\alpha\hay^2}\big)} ~.
\end{equation*}
A second solution is given by taking $\eta$ to be the negative square root.
}
\else
In contrast to the trigonometric deformation, the action and Lax connection of the YB deformation are only real for $\alpha \leq 1$.

Noting that
\begin{equation}\label{eq:si}
\alpha\mathcal{A}_\alpha = 1 + (1-\alpha)\mathcal{R}^2 ~, \qquad \mathcal{R}^3 = -\mathcal{R} ~,
\qquad \mathcal{R} \mathcal{A}_\alpha = \mathcal{A}_\alpha \mathcal{R} = \mathcal{R} ~,
\end{equation}
we observe that we can write
\begin{equation}
\alpha \mathcal{A}_\alpha + \sqrt{\alpha(1-\alpha)}\mathcal{R} = \frac{1}{\gamma_0 + \gamma_1 \mathcal{R} + \gamma_2 \mathcal{R}^2} ~,
\end{equation}
where, acting on both sides with $\gamma_0 + \gamma_1 \mathcal{R} + \gamma_2 \mathcal{R}^2$, the parameters $\gamma_0$, $\gamma_1$ and $\gamma_2$ are determined to be
\begin{equation}
\gamma_0 = 1 ~, \qquad \gamma_1 = - \sqrt{\frac{1-\alpha}{\alpha}} = -\eta ~, \qquad \gamma_2 = 0 ~.
\end{equation}
Therefore, the YB deformation of the $\grp{SU}(2)$ PCM~\eqref{eq:ybsu2} can be written in the form
\begin{equation}\label{eq:ybsu2alt}
\Act_\indrm{YB} = - \frac{\hay}{2} \int d^2 x \, \tr \big(j_+ \frac{1}{1-\eta \mathcal{R}_g} j_- \big) ~.
\end{equation}
The range $\alpha \in [1,0)$ corresponds to $\eta \in [0,\infty)$.
Allowing $\eta$ to take negative values we can simultaneously cover the second solution to eq.~\eqref{eq:eqstosol}.
As we will see in \secref{sec:ybpcm}, this form of the action is particularly natural for studying integrable deformations of the PCM for general simple Lie group $\grp{G}$.

\subsection{The Yang-Baxter deformation of the \texorpdfstring{$\grp{SU}(2)$}{SU(2)} PCWZM}

A similar approach can be used to construct the YB deformation of the $\grp{SU}(2)$ PCWZM starting from the ansatz
\begin{equation}\label{eq:su2ybwz}
\Act_\indrm{YB-PCWZM} = - \frac{\hay}{2} \int d^2x \, \tr\big(j_+ ( \alpha\mathcal{A}_{\alpha g} + \alpha \eta \mathcal{R}_g) j_-\big) + \frac{\kay}{6} \int d^3 x\, \epsilon^{ijk} \tr\big(j_i[j_j,j_k]\big) ~.
\end{equation}
Given that the model is invariant under a $\grp{SU}(2)$ global symmetry acting as $g \to g g_\ind{R}$, there is an associated conserved current, which we can compute by considering the infinitesimal variation $g \to g e^{\varepsilon} \sim g(1+\varepsilon)$.
As expected, the resulting equations of motion take the form of a conservation equation
\begin{equation}\label{eq:ccc}
\partial_+ J_- + \partial_- J_+ = 0 ~, \qquad
J_\pm = \frac{1}{\xi}(\mathcal{A}_{\alpha g} \mp \eta \mathcal{R}_g \mp \frac{\kay}{\alpha\hay}) j_\pm ~,
\end{equation}
where $\xi$ is again a free constant.
Since $J_{\mathcal{R}\pm} = \mp \mathcal{R}_g j_\pm$ is conserved off-shell for $\grp{G} = \grp{SU}(2)$, it follows that $J_{\mathcal{A}\pm} = J_\pm - \frac{\eta}{\xi} J_{\mathcal{R}\pm} = \frac{1}{\xi}\big(\mathcal{A}_{\alpha g} \mp \frac{\kay}{\alpha\hay}\big) j_\pm$ is also conserved on-shell.
Now computing $\partial_+ J_- - \partial_- J_+ + [J_+,J_-]$, and using the conservation of $J_{\mathcal{A}\pm}$ on-shell and the flatness of $j_\pm$ to replace $\partial_+ j_- + \partial_- j_+$ by $\mathcal{A}_{\alpha g}^{-1}([j_+,\mathcal{A}_{\alpha g}j_-] -[\mathcal{A}_{\alpha g}j_+,j_-] + \frac{\kay}{\alpha\hay}[j_+,j_-])$, we find that $J_\pm$ is flat if
\begin{equation}\begin{split}
& \frac{1}{\xi}\Big( \mathcal{A}_{\alpha} [X,Y]
+ (\eta\mathcal{R} + \frac{\kay}{\alpha\hay} )\mathcal{A}_{\alpha}^{-1}\big([X,\mathcal{A}_{\alpha}Y] -[\mathcal{A}_{\alpha}X,Y] + \frac{\kay}{\alpha\hay}[X,Y]\big)
- [X,(\mathcal{A}_\alpha + \eta \mathcal{R}) Y]
\\ & \hspace{15pt}
- [(\mathcal{A}_\alpha - \eta \mathcal{R}) X,Y]
+ \frac{1}{\xi} [(\mathcal{A}_\alpha - \eta \mathcal{R} - \frac{\kay}{\alpha\hay}) X,(\mathcal{A}_\alpha + \eta \mathcal{R}+\frac{\kay}{\alpha\hay}) Y] \Big)
= 0 ~, \qquad X,Y\in\alg{su}(2)~.
\end{split}\end{equation}
Substituting in the four cases:
(i) $X=T_{\bar{a}}$, $Y=T_{\bar{b}}$; (ii)  $X=T_{\bar{a}}$, $Y=T_3$, (iii) $X=T_3$, $Y = T_{\bar{a}}$ and (iv) $X=T_3$, $Y=T_3$, the resulting set of equations can be solved for $\xi$ and $\eta$ by setting
\begin{equation}\label{eq:xieta}
\xi = 1 ~, \qquad \eta = \sqrt{\frac{1-\alpha}{\alpha}\big(1 - \frac{\kay^2}{\alpha\hay^2}\big)} ~.
\end{equation}
A second solution is given by taking $\eta$ to be the negative square root.
Therefore, for $\xi$ and $\eta$ given by eq.~\eqref{eq:xieta}, the conserved current~\eqref{eq:ccc} associated to the right-acting $\grp{SU}(2)$ global symmetry of the deformed action~\eqref{eq:su2ybwz} is flat on-shell and we can write down the Lax connection of this model.

Introducing the parameters $\chi$ and $\rho$ in place of $\hay$ and $\alpha$
\begin{equation}
\hay = \kay \coth \frac{\chi}{2} ~, \qquad
\alpha = \frac{\cosh\chi-1}{\cosh\chi-\cos\rho} ~,
\qquad
\eta = \sqrt{\frac{1-\alpha}{\alpha}\big(1- \frac{\kay^2}{\alpha\hay^2}\big)} = \frac{\sin\rho}{\sinh\chi} ~,
\end{equation}
and again using the identities~\eqref{eq:si}, we can show that
\unskip\footnote{The operator $e^{\rho \mathcal{R}}$ satisfies the following differential equation and boundary condition
\begin{equation*}
\frac{d}{d\rho} e^{\rho \mathcal{R}} = \mathcal{R} e^{\rho \mathcal{R}} ~, \qquad
e^{\rho \mathcal{R}}\big|_{\rho = 0} = 1 ~.
\end{equation*}
Substituting in $e^{\rho \mathcal{R}} = \gamma_0 + \gamma_1 \mathcal{R} + \gamma_2 \mathcal{R}^2$ and using $\mathcal{R}^3 = -\mathcal{R}$ we find
\begin{equation*}\begin{aligned}
\frac{d\gamma_0}{d\rho} = 0 ~, \qquad
\frac{d\gamma_1}{d\rho} = \gamma_0-\gamma_2 ~, \qquad
\frac{d\gamma_2}{d\rho} = \gamma_1 ~, \qquad
\gamma_0\big|_{\rho = 0} = 1 ~, \qquad
\gamma_1\big|_{\rho = 0} = 0 ~, \qquad
\gamma_2\big|_{\rho = 0} = 0 ~,
\end{aligned}\end{equation*}
which is solved by
\begin{equation*}
\gamma_1 = 1 ~, \qquad \gamma_2 = \sin\rho ~, \qquad \gamma_3 = 1- \cos\rho ~.
\end{equation*}
Using this expansion of $e^{\rho \mathcal{R}}$ it is then straightforward to check that
\begin{equation*}
\kay\frac{e^\chi + e^{\rho \mathcal{R}} }{e^\chi - e^{\rho \mathcal{R}}} =
\kay \coth \frac{\chi}{2} + \frac{\kay\sin\rho}{\cosh\chi-\cos\rho}\mathcal{R}
+ \kay \coth \frac{\chi}{2} \frac{1-\cos\rho}{\cosh\chi - \cos\rho} \mathcal{R}^2
= \hay(\alpha\mathcal{A}_\alpha + \alpha \eta \mathcal{R}) ~.
\end{equation*}}
\begin{equation}
\hay(\alpha\mathcal{A}_\alpha + \alpha \eta \mathcal{R}) =
\kay\frac{e^\chi + e^{\rho \mathcal{R}} }{e^\chi - e^{\rho \mathcal{R}}} ~.
\end{equation}
Therefore, the YB deformation of the $\grp{SU}(2)$ PCWZM~\eqref{eq:su2ybwz} can be written in the form
\begin{equation}
\Act_\indrm{YB-PCWZM} = -\frac{\kay}{2} \int d^2x \, \tr\big(j_+ \frac{e^\chi + e^{\rho \mathcal{R}_g} }{e^\chi - e^{\rho \mathcal{R}_g}} j_-\big)
+ \frac{\kay}{6} \int d^3 x\, \epsilon^{ijk} \tr\big(j_i[j_j,j_k]\big) ~.
\end{equation}
It turns out that this form is particularly natural for studying integrable deformations of the PCWZM for general simple Lie group $\grp{G}$.
\fi

\section{The Yang-Baxter deformation of the PCM}\label{sec:ybpcm}

Now that we have explored integrable deformations of the $\grp{SU}(2)$ PCM and PCWZM in some detail, we generalise to a general simple Lie group $\grp{G}$.
From this point on we will restrict ourselves to those integrable deformations for which the Lax connection can be built from a conserved and flat current.
This means that the methods for constructing conserved charges discussed in \secref{sec:pcmwz} can be straightforwardly applied to these models.

Motivated by our discussion of the Yang-Baxter (YB) deformation of the $\grp{SU}(2)$ PCM we consider the following action as our starting point
\begin{equation}\label{eq:ybdefgen}
\Act_\indrm{YB} = -\frac{\hay}{2} \int d^2x \, \tr\big(j_+ \frac{1}{1 - \eta \mathcal{R}_g} j_-\big) ~,
\qquad \mathcal{R}_g = \Ad_g^{-1} \mathcal{R} \Ad_g ~,
\end{equation}
where $\mathcal{R}$ is a general constant linear operator on $\alg{g}$.
For now we do not assume that $\mathcal{R}$ satisfies any particular symmetry properties.
However, we do assume that the operator $1 - \eta \mathcal{R}$ is invertible on $\alg{g}$ for all $\eta$.
This may not always be the case, and typically leads to singularities in the target-space metric and B-field.

The action~\eqref{eq:ybdefgen} defines a deformation of the PCM with the undeformed limit given by $\eta \to 0$.
The deformation preserves the right-acting $\grp{G}$ global symmetry, but breaks the left-acting symmetry to the subgroup $\grp{G}_0$ of $\grp{G}$ that commutes with the operator $\mathcal{R}$
\begin{equation}\label{eq:symg0}
\Ad_{g_0}^{-1} \mathcal{R} \Ad_{g_0} = \mathcal{R} ~, \qquad g_0 \in \grp{G}_0 ~.
\end{equation}
We say that elements of $\grp{G}_0$ are symmetries of the operator $\mathcal{R}$.

The equations of motion for the action~\eqref{eq:ybdefgen} can be written as the conservation equation associated to the right-acting $\grp{G}$ global symmetry.
Considering the infinitesimal variation $g \to ge^{\varepsilon} \sim g(1+\varepsilon)$ we find
\begin{equation}\label{eq:jpm}
\partial_+ J_- + \partial_- J_+ = 0 ~, \qquad J_- = \frac{1}{\xi} \frac{1}{1-\eta \mathcal{R}_g} j_- ~,
\qquad J_+ = \frac{1}{\xi} \frac{1}{1-\eta \mathcal{R}^t_g} j_+ ~,
\end{equation}
where $\mathcal{R}^t$ is the transpose of $\mathcal{R}$ with respect to the invariant bilinear form, that is
\begin{equation}
\tr\big(X \mathcal{R} Y\big) = \tr\big((\mathcal{R}^t X) Y\big) ~, \qquad X,Y\in\alg{g}~,
\end{equation}
and $\xi$ is a free parameter.
In order to construct integrable deformations of the PCM, we follow the strategy of requiring that the conserved current $J_\pm$ is also flat on-shell.
While this will lead to a large class of integrable deformations known as the YB deformations, it should be noted that not all integrable deformations of sigma models, for example, the elliptic deformation of the $\grp{SU}(2)$ PCM, are of this type.

From eq.~\eqref{eq:jpm} we have
\begin{equation}
j_- = \xi(1-\eta\mathcal{R}_g) J_- ~, \qquad j_+ = \xi(1-\eta \mathcal{R}_g^t) J_+ ~.
\end{equation}
Substituting into the flatness condition~\eqref{eq:flatness} we find
\begin{equation}\begin{split}
0 & = \xi \partial_+ ( (1-\eta \mathcal{R}_g) J_-) - \xi \partial_- ((1 - \eta \mathcal{R}^t_g) J_+) + \xi^2 [(1 - \eta \mathcal{R}^t_g) J_+,(1 - \eta \mathcal{R}_g) J_-]
\\
& = \xi (\partial_+ J_- - \partial_- J_+)
- \xi^2 \eta (\mathcal{R}_g + \mathcal{R}_g^t)[J_+,J_-]
\\ & \qquad
- \xi^2\eta^2([\mathcal{R}_g^t J_+,\mathcal{R}_g J_-] - \mathcal{R}_g^t[J_+,\mathcal{R}_g J_-] - \mathcal{R}_g [\mathcal{R}_g^t J_+,J_-] ) + \xi^2 [J_+,J_-] ~,
\end{split}\end{equation}
where we have used that the current $J_\pm$ is conserved.
It therefore follows that $J_\pm$ is also flat on-shell if the operator $\mathcal{R}$ satisfies
\begin{equation}\label{eq:eqts}
\xi \eta (\mathcal{R} + \mathcal{R}^t)[X,Y]
+ \xi\eta^2([\mathcal{R}^t X,\mathcal{R} Y] - \mathcal{R}^t[X,\mathcal{R} Y] - \mathcal{R} [\mathcal{R}^t X,Y] ) + (1-\xi) [X,Y] = 0 ~, \qquad X,Y \in \alg{g} ~.
\end{equation}
It is possible to show with some simple manipulations that $\mathcal{R} + \mathcal{R}^t$, that is the symmetric part of $\mathcal{R}$, must be proportional to the identity operator.
\unskip\footnote{To show this we first note that eq.~\eqref{eq:eqts} is equivalent to
\begin{equation*}
\tr\big(Z(\xi \eta (\mathcal{R} + \mathcal{R}^t)[X,Y]
+ \xi\eta^2([\mathcal{R}^t X,\mathcal{R} Y] - \mathcal{R}^t[X,\mathcal{R} Y] - \mathcal{R} [\mathcal{R}^t X,Y] ) + (1-\xi) [X,Y])\big) = 0 ~, \qquad X,Y,Z \in \alg{g} ~.
\end{equation*}
Using the invariance of the bilinear form this can be rewritten as
\begin{equation*}
\tr\big((\xi\eta[Y,(\mathcal{R}+\mathcal{R}^t)] + \xi\eta^2(\mathcal{R}[\mathcal{R}Y,Z] - [\mathcal{R}Y,\mathcal{R}Z]
-\mathcal{R}[Y,\mathcal{R}^tZ]) + (1-\xi)[Y,Z])X\big) = 0 ~, \qquad X,Y,Z \in \alg{g} ~.
\end{equation*}
Adding this equation to itself with $Y$ and $Z$ interchanged we arrive at
\begin{equation*}
\xi\eta(1-\eta \mathcal{R})([Y,(\mathcal{R}+\mathcal{R}^t)Z] - [(\mathcal{R}+\mathcal{R}^t)Y,Z]) = 0 ~,
\qquad Y,Z \in \alg{g} ~.
\end{equation*}
Given that we assume $1-\eta\mathcal{R}$ is invertible and that $\alg{g}$ is a simple Lie algebra, it follows from Schur's lemma that $\mathcal{R} + \mathcal{R}^t$ is proportional to the identity operator.}
Redefining $\hay$ and $\eta$ in \eqref{eq:ybdefgen} we can always remove a term proportional to the identity in $\mathcal{R}$.
Combining these two facts, we take the symmetric part of $\mathcal{R}$ to vanish without loss of generality.
We are then left with the following two equations for $\mathcal{R}$
\begin{equation}\begin{split}\label{eq:ybe}
& [\mathcal{R} X,\mathcal{R} Y] - \mathcal{R}[X,\mathcal{R} Y] - \mathcal{R} [\mathcal{R} X,Y] + c^2 [X,Y] = 0 ~,
\qquad \tr\big(X\mathcal{R}Y\big) + \tr\big((\mathcal{R}X)Y\big) = 0 ~, \qquad X,Y \in \alg{g} ~,
\end{split}\end{equation}
where we have set
\begin{equation}
\xi = \frac{1}{1-c^2\eta^2} ~.
\end{equation}
The first of these equations is known as the (modified) classical Yang-Baxter equation ((m)cYBe), where modified refers to the case $c \neq 0$.
That the operator $\mathcal{R}$ is required to satisfy this equation is the reason why these models are named YB deformations.

To summarise, we have found that the action~\eqref{eq:ybdefgen} defines an integrable deformation of the PCM for simple Lie group $\grp{G}$ when the constant linear operator $\mathcal{R}$ is an antisymmetric solution to the (m)cYBe~\eqref{eq:ybe}.
In particular, the current
\begin{equation}
J_\pm = \frac{1-c^2\eta^2}{1\pm\eta \mathcal{R}_g} j_\pm ~,
\end{equation}
is both conserved and flat on-shell.
Therefore, the Lax connection is given by
\begin{equation}
\Lax_\pm = \frac{J_\pm}{1\mp z} ~.
\end{equation}
Given that $\mathcal{R}$ always appears with the real deformation parameter $\eta$, we have the freedom to rescale $\mathcal{R}$ by a real number.
Noting that we have $c^2 \in \Real$, we can use this freedom to fix $c$ to take one of the following three values:
\begin{itemize}
\item $c = 0$: this is known as the homogeneous case and $\mathcal{R}$ satisfies the cYBe.
\item $c = i$: this is known as the non-split inhomogeneous case and $\mathcal{R}$ satisfies the non-split mcYBe;
\item $c = 1$: this is known as the split inhomogeneous case and $\mathcal{R}$ satisfies the split mcYBe.
\end{itemize}

Before we explore each of these three cases in more detail, let us relate the operator form of the (m)cYBe~\eqref{eq:ybe} to its more familiar matrix form.
Any antisymmetric constant linear operator $\mathcal{R}$ on $\alg{g}$ can be defined in terms of a matrix $r \in \alg{g} \wedge \alg{g}$
\begin{equation}\begin{split}
r + P(r) = 0 ~, \qquad
& \mathcal{R} X = \tr_{\mathbf{2}}(r (1\otimes X)) ~, \qquad X\in \alg{g} ~,
\end{split}\end{equation}
known as the r-matrix, where the subscript indicates that the trace is taken over the second entry in the tensor product and $P$ permutes the entries in the tensor product.
In terms of the r-matrix, the (m)cYBe is given by the following equation in $\alg{g} \wedge \alg{g} \wedge \alg{g}$
\begin{equation}\label{eq:ybemat}
[r_{\mathbf{12}},r_{\mathbf{13}}] + [r_{\mathbf{12}},r_{\mathbf{23}}] + [r_{\mathbf{13}},r_{\mathbf{23}}] + c^2 \Omega = 0 ~,
\end{equation}
where $r_{\mathbf{12}} = r \otimes 1$, $r_{\mathbf{13}} = P_{\mathbf{23}} (r_{\mathbf{12}})$, $r_{\mathbf{23}} = P_{\mathbf{12}}(r_{\mathbf{13}})$ and $\Omega$ is the canonical invariant 3-form in $\alg{g} \wedge \alg{g} \wedge \alg{g}$.
\unskip\footnote{Let us introduce generators $T_a$, $a=1,\dots,\dim\alg{g}$, of the Lie algebra $\alg{g}$ with $[T_a,T_b] = f_{ab}{}^c T_c$, $\tr\big(T_aT_b\big) = \kappa_{ab}$ and $\kappa^{ab}\kappa_{bc} = \delta^a_c$.
Using $\kappa_{ab}$ and its inverse $\kappa^{ab}$ to lower and raise indices, we have $\Omega = f^{abc} T_a \otimes T_b \otimes T_c$.
$\Omega$ is valued in $\alg{g} \wedge \alg{g} \wedge \alg{g}$ since the structure constants with all indices raised are totally antisymmetric.}
Here $P_{\mathbf{ij}}$ permutes entries $i$ and $j$ in the tensor product.
Acting with eq.~\eqref{eq:ybemat} on $1\otimes X \otimes Y$ and taking the trace over the second and third entries in the tensor product we find the (m)cYBe in its operator form~\eqref{eq:ybe}.

\subsection{Homogeneous YB deformations}

Let us start by analysing the case of homogeneous YB deformations.
In this case the antisymmetric operator $\mathcal{R}$ satisfies the cYBe
\begin{equation}\begin{split}\label{eq:cybe}
& [\mathcal{R} X,\mathcal{R} Y] - \mathcal{R}[X,\mathcal{R} Y] - \mathcal{R} [\mathcal{R} X,Y] = 0 ~,
\qquad \tr\big(X\mathcal{R}Y\big) + \tr\big((\mathcal{R}X)Y\big) = 0 ~, \qquad X,Y \in \alg{g} ~.
\end{split}\end{equation}
When discussing solutions of the cYBe, it is often more convenient to work in terms of the r-matrix, which satisfies
\begin{equation}\label{eq:cybemat}
[r_{\mathbf{12}},r_{\mathbf{13}}] + [r_{\mathbf{12}},r_{\mathbf{23}}] + [r_{\mathbf{13}},r_{\mathbf{23}}] = 0 ~,
\qquad
r + P(r) = 0 ~.
\end{equation}
We also introduce generators $T_a$, $a=1,\dots,\dim\alg{g}$, of the Lie algebra $\alg{g}$.

\paragraph{Abelian r-matrices and TsT transformations.}
An important class of solutions of the cYBe~\eqref{eq:cybemat} are the abelian r-matrices, which take the form
\begin{equation}
r = T_1 \wedge T_2 ~, \qquad [T_1,T_2] = 0 ~.
\end{equation}
Given that $T_1$ and $T_2$ commute it immediately follows that such r-matrices satisfy~\eqref{eq:cybemat}.
To explore the resulting deformed models, let us assume that
\begin{equation}\label{eq:simplifying}
\tr\big(T_aT_b\big) = - \delta_{ab} ~, \qquad a = 1,\dots,\dim\alg{g} ~,
\end{equation}
so that
\begin{equation}
\mathcal{R} T_{\bar{a}} = \epsilon_{\bar{a}}{}^{\bar{b}} T_{\bar{b}} ~, \qquad \mathcal{R} T_{\hat{a}} = 0 ~,
\end{equation}
where the index $\bar{a} = 1,2$ and the index $\hat{a}$ runs over the remaining generators of the Lie algebra $\alg{g}$.
Under these assumptions we have that $\mathcal{R}^3 = - \mathcal{R}$ and therefore
\begin{equation}\label{eq:rex}
\frac{1}{1-\eta \mathcal{R}} = (1 + \mathcal{R}^2) + \frac{\eta}{1+\eta^2} \mathcal{R} - \frac{1}{1+\eta^2} \mathcal{R}^2 ~,
\end{equation}
where
\begin{equation}\begin{aligned}
\mathcal{R}^2 T_{\bar{a}} & = - T_{\bar{a}} ~, \qquad
& (1+\mathcal{R}^2) T_{\bar{a}} & = 0 ~,
\\
\mathcal{R}^2 T_{\hat{a}} & = 0 ~, \qquad
& (1+\mathcal{R}^2) T_{\hat{a}} & = T_{\hat{a}} ~.
\end{aligned}\end{equation}
Now writing
\begin{equation}
k_\pm = - \partial_\pm g g^{-1} = k_\pm^{\bar{a}} T_{\bar{a}} + k_\pm^\perp ~,
\qquad k_\pm^\perp = k_\pm^{\hat{a}} T_{\hat{a}} ~,
\end{equation}
the action of the YB deformation~\eqref{eq:ybdefgen} for such abelian r-matrices becomes
\begin{equation}\label{eq:ybabe}
\Act_{\indrm{YB}} = \frac{\hay}{2} \int d^2 x \, \big( - \tr\big(k_+^\perp k_-^\perp\big) + \frac{1}{1+\eta^2} \delta_{\bar{a}\bar{b}} k_+^{\bar{a}} k_-^{\bar{b}} - \frac{\eta}{1+\eta^2} \epsilon_{\bar{a}\bar{b}} k_+^{\bar{a}} k_-^{\bar{b}} \big) ~.
\end{equation}
It turns out that this is nothing but the well-known TsT transformation of the PCM.
In a model with two commuting isometries, we can always choose coordinates on the target space such that these isometries are realised as shift symmetries
\begin{equation}
\Theta^{\bar{a}} \to \Theta^{\bar{a}} + \gamma^{\bar{a}} ~.
\end{equation}
A TsT transformation then amounts to first T-dualising $\Theta^1 \to \tilde\Theta_1$, shifting $\Theta^2 \to \Theta^2 - \eta \tilde\Theta_1$ and finally T-dualising back $\tilde\Theta_1 \to \Theta^1$.
Given that T-dualising twice is the identity map, for $\eta = 0$ the TsT transformation has no effect, hence for $\eta \neq 0$ it defines a deformation of the original model.
Since T-duality is a canonical transformation on phase space, it preserves the integrability of a model, in agreement with the fact that the YB deformations are integrable deformations of the PCM.
\iflectures
This result actually holds in full generality.
That is, any YB deformation constructed from an abelian r-matrix is equivalent to a TsT transformation.
This is our first encounter of the close relationship between integrability, deformations and dualities.

\question{Using the same simplifying assumptions as in eq.~\eqref{eq:simplifying} and parametrising
\begin{equation*}
g = e^{-\Theta} \tilde{g} ~, \qquad \Theta = \Theta^{\bar{a}} T_{\bar{a}} ~,
\end{equation*}
show that the TsT transformation of the PCM in $\Theta^1$ and $\Theta^2$ can be written in the form of the YB deformation~\eqref{eq:ybdefgen}.
You should first T-dualise $\Theta^1 \to \tilde\Theta_1$, then shift $\Theta^2 \to \Theta^2 - \eta \tilde\Theta_1$ and finally T-dualise $\tilde\Theta_1 \to \Theta^1$.
}

\answer{Parametrising $g = e^{-\Theta} \tilde{g}$ we find
\begin{equation*}
k_\pm = - \partial_\pm g g^{-1} = e^{-\Theta}( \tilde{k}_\pm + \partial_\pm \Theta ) e^{\Theta} ~, \qquad \tilde{k}_\pm = - \partial_\pm \tilde{g} \tilde{g}^{-1}  ~,
\end{equation*}
where we have used that $\Theta$ commutes with itself and its derivatives.
Let us also denote
\begin{equation*}
\tilde{k}_\pm = \tilde{k}_\pm^{\bar{a}} T_{\bar{a}} + \tilde{k}_\pm^\perp ~,
\qquad \tilde{k}_\pm^\perp = \tilde{k}_\pm^{\hat{a}} T_{\hat{a}} ~.
\end{equation*}
Substituting into the action of the PCM~\eqref{eq:pcmact} gives
\begin{equation*}
\Act_{\indrm{PCM}} = \frac{\hay}{2} \int d^2x \, \big( - \tr\big( \tilde{k}_+^\perp \tilde{k}_-^\perp\big)
+ (\tilde{k}_+^1 + \partial_+ \Theta^1)( \tilde{k}_-^1 + \partial_- \Theta^1)
+ (\tilde{k}_+^2 + \partial_+ \Theta^2)( \tilde{k}_-^2 + \partial_- \Theta^2) \big) ~.
\end{equation*}
To T-dualise in $\Theta^1$ we first gauge the shift symmetry $\Theta^1 \to \Theta^1 + \gamma^1$, add a Lagrange multiplier $\tilde \Theta_1$ enforcing that the field strength of the abelian gauge field is zero and gauge fix $\Theta^1 = 0$
\begin{equation*}\begin{split}
\Act_{\indrm{TD-INT-PCM}} = \frac{\hay}{2} \int d^2x \, \big( & - \tr\big( \tilde{k}_+^\perp \tilde{k}_-^\perp\big)
+ (\tilde{k}_+^1 + A_+)( \tilde{k}_-^1 + A_-)
\\ & + (\tilde{k}_+^2 + \partial_+ \Theta^2)( \tilde{k}_-^2 + \partial_- \Theta^2) + \tilde\Theta_1 (\partial_+ A_- - \partial_- A_+) \big) ~.
\end{split}\end{equation*}
The T-dual model is then given by integrating out the gauge field $A_\pm$
\begin{equation*}
\Act_{\indrm{TD-PCM}} = \frac{\hay}{2} \int d^2x \, \big( - \tr\big( \tilde{k}_+^\perp \tilde{k}_-^\perp\big)
+ \partial_+\tilde\Theta_1 \partial_-\tilde\Theta_1
- \tilde{k}_+^1 \partial_- \tilde\Theta_1
+ \partial_+ \tilde\Theta_1 \tilde k_-^1
+ (\tilde{k}_+^2 + \partial_+ \Theta^2)( \tilde{k}_-^2 + \partial_- \Theta^2) \big) ~.
\end{equation*}
We next shift $\Theta^2 \to \Theta^2 - \eta \tilde\Theta_1$ in the T-dual model
\begin{equation*}\begin{split}
\Act_{\indrm{TD-PCM}} & = \frac{\hay}{2} \int d^2x \, \big( - \tr\big( \tilde{k}_+^\perp \tilde{k}_-^\perp\big)
+ \partial_+\tilde\Theta_1 \partial_-\tilde\Theta_1
- \tilde{k}_+^1 \partial_- \tilde\Theta_1
+ \partial_+ \tilde\Theta_1 \tilde k_-^1
\\ & \hphantom{ = \frac{\hay}{2} \int d^2x \, \big( ~ }
+ (\tilde{k}_+^2 + \partial_+ \Theta^2 - \eta \partial_+ \tilde\Theta_1)( \tilde{k}_-^2 + \partial_- \Theta^2 - \eta \partial_- \tilde \Theta_1) \big) ~,
\end{split}\end{equation*}
and T-dualise in $\tilde\Theta_1$ to find the TsT transformation of the PCM
\begin{equation*}\begin{split}
\Act_{\indrm{TsT-PCM}} & = \frac{\hay}{2} \int d^2x \, \big( - \tr\big( \tilde{k}_+^\perp \tilde{k}_-^\perp\big)
+\frac{1}{1+\eta^2}\delta_{\bar{a}\bar{b}} (\tilde k_+^{\bar{a}} + \partial_+\Theta^{\bar{a}})(\tilde k_-^{\bar{b}} + \partial_-\Theta^{\bar{b}})
\\ & \hphantom{ = \frac{\hay}{2} \int d^2x \, \big( - \tr\big( \tilde{k}_+^\perp \tilde{k}_-^\perp\big) ~ }
- \frac{\eta}{1+\eta^2} \epsilon_{\bar{a}\bar{b}} (\tilde k_+^{\bar{a}} + \partial_+\Theta^{\bar{a}})(\tilde k_-^{\bar{b}} + \partial_-\Theta^{\bar{b}})\big)
\\ & = - \frac{\hay}{2} \int d^2x \, \tr\big((\tilde k_+ + \partial_+ \Theta) \frac{1}{1-\eta\mathcal{R}} (\tilde k_- + \partial_- \Theta)\big) ~,
\end{split}\end{equation*}
where we have used the identity~\eqref{eq:rex}.
Finally, using that
\begin{equation*}
\Ad_{e^{-\Theta}} \mathcal{R} \Ad_{e^\Theta} = \mathcal{R} ~,
\end{equation*}
that is $T_1$ and $T_2$ are symmetries of the operator $\mathcal{R}$, which can be seen from the equivalent form
\begin{equation*}
(\Ad_{e^{-\Theta}}\otimes\Ad_{e^{-\Theta}})r = r ~,
\end{equation*}
we can replace $\tilde k_\pm + \partial_\pm \Theta$ by $e^{-\Theta}(\tilde k_\pm + \partial_\pm \Theta) e^{\Theta} = k_\pm$, and write the TsT transformation of the PCM in the form of the YB deformation
\begin{equation*}
\Act_{\indrm{TsT-PCM}} = - \frac{\hay}{2} \int d^2x \, \tr\big(k_+ \frac{1}{1-\eta\mathcal{R}} k_-\big) =
- \frac{\hay}{2} \int d^2x \, \tr\big(j_+ \frac{1}{1-\eta\mathcal{R}_g} j_-\big) = \Act_{\indrm{YB}} ~.
\end{equation*}
}
\else

To see the relation to TsT transformations, we use the same simplifying assumptions as in eq.~\eqref{eq:simplifying} and parametrise
\begin{equation}
g = e^{-\Theta} \tilde{g} ~, \qquad \Theta = \Theta^{\bar{a}} T_{\bar{a}} ~.
\end{equation}
It follows that
\begin{equation}
k_\pm = - \partial_\pm g g^{-1} = e^{-\Theta}( \tilde{k}_\pm + \partial_\pm \Theta ) e^{\Theta} ~, \qquad \tilde{k}_\pm = - \partial_\pm \tilde{g} \tilde{g}^{-1}  ~,
\end{equation}
where we have used that $\Theta$ commutes with itself and its derivatives.
Let us also denote
\begin{equation}
\tilde{k}_\pm = \tilde{k}_\pm^{\bar{a}} T_{\bar{a}} + \tilde{k}_\pm^\perp ~,
\qquad \tilde{k}_\pm^\perp = \tilde{k}_\pm^{\hat{a}} T_{\hat{a}} ~.
\end{equation}
Substituting into the action of the PCM~\eqref{eq:pcmact} gives
\begin{equation}
\Act_{\indrm{PCM}} = \frac{\hay}{2} \int d^2x \, \big( - \tr\big( \tilde{k}_+^\perp \tilde{k}_-^\perp\big)
+ (\tilde{k}_+^1 + \partial_+ \Theta^1)( \tilde{k}_-^1 + \partial_- \Theta^1)
+ (\tilde{k}_+^2 + \partial_+ \Theta^2)( \tilde{k}_-^2 + \partial_- \Theta^2) \big) ~.
\end{equation}
To T-dualise in $\Theta^1$ we first gauge the shift symmetry $\Theta^1 \to \Theta^1 + \gamma^1$, add a Lagrange multiplier $\tilde \Theta_1$ enforcing that the field strength of the abelian gauge field is zero, and gauge fix $\Theta^1 = 0$
\begin{equation}\begin{split}
\Act_{\indrm{TD-INT-PCM}} = \frac{\hay}{2} \int d^2x \, \big( & - \tr\big( \tilde{k}_+^\perp \tilde{k}_-^\perp\big)
+ (\tilde{k}_+^1 + A_+)( \tilde{k}_-^1 + A_-)
\\ & + (\tilde{k}_+^2 + \partial_+ \Theta^2)( \tilde{k}_-^2 + \partial_- \Theta^2) + \tilde\Theta_1 (\partial_+ A_- - \partial_- A_+) \big) ~.
\end{split}\end{equation}
The T-dual model is then given by integrating out the gauge field $A_\pm$
\begin{equation}
\Act_{\indrm{TD-PCM}} = \frac{\hay}{2} \int d^2x \, \big( - \tr\big( \tilde{k}_+^\perp \tilde{k}_-^\perp\big)
+ \partial_+\tilde\Theta_1 \partial_-\tilde\Theta_1
- \tilde{k}_+^1 \partial_- \tilde\Theta_1
+ \partial_+ \tilde\Theta_1 \tilde k_-^1
+ (\tilde{k}_+^2 + \partial_+ \Theta^2)( \tilde{k}_-^2 + \partial_- \Theta^2) \big) ~.
\end{equation}
We next shift $\Theta^2 \to \Theta^2 - \eta \tilde\Theta_1$ in the T-dual model
\begin{equation}\begin{split}
\Act_{\indrm{TD-PCM}} & = \frac{\hay}{2} \int d^2x \, \big( - \tr\big( \tilde{k}_+^\perp \tilde{k}_-^\perp\big)
+ \partial_+\tilde\Theta_1 \partial_-\tilde\Theta_1
- \tilde{k}_+^1 \partial_- \tilde\Theta_1
+ \partial_+ \tilde\Theta_1 \tilde k_-^1
\\ & \hphantom{= \frac{\hay}{2} \int d^2x \, \big( ~ }
+ (\tilde{k}_+^2 + \partial_+ \Theta^2 - \eta \partial_+ \tilde\Theta_1)( \tilde{k}_-^2 + \partial_- \Theta^2 - \eta \partial_- \tilde \Theta_1) \big) ~,
\end{split}\end{equation}
and T-dualise in $\tilde\Theta_1$ to find the TsT transformation of the PCM
\begin{equation}\begin{split}
\Act_{\indrm{TsT-PCM}} & = \frac{\hay}{2} \int d^2x \, \big( - \tr\big( \tilde{k}_+^\perp \tilde{k}_-^\perp\big)
+\frac{1}{1+\eta^2}\delta_{\bar{a}\bar{b}} (\tilde k_+^{\bar{a}} + \partial_+\Theta^{\bar{a}})(\tilde k_-^{\bar{b}} + \partial_-\Theta^{\bar{b}})
\\ & \hphantom{= \frac{\hay}{2} \int d^2x \, \big( - \tr\big( \tilde{k}_+^\perp \tilde{k}_-^\perp\big) ~ }
- \frac{\eta}{1+\eta^2} \epsilon_{\bar{a}\bar{b}} (\tilde k_+^{\bar{a}} + \partial_+\Theta^{\bar{a}})(\tilde k_-^{\bar{b}} + \partial_-\Theta^{\bar{b}})\big)
\\ & = - \frac{\hay}{2} \int d^2x \, \tr\big((\tilde k_+ + \partial_+ \Theta) \frac{1}{1-\eta\mathcal{R}} (\tilde k_- + \partial_- \Theta)\big) ~,
\end{split}\end{equation}
where we have used the identity~\eqref{eq:rex}.
Finally, using that
\begin{equation*}
\Ad_{e^{-\Theta}} \mathcal{R} \Ad_{e^\Theta} = \mathcal{R} ~,
\end{equation*}
that is $T_1$ and $T_2$ are symmetries of the operator $\mathcal{R}$, which can be seen from the equivalent form
\begin{equation*}
(\Ad_{e^{-\Theta}}\otimes\Ad_{e^{-\Theta}})r = r ~,
\end{equation*}
we can replace $\tilde k_\pm + \partial_\pm \Theta$ by $e^{-\Theta}(\tilde k_\pm + \partial_\pm \Theta) e^{\Theta} = k_\pm$.
Therefore, we indeed find that the TsT transformation of the PCM takes the form of the YB deformation constructed from an abelian r-matrix
\begin{equation}
\Act_{\indrm{TsT-PCM}} = - \frac{\hay}{2} \int d^2x \, \tr\big(k_+ \frac{1}{1-\eta\mathcal{R}} k_-\big) =
- \frac{\hay}{2} \int d^2x \, \tr\big(j_+ \frac{1}{1-\eta\mathcal{R}_g} j_-\big) = \Act_{\indrm{YB}} ~.
\end{equation}
This result holds in full generality, without the simplifying assumptions in eq.~\eqref{eq:simplifying}.
That is, any YB deformation constructed from an abelian r-matrix is equivalent to a TsT transformation.
This is our first encounter of the close relationship between integrability, deformations and dualities.
\fi

\paragraph{General r-matrices and non-abelian T-duality.}
The classification of homogeneous r-matrices for a given Lie algebra $\alg{g}$ can be a rather involved exercise.
We can make progress by noting that the image of the operator $\mathcal{R}$ is a subalgebra of $\alg{g}$.
We denote $\alg{h} = \im \mathcal{R}$.
This can be easily seen by writing the cYBe~\eqref{eq:cybe} as
\begin{equation}
[\mathcal{R}X,\mathcal{R}Y] = \mathcal{R}([X,\mathcal{R}Y] + [\mathcal{R}X,Y]) ~, \qquad X,Y \in \alg{g} ~,
\end{equation}
that is, the commutator of two elements of $\alg{h}$ is in the image of $\mathcal{R}$, hence is also an element of $\alg{h}$.
It follows that the r-matrix is valued in $\alg{h} \wedge \alg{h}$.
Moreover, writing
\begin{equation}\label{eq:rmat}
r = r^{\bar{a}\bar{b}} T_{\bar{a}}\wedge T_{\bar{b}} ~, \qquad r^{\bar{a}\bar{b}} = - r^{\bar{b}\bar{a}} ~,
\end{equation}
where $T_{\bar{a}}$ are generators of $\alg{h}$, $r^{\bar{a}\bar{b}}$ is invertible as a $\dim \alg{h} \times \dim \alg{h}$ matrix.
Denoting the inverse by $\omega_{\bar{a}\bar{b}}$, we can use it to define a 2-cochain on $\alg{h}$
\begin{equation}\label{eq:2cocycle}
\omega\big(T_{\bar{a}},T_{\bar{b}}\big) = \omega_{\bar{a}\bar{b}} ~, \qquad \omega_{\bar{a}\bar{b}} = - \omega_{\bar{b}\bar{a}} ~.
\end{equation}
An r-matrix solving the cYBe~\eqref{eq:cybemat} is then equivalent to $\omega$ being a 2-cocycle, that is a closed 2-cochain
\begin{equation}
\omega\big(X,[Y,Z]\big) + \omega\big(Y,[Z,X]\big) + \omega\big(Z,[X,Y]\big) = 0~, \qquad X,Y,Z \in \alg{h} ~.
\end{equation}
\iflectures
It follows that $\alg{h}$ is a quasi-Frobenius subalgebra of $\alg{g}$, that is a subalgebra equipped with a non-degenerate 2-cocycle $\omega$.
When $\omega$ is also exact
\begin{equation}
\omega\big(X,Y\big) = \tilde\omega\big([X,Y]\big) ~, \qquad X,Y \in \alg{h} ~,
\end{equation}
the subalgebra $\alg{h}$ is Frobenius.
Therefore, the classification of homogeneous r-matrices for a given Lie algebra $\alg{g}$ is equivalent to the classification of quasi-Frobenius subalgebras and non-degenerate 2-cocycles.

\question{Show that if the r-matrix~\eqref{eq:rmat} solves the cYBe~\eqref{eq:cybemat}, then the 2-cochain~\eqref{eq:2cocycle} is closed, where $\omega_{\bar{a}\bar{b}}$ is the inverse of $r^{\bar{a}\bar{b}}$.
}

\answer{For the r-matrix~\eqref{eq:rmat}, the cYBe~\eqref{eq:cybemat} can be written as
\begin{equation*}
r^{\bar{a}\bar{b}}r^{\bar{c}\bar{d}}\big([T_{\bar{a}},T_{\bar{c}}] \wedge T_{\bar{b}} \wedge T_{\bar{d}}
+ T_{\bar{a}}\wedge [T_{\bar{b}},T_{\bar{c}}] \wedge T_{\bar{d}}
+ T_{\bar{a}}\wedge T_{\bar{c}} \wedge [T_{\bar{b}},T_{\bar{d}}]\big) = 0 ~,
\end{equation*}
which implies
\begin{equation*}
\big(f_{\bar{d}\bar{e}}{}^{\bar{a}} r^{\bar{b}\bar{d}}r^{\bar{c}\bar{e}}
+f_{\bar{d}\bar{e}}{}^{\bar{b}}  r^{\bar{c}\bar{d}} r^{\bar{a}\bar{e}}
+f_{\bar{d}\bar{e}}{}^{\bar{c}} r^{\bar{a}\bar{d}}r^{\bar{b}\bar{e}}\big)
T_{\bar{a}}\wedge T_{\bar{b}} \wedge T_{\bar{c}} = 0 ~,
\end{equation*}
or equivalently
\begin{equation*}
f_{\bar{d}\bar{e}}{}^{\bar{a}} r^{\bar{b}\bar{d}}r^{\bar{c}\bar{e}}
+f_{\bar{d}\bar{e}}{}^{\bar{b}}  r^{\bar{c}\bar{d}} r^{\bar{a}\bar{e}}
+f_{\bar{d}\bar{e}}{}^{\bar{c}} r^{\bar{a}\bar{d}}r^{\bar{b}\bar{e}} = 0 ~.
\end{equation*}
Contracting this identity with $\omega_{\bar{f}\bar{a}}\omega_{\bar{g}\bar{b}}\omega_{\bar{h}\bar{c}}$ and using that $\omega_{\bar{f}\bar{a}}r^{\bar{a}\bar{d}} = \delta_{\bar{f}}^{\bar{d}}$ we find
\begin{equation*}
\omega_{\bar{f}\bar{a}}f_{\bar{g}\bar{h}}{}^{\bar{a}}
+\omega_{\bar{g}\bar{b}}f_{\bar{h}\bar{f}}{}^{\bar{b}}
+\omega_{\bar{h}\bar{c}}f_{\bar{f}\bar{g}}{}^{\bar{c}}= 0 ~.
\end{equation*}
It is then straightforward to see that this is equivalent to the closure of the 2-cochain~\eqref{eq:2cocycle}
\begin{equation*}
\omega\big(T_{\bar{f}},[T_{\bar{g}},T_{\bar{h}}]\big)+
\omega\big(T_{\bar{g}},[T_{\bar{h}},T_{\bar{f}}]\big)+
\omega\big(T_{\bar{h}},[T_{\bar{f}},T_{\bar{g}}]\big) = 0 ~.
\end{equation*}
}
\else
To see this explicitly, we note that for the r-matrix~\eqref{eq:rmat} the cYBe~\eqref{eq:cybemat} can be written as
\begin{equation}
r^{\bar{a}\bar{b}}r^{\bar{c}\bar{d}}\big([T_{\bar{a}},T_{\bar{c}}] \wedge T_{\bar{b}} \wedge T_{\bar{d}}
+ T_{\bar{a}}\wedge [T_{\bar{b}},T_{\bar{c}}] \wedge T_{\bar{d}}
+ T_{\bar{a}}\wedge T_{\bar{c}} \wedge [T_{\bar{b}},T_{\bar{d}}]\big) = 0 ~,
\end{equation}
which implies
\begin{equation}
\big(f_{\bar{d}\bar{e}}{}^{\bar{a}} r^{\bar{b}\bar{d}}r^{\bar{c}\bar{e}}
+f_{\bar{d}\bar{e}}{}^{\bar{b}}  r^{\bar{c}\bar{d}} r^{\bar{a}\bar{e}}
+f_{\bar{d}\bar{e}}{}^{\bar{c}} r^{\bar{a}\bar{d}}r^{\bar{b}\bar{e}}\big)
T_{\bar{a}}\wedge T_{\bar{b}} \wedge T_{\bar{c}} = 0 ~,
\end{equation}
or equivalently
\begin{equation}
f_{\bar{d}\bar{e}}{}^{\bar{a}} r^{\bar{b}\bar{d}}r^{\bar{c}\bar{e}}
+f_{\bar{d}\bar{e}}{}^{\bar{b}}  r^{\bar{c}\bar{d}} r^{\bar{a}\bar{e}}
+f_{\bar{d}\bar{e}}{}^{\bar{c}} r^{\bar{a}\bar{d}}r^{\bar{b}\bar{e}} = 0 ~.
\end{equation}
Contracting this identity with $\omega_{\bar{f}\bar{a}}\omega_{\bar{g}\bar{b}}\omega_{\bar{h}\bar{c}}$ and using that $\omega_{\bar{f}\bar{a}}r^{\bar{a}\bar{d}} = \delta_{\bar{f}}^{\bar{d}}$ we find
\begin{equation}
\omega_{\bar{f}\bar{a}}f_{\bar{g}\bar{h}}{}^{\bar{a}}
+\omega_{\bar{g}\bar{b}}f_{\bar{h}\bar{f}}{}^{\bar{b}}
+\omega_{\bar{h}\bar{c}}f_{\bar{f}\bar{g}}{}^{\bar{c}}= 0 ~.
\end{equation}
It is then straightforward to see that this is equivalent to the closure of the 2-cochain~\eqref{eq:2cocycle}
\begin{equation}
\omega\big(T_{\bar{f}},[T_{\bar{g}},T_{\bar{h}}]\big)+
\omega\big(T_{\bar{g}},[T_{\bar{h}},T_{\bar{f}}]\big)+
\omega\big(T_{\bar{h}},[T_{\bar{f}},T_{\bar{g}}]\big) = 0 ~.
\end{equation}

It follows that $\alg{h}$ is a quasi-Frobenius subalgebra of $\alg{g}$, that is a subalgebra equipped with a non-degenerate 2-cocycle $\omega$.
When $\omega$ is also exact
\begin{equation}
\omega\big(X,Y\big) = \tilde\omega\big([X,Y]\big) ~, \qquad X,Y \in \alg{h} ~,
\end{equation}
the subalgebra $\alg{h}$ is Frobenius.
Therefore, the classification of homogeneous r-matrices for a given Lie algebra $\alg{g}$ is equivalent to the classification of quasi-Frobenius subalgebras and non-degenerate 2-cocycles.
\fi

It turns out that all homogeneous YB deformations can be related to duality transformations.
As an example of this we consider the following simplified setup.
We assume that we have a basis of generators $T_a$, $a=1,\dots,\dim\alg{g}$, of $\alg{g}$ that we split into $T_{\bar{a}}$, $\bar{a} = 1,\dots,\dim\alg{h}$, which are generators of $\alg{h}$, and $T_{\hat{a}}$, $\hat{a} = \dim\alg{h}+1,\dots,\dim\alg{g}$, such that $\kappa_{ab} = \tr\big(T_aT_b\big)$ is block diagonal, that is $\kappa_{\bar{a}\hat{b}} = \kappa_{\hat{a}\bar{b}} = 0$.
In particular, this means that $\kappa_{\bar{a}\bar{b}} = \tr\big(T_{\bar{a}}T_{\bar{b}}\big)$ is invertible, with the inverse denoted by $\kappa^{\bar{a}\bar{b}}$, and we can use the invariant bilinear form to identify the dual of $\alg{h}$ with itself.
While this is quite a severe restriction, the following construction can be generalised to any homogeneous YB deformation.

For the r-matrix~\eqref{eq:rmat}, we find that in this setup
\begin{equation}
\mathcal{R} T_{\bar{a}} = \mathcal{R}^{\bar{b}}{}_{\bar{a}} T_{\bar{b}} = 2 r^{\bar{b}\bar{c}} \kappa_{\bar{c}\bar{a}} T_{\bar{b}} ~, \qquad \mathcal{R} T_{\hat{a}} = 0 ~.
\end{equation}
It follows that the restriction of $\mathcal{R}$ to $\alg{h}$ is invertible.
We can also use the 2-cocycle $\omega$ to define an antisymmetric map $\Omega : \alg{h} \to \alg{h}$ as
\begin{equation}
\omega\big(X,Y\big) = 2\tr\big(X\Omega Y\big) = - 2\tr\big((\Omega X) Y\big) ~, \qquad X,Y \in \alg{h} ~,
\end{equation}
such that
\begin{equation}
\omega_{\bar{a}\bar{b}} = 2 \kappa_{\bar{a}\bar{c}} \Omega^{\bar{c}}{}_{\bar{b}} ~.
\end{equation}
Given that $\omega_{\bar{a}\bar{b}}$ is the inverse of $r^{\bar{a}\bar{b}}$, it follows that $\Omega$ is the inverse of the restriction of $\mathcal{R}$ to $\alg{h}$.
It is also useful to note that in terms of $\Omega$ the closure of the 2-cocycle $\omega$ is
\begin{equation}\label{eq:derivation}
\Omega[X,Y] = [X,\Omega Y] + [\Omega X,Y] ~, \qquad X,Y \in \alg{h} ~,
\end{equation}
that is $\Omega$ acts as a derivation on commutators of $\alg{h}$.

Let us parametrise the group-valued field of the PCM as
\begin{equation}\label{eq:param}
g = h \tilde{g} ~, \qquad h \in \grp{H} ~, \qquad \tilde{g} \in \grp{G} ~,
\end{equation}
where $\Lie\grp{H} = \alg{h}$, such that
\begin{equation}\label{eq:jsub}
j_\pm = g^{-1} \partial_\pm g = \tilde{g}^{-1} (h^{-1} \partial_\pm h - \tilde k_\pm ) \tilde{g} ~, \qquad
\tilde k_\pm = -\partial_\pm \tilde{g} \tilde{g}^{-1} ~.
\end{equation}
Given that $h^{-1} \partial_\pm h \in \alg{h}$, we can consider the following modification of the PCM
\begin{equation}\label{eq:pcmactmod}
\Act_{\indrm{PCM}_\zeta} = -\frac{\hay}{2} \int d^2x \, \tr\big(j_+j_-\big) + \frac{\hay\zeta}{4} \int d^2 x \, \omega\big(h^{-1}\partial_+ h,h^{-1}\partial_- h\big) ~.\end{equation}
Since $\omega$ is a 2-cocycle on $\alg{h}$, hence it is closed, it follows that locally $\omega\big(h^{-1}\partial_+ h, h^{-1}\partial_- h\big)$ is a total derivative.
Therefore, adding this term does not change the equations of motion.
We will call such a term a closed term.

We now claim that, up to a closed term, the non-abelian T-dual of the modified action~\eqref{eq:pcmactmod} gives the YB deformation of the PCM~\eqref{eq:ybdefgen}.
To perform the non-abelian T-duality transformation we start by substituting in for $j_\pm$ using eq.~\eqref{eq:jsub} in the modified action~\eqref{eq:pcmactmod}.
We then gauge the left-acting global $H$ symmetry, $h \to h_\ind{L} h$, gauge-fix $h=1$, and add a Lagrange multiplier $v \in \alg{h}$ enforcing that the gauge field is flat.
After doing this we find
\begin{equation}\begin{split}
\Act_{\indrm{INT}_\zeta} & = - \frac{\hay}{2} \int d^2x \, \tr\big((A_+ - \tilde{k}_+)(A_- - \tilde{k}_-)\big)
+ \frac{\hay\zeta}{4} \int d^2x \, \omega\big(A_+,A_-\big)
\\ & \qquad
- \frac{\hay}{2}\int d^2x \, \tr\big(v(\partial_+ A_- - \partial_- A_+ + [A_+,A_-])\big)
~, \qquad v, A_\pm \in \alg{h} ~.
\end{split}\end{equation}
The non-abelian T-dual model is then given by integrating out the gauge field $A_\pm$
\begin{equation}\label{eq:c1}
\Act_{\indrm{NATD}_\zeta} = -\frac{\hay}{2} \int d^2 x \,
\tr\big((\partial_+ v + P \tilde{k}_+) \frac{1}{1-\zeta \Omega - \ad_v} (\partial_-v - P \tilde{k}_-)
+ \tilde{k}_+\tilde{k}_- \big) ~,
\end{equation}
where $P$ is the orthogonal projector onto $\alg{h}$, that is
\begin{equation}
P T_{\bar{a}} = T_{\bar{a}} ~, \qquad P T_{\hat{a}} = 0 ~.
\end{equation}

In order to show that this is equivalent to the YB deformation up to a closed term, we substitute the parametrisation \eqref{eq:param} into the action~\eqref{eq:ybdefgen} to give
\begin{equation}\label{eq:c2}
\Act_{\indrm{YB}} = - \frac{\hay}{2} \int d^2 x \, \tr \big((h^{-1}\partial_+ h - \tilde{k}_+) \frac{1}{1-\eta \mathcal{R}_h} (h^{-1}\partial_- h - \tilde{k}_-)\big) ~, \qquad \mathcal{R}_h = \Ad_h^{-1} \mathcal{R} \Ad_h ~.
\end{equation}
Comparing the actions~\eqref{eq:c1} and~\eqref{eq:c2} we would like to solve the following equations
\begin{equation}\label{eq:ts1}
1 - P \frac{1}{1-\zeta \Omega - \ad_v} P = \frac{1}{1-\eta \mathcal{R}_h} ~,
\qquad
\pm \frac{1}{1\pm\zeta \Omega \pm \ad_v} \partial_\pm v = \frac{1}{1\pm\eta \mathcal{R}_h} h^{-1}\partial_\pm h ~.
\end{equation}
Rearranging and using that, restricted to $\alg{h}$, $\mathcal{R}^{-1} = \Omega$, we find
\begin{equation}\label{eq:ts2}
\ad_v = \eta^{-1} \Omega_h - \zeta \Omega ~, \qquad
\partial_\pm v = \eta^{-1} \Omega_h(h^{-1} \partial_\pm h) ~, \qquad \Omega_h = \Ad_h^{-1} \Omega \Ad_h ~,
\end{equation}
where the first equation should be understood as an equation on $\alg{h}$.
It is natural to expect that $h=1$ is equivalent to $v=0$, in which case the first equation tells us that $\zeta = \eta^{-1}$.
The equations can then be solved by setting
\begin{equation}
h = e^{\bar v} ~,
\qquad v = \eta^{-1} \frac{1-e^{-ad_{\bar{v}}}}{\ad_{\bar{v}}} \Omega \bar v ~,
\qquad \bar v \in \alg{h} ~.
\end{equation}
This follows from the fact that $\Omega$ acts as a derivation on commutators of $\alg{h}$~\eqref{eq:derivation}.
\iflectures

Finally, having solved eq.~\eqref{eq:ts1}, we observe that
\begin{equation}
\partial_+ v  \frac{1}{1-\zeta \Omega - \ad_v} \partial_- v
= h^{-1}\partial_+ h \frac{1}{1-\eta\mathcal{R}_h} h^{-1}\partial_- h + \eta^{-1} h^{-1}\partial_+ h \Omega_h h^{-1}\partial_- h ~,
\end{equation}
hence
\begin{equation}
\Act_{\indrm{NATD}_\zeta} = \Act_{\indrm{YB}} - \frac{\hay\eta^{-1}}{4} \int d^2x \, \omega\big(\partial_+ h h^{-1},\partial_- h h^{-1}\big) ~.
\end{equation}
Therefore, we indeed find that the non-abelian T-dual of the modified action~\eqref{eq:c1} gives the YB deformation of the PCM up to a closed term.
As mentioned above, this construction can be generalised to any homogeneous YB deformation.
That is, up to a closed term, any homogeneous YB deformation is the non-abelian T-dual of the PCM modified by a closed term.
It is interesting to note that when $\omega$ is exact the closed terms become boundary terms and the parameter $\eta$ can be eliminated by a coordinate redefinition.
Therefore, in these cases the YB deformation is not strictly speaking a deformation.

\question{Setting $\zeta = \eta^{-1}$, show that
\begin{equation*}
h = e^{\bar{v}} ~, \qquad v = \eta^{-1} \frac{1-e^{-ad_{\bar{v}}}}{\ad_{\bar{v}}} \Omega \bar v ~,
\end{equation*}
solves eq.~\eqref{eq:ts2}.
You should assume that $\Omega$ can be extended to a derivation of the universal enveloping algebra of $\alg{h}$.
}

\answer{For any derivation $\delta$ of the universal enveloping algebra of $\alg{h}$
\begin{equation*}
h^{-1} \delta h = \frac{1-e^{-ad_{\bar{v}}}}{\ad_{\bar{v}}} \delta \bar v ~, \qquad
\delta h h^{-1} = \frac{e^{ad_{\bar{v}}}-1}{\ad_{\bar{v}}} \delta \bar v ~,
\end{equation*}
where $h = e^{\bar v}$.
Moreover, if we have two such derivations $\delta_1$ and $\delta_2$ that commute then
\begin{equation*}\begin{split}
& \delta_1(h^{-1}\delta_2 h) - \delta_2(h^{-1}\delta_1h) + [h^{-1}\delta_1 h,h^{-1}\delta_2 h] = 0 ~,
\\ &
\delta_1(\delta_2 hh^{-1}) - \delta_2(\delta_1hh^{-1}) - [h^{-1}\delta_1 h,h^{-1}\delta_2 h] = 0 ~.
\end{split}\end{equation*}

To solve the first equation in eq.~\eqref{eq:ts2}, we observe that
\begin{equation*}
\Ad_h^{-1} \delta \Ad_h - \delta = \ad_{h^{-1}\delta h} ~.
\end{equation*}
Indeed, acting with the left-hand side on $X \in \alg{h}$ we find
\begin{equation*}
h^{-1} \delta(hXh^{-1}) h - \delta X = h^{-1}\delta h X + \delta X - X h^{-1}\delta h - \delta X = [h^{-1}\delta h,X] ~.
\end{equation*}
Therefore, setting $\delta = \Omega$, it follows that the first equation in eq.~\eqref{eq:ts2} is solved by
\begin{equation*}
v = \eta^{-1} (h^{-1}\Omega h) = \eta^{-1} \frac{1-e^{-ad_{\bar{v}}}}{\ad_{\bar{v}}} \Omega \bar v ~.
\end{equation*}
as claimed.

It remains to show that the second equation in eq.~\eqref{eq:ts2} is solved by this expression for $v$.
Using that $\partial_\pm$ and $\Omega$ commute, we have
\begin{equation*}\begin{split}
\partial_\pm v & = \eta^{-1} \partial_\pm (h^{-1}\Omega h)
\\
& = \eta^{-1} \Omega(h^{-1}\partial_\pm h) + \eta^{-1} [h^{-1}\Omega h, h^{-1}\partial_\pm h]
\\
& = \eta^{-1} (\Omega + \ad_{h^{-1}\Omega h}) h^{-1}\partial_\pm h
\\
& = \eta^{-1} \Omega_h (h^{-1}\partial_\pm h) ~,
\end{split}\end{equation*}
as required.
}
\else
\unskip\footnote{
For any derivation $\delta$ of the universal enveloping algebra of $\alg{h}$
\begin{equation*}
h^{-1} \delta h = \frac{1-e^{-ad_{\bar{v}}}}{\ad_{\bar{v}}} \delta \bar v ~, \qquad
\delta h h^{-1} = \frac{e^{ad_{\bar{v}}}-1}{\ad_{\bar{v}}} \delta \bar v ~,
\end{equation*}
where $h = e^{\bar v}$.
Moreover, if we have two such derivations $\delta_1$ and $\delta_2$ that commute then
\begin{equation*}\begin{split}
& \delta_1(h^{-1}\delta_2 h) - \delta_2(h^{-1}\delta_1h) + [h^{-1}\delta_1 h,h^{-1}\delta_2 h] = 0 ~,
\qquad
\delta_1(\delta_2 hh^{-1}) - \delta_2(\delta_1hh^{-1}) - [h^{-1}\delta_1 h,h^{-1}\delta_2 h] = 0 ~.
\end{split}\end{equation*}
To solve the first equation in eq.~\eqref{eq:ts2}, we observe that
$\Ad_h^{-1} \delta \Ad_h - \delta = \ad_{h^{-1}\delta h}$
Indeed, acting with the left-hand side on $X \in \alg{h}$ we find
$h^{-1} \delta(hXh^{-1}) h - \delta X = h^{-1}\delta h X + \delta X - X h^{-1}\delta h - \delta X = [h^{-1}\delta h,X]$.
Therefore, setting $\delta = \Omega$ and assuming that $\Omega$ can be extended to a derivation of the universal enveloping algebra of $\alg{h}$, it follows that the first equation in eq.~\eqref{eq:ts2} is solved by
\begin{equation*}
v = \eta^{-1} (h^{-1}\Omega h) = \eta^{-1} \frac{1-e^{-ad_{\bar{v}}}}{\ad_{\bar{v}}} \Omega \bar v ~,
\end{equation*}
as claimed.
It remains to show that the second equation in eq.~\eqref{eq:ts2} is solved by this expression for $v$.
Using that $\partial_\pm$ and $\Omega$ commute, we have
\begin{equation*}\begin{split}
\partial_\pm v & = \eta^{-1} \partial_\pm (h^{-1}\Omega h)
= \eta^{-1} \Omega(h^{-1}\partial_\pm h) + \eta^{-1} [h^{-1}\Omega h, h^{-1}\partial_\pm h]
= \eta^{-1} (\Omega + \ad_{h^{-1}\Omega h}) h^{-1}\partial_\pm h
= \eta^{-1} \Omega_h (h^{-1}\partial_\pm h) ~,
\end{split}\end{equation*}
as required.
}

Finally, having solved eq.~\eqref{eq:ts1}, we observe that
\begin{equation}
\partial_+ v  \frac{1}{1-\zeta \Omega - \ad_v} \partial_- v
= h^{-1}\partial_+ h \frac{1}{1-\eta\mathcal{R}_h} h^{-1}\partial_- h + \eta^{-1} h^{-1}\partial_+ h \Omega_h h^{-1}\partial_- h ~,
\end{equation}
hence
\begin{equation}
\Act_{\indrm{NATD}_\zeta} = \Act_{\indrm{YB}} - \frac{\hay\eta^{-1}}{4} \int d^2x \, \omega\big(\partial_+ h h^{-1},\partial_- h h^{-1}\big) ~.
\end{equation}
Therefore, we indeed find that the non-abelian T-dual of the modified action~\eqref{eq:c1} gives the YB deformation of the PCM up to a closed term.
As mentioned above, this construction can be generalised to any homogeneous YB deformation.
That is, up to a closed term, any homogeneous YB deformation is the non-abelian T-dual of the PCM modified by a closed term.
It is interesting to note that when $\omega$ is exact the closed terms become boundary terms and the parameter $\eta$ can be eliminated by a coordinate redefinition.
Therefore, in these cases the YB deformation is not strictly speaking a deformation.
\fi

\paragraph{Examples based on $\grp{SL}(2)$.}
We finish our discussion of homogeneous deformations with an explicit example of a non-abelian r-matrix.
It turns out there is no solution to the cYBe on $\alg{su}(2)$.
Therefore, we instead consider the non-compact real form $\alg{sl}(2,\Real)$ and the PCM for $\grp{SL}(2,\Real) \cong \AdS_3$.
Recall that for $\grp{G} = \grp{SL}(2,\Real)$ we reverse the overall sign of the action to ensure that the target-space metric has Lorentzian signature.
We use the following matrix representation of $\alg{sl}(2,\Real)$
\begin{equation}\label{eq:sl2gen}
S_0 = \begin{pmatrix} 1 & 0 \\ 0 & -1 \end{pmatrix} ~, \qquad
S_+ = \begin{pmatrix} 0 & 1 \\ 0 & 0 \end{pmatrix} ~, \qquad
S_- = \begin{pmatrix} 0 & 0 \\ 1 & 0 \end{pmatrix} ~.
\end{equation}
In this representation we take $\tr = \Tr$, where $\Tr$ is the standard matrix trace.
Therefore, we have
\begin{equation}
\tr\big(S_0^2\big) = 2 ~, \qquad \tr\big(S_+S_-\big) = 1 ~, \qquad [S_0,S_\pm] = \pm 2 S_\pm ~, \qquad
[S_+,S_-] = S_0 ~.
\end{equation}
Parametrising the group-valued field as
\begin{equation}\label{eq:gparamads}
g = \exp\big(y_+ S_+\big)\exp\big(\log z \, S_0\big)\exp\big(y_- S_-\big) ~,
\end{equation}
and substituting into the PCM action~\eqref{eq:pcmact} with the overall sign reversed, we find
\begin{equation}
\Act_{\indrm{SL(2,\Real)-PCM}} = \hay \int d^2x \, \big(\frac{2\partial_+ z \partial_- z + \partial_+ y_+ \partial_-y_- + \partial_+ y_- \partial_-y_+}{2z^2} \big) ~.
\end{equation}
The explicit form of the target-space metric in local coordinates is thus
\begin{equation}
G = 2\hay\big(\frac{dz^2 + dy_+ dy_-}{z^2}\big)~,
\end{equation}
which is the familiar Poincar\'e metric for $\AdS_3$.

Up to automorphisms, there is only one solution to the cYBe on $\alg{sl}(2,\Real)$.
This is the jordanian solution and is given by
\begin{equation}
r = S_0 \wedge S_+ ~.
\end{equation}
In terms of the operator $\mathcal{R}$ we have
\begin{equation}\label{eq:jord}
\mathcal{R} S_- = S_0 ~, \qquad \mathcal{R} S_0 = - 2 S_+ ~, \qquad \mathcal{R} S_+ = 0 ~.
\end{equation}
Therefore, $\mathcal{R}^3 = 0$ and
\begin{equation}
\frac{1}{1-\eta \mathcal{R}} = 1 + \eta \mathcal{R} + \eta^2 \mathcal{R}^2 ~.
\end{equation}
Parametrising $g$ as in \eqref{eq:gparamads} and substituting into the action of the deformed model~\eqref{eq:ybdefgen} with the overall sign reversed, we can read off the explicit form of the deformed target-space metric and B-field in local coordinates
\begin{equation}\label{eq:homback}
G= 2\hay\big(\frac{dz^2 + dy_+ dy_- - \eta^2 z^2 dy_-^2}{z^2}\big) ~,
\qquad
B = 2\hay \eta \frac{dz}{z^3} \wedge dy_- ~.
\end{equation}
Let us note that strictly speaking this is not a deformation since we can always set $\eta = 1$ by rescaling
\begin{equation}
y_\pm \to \eta^{\pm 1} y_\pm ~.
\end{equation}
In this example, the Lie algebra $\alg{h}$ is generated by $S_0$ and $S_+$, and the 2-cocycle $\omega$ is given by
\begin{equation}
\omega\big(S_0,S_+\big) = - \omega\big(S_+,S_0\big) = 2 ~.
\end{equation}
This 2-cocycle is exact since $\omega\big(X,Y\big)$ can be written as $\tilde\omega\big([X,Y]\big)$ where
\begin{equation}
\tilde\omega\big(S_0\big) = 0 ~, \quad \tilde\omega\big(S_+\big) = 1 ~.
\end{equation}

\subsection{Inhomogeneous YB deformations}

Let us now turn to the case of inhomogeneous YB deformations.
In this case the antisymmetric operator $\mathcal{R}$ satisfies the mcYBe
\begin{equation}\label{eq:mcybe}
[\mathcal{R} X, \mathcal{R}Y] - \mathcal{R}[X,\mathcal{R}Y] - \mathcal{R}[\mathcal{R}X,Y] + c^2 [X,Y] = 0 ~,
\qquad
\tr\big(X\mathcal{R}Y\big) + \tr\big((\mathcal{R} X) Y\big)  = 0 ~,
\qquad
X,Y \in \alg{g} ~.
\end{equation}
The mcYBe can be rewritten as
\begin{equation}
[(\mathcal{R}\pm c) X, (\mathcal{R}\pm c) Y] = (\mathcal{R}\pm c) ([X,\mathcal{R}Y] + [\mathcal{R} X,Y]) ~, \qquad
X,Y \in \alg{g} ~,
\end{equation}
from which we see that the vector spaces $\alg{h}_\pm = \im(\mathcal{R} \pm c)$ are Lie algebras.

\paragraph{The Drinfel'd-Jimbo solution.}
Antisymmetric solutions to the mcYBe, that is with $c \neq 0$, have been classified for simple Lie algebras over the complex numbers.
Here we focus on the simplest, Drinfel'd-Jimbo, solution to the mcYBe.
To define this solution we introduce a Cartan-Weyl basis for the complexified Lie algebra $\alg{g}^{\Complex}$.
We denote the Cartan generators by $H_i$, $i = 1,\dots\rank\alg{g}^{\Complex}$, and the positive and negative roots by $E_\alpha$ and $F_\alpha$ respectively, $\alpha = 1,\dots,\frac12(\dim_{\Complex}\alg{g}^{\Complex} - \rank\alg{g}^{\Complex})$.
\iflectures
In operator form, the Drinfel'd-Jimbo solution to the mcYBe is then given by
\begin{equation}\label{eq:djcomplex}
\mathcal{R} H_i = 0 ~, \qquad
\mathcal{R} E_\alpha = -c E_\alpha ~, \qquad
\mathcal{R} F_\alpha = c F_\alpha ~.
\end{equation}

\question{Show that the Drinfel'd-Jimbo solution~\eqref{eq:djcomplex} solves the mcYBe~\eqref{eq:mcybe}.
Let $\Pi = 1 - c^{-2} \mathcal{R}^2$.
Show that $\Pi$ projects onto the Cartan subalgebra and that
\begin{equation*}
\Pi([X,\mathcal{R}Y] + [\mathcal{R}X,Y]) = 0 ~, \qquad X,Y \in \alg{g} ~.
\end{equation*}
}

\answer{Using that
\begin{equation*}
\mathcal{R} H_i = 0 ~, \qquad
\mathcal{R} E_\alpha = -c E_\alpha ~, \qquad
\mathcal{R} F_\alpha = c F_\alpha ~,
\end{equation*}
and working through the various cases we have
\begin{equation*}
[\mathcal{R}H_i,\mathcal{R}H_j] - \mathcal{R}[H_i,\mathcal{R} H_j] - \mathcal{R}[\mathcal{R}H_i,H_j] + c^2 [H_i,H_j] = 0 ~,
\end{equation*}
since $H_i$ are Cartan generators and thus commute,
\begin{equation*}\begin{split}
[\mathcal{R}H_i,\mathcal{R}E_\alpha] - \mathcal{R}[H_i,\mathcal{R}E_\alpha] - \mathcal{R}[\mathcal{R}H_i,E_\alpha] + c^2 [H_i,E_\alpha]
& =
- \mathcal{R} [H_i,\mathcal{R}E_\alpha] + c^2 [H_i,E_\alpha]
\\ & = - c^2 [H_i,E_\alpha] + c^2 [H_i,E_\alpha]
= 0 ~,
\end{split}\end{equation*}
since $[H_i,E_\alpha]$ is a sum of positive roots, hence $\mathcal{R} [H_i,E_\alpha] = -c [H_i,E_\alpha]$,
\begin{equation*}\begin{split}
[\mathcal{R}H_i,\mathcal{R}F_\alpha] - \mathcal{R}[H_i,\mathcal{R}F_\alpha]  - \mathcal{R}[\mathcal{R} H_i,F_\alpha] + c^2 [H_i,F_\alpha]
& =
- \mathcal{R} [H_i,\mathcal{R}F_\alpha] + c^2 [H_i,F_\alpha]
\\ & = - c^2 [H_i,F_\alpha] + c^2 [H_i,F_\alpha]
= 0 ~,
\end{split}\end{equation*}
since $[H_i,F_\alpha]$ is a sum of negative roots, hence $\mathcal{R} [H_i,F_\alpha] = c [H_i,F_\alpha]$,
\begin{equation*}\begin{split}
& [\mathcal{R}E_\alpha,\mathcal{R}E_\beta] - \mathcal{R}[E_\alpha,\mathcal{R}E_\beta] - \mathcal{R}[\mathcal{R}E_\alpha,E_\beta] + c^2 [E_\alpha,E_\beta]
= c^2[E_\alpha,E_\beta] + 2c\mathcal{R}[E_\alpha,E_\beta] + c^2 [E_\alpha,E_\beta] = 0 ~,
\end{split}\end{equation*}
since $[E_\alpha,E_\beta]$ is a positive root, hence $\mathcal{R} [E_\alpha,E_\beta] = -c [E_\alpha,E_\beta]$,
\begin{equation*}\begin{split}
& [\mathcal{R}F_\alpha,\mathcal{R}F_\beta] - \mathcal{R}[F_\alpha,\mathcal{R}F_\beta] - \mathcal{R}[\mathcal{R}F_\alpha,F_\beta] + c^2[F_\alpha,F_\beta]
=
c^2[F_\alpha,F_\beta] - 2c\mathcal{R}[F_\alpha,F_\beta] + c^2 [F_\alpha,F_\beta] = 0 ~,
\end{split}\end{equation*}
since $[F_\alpha,F_\beta]$ is a negative root, hence $\mathcal{R} [F_\alpha,F_\beta] = c [F_\alpha,F_\beta]$, and
\begin{equation*}\begin{split}
& [\mathcal{R}E_\alpha,\mathcal{R}F_\beta] - \mathcal{R}[E_\alpha,\mathcal{R}F_\beta] - \mathcal{R}[\mathcal{R}E_\alpha,F_\beta] + c^2 [E_\alpha,F_\beta]
\\ & \quad =
-c^2[E_\alpha,F_\beta] - \mathcal{R}(c[E_\alpha,F_\beta]-c[E_\alpha,F_\beta]) +c^2 [E_\alpha,F_\beta] = 0 ~.
\end{split}\end{equation*}
Therefore, the Drinfel'd-Jimbo solution~\eqref{eq:djcomplex} indeed solves the mcYBe~\eqref{eq:mcybe}.

Defining $\Pi = 1 - c^{-2} \mathcal{R}^2$ we have
\begin{equation*}
\Pi H_i = H_i ~, \qquad \Pi E_\alpha = E_\alpha - E_\alpha = 0 ~, \qquad
\Pi F_\alpha = F_\alpha - F_\alpha = 0 ~,
\end{equation*}
hence $\Pi$ projects onto the Cartan subalgebra.
Working through the various cases we have
\begin{equation*}\begin{aligned}
& \Pi([H_i,\mathcal{R}H_j] + [\mathcal{R}H_i,H_j]) = 0 ~,
\\ & \Pi([H_i,\mathcal{R}E_\alpha] + [\mathcal{R}H_i,E_\alpha]) = \Pi[H_i,\mathcal{R}E_\alpha] = - c \Pi [H_i,E_\alpha] = 0 ~,
\\ & \Pi([H_i,\mathcal{R}F_\alpha]  +[\mathcal{R} H_i,F_\alpha]) = \Pi[H_i,\mathcal{R}F_\alpha] = c \Pi [H_i,F_\alpha] = 0 ~,
\\ & \Pi([E_\alpha,\mathcal{R}F_\beta] + [\mathcal{R}E_\alpha,F_\beta]) = \Pi(c[E_\alpha,F_\beta]-c[E_\alpha,F_\beta]) = 0 ~,
\\ & \Pi ([E_\alpha,\mathcal{R}E_\beta] + [\mathcal{R}E_\alpha,E_\beta] ) =-2c\Pi[E_\alpha,E_\beta] = 0 ~,
\\ & \Pi ([F_\alpha,\mathcal{R}F_\beta] + [\mathcal{R}F_\alpha,F_\beta]) =2c\Pi[F_\alpha,F_\beta] = 0 ~,
\end{aligned}\end{equation*}
and we indeed find that
\begin{equation*}
\Pi([X,\mathcal{R}Y]+ [\mathcal{R}X,Y]) = 0 ~, \qquad X,Y \in \alg{g} ~.
\end{equation*}
}

\else
In operator form, the Drinfel'd-Jimbo solution to the mcYBe is then given by
\unskip\footnote{Using that the Cartan generators $H_i$ commute, that $[H_i,E_\alpha]$ and $[E_\alpha,E_\beta]$ are sums of positive roots, and that $[H_i,F_\alpha]$ and $[F_\alpha,F_\beta]$ are sums of negative roots, working through the various cases we have
\begin{equation*}\begin{split}
& [\mathcal{R}H_i,\mathcal{R}H_j] - \mathcal{R}[H_i,\mathcal{R} H_j] - \mathcal{R}[\mathcal{R}H_i,H_j] + c^2 [H_i,H_j] = 0 ~,
\\ & [\mathcal{R}E_\alpha,\mathcal{R}E_\beta] - \mathcal{R}[E_\alpha,\mathcal{R}E_\beta] - \mathcal{R}[\mathcal{R}E_\alpha,E_\beta] + c^2 [E_\alpha,E_\beta]
= c^2[E_\alpha,E_\beta] + 2c\mathcal{R}[E_\alpha,E_\beta] + c^2 [E_\alpha,E_\beta] = 0 ~,
\\ & [\mathcal{R}F_\alpha,\mathcal{R}F_\beta] - \mathcal{R}[F_\alpha,\mathcal{R}F_\beta] - \mathcal{R}[\mathcal{R}F_\alpha,F_\beta] + c^2[F_\alpha,F_\beta]
=
c^2[F_\alpha,F_\beta] - 2c\mathcal{R}[F_\alpha,F_\beta] + c^2 [F_\alpha,F_\beta] = 0 ~,
\\ & [\mathcal{R}H_i,\mathcal{R}E_\alpha] - \mathcal{R}[H_i,\mathcal{R}E_\alpha] - \mathcal{R}[\mathcal{R}H_i,E_\alpha] + c^2 [H_i,E_\alpha]
=
- \mathcal{R} [H_i,\mathcal{R}E_\alpha] + c^2 [H_i,E_\alpha]
= - c^2 [H_i,E_\alpha] + c^2 [H_i,E_\alpha]
= 0 ~,
\\ & [\mathcal{R}H_i,\mathcal{R}F_\alpha] - \mathcal{R}[H_i,\mathcal{R}F_\alpha]  - \mathcal{R}[\mathcal{R} H_i,F_\alpha] + c^2 [H_i,F_\alpha]
=
- \mathcal{R} [H_i,\mathcal{R}F_\alpha] + c^2 [H_i,F_\alpha]
= - c^2 [H_i,F_\alpha] + c^2 [H_i,F_\alpha]
= 0 ~,
\\ & [\mathcal{R}E_\alpha,\mathcal{R}F_\beta] - \mathcal{R}[E_\alpha,\mathcal{R}F_\beta] - \mathcal{R}[\mathcal{R}E_\alpha,F_\beta] + c^2 [E_\alpha,F_\beta]
=
-c^2[E_\alpha,F_\beta] - \mathcal{R}(c[E_\alpha,F_\beta]-c[E_\alpha,F_\beta]) +c^2 [E_\alpha,F_\beta] = 0 ~.
\end{split}\end{equation*}}
\begin{equation}\label{eq:djcomplex}
\mathcal{R} H_i = 0 ~, \qquad
\mathcal{R} E_\alpha = -c E_\alpha ~, \qquad
\mathcal{R} F_\alpha = c F_\alpha ~.
\end{equation}
It is also useful to introduce the operator
\begin{equation}
\Pi = 1 - c^{-2} \mathcal{R}^2 ~,
\end{equation}
which projects onto the Cartan subalgebra
\begin{equation}
\Pi H_i = H_i ~, \qquad \Pi E_\alpha = 0 ~, \qquad
\Pi F_\alpha = 0 ~,
\end{equation}
and satisfies the identity
\unskip\footnote{Working through the various cases we have
\begin{equation*}\begin{aligned}
& \Pi([H_i,\mathcal{R}H_j] + [\mathcal{R}H_i,H_j]) = 0 ~,
&& \Pi([H_i,\mathcal{R}E_\alpha] + [\mathcal{R}H_i,E_\alpha]) = \Pi[H_i,\mathcal{R}E_\alpha] = - c \Pi [H_i,E_\alpha] = 0 ~,
\\ & \Pi ([E_\alpha,\mathcal{R}E_\beta] + [\mathcal{R}E_\alpha,E_\beta] ) =-2c\Pi[E_\alpha,E_\beta] = 0 ~,
&& \Pi([H_i,\mathcal{R}F_\alpha]  +[\mathcal{R} H_i,F_\alpha]) = \Pi[H_i,\mathcal{R}F_\alpha] = c \Pi [H_i,F_\alpha] = 0 ~,
\\
& \Pi ([F_\alpha,\mathcal{R}F_\beta] + [\mathcal{R}F_\alpha,F_\beta]) =2c\Pi[F_\alpha,F_\beta] = 0 ~,
&& \Pi([E_\alpha,\mathcal{R}F_\beta] + [\mathcal{R}E_\alpha,F_\beta]) = \Pi(c[E_\alpha,F_\beta]-c[E_\alpha,F_\beta]) = 0 ~.
\end{aligned}\end{equation*}}
\begin{equation}
\Pi([X,\mathcal{R}Y]+ [\mathcal{R}X,Y]) = 0 ~, \qquad X,Y \in \alg{g} ~.
\end{equation}
\fi

When we consider real forms of the complexified Lie algebra $\alg{g}^\Complex$ we require that $c^2\in\Real$, and the distinction between non-split, $c=i$, and split, $c=1$, becomes relevant.
For a given real form there may be no, one or more than one inequivalent Drinfel'd-Jimbo solutions to the non-split or split mcYBe.
Typically, a careful analysis of the particular Lie algebra under consideration is needed to classify these.
Nevertheless, we can still make some universal statements.
In particular, let us focus on two real forms that can be defined for every simple Lie algebra.
These are the compact and split real forms.
The compact real form is generated by $\{iH_i,i(E_\alpha+F_\alpha),(E_\alpha-F_\alpha)\}$, while the split real form is generated by $\{H_i,E_\alpha,F_\alpha\}$.
For example, for $\alg{g}^\Complex = \alg{sl}(N,\Complex)$ the compact real form is $\alg{su}(N)$, while the split real form is $\alg{sl}(N,\Real)$.

The Drinfel'd-Jimbo solution~\eqref{eq:djcomplex} allows us to write down solutions to the mcYBe with $c=i$ and $c=1$
\begin{align}\label{eq:nssol}
&c=i ~,\qquad
&&\mathcal{R} H_i = 0 ~, \qquad
&&\mathcal{R} E_\alpha = - i E_\alpha ~, \qquad
&&\mathcal{R} F_\alpha = i F_\alpha ~,
\\\label{eq:ssol}
&c=1 ~,\qquad
&&\mathcal{R} H_i = 0 ~, \qquad
&&\mathcal{R} E_\alpha = - E_\alpha ~, \qquad
&&\mathcal{R} F_\alpha = F_\alpha ~.
\end{align}
We would now like to ask whether these solutions preserve the compact and split real forms, that is whether the image of $\mathcal{R}$ restricted to the real form is also valued in the real form.
It is straightforward to check that the $c=i$ solution preserves the compact real form, hence we have a solution to the non-split mcYBe, while the $c=1$ solution preserves the split real form and we have a solution to the split mcYBe.

The Drinfel'd-Jimbo solution to the mcYBe~\eqref{eq:djcomplex} can be extended to include $\frac12 \rank\alg{g}(\rank\alg{g} - 1)$ free parameters by setting
\begin{equation}\begin{split}\label{eq:djext}
\mathcal{R} H_i = \beta^j{}_i H_j ~, \qquad
\mathcal{R} E_\alpha = -c E_\alpha ~, \qquad
\mathcal{R} F_\alpha = c F_\alpha ~, \qquad
\beta^j{}_i \kappa^{ik} = - \beta^k{}_i \kappa^{ij} ~,
\end{split}\end{equation}
where $\kappa_{ij} = \tr\big(H_i H_j\big)$ and $\kappa^{ij}\kappa_{jk} = \delta^i_k$.
To preserve the compact and split real forms the parameters $\beta^j{}_i$ should be real.
More generally, the reality conditions on the parameters $\beta^j{}_i$ are determined by the particular real form being considered.
Since the Cartan generators commute, this extension can be understood as combining an abelian solution to the cYBe with the Drinfel'd-Jimbo solution to the mcYBe.
Indeed, the Cartan subgroup of $\grp{G}^\Complex$ generated by $H_i$ is the symmetry of both the Drinfel'd-Jimbo solution~\eqref{eq:djcomplex} and its extension~\eqref{eq:djext}
\begin{equation}
\ad_{H_i} \mathcal{R} = \mathcal{R} \ad_{H_i} ~,
\end{equation}
which is the infinitesimal version of eq.~\eqref{eq:symg0}.
This means that the Cartan subgroup of the left-acting $\grp{G}$ global symmetry is preserved by YB deformations based on Drinfel'd-Jimbo solutions.
The extension~\eqref{eq:djext} is then equivalent to performing additional TsT transformations in the corresponding coordinates on target space.

Note that, apart from the case of the rank 1 algebra $\alg{sl}(2,\Complex)$, the extended Drinfel'd-Jimbo solution~\eqref{eq:djext} is not the unique solution to the mcYBe for simple Lie algebras over $\Complex$.
However, for the compact real form it is the only solution to the non-split mcYBe and there is no solution to the split mcYBe.
For the split real form, and for non-compact real forms more generally, the situation is typically more complicated.
Indeed, as we will see shortly, the split real form $\alg{sl}(2,\Real)$ has a solution to both the split and non-split mcYBe.

\iflectures
\question{Consider the following deformation of the PCWZM for general Lie group $\grp{G}$
\begin{equation*}
\Act_\indrm{YB-PCWZM} = - \frac{\hay}{2} \int d^2x \, \tr\big(j_+ (1 + \alpha \eta \mathcal{R}_g + (1-\alpha)\mathcal{R}^2_g) j_-\big) + \frac{\kay}{6} \int d^3 x\, \epsilon^{ijk} \tr\big(j_i[j_j,j_k]\big) ~,
\end{equation*}
with
\begin{equation*}
\eta = \sqrt{\frac{1-\alpha}{\alpha}\big(1- \frac{\kay^2}{\alpha\hay^2}\big)} ~.
\end{equation*}
Taking $\mathcal{R}$ to be the Drinfel'd-Jimbo solution to the mcYBe~\eqref{eq:nssol} show that the conserved current
\begin{equation*}
J_\pm = \alpha^{-1} (1 + (1-\alpha)\mathcal{R}_g^2 \mp (\alpha \eta \mathcal{R}_g + \frac{\kay}{\hay})) j_\pm ~,
\end{equation*}
is flat on-shell.
You will need to use the identities $\mathcal{R}^3 = -\mathcal{R}$ and $\Pi([X,\mathcal{R}Y]+ [\mathcal{R}X,Y]) = 0$ for $X,Y\in\alg{g}$ where $\Pi = 1+\mathcal{R}^2$.

Again using $\mathcal{R}^3 = -\mathcal{R}$ show that this action can be rewritten in the compact form
\begin{equation*}
\Act_\indrm{YB-PCWZM} = -\frac{\kay}{2} \int d^2x \, \tr\big(j_+ \frac{e^\chi + e^{\rho \mathcal{R}_g} }{e^\chi - e^{\rho \mathcal{R}_g}} j_-\big)
+ \frac{\kay}{6} \int d^3 x\, \epsilon^{ijk} \tr\big(j_i[j_j,j_k]\big) ~,
\end{equation*}
where you should identify how the parameters $\rho$ and $\chi$ are related to $\hay$ and $\eta$.}

\answer{We start by writing
\begin{equation*}
j_\pm = \frac{\alpha}{1 + (1-\alpha)\mathcal{R}_g^2 \mp ( \alpha\eta \mathcal{R}_g + \frac{\kay}{\hay})} J_\pm
= \frac{1}{(1\mp\frac{\kay}{\hay})^2} (\alpha \mp  \frac{\kay}{\hay} \pm \alpha\eta \mathcal{R}_g \pm \frac{\kay}{\hay}(1-\alpha) \Pi_g )J_\pm ~,
\end{equation*}
where we have used that $\mathcal{R}^3 = -\mathcal{R}$, which implies that $\mathcal{R}_g^3 = -\mathcal{R}_g$,
$\eta = \sqrt{\frac{1-\alpha}{\alpha}\big(1 - \frac{\kay^2}{\alpha\hay^2}\big)}$,
and $\Pi_g = \Ad_g^{-1} \Pi \Ad_g$ with $\Pi = 1+ \mathcal{R}^2$.
Writing $j_\pm = \mathcal{Q}_{\pm g} J_\pm$ the flatness condition~\eqref{eq:flatness} becomes
\begin{equation*}\begin{split}
& \partial_+ j_- - \partial_- j_+ + [j_+,j_-] \\ & =
\frac12 (\mathcal{Q}_{+g} + \mathcal{Q}_{-g}) (\partial_+ J_- - \partial_- J_+)
+\mathcal{Q}_{+g}[J_+,\mathcal{Q}_{-g} J_-] + \mathcal{Q}_{-g}[\mathcal{Q}_{+g} J_+,J_-] - [\mathcal{Q}_{+g} J_+,\mathcal{Q}_{-g} J_-] ~,
\end{split}\end{equation*}
where we have used that the current $J_\pm$ is conserved.
By repeatedly applying the mcYBe~\eqref{eq:mcybe} with $c=i$ we can prove the following identities
\begin{equation*}\begin{split}
& \Pi [X,\Pi Y] + \Pi [\Pi X,Y] - [\Pi X,\Pi Y] = 0 ~,
\\ & \Pi [X,\mathcal{R} Y] + \mathcal{R} [\Pi X,Y] - [\Pi X,\mathcal{R} Y] = -\Pi[\mathcal{R}X,Y] ~,
\\ & \mathcal{R} [X,\Pi Y] + \Pi [\mathcal{R} X,Y] - [\mathcal{R} X,\Pi Y] = -\Pi[X,\mathcal{R}Y] ~,
\end{split}\end{equation*}
for $X,Y \in \alg{g}$.
Therefore, for $\mathcal{Q}_{\pm g} = \frac{1}{(1\mp\frac{\kay}{\hay})^2} (\alpha \mp  \frac{\kay}{\hay} \pm \alpha \eta \mathcal{R}_g \pm \frac{\kay}{\hay}(1-\alpha) \Pi_g )$ we find that
\begin{equation*}\begin{split}
& \mathcal{Q}_{+g}[J_+,\mathcal{Q}_{-g} J_-] + \mathcal{Q}_{-g}[\mathcal{Q}_{+g} J_+,J_-] - [\mathcal{Q}_{+g} J_+,\mathcal{Q}_{-g} J_-]
\\ & = \frac{1}{(1-\frac{\kay^2}{\hay^2})^2}\big((\alpha - \frac{\kay^2}{\hay^2}(2-\alpha) + \frac{2\alpha\eta\kay}{\hay}\mathcal{R}_g + \frac{2\kay^2}{\hay^2}(1-\alpha)\Pi_g)[J_+,J_-] \\ & \hspace{124pt} - \frac{\eta\kay}{\hay}\alpha(1-\alpha)\Pi_g ([J_+,\mathcal{R}_gJ_-]+ [\mathcal{R}_gJ_+,J_-])\big)
\\ & = \frac{1}{(1-\frac{\kay^2}{\hay^2})^2}(\alpha - \frac{\kay^2}{\hay^2}(2-\alpha) + \frac{2\alpha\eta\kay}{\hay}\mathcal{R}_g + \frac{2\kay^2}{\hay^2}(1-\alpha)\Pi_g)[J_+,J_-] ~,
\end{split}\end{equation*}
where we have used the identity $\Pi([X,\mathcal{R}Y]+ [\mathcal{R}X,Y]) = 0$ for $X,Y\in\alg{g}$ and $\eta = \sqrt{\frac{1-\alpha}{\alpha}\big(1 - \frac{\kay^2}{\alpha\hay^2}\big)}$.
On the other hand we have
\begin{equation*}
\frac12 (\mathcal{Q}_{+g} + \mathcal{Q}_{-g}) = \frac{1}{(1-\frac{\kay^2}{\hay^2})^2}(\alpha - \frac{\kay^2}{\hay^2}(2-\alpha) + \frac{2\alpha\eta\kay}{\hay}\mathcal{R}_g + \frac{2\kay^2}{\hay^2}(1-\alpha)\Pi_g) ~.
\end{equation*}
Therefore,
\begin{equation*}\begin{split}
& \partial_+ j_- - \partial_- j_+ + [j_+,j_-] \\ & =
\frac{1}{(1-\frac{\kay^2}{\hay^2})^2}(\alpha - \frac{\kay^2}{\hay^2}(2-\alpha) + \frac{2\alpha\eta\kay}{\hay}\mathcal{R}_g + \frac{2\kay^2}{\hay^2}(1-\alpha)\Pi_g)(\partial_+ J_- - \partial_- J_+ + [J_+,J_-]) ~,
\end{split}\end{equation*}
and we find that the flatness of $j_\pm$ implies that $J_\pm$ is also flat on-shell.

In order to rewrite the action in the required form we compare the two actions and see that we would like to show that
\begin{equation*}
\kay\frac{e^\chi + e^{\rho \mathcal{R}_g} }{e^\chi - e^{\rho \mathcal{R}_g}}
= \hay(1 + \alpha \eta \mathcal{R}_g + (1-\alpha)\mathcal{R}^2_g) ~.
\end{equation*}
We start by observing that, since $\mathcal{R}^3 = -\mathcal{R}$, we can write
\begin{equation*}
e^{\rho \mathcal{R}_g} = \gamma_1 + \gamma_2 \mathcal{R}_g + \gamma_3 \mathcal{R}^2_g ~.
\end{equation*}
The operator $e^{\rho \mathcal{R}_g}$ satisfies the following differential equation and boundary condition
\begin{equation*}
\frac{d}{d\rho} e^{\rho \mathcal{R}_g} = \mathcal{R}_g e^{\rho \mathcal{R}_g} ~, \qquad
e^{\rho \mathcal{R}_g}\big|_{\rho = 0} = 1 ~.
\end{equation*}
Substituting in $e^{\rho \mathcal{R}_g} = \gamma_1 + \gamma_2 \mathcal{R}_g + \gamma_3 \mathcal{R}^2_g$ and using $\mathcal{R}^3 = -\mathcal{R}$ we find
\begin{equation*}\begin{aligned}
\frac{d\gamma_1}{d\rho} = 0 ~, \qquad
\frac{d\gamma_2}{d\rho} = \gamma_1-\gamma_3 ~, \qquad
\frac{d\gamma_3}{d\rho} = \gamma_2 ~, \qquad
\gamma_1\big|_{\rho = 0} = 1 ~, \qquad
\gamma_2\big|_{\rho = 0} = 0 ~, \qquad
\gamma_3\big|_{\rho = 0} = 0 ~,
\end{aligned}\end{equation*}
which is solved by
\begin{equation*}
\gamma_1 = 1 ~, \qquad \gamma_2 = \sin\rho ~, \qquad \gamma_3 = 1- \cos\rho ~.
\end{equation*}
Using this expansion of $e^{\rho \mathcal{R}_g}$ it is then straightforward to show that
\begin{equation*}
\kay\frac{e^\chi + e^{\rho \mathcal{R}_g} }{e^\chi - e^{\rho \mathcal{R}_g}} =
\kay \coth \frac{\chi}{2} + \frac{\kay\sin\rho}{\cosh\chi-\cos\rho}\mathcal{R}_g
+ \kay \coth \frac{\chi}{2} \frac{1-\cos\rho}{\cosh\chi - \cos\rho} \mathcal{R}_g^2 ~.
\end{equation*}
On the other hand, setting
\begin{equation*}
\hay = \kay \coth \frac{\chi}{2} ~, \qquad
\alpha = \frac{\cosh\chi-1}{\cosh\chi-\cos\rho} ~,
\end{equation*}
implies that
\begin{equation*}
\eta = \sqrt{\frac{1-\alpha}{\alpha}\big(1- \frac{\kay^2}{\alpha\hay^2}\big)} = \frac{\sin\rho}{\sinh\chi} ~,
\end{equation*}
hence
\begin{equation*}
\hay(1 + \alpha \eta \mathcal{R}_g + (1-\alpha)\mathcal{R}^2_g) =
\kay \coth \frac{\chi}{2} + \frac{\kay\sin\rho}{\cosh\chi-\cos\rho}\mathcal{R}_g
+ \kay \coth \frac{\chi}{2} \frac{1-\cos\rho}{\cosh\chi - \cos\rho} \mathcal{R}_g^2 ~,
\end{equation*}
as required.
}
\fi

\paragraph{Examples based on $\grp{SL}(2)$.}
Let us now give a few explicit examples of inhomogeneous YB deformations of the PCM based on different real forms of $\grp{G} = \grp{SL}(2,\Complex)$.
We identify the Cartan generator $H$ with $S_0$, the positive root $E$ with $S_+$ and the negative root $F$ with $S_-$ where $S_0$, $S_\pm$ are the generators of the split real form $\alg{sl}(2,\Real)$ defined in eq.~\eqref{eq:sl2gen}.
This then implies that $iH = T_3$, $i(E+F) = T_1$ and $E-F = T_2$ where $T_1$, $T_2$ and $T_3$ are the generators of the compact real form $\alg{su}(2)$ defined in eq.~\eqref{eq:su2gen}.

We start by considering inhomogeneous YB deformations of the $\grp{SU}(2)$ PCM.
Since $\alg{su}(2)$ is a compact real form it has a solution~\eqref{eq:nssol} to the non-split mcYBe.
In terms of the generators $T_1$, $T_2$ and $T_3$ this is given by
\begin{equation}\label{eq:su2sol}
\mathcal{R} T_1 = T_2 ~, \qquad \mathcal{R} T_2 = - T_1 ~, \qquad \mathcal{R} T_3 = 0 ~.
\end{equation}
This precisely coincides with the operator $\mathcal{R}$ that we encountered previously in \secref{sec:su2pcm} in eq.~\eqref{eq:firstr}.
Therefore, the deformed target-space metric will be given by eq.~\eqref{eq:su2def} and the deformed B-field is closed.
Noting that $\mathcal{R}^3 = -\mathcal{R}$ we have
\begin{equation}
\frac{1}{1-\eta \mathcal{R}} = 1 + \frac{\eta}{1+\eta^2} \mathcal{R} + \frac{\eta^2}{1+\eta^2} \mathcal{R}^2 ~.
\end{equation}
Parametrising $g$ as in~\eqref{eq:gparam} and substituting into the action of the deformed model~\eqref{eq:ybdefgen} with $\tr = \Tr$, we can read off the explicit form of the deformed target-space metric and B-field in local coordinates
\begin{equation}\begin{split}
G & = \frac{2\hay}{1+\eta^2}\big(d\theta^2 + \cos^2\theta(1 + \eta^2\cos^2\theta) d\varphi^2
+ \sin^2\theta(1+\eta^2\sin^2\theta) d\phi^2
+ 2\eta^2\sin^2\theta\cos^2\theta d\varphi d\phi\big) ~,
\\
B & = -\frac{2\hay\eta}{1+\eta^2} \sin\theta\cos\theta d\theta \wedge d(\varphi - \phi) ~.
\end{split}\end{equation}

We now turn to inhomogeneous YB deformations of the $\grp{SL}(2,\Real)$ PCM.
Recall that for $\grp{G} = \grp{SL}(2,\Real)$ we reverse the overall sign of the action to ensure that the target-space metric has Lorentzian signature.
Given that $\alg{sl}(2,\Real)$ is a split real form it has a solution~\eqref{eq:ssol} to the split mcYBe.
In terms of the generators $S_0$ and $S_\pm$ this is given by
\begin{equation}\label{eq:sl2rsplit}
\mathcal{R} S_0 = 0 ~, \qquad \mathcal{R} S_\pm = \mp S_\pm ~.
\end{equation}
Noting that $\mathcal{R}^3 = \mathcal{R}$ we have
\begin{equation}
\frac{1}{1-\eta \mathcal{R}} = 1 + \frac{\eta}{1-\eta^2} \mathcal{R} + \frac{\eta^2}{1-\eta^2} \mathcal{R}^2 ~.
\end{equation}
Parametrising the group-valued field as
\begin{equation}\label{eq:param2}
g = \exp\big(\frac{y_1 + y_0}{2} S_0\big) \exp\big(u \, (S_+ + S_-)\big) \exp\big(\frac{y_1 - y_0}{2} S_0\big) ~,
\end{equation}
and substituting into the action of the deformed model~\eqref{eq:ybdefgen} with $\tr = \Tr$ and the overall sign reversed, we find that the deformed target-space metric and B-field are given by
\begin{equation}\begin{split}
G & = \frac{2\hay}{1-\eta^2} \big(du^2
- \sinh^2 u(1+\eta^2 \sinh^2 u)dy_0^2
+ \cosh^2 u(1-\eta^2 \cosh^2 u)dy_1^2
+ 2\eta^2\sinh^2u\cosh^2u dy_0 dy_1\big) ~,
\\
B & = \frac{2\hay\eta}{1-\eta^2} \sinh u \cosh u du \wedge d(y_1 - y_0) ~.
\end{split}\end{equation}
This deformed background has a singularity at $\eta = 1$, which is related to the fact that the operator $1-\eta\mathcal{R}$ is not invertible at this point.
For our purposes, it will be sufficient to restrict $\eta$ to lie in the range $[0,1)$.
It is interesting to observe that if we consider the coordinate redefinition
\begin{equation}\label{eq:coordredf}
u = \log(\gamma z) ~, \qquad y_0 \to - \gamma (y_+ - y_-) ~, \qquad y_1 \to - \gamma (y_+ + y_-) ~,
\end{equation}
where $\gamma$ is a constant parameter, redefine $\eta \to 2\gamma \eta$ and take $\gamma \to 0$, then this deformed background becomes that of the homogeneous YB deformation~\eqref{eq:homback}.
This is not an accident.
Indeed, let us consider the field redefinition
\begin{equation}\begin{split}
g \to g_\ind{L} g g_\ind{R} ~, \qquad
g_\ind{L} & = \exp\big(\frac{\pi}{4} (S_+ - S_-)\big) \exp\big(-\frac12\log\gamma\,(S_++S_-)\big) ~, \qquad
\\
g_\ind{R} &= \exp\big(-\frac12\log\gamma\,(S_++S_-)\big) \exp\big(-\frac{\pi}{4} (S_+ - S_-)\big) ~,
\end{split}\end{equation}
together with the coordinate redefinition~\eqref{eq:coordredf} in the action~\eqref{eq:ybdefgen}.
The form of the action is the same except that the operator $\mathcal{R}$ is replaced by the operator
$\Ad_{g_\ind{L}} \mathcal{R} \Ad_{g_\ind{L}}^{-1}$.
We can now check that
\begin{equation}\begin{split}
& \lim_{\gamma \to 0} \ \big( g_\ind{L} \exp\big(-\gamma y_+ S_0\big) g_\ind{L}^{-1}\big) = \exp\big(y_+ S_+\big) ~,
\\ &
\lim_{\gamma \to 0} \ \big( g_\ind{L} \exp\big(\log(\gamma z)(S_+ + S_-)\big) g_\ind{R}\big) = \exp\big(\log z \, S_0\big) ~,
\\ &
\lim_{\gamma \to 0} \ \big( g_\ind{R}^{-1} \exp\big(-\gamma y_- S_0) g_\ind{R}\big) = \exp\big(y_- S_-\big) ~,
\end{split}\end{equation}
while $2\gamma\eta \Ad_{g_\ind{L}} \mathcal{R} \Ad_{g_\ind{L}}^{-1}$ becomes~\eqref{eq:jord} in the limit $\gamma \to 0$.

We now turn to our final example of an inhomogeneous YB deformation.
Even though $\alg{sl}(2,\Real)$ is a split real form, it also has a solution to the non-split mcYBe, which is given by
\begin{equation}\label{eq:sl2rns}
\mathcal{R} S_0 = S_+ + S_- ~, \qquad \mathcal{R} S_\pm = - \frac{1}{2} S_0 ~.
\end{equation}
This solution takes the form~\eqref{eq:nssol} if we use the isomorphism $\alg{sl}(2,\Real) \cong \alg{su}(1,1)$.
\unskip\footnote{
Explicitly, we can take $\tilde{S}_1 = i T_2 = i(E-F)$, $\tilde{S}_2 = -i T_1 = E+F$ and $\tilde{S}_3 = T_3 = i H$ as generators of $\alg{su}(1,1)$.
The solution~\eqref{eq:nssol} then takes the form
\begin{equation*}
\mathcal{R} \tilde{S}_1 = \tilde{S}_2 ~, \qquad \mathcal{R} \tilde{S}_2 = - \tilde{S}_1 ~,
\qquad
\mathcal{R} \tilde{S}_3 = 0 ~.
\end{equation*}
Using the isomorphism $\alg{su}(1,1) \cong \alg{sl}(2,\Real)$ given by $\tilde{S}_1 \to S_0$, $\tilde{S}_2 \to S_+ + S_-$ and $\tilde{S}_3 \to S_+ - S_-$, we see that this solution is equivalent to the solution~\eqref{eq:sl2rns}.}
Note that the solution~\eqref{eq:sl2rsplit} to the split mcYBe satisfies $\mathcal{R}^3 = \mathcal{R}$ and annihilates the non-compact generator $S_0$, while the solution~\eqref{eq:sl2rns} to the non-split mcYBe satisfies $\mathcal{R}^3 = -\mathcal{R}$ and annihilates the compact generator $S_+ - S_-$, the same as for the solution~\eqref{eq:su2sol} to the non-split mcYBe on $\alg{su}(2)$.
Parametrising the group-valued field as
\begin{equation}\label{eq:param3}
g = \exp\big(\frac{t+\psi}{2} (S_+ - S_-)\big) \exp \big(\rho S_0\big) \exp \big(\frac{t-\psi}{2} (S_+ - S_-) \big)~,
\end{equation}
and substituting into the action of the deformed model~\eqref{eq:ybdefgen} with $\tr = \Tr$ and the overall sign reversed, we find that the deformed target-space metric and B-field are given by
\begin{equation}\begin{split}
G & = \frac{2\hay}{1+\eta^2} \big(d\rho^2 - \cosh^2\rho(1 +\eta^2 \cosh^2\rho) dt^2
+ \sinh^2\rho(1- \eta^2\sinh^2\rho) d\psi^2
+2\eta^2\sinh^2\rho\cosh^2\rho dt d\psi \big) ~,
\\
B & = - \frac{2\hay\eta}{1+\eta^2} \sinh\rho \cosh\rho d\rho \wedge d(t-\psi) ~.
\end{split}\end{equation}

It is worth clarifying that we have used three different coordinate systems for $\grp{SL}(2,\Real) \cong \AdS_3$, based on the three parametrisations~\eqref{eq:gparamads},~\eqref{eq:param2} and~\eqref{eq:param3}.
The corresponding metrics are
\unskip\footnote{
In terms of embedding coordinates we have that $\AdS_3$ with unit radius is given by the locus of all points $(Z_0,Z_1,Z_2,Z_3) \in \Real^{2,2}$ satisfying $-Z_0^2 + Z_1^2 + Z_2^2 - Z_3^2 = - 1$.
Solving this constraint equation by setting
\begin{equation*}\begin{aligned}
Z_0 & = \frac{y_+ - y_-}{2z} ~, \qquad &
Z_1 & = \frac{y_+ + y_-}{2z} ~, \qquad &
Z_2 & = \frac{z}{2} - \frac{1}{2z} (1 - y_+ y_-)~, \qquad &
Z_3 & = \frac{z}{2} + \frac{1}{2z} (1 + y_+ y_-)~,
\\
Z_0 & = \sinh u \sinh y_0 ~, \qquad &
Z_1 & = \cosh u \sinh y_1 ~, \qquad &
Z_2 & = \sinh u \cosh y_0 ~, \qquad &
Z_3 & = \cosh u \cosh y_1 ~,
\\
Z_0 & = \cosh\rho \cos t ~, \qquad &
Z_1 & = \sinh \rho \cos \psi ~, \qquad &
Z_2 & = \sinh \rho \sin \psi ~, \qquad &
Z_3 & = \cosh \rho \sin t ~,
\end{aligned}\end{equation*}
and substituting into the metric on $\Real^{2,2}$, $G = -dZ_0^2 + dZ_1^2 + dZ_2^2 - dZ_3^2$, the induced metrics on $\AdS_3$ are given by~\eqref{eq:metrics} up to the factor of $2\hay$.}
\begin{equation}\begin{split}\label{eq:metrics}
G & = 2\hay\big(\frac{dz^2 + dy^+ dy^-}{z^2}\big) ~,
\\
G & = 2\hay \big(du^2 - \sinh^2 u dy_0^2 + \cosh^2u dy_1^2 \big) ~,
\\
G & = 2\hay \big(d\rho^2 - \cosh^2\rho dt^2 + \sinh^2\rho d\psi^2 \big) ~.
\end{split}\end{equation}
The first set are known as Poincar\'e coordinates, which cover the Poincar\'e patch of $\AdS_3$.
The second set also only cover a patch of $\AdS_3$, while the third set are known as global coordinates and cover the whole of $\AdS_3$.
The reason for introducing three different coordinate systems is that they are each adapted to the YB deformation for which we used them.
In particular, the isometries preserved by the deformation are manifest.

\section{Drinfel'd doubles, the \texorpdfstring{$\mathcal{E}$}{E}-model and Poisson-Lie T-duality}\label{sec:dd}

Thus far in our discussion of YB deformations we have used the fact that the right-acting $\grp{G}$ global symmetry is preserved and exploited the existence of an associated conserved current to construct a Lax connection.
It is also instructive to explore what happens to the broken left-acting symmetry.
Recalling that $\mathcal{R}$ is an antisymmetric solution to the (m)cYBe, varying the action~\eqref{eq:ybdefgen} under the infinitesimal variation $g \to e^{-\varepsilon} g \sim (1-\varepsilon) g$ we find the equations of motion
\begin{equation}\label{eq:eomk}
\partial_+ K_- + \partial_- K_+ + \eta [K_+,K_-]_{\mathcal{R}} = 0 ~, \qquad K_\pm = \frac{1}{1\pm \eta \mathcal{R}} k_\pm ~,
\end{equation}
where
\begin{equation}\label{eq:rbracket}
[X,Y]_{\mathcal{R}} = [X,\mathcal{R}Y] + [\mathcal{R}X,Y] ~, \qquad X,Y \in \alg{g} ~,
\end{equation}
is known as the R-bracket and defines a second Lie bracket on the vector space $\alg{g}$.
\iflectures
We denote the corresponding Lie algebra by $\alg{g}_{\mathcal{R}}$.
Note that in the undeformed $\eta \to 0$ limit we recover the conservation equation associated to the left-acting $\grp{G}$ global symmetry.

\question{Show that $[X,Y]_{\mathcal{R}}$ defines a Lie bracket on $\alg{g}$ assuming that $\mathcal{R}$ solves the (m)cYBe.}

\answer{We need to check that the Lie bracket satisfies the Jacobi identity
\begin{equation*}
[X,[Y,Z]_{\mathcal{R}}]_{\mathcal{R}} + [Y,[Z,X]_{\mathcal{R}}]_{\mathcal{R}} + [Z,[X,Y]_{\mathcal{R}}]_{\mathcal{R}} = 0~.
\end{equation*}
Just considering the first term we have
\begin{equation*}\begin{split}
[X,[Y,Z]_{\mathcal{R}}]_{\mathcal{R}} &= [\mathcal{R} X,[Y,Z]_{\mathcal{R}}] + [X,\mathcal{R}[Y,Z]_{\mathcal{R}}]
\\ & = [\mathcal{R} X,[Y,\mathcal{R}Z]] + [\mathcal{R}X,[\mathcal{R}Y,Z]] + [X,[\mathcal{R}Y,\mathcal{R}Z]] + c^2 [X,[Y,Z]] ~,
\end{split}\end{equation*}
where we have used the (m)cYBe.
Now combining this equation with its cyclic permutations we find
\begin{equation*}\begin{split}
& [X,[Y,Z]_{\mathcal{R}}]_{\mathcal{R}} + [Y,[Z,X]_{\mathcal{R}}]_{\mathcal{R}} + [Z,[X,Y]_{\mathcal{R}}]_{\mathcal{R}}
\\
& = [\mathcal{R} X,[Y,\mathcal{R}Z]] + [\mathcal{R}X,[\mathcal{R}Y,Z]] + [X,[\mathcal{R}Y,\mathcal{R}Z]] + c^2 [X,[Y,Z]]
\\ & \quad +
[\mathcal{R} Y,[Z,\mathcal{R}X]] + [\mathcal{R}Y,[\mathcal{R}Z,X]] + [Y,[\mathcal{R}Z,\mathcal{R}X]]+ c^2 [Y,[Z,X]]
\\ & \quad +
[\mathcal{R} Z,[X,\mathcal{R}Y]] + [\mathcal{R}Z,[\mathcal{R}X,Y]] + [Z,[\mathcal{R}X,\mathcal{R}Y]] + c^2 [ Z,[X,Y]]
\\
& = [\mathcal{R} X,[Y,\mathcal{R}Z]]
+ [Y,[\mathcal{R}Z,\mathcal{R}X]]
+ [\mathcal{R}Z,[\mathcal{R}X,Y]]
\\ & \quad
+ [\mathcal{R}X,[\mathcal{R}Y,Z]]
+ [\mathcal{R} Y,[Z,\mathcal{R}X]]
+ [Z,[\mathcal{R}X,\mathcal{R}Y]]
\\ & \quad
+ [X,[\mathcal{R}Y,\mathcal{R}Z]]
+ [\mathcal{R}Y,[\mathcal{R}Z,X]]
+ [\mathcal{R} Z,[X,\mathcal{R}Y]]
\\ & \quad
+c^2([X,[Y,Z]] + [Y,[Z,X]] + [Z,[X,Y]])
= 0 ~,
\end{split}\end{equation*}
where each line vanishes by the Jacobi identity for the Lie bracket $[X,Y]$.}
\else
\unskip\footnote{
Using the fact that $\mathcal{R}$ satisfies the (m)cYBE we have
\begin{equation*}\begin{split}
[X,[Y,Z]_{\mathcal{R}}]_{\mathcal{R}} &= [\mathcal{R} X,[Y,Z]_{\mathcal{R}}] + [X,\mathcal{R}[Y,Z]_{\mathcal{R}}]
= [\mathcal{R} X,[Y,\mathcal{R}Z]] + [\mathcal{R}X,[\mathcal{R}Y,Z]] + [X,[\mathcal{R}Y,\mathcal{R}Z]] + c^2 [X,[Y,Z]] ~.
\end{split}\end{equation*}
Combining this equation with its cyclic permutations we find
\begin{equation*}\begin{split}
[X,[Y,Z]_{\mathcal{R}}]_{\mathcal{R}} + [Y,[Z,X]_{\mathcal{R}}]_{\mathcal{R}} + [Z,[X,Y]_{\mathcal{R}}]_{\mathcal{R}}
& = [\mathcal{R} X,[Y,\mathcal{R}Z]] + [\mathcal{R}X,[\mathcal{R}Y,Z]] + [X,[\mathcal{R}Y,\mathcal{R}Z]] + c^2 [X,[Y,Z]]
\\ & \quad +
[\mathcal{R} Y,[Z,\mathcal{R}X]] + [\mathcal{R}Y,[\mathcal{R}Z,X]] + [Y,[\mathcal{R}Z,\mathcal{R}X]]+ c^2 [Y,[Z,X]]
\\ & \quad +
[\mathcal{R} Z,[X,\mathcal{R}Y]] + [\mathcal{R}Z,[\mathcal{R}X,Y]] + [Z,[\mathcal{R}X,\mathcal{R}Y]] + c^2 [ Z,[X,Y]]
= 0 ~,
\end{split}\end{equation*}
which vanishes by the Jacobi identity for the Lie bracket $[X,Y]$.
Therefore, the R-bracket also satisfies the Jacobi identity, hence defines a Lie bracket.
}
We denote the corresponding Lie algebra by $\alg{g}_{\mathcal{R}}$.
Note that in the undeformed $\eta \to 0$ limit we recover the conservation equation associated to the left-acting $\grp{G}$ global symmetry.
\fi

\subsection{Drinfel'd doubles and the \texorpdfstring{$\mathcal{E}$}{E}-model}

The appearance of the R-bracket is our first glimpse of a deeper underlying algebraic structure.
In addition to $\alg{g}$, let us introduce a vector space $\tilde{\alg{g}}$ such that $\dim\tilde{\alg{g}} = \dim\alg{g}$, their direct sum (as vector spaces)
\begin{equation}\label{eq:dd}
\alg{d} = \alg{g} \dotplus \tilde{\alg{g}}~,
\end{equation}
and an invertible linear map $\sigma : \alg{g} \to \tilde{\alg{g}}$.
We denote an element of $\alg{d}$ by $X + \sigma Y$, $X,Y\in\alg{g}$.
Given an antisymmetric solution to the (m)cYBe~\eqref{eq:ybe}, we can define the following Lie bracket $[[\cdot,\cdot]]$ on the vector space $\alg{d}$
\begin{equation}\begin{split}\label{eq:lb1}
[[X_1+\sigma Y_1,X_2+\sigma Y_2]] & = [X_1,X_2]
+ [X_1, \mathcal{R} Y_2] - \mathcal{R} [X_1,Y_2]
+ [\mathcal{R}Y_1, X_2] - \mathcal{R}[Y_1,X_2]
\\ & \quad
+ \sigma ([Y_1,Y_2]_{\mathcal{R}} + [X_1,Y_2] + [Y_1,X_2]) ~,
\end{split}\end{equation}
which satisfies the Jacobi identity as a consequence of the (m)cYBe.
The two subalgebras $\alg{g}$ and $\tilde{\alg{g}} \cong \alg{g}_{\mathcal{R}}$, where $\alg{g}_{\mathcal{R}}$ is the Lie algebra whose Lie bracket is the R-bracket~\eqref{eq:rbracket}, are maximally isotropic or Lagrangian with respect to the following invariant bilinear form
\begin{equation}\label{eq:bilinear}
\lang X_1+\sigma Y_1,X_2+\sigma Y_2\rang = \tr\big(X_1Y_2\big) + \tr\big(Y_1X_2\big) ~.
\end{equation}
The invariance of this bilinear form follows from the antisymmetry of $\mathcal{R}$.
Altogether, this means that $\alg{d}$ has the structure of a Drinfel'd double.
A Drinfel'd double is an even-dimensional Lie algebra $\alg{d}$ that admits a Manin triple $\{\alg{g},\tilde{\alg{g}},\lang\cdot,\cdot\rang\}$.
A Manin triple is comprised of two subalgebras $\alg{g}$ and $\tilde{\alg{g}}$ of $\alg{d}$, such that $\alg{d} = \alg{g} \dotplus \tilde{\alg{g}}$, together with an invariant bilinear form $\lang\cdot,\cdot\rang$ with respect to which these two subalgebras are Lagrangian.

If we define the injective map $\iota : \alg{g} \to \alg{d}$ that acts as $\iota X = \sigma X - \mathcal{R}X$, $X\in\alg{g}$, then $X + \sigma Y = X' + \iota Y$, where $X,Y\in\alg{g}$ and $X' = X+\mathcal{R}Y \in \alg{g}$.
Therefore, an element of $\alg{d}$ can be equivalently written in the form $X + \iota Y$, $X,Y \in \alg{g}$.
In this basis, the Lie bracket \eqref{eq:lb1} and invariant bilinear form \eqref{eq:bilinear} are then given by
\begin{equation}\label{eq:comiota}
\begin{split}
[[X_1+\iota Y_1,X_2+\iota Y_2]] & = [X_1,X_2] + c^2[Y_1,Y_2] + \iota([X_1,Y_2]+[Y_1,X_2]) ~,
\\
\lang X_1+\iota Y_1,X_2+\iota Y_2\rang & = \tr\big(X_1Y_2\big) + \tr\big(Y_1X_2\big) ~,
\end{split}
\end{equation}
where we have used that $\mathcal{R}$ is an antisymmetric solution to the (m)cYBe.
Therefore, we find that $\alg{d}$ is isomorphic to the real double $\alg{g} \oplus \alg{g}$, the complex double $\alg{g}^\Complex$, or the semi-abelian double $\alg{g} \ltimes \alg{g}^{\text{ab}}$, for $c = 1$, $c=i$ and $c=0$ respectively.
Explicitly, given $X+\iota Y \in \alg{d}$, we have that $\big(X-Y,X+Y\big) \in \alg{g} \oplus \alg{g}$, $X-iY \in \alg{g}^\Complex$ and $\big(X,-Y\big) \in \alg{g} \ltimes \alg{g}^{\text{ab}}$, for $c = 1$, $c=i$ and $c=0$ respectively.

The YB deformation~\eqref{eq:ybdefgen} can be rewritten as a first-order model, that is first-order in time derivatives, on the Drinfel'd double.
This model takes the form of an $\mathcal{E}$-model and its action is given by
\begin{equation}\begin{split}\label{eq:doubact}
\Act_{{\mathcal{E}}} & = N \Big(\int d^2 x \, \lang \gdsl^{-1} \partial_t \gdsl, \gdsl^{-1}\partial_x \gdsl \rang
+ \frac{1}{6} \int d^3 x \, \epsilon^{ijk} \lang \gdsl^{-1}\partial_i\gdsl,[[\gdsl^{-1}\partial_j\gdsl, \gdsl^{-1} \partial_k\gdsl]] \rang
\\ & \qquad \quad
- \int d^2 x\; \lang \gdsl^{-1} \partial_x \gdsl, \mathcal{E} \gdsl^{-1} \partial_x \gdsl \rang \Big) ~.
\end{split}\end{equation}
Here $\gdsl$ is a field valued in the Drinfel'd double $\grp{D}$ where $\Lie\grp{D} = \alg{d}$ and $\lang\cdot,\cdot\rang$ is the invariant bilinear form on $\alg{d}$.
\unskip\footnote{Note that the invariant bilinear form $\lang\cdot,\cdot\rang$ could be replaced by any invariant bilinear form on $\alg{d}$.
In particular, it does not need part of a Manin triple.
However, we do require that $\alg{d}$ has at least one Lagrangian subalgebra with respect to this bilinear form in order to integrate out the corresponding degrees of freedom and recover a relativistic second-order model.}
$N$ is an overall normalisation to be fixed later and $\mathcal{E} : \alg{d} \to \alg{d}$ is a constant linear operator that squares to the identity, $\mathcal{E}^2 = 1$, and satisfies
\begin{equation}
\lang \mathcal{E}\mathscr{X},\mathscr{Y} \rang = \lang \mathscr{X},\mathcal{E}\mathscr{Y} \rang ~, \qquad
\mathscr{X},\mathscr{Y} \in \alg{d} ~,
\end{equation}
or equivalently $\mathcal{E}^t = \mathcal{E}$.
Let us take any Lagrangian subalgebra of $\alg{d}$, which we denote $\alg{b}$, that is $\dim\alg{b} = \frac12\dim\alg{d}$ and $\lang\alg{b},\alg{b}\rang = 0$.
\iflectures
Redefining
\begin{equation}\label{eq:redefgb}
\gdsl \to b \gdsl, \qquad b \in \grp{B} ~,
\end{equation}
where $\Lie\grp{B} = \alg{b}$, we find that the action~\eqref{eq:doubact} only depends on $b$ through $b^{-1}\partial_x b \in \alg{b}$.
If $\mathcal{E}$ is such that $\Ad_\gdsl^{-1} \alg{b}$ and $\mathcal{E} \Ad_\gdsl^{-1} \alg{b}$ have trivial intersection, then we can integrate out the degrees of freedom in $b$ to obtain the action
\begin{equation}\begin{split}\label{eq:doubact1}
\Act_{\mathcal{E}_{\grp{B}}} & = N \Big(\frac12\int d^2 x \, \big(\lang \gdsl^{-1} \partial_+ \gdsl, \mathcal{E}\mathcal{P}(\mathcal{E}+1) \gdsl^{-1}\partial_- \gdsl \rang
- \lang \gdsl^{-1} \partial_- \gdsl, \mathcal{E}\mathcal{P}(\mathcal{E}-1) \gdsl^{-1}\partial_+ \gdsl \rang\big)
\\ & \qquad \quad
+ \frac{1}{6} \int d^3 x \, \epsilon^{ijk} \lang \gdsl^{-1}\partial_i \gdsl,[[\gdsl^{-1}\partial_j \gdsl, \gdsl^{-1} \partial_k \gdsl]] \rang\Big) ~,
\end{split}\end{equation}
where $\mathcal{P}$ is the projector with $\im \mathcal{P} = \mathcal{E} \Ad_\gdsl^{-1} \alg{b}$ and $\ker \mathcal{P} = \Ad_\gdsl^{-1} \alg{b}$.
It follows that the operators $\mathcal{E}\mathcal{P}(\mathcal{E}\pm1)$ are projectors with $\im \mathcal{E}\mathcal{P}(\mathcal{E}\pm1) = \Ad_\gdsl^{-1} \alg{b}$ and $\ker \mathcal{E}\mathcal{P}(\mathcal{E}\pm1) = \alg{e}_\mp$ where $\alg{e}_\pm$ are the eigenspaces of $\mathcal{E}$ with eigenvalues $\pm 1$.
To compensate the additional degrees of freedom that the redefinition \eqref{eq:redefgb} introduces, the action~\eqref{eq:doubact1} has a $\grp{B}$ gauge symmetry
\begin{equation}\label{eq:gsym}
\gdsl \to b \gdsl , \qquad b(t,x) \in \grp{B} ~,
\end{equation}
hence describes a relativistic second-order model on $\grp{B} \backslash \grp{D}$.

\question{Derive the relativistic second-order model~\eqref{eq:doubact1} from the $\mathcal{E}$-model~\eqref{eq:doubact}.}

\answer{Redefining $\gdsl \to b \gdsl$ in the action~\eqref{eq:doubact} we find
\begin{equation*}\begin{split}
\Act_{\mathcal{E}} & = N\Big(\int d^2x \, \big( \lang \gdsl^{-1}\partial_t \gdsl,\gdsl^{-1}\partial_x \gdsl\rang
+ \lang \Ad_{\gdsl}^{-1} b^{-1}\partial_t b, \gdsl^{-1} \partial_x \gdsl \rang
+ \lang \Ad_{\gdsl}^{-1} b^{-1} \partial_x b ,\gdsl^{-1} \partial_t \gdsl \rang \big)
\\
& \qquad\quad + \frac{1}{6} \int d^3 x \, \epsilon^{ijk} \big(\lang \gdsl^{-1}\partial_i \gdsl,[[\gdsl^{-1}\partial_j \gdsl, \gdsl^{-1} \partial_k \gdsl]] \rang
- 6 \partial_i \lang \Ad_{\gdsl}^{-1} b^{-1}\partial_j b, \gdsl^{-1} \partial_k \gdsl \rang \big)
\\
& \qquad\quad - \int d^2x \,
\lang (\Ad_{\gdsl}^{-1}b^{-1} \partial_x b + \gdsl^{-1} \partial_x \gdsl ), \mathcal{E}(\Ad_{\gdsl}^{-1} b^{-1}\partial_x b + \gdsl^{-1} \partial_x \gdsl) \rang \Big)
\\
& = N\Big(\int d^2x \, \big( \lang \gdsl^{-1}\partial_t \gdsl,\gdsl^{-1}\partial_x \gdsl\rang
+ \lang \Ad_{\gdsl}^{-1} b^{-1}\partial_t b, \gdsl^{-1} \partial_x \gdsl \rang
+ \lang \Ad_{\gdsl}^{-1} b^{-1} \partial_x b ,\gdsl^{-1} \partial_t \gdsl \rang \big)
\\ & \qquad\quad
- \int d^2 x \, \big(
\lang \Ad_{\gdsl}^{-1} b^{-1}\partial_t b, \gdsl^{-1} \partial_x \gdsl \rang
- \lang \Ad_{\gdsl}^{-1} b^{-1} \partial_x b , \gdsl^{-1} \partial_t \gdsl \rang \big)
\\
& \qquad\quad - \int d^2x \,
\lang (\Ad_{\gdsl}^{-1}b^{-1} \partial_x b + \gdsl^{-1} \partial_x \gdsl ), \mathcal{E}(\Ad_{\gdsl}^{-1} b^{-1}\partial_x b + \gdsl^{-1} \partial_x \gdsl) \rang
\\
& \qquad\quad + \frac{1}{6} \int d^3 x \, \epsilon^{ijk} \lang \gdsl^{-1}\partial_i \gdsl,[[\gdsl^{-1}\partial_j \gdsl, \gdsl^{-1} \partial_k \gdsl]] \rang\Big)
\\
& = N\Big(\int d^2x \, \big(
\lang \gdsl^{-1} \partial_t \gdsl, \mathcal{E} \gdsl^{-1} \partial_t \gdsl\rang
- \lang \gdsl^{-1} \partial_t \gdsl,\gdsl^{-1}\partial_x \gdsl\rang \big)
\\
& \qquad\quad - \int d^2x \,
\lang (\Ad_{\gdsl}^{-1} b^{-1} \partial_x b + \gdsl^{-1} \partial_x \gdsl - \mathcal{E} \gdsl^{-1} \partial_t \gdsl  ), \mathcal{E} (\Ad_{\gdsl}^{-1} b^{-1}\partial_x b + \gdsl^{-1}  \partial_x \gdsl- \mathcal{E} \gdsl^{-1} \partial_t \gdsl) \rang
\\
& \qquad\quad + \frac{1}{6} \int d^3 x \, \epsilon^{ijk} \lang \gdsl^{-1}\partial_i \gdsl,[[\gdsl^{-1}\partial_j \gdsl, \gdsl^{-1} \partial_k \gdsl]] \rang\Big) ~,
\end{split}\end{equation*}
where we have used $\mathcal{E}^2 = 1$ and $\mathcal{E}^t = \mathcal{E}$, and that $\alg{b}$ is a Lagrangian subalgebra, hence
\begin{equation*}
\lang b^{-1} \partial_t b , b^{-1} \partial_x b \rang =
\lang b^{-1} \partial_i b , [[ b^{-1} \partial_j b,b^{-1}\partial_k b]] \rang = 0 ~.
\end{equation*}
Therefore, we find that the action only depends on $b$ through $b^{-1}\partial_x b \in \alg{b}$.

If $\mathcal{E}$ is such that $\Ad_\gdsl^{-1} \alg{b}$ and $\mathcal{E} \Ad_\gdsl^{-1} \alg{b}$ have trivial intersection, then we can integrate out the degrees of freedom in $b$.
To this end, we introduce the projector $\mathcal{P}$ with $\im \mathcal{P} = \mathcal{E} \Ad_\gdsl^{-1} \alg{b}$ and $\ker \mathcal{P} = \Ad_\gdsl^{-1} \alg{b}$.
The projector $1-\mathcal{P}$ therefore has $\im(1-\mathcal{P}) = \ker \mathcal{P} = \Ad_\gdsl^{-1} \alg{b}$ and $\ker(1-\mathcal{P}) = \im\mathcal{P} = \mathcal{E} \Ad_\gdsl^{-1} \alg{b}$.
These projectors also satisfy
\begin{equation*}
\lang\mathcal{P}\mathscr{X},\mathcal{P}\mathscr{Y}\rang = \lang(1-\mathcal{P})\mathscr{X},(1-\mathcal{P})\mathscr{Y}\rang = 0 ~, \qquad \mathscr{X},\mathscr{Y} \in \alg{d} ~,
\end{equation*}
which follows from $\mathcal{E}^2 = 1$, $\mathcal{E}^t = \mathcal{E}$, the invariance of the bilinear form $\lang\cdot,\cdot\rang$ and the fact that $\alg{b}$ is a Lagrangian subalgebra.
Defining $\tilde{X} = \Ad_{\gdsl}^{-1} b^{-1} \partial_x b + (1-\mathcal{P})(\gdsl^{-1} \partial_x \gdsl - \mathcal{E} \gdsl^{-1} \partial_t \gdsl) \in \Ad_\gdsl^{-1} \alg{b}$, we have
\begin{equation*}\begin{split}
& \lang (\Ad_{\gdsl}^{-1} b^{-1} \partial_x b + \gdsl^{-1} \partial_x \gdsl - \mathcal{E} \gdsl^{-1} \partial_t \gdsl  ), \mathcal{E} (\Ad_{\gdsl}^{-1} b^{-1}\partial_x b + \gdsl^{-1}  \partial_x \gdsl- \mathcal{E} \gdsl^{-1} \partial_t \gdsl) \rang
\\
& = \lang \tilde{X} +\mathcal{P}( \gdsl^{-1} \partial_x \gdsl - \mathcal{E} \gdsl^{-1} \partial_t \gdsl  ), \mathcal{E} (\tilde X + \mathcal{P} (\gdsl^{-1}  \partial_x \gdsl- \mathcal{E} \gdsl^{-1} \partial_t \gdsl)) \rang
\\
& = \lang \tilde{X} , \mathcal{E} \tilde{X} \rang  + \lang \mathcal{P}( \gdsl^{-1} \partial_x \gdsl - \mathcal{E} \gdsl^{-1} \partial_t \gdsl  ), \mathcal{E} \mathcal{P} (\gdsl^{-1}  \partial_x \gdsl- \mathcal{E} \gdsl^{-1} \partial_t \gdsl) \rang
+2 \lang \tilde{X},\mathcal{E} \mathcal{P} (\gdsl^{-1}  \partial_x \gdsl- \mathcal{E} \gdsl^{-1} \partial_t \gdsl)\rang
\\
& = \lang \tilde{X} , \mathcal{E} \tilde{X} \rang  + \lang \mathcal{P}( \gdsl^{-1} \partial_x \gdsl - \mathcal{E} \gdsl^{-1} \partial_t \gdsl  ), \mathcal{E} \mathcal{P} (\gdsl^{-1}  \partial_x \gdsl- \mathcal{E} \gdsl^{-1} \partial_t \gdsl) \rang ~,
\end{split}\end{equation*}
where we have used that $\mathcal{E} \mathcal{P} (\gdsl^{-1}  \partial_x \gdsl- \mathcal{E} \gdsl^{-1} \partial_t \gdsl) \in \Ad_\gdsl^{-1} \alg{b}$, hence $\lang \tilde{X},\mathcal{E} (\mathcal{P} (\gdsl^{-1}  \partial_x \gdsl- \mathcal{E} \gdsl^{-1} \partial_t \gdsl)\rang=0$.
Integrating out the degrees of freedom in $b$ is then equivalent to integrating out $\tilde X$, which gives the action
\begin{equation*}\begin{split}
\Act_{\mathcal{E}_{\grp{B}}} & = N\Big(\int d^2x \, \big(
\lang \gdsl^{-1} \partial_t \gdsl, \mathcal{E} \gdsl^{-1} \partial_t \gdsl\rang
- \lang \gdsl^{-1} \partial_t \gdsl,\gdsl^{-1}\partial_x \gdsl\rang \big)
\\
& \qquad\quad - \int d^2x \,
\lang \mathcal{P} (\gdsl^{-1} \partial_x \gdsl - \mathcal{E} \gdsl^{-1} \partial_t \gdsl  ), \mathcal{E} \mathcal{P} (\gdsl^{-1}  \partial_x \gdsl- \mathcal{E} \gdsl^{-1} \partial_t \gdsl) \rang
\\
& \qquad\quad + \frac{1}{6} \int d^3 x \, \epsilon^{ijk} \lang \gdsl^{-1}\partial_i \gdsl,[[\gdsl^{-1}\partial_j \gdsl, \gdsl^{-1} \partial_k \gdsl]] \rang\Big)
\\
& = N\Big(\frac{1}{4}\int d^2x \, \big(
\lang \gdsl^{-1} \partial_+ \gdsl, ((\mathcal{E}-1) - (\mathcal{E}-1) \mathcal{P}^t \mathcal{E} \mathcal{P} (\mathcal{E}-1)) \gdsl^{-1} \partial_+ \gdsl\rang
\\ & \qquad \qquad \qquad \qquad +
\lang \gdsl^{-1} \partial_- \gdsl, ((\mathcal{E}+1) - (\mathcal{E}+1) \mathcal{P}^t \mathcal{E} \mathcal{P} (\mathcal{E}+1)) \gdsl^{-1} \partial_- \gdsl\rang
\\ & \qquad \qquad \qquad \qquad +
\lang \gdsl^{-1} \partial_+ \gdsl, ((\mathcal{E}+1) - (\mathcal{E}-1) \mathcal{P}^t \mathcal{E} \mathcal{P} (\mathcal{E}+1)) \gdsl^{-1} \partial_- \gdsl\rang
\\ & \qquad \qquad \qquad \qquad +
\lang \gdsl^{-1} \partial_- \gdsl, ((\mathcal{E}-1) - (\mathcal{E}+1) \mathcal{P}^t \mathcal{E} \mathcal{P} (\mathcal{E}-1)) \gdsl^{-1} \partial_+ \gdsl\rang \big)
\\
& \qquad\quad + \frac{1}{6} \int d^3 x \, \epsilon^{ijk} \lang \gdsl^{-1}\partial_i \gdsl,[[\gdsl^{-1}\partial_j \gdsl, \gdsl^{-1} \partial_k \gdsl]] \rang\Big) ~,
\end{split}\end{equation*}
where
\begin{equation*}
\lang \mathcal{P}^t \mathscr{X} ,\mathscr{Y} \rang  = \lang \mathscr{X},\mathcal{P}\mathscr{Y}\rang ~, \qquad
\mathscr{X},\mathscr{Y} \in \alg{d} ~,
\end{equation*}
and $\gdsl^{-1}\partial_\pm \gdsl = \gdsl^{-1}\partial_t \gdsl \pm \gdsl^{-1}\partial_x \gdsl$.
From $\lang\mathcal{P}\mathscr{X},\mathcal{P}\mathscr{Y}\rang = 0$ we have $\mathcal{P}^t\mathcal{P} = 0$.
In addition, since $\mathcal{E}^2 = 1$, $(\mathcal{E}\mathcal{P} \mathcal{E})^2 = \mathcal{E}\mathcal{P} \mathcal{E}$, with $\ker \mathcal{E}\mathcal{P} \mathcal{E} = \im \mathcal{P}$ and $\im \mathcal{E}\mathcal{P}\mathcal{E} = \ker \mathcal{P}$, which implies that the projector $\mathcal{E}\mathcal{P}\mathcal{E} = 1- \mathcal{P}$.
Combining these identities, and using $\mathcal{E}^2 = 1$  and $\mathcal{E}^t = \mathcal{E}$, it is straightforward to show that
\begin{equation*}\begin{split}
(\mathcal{E}\pm1) - (\mathcal{E}\pm1) \mathcal{P}^t \mathcal{E} \mathcal{P} (\mathcal{E}\pm1) & = 0 ~,
\\
(\mathcal{E}\pm1) - (\mathcal{E}\mp1) \mathcal{P}^t \mathcal{E} \mathcal{P} (\mathcal{E}\pm1) & = \pm 2\mathcal{E}\mathcal{P}(\mathcal{E}\pm1) ~.
\end{split}\end{equation*}
Substituting into the action $\Act_{\mathcal{E}_{\grp{B}}}$ above we indeed recover eq.~\eqref{eq:doubact1} as required.
}
\else
Redefining
\begin{equation}\label{eq:redefgb}
\gdsl \to b \gdsl, \qquad b \in \grp{B} ~,
\end{equation}
where $\Lie\grp{B} = \alg{b}$, in the action~\eqref{eq:doubact} we find
\begin{equation}\begin{split}
\Act_{\mathcal{E}} & = N\Big(\int d^2x \, \big( \lang \gdsl^{-1}\partial_t \gdsl,\gdsl^{-1}\partial_x \gdsl\rang
+ \lang \Ad_{\gdsl}^{-1} b^{-1}\partial_t b, \gdsl^{-1} \partial_x \gdsl \rang
+ \lang \Ad_{\gdsl}^{-1} b^{-1} \partial_x b ,\gdsl^{-1} \partial_t \gdsl \rang \big)
\\
& \qquad\quad + \frac{1}{6} \int d^3 x \, \epsilon^{ijk} \big(\lang \gdsl^{-1}\partial_i \gdsl,[[\gdsl^{-1}\partial_j \gdsl, \gdsl^{-1} \partial_k \gdsl]] \rang
- 6 \partial_i \lang \Ad_{\gdsl}^{-1} b^{-1}\partial_j b, \gdsl^{-1} \partial_k \gdsl \rang \big)
\\
& \qquad\quad - \int d^2x \,
\lang (\Ad_{\gdsl}^{-1}b^{-1} \partial_x b + \gdsl^{-1} \partial_x \gdsl ), \mathcal{E}(\Ad_{\gdsl}^{-1} b^{-1}\partial_x b + \gdsl^{-1} \partial_x \gdsl) \rang \Big)
\\
& = N\Big(\int d^2x \, \big( \lang \gdsl^{-1}\partial_t \gdsl,\gdsl^{-1}\partial_x \gdsl\rang
+ \lang \Ad_{\gdsl}^{-1} b^{-1}\partial_t b, \gdsl^{-1} \partial_x \gdsl \rang
+ \lang \Ad_{\gdsl}^{-1} b^{-1} \partial_x b ,\gdsl^{-1} \partial_t \gdsl \rang \big)
\\ & \qquad\quad
- \int d^2 x \, \big(
\lang \Ad_{\gdsl}^{-1} b^{-1}\partial_t b, \gdsl^{-1} \partial_x \gdsl \rang
- \lang \Ad_{\gdsl}^{-1} b^{-1} \partial_x b , \gdsl^{-1} \partial_t \gdsl \rang \big)
\\
& \qquad\quad - \int d^2x \,
\lang (\Ad_{\gdsl}^{-1}b^{-1} \partial_x b + \gdsl^{-1} \partial_x \gdsl ), \mathcal{E}(\Ad_{\gdsl}^{-1} b^{-1}\partial_x b + \gdsl^{-1} \partial_x \gdsl) \rang
\\
& \qquad\quad + \frac{1}{6} \int d^3 x \, \epsilon^{ijk} \lang \gdsl^{-1}\partial_i \gdsl,[[\gdsl^{-1}\partial_j \gdsl, \gdsl^{-1} \partial_k \gdsl]] \rang\Big)
\\
& = N\Big(\int d^2x \, \big(
\lang \gdsl^{-1} \partial_t \gdsl, \mathcal{E} \gdsl^{-1} \partial_t \gdsl\rang
- \lang \gdsl^{-1} \partial_t \gdsl,\gdsl^{-1}\partial_x \gdsl\rang \big)
\\
& \qquad\quad - \int d^2x \,
\lang (\Ad_{\gdsl}^{-1} b^{-1} \partial_x b + \gdsl^{-1} \partial_x \gdsl - \mathcal{E} \gdsl^{-1} \partial_t \gdsl  ), \mathcal{E} (\Ad_{\gdsl}^{-1} b^{-1}\partial_x b + \gdsl^{-1}  \partial_x \gdsl- \mathcal{E} \gdsl^{-1} \partial_t \gdsl) \rang
\\
& \qquad\quad + \frac{1}{6} \int d^3 x \, \epsilon^{ijk} \lang \gdsl^{-1}\partial_i \gdsl,[[\gdsl^{-1}\partial_j \gdsl, \gdsl^{-1} \partial_k \gdsl]] \rang\Big) ~,
\end{split}\end{equation}
where we have used $\mathcal{E}^2 = 1$ and $\mathcal{E}^t = \mathcal{E}$, and that $\alg{b}$ is a Lagrangian subalgebra, hence
\begin{equation}
\lang b^{-1} \partial_t b , b^{-1} \partial_x b \rang =
\lang b^{-1} \partial_i b , [[ b^{-1} \partial_j b,b^{-1}\partial_k b]] \rang = 0 ~.
\end{equation}
Therefore, we find that the action~\eqref{eq:doubact} only depends on $b$ through $b^{-1}\partial_x b \in \alg{b}$.

If $\mathcal{E}$ is such that $\Ad_\gdsl^{-1} \alg{b}$ and $\mathcal{E} \Ad_\gdsl^{-1} \alg{b}$ have trivial intersection, then we can integrate out the degrees of freedom in $b$.
To this end, we introduce the projector $\mathcal{P}$ with $\im \mathcal{P} = \mathcal{E} \Ad_\gdsl^{-1} \alg{b}$ and $\ker \mathcal{P} = \Ad_\gdsl^{-1} \alg{b}$.
The projector $1-\mathcal{P}$ therefore has $\im(1-\mathcal{P}) = \ker \mathcal{P} = \Ad_\gdsl^{-1} \alg{b}$ and $\ker(1-\mathcal{P}) = \im\mathcal{P} = \mathcal{E} \Ad_\gdsl^{-1} \alg{b}$.
These projectors also satisfy
\begin{equation}
\lang\mathcal{P}\mathscr{X},\mathcal{P}\mathscr{Y}\rang = \lang(1-\mathcal{P})\mathscr{X},(1-\mathcal{P})\mathscr{Y}\rang = 0 ~, \qquad \mathscr{X},\mathscr{Y} \in \alg{d} ~,
\end{equation}
which follows from $\mathcal{E}^2 = 1$, $\mathcal{E}^t = \mathcal{E}$, the invariance of the bilinear form $\lang\cdot,\cdot\rang$ and the fact that $\alg{b}$ is a Lagrangian subalgebra.
Defining $\mathscr{B} = \Ad_{\gdsl}^{-1} b^{-1} \partial_x b + (1-\mathcal{P})(\gdsl^{-1} \partial_x \gdsl - \mathcal{E} \gdsl^{-1} \partial_t \gdsl) \in \Ad_\gdsl^{-1} \alg{b}$, we have
\begin{equation}\begin{split}
& \lang (\Ad_{\gdsl}^{-1} b^{-1} \partial_x b + \gdsl^{-1} \partial_x \gdsl - \mathcal{E} \gdsl^{-1} \partial_t \gdsl  ), \mathcal{E} (\Ad_{\gdsl}^{-1} b^{-1}\partial_x b + \gdsl^{-1}  \partial_x \gdsl- \mathcal{E} \gdsl^{-1} \partial_t \gdsl) \rang
\\
& = \lang \mathscr{B} +\mathcal{P}( \gdsl^{-1} \partial_x \gdsl - \mathcal{E} \gdsl^{-1} \partial_t \gdsl  ), \mathcal{E} (\mathscr{B} + \mathcal{P} (\gdsl^{-1}  \partial_x \gdsl- \mathcal{E} \gdsl^{-1} \partial_t \gdsl)) \rang
\\
& = \lang \mathscr{B} , \mathcal{E} \mathscr{B} \rang  + \lang \mathcal{P}( \gdsl^{-1} \partial_x \gdsl - \mathcal{E} \gdsl^{-1} \partial_t \gdsl  ), \mathcal{E} \mathcal{P} (\gdsl^{-1}  \partial_x \gdsl- \mathcal{E} \gdsl^{-1} \partial_t \gdsl) \rang ~,
\end{split}\end{equation}
where we have used that $\mathcal{E} \mathcal{P} (\gdsl^{-1}  \partial_x \gdsl- \mathcal{E} \gdsl^{-1} \partial_t \gdsl) \in \Ad_\gdsl^{-1} \alg{b}$, hence $\lang \mathscr{B},\mathcal{E} (\mathcal{P} (\gdsl^{-1}  \partial_x \gdsl- \mathcal{E} \gdsl^{-1} \partial_t \gdsl)\rang=0$.
Integrating out the degrees of freedom in $b$ is then equivalent to integrating out $\mathscr{B}$, which gives the action
\begin{equation}\begin{split}\label{eq:acteb1}
\Act_{\mathcal{E}_{\grp{B}}} & = N\Big(\int d^2x \, \big(
\lang \gdsl^{-1} \partial_t \gdsl, \mathcal{E} \gdsl^{-1} \partial_t \gdsl\rang
- \lang \gdsl^{-1} \partial_t \gdsl,\gdsl^{-1}\partial_x \gdsl\rang \big)
\\
& \qquad\quad - \int d^2x \,
\lang \mathcal{P} (\gdsl^{-1} \partial_x \gdsl - \mathcal{E} \gdsl^{-1} \partial_t \gdsl  ), \mathcal{E} \mathcal{P} (\gdsl^{-1}  \partial_x \gdsl- \mathcal{E} \gdsl^{-1} \partial_t \gdsl) \rang
\\
& \qquad\quad + \frac{1}{6} \int d^3 x \, \epsilon^{ijk} \lang \gdsl^{-1}\partial_i \gdsl,[[\gdsl^{-1}\partial_j \gdsl, \gdsl^{-1} \partial_k \gdsl]] \rang\Big)
\\
& = N\Big(\frac{1}{4}\int d^2x \, \big(
\lang \gdsl^{-1} \partial_+ \gdsl, ((\mathcal{E}-1) - (\mathcal{E}-1) \mathcal{P}^t \mathcal{E} \mathcal{P} (\mathcal{E}-1)) \gdsl^{-1} \partial_+ \gdsl\rang
\\ & \qquad \qquad \qquad \qquad +
\lang \gdsl^{-1} \partial_- \gdsl, ((\mathcal{E}+1) - (\mathcal{E}+1) \mathcal{P}^t \mathcal{E} \mathcal{P} (\mathcal{E}+1)) \gdsl^{-1} \partial_- \gdsl\rang
\\ & \qquad \qquad \qquad \qquad +
\lang \gdsl^{-1} \partial_+ \gdsl, ((\mathcal{E}+1) - (\mathcal{E}-1) \mathcal{P}^t \mathcal{E} \mathcal{P} (\mathcal{E}+1)) \gdsl^{-1} \partial_- \gdsl\rang
\\ & \qquad \qquad \qquad \qquad +
\lang \gdsl^{-1} \partial_- \gdsl, ((\mathcal{E}-1) - (\mathcal{E}+1) \mathcal{P}^t \mathcal{E} \mathcal{P} (\mathcal{E}-1)) \gdsl^{-1} \partial_+ \gdsl\rang \big)
\\
& \qquad\quad + \frac{1}{6} \int d^3 x \, \epsilon^{ijk} \lang \gdsl^{-1}\partial_i \gdsl,[[\gdsl^{-1}\partial_j \gdsl, \gdsl^{-1} \partial_k \gdsl]] \rang\Big) ~,
\end{split}\end{equation}
where
\begin{equation*}
\lang \mathcal{P}^t \mathscr{X} ,\mathscr{Y} \rang  = \lang \mathscr{X},\mathcal{P}\mathscr{Y}\rang ~, \qquad
\mathscr{X},\mathscr{Y} \in \alg{d} ~,
\end{equation*}
and $\gdsl^{-1}\partial_\pm \gdsl = \gdsl^{-1}\partial_t \gdsl \pm \gdsl^{-1}\partial_x \gdsl$.
From $\lang\mathcal{P}\mathscr{X},\mathcal{P}\mathscr{Y}\rang = 0$ we have $\mathcal{P}^t\mathcal{P} = 0$.
In addition, since $\mathcal{E}^2 = 1$, $(\mathcal{E}\mathcal{P} \mathcal{E})^2 = \mathcal{E}\mathcal{P} \mathcal{E}$, with $\ker \mathcal{E}\mathcal{P} \mathcal{E} = \im \mathcal{P}$ and $\im \mathcal{E}\mathcal{P}\mathcal{E} = \ker \mathcal{P}$, which implies that the projector $\mathcal{E}\mathcal{P}\mathcal{E} = 1- \mathcal{P}$.
Combining these identities, and using $\mathcal{E}^2 = 1$  and $\mathcal{E}^t = \mathcal{E}$, it is straightforward to show that
\begin{equation*}\begin{split}
(\mathcal{E}\pm1) - (\mathcal{E}\pm1) \mathcal{P}^t \mathcal{E} \mathcal{P} (\mathcal{E}\pm1) & = 0 ~,
\\
(\mathcal{E}\pm1) - (\mathcal{E}\mp1) \mathcal{P}^t \mathcal{E} \mathcal{P} (\mathcal{E}\pm1) & = \pm 2\mathcal{E}\mathcal{P}(\mathcal{E}\pm1) ~.
\end{split}\end{equation*}
Substituting into the action~\eqref{eq:acteb1} we arrive at the relativistic second-order action
\begin{equation}\begin{split}\label{eq:doubact1}
\Act_{\mathcal{E}_{\grp{B}}} & = N \Big(\frac12\int d^2 x \, \big(\lang \gdsl^{-1} \partial_+ \gdsl, \mathcal{E}\mathcal{P}(\mathcal{E}+1) \gdsl^{-1}\partial_- \gdsl \rang
- \lang \gdsl^{-1} \partial_- \gdsl, \mathcal{E}\mathcal{P}(\mathcal{E}-1) \gdsl^{-1}\partial_+ \gdsl \rang\big)
\\ & \qquad \quad
+ \frac{1}{6} \int d^3 x \, \epsilon^{ijk} \lang \gdsl^{-1}\partial_i \gdsl,[[\gdsl^{-1}\partial_j \gdsl, \gdsl^{-1} \partial_k \gdsl]] \rang\Big) ~.
\end{split}\end{equation}
The operators $\mathcal{E}\mathcal{P}(\mathcal{E}\pm1)$ are projectors with $\im \mathcal{E}\mathcal{P}(\mathcal{E}\pm1) = \Ad_\gdsl^{-1} \alg{b}$ and $\ker \mathcal{E}\mathcal{P}(\mathcal{E}\pm1) = \alg{e}_\mp$ where $\alg{e}_\pm$ are the eigenspaces of $\mathcal{E}$ with eigenvalues $\pm 1$.
To compensate the additional degrees of freedom that the redefinition \eqref{eq:redefgb} introduces, the action~\eqref{eq:doubact1} has a $\grp{B}$ gauge symmetry
\begin{equation}\label{eq:gsym}
\gdsl \to b \gdsl , \qquad b(t,x) \in \grp{B} ~,
\end{equation}
hence describes a relativistic second-order model on $\grp{B} \backslash \grp{D}$.
\fi

To show that the YB deformation~\eqref{eq:ybdefgen} can be written as an $\mathcal{E}$-model we set
\begin{equation}
\grp{B} = \widetilde{\grp{G}} ~, \qquad \alg{b} = \tilde{\alg{g}} \cong \alg{g}_{\mathcal{R}} ~,
\end{equation}
and, writing a general element of the Drinfel'd double $\alg{d}$ as $X + \iota Y$, $X,Y\in\alg{g}$, define the operator $\mathcal{E}$ to act as
\begin{equation}\label{eq:eop}
\mathcal{E}(X + \iota Y) = - \eta^{-1} Y - \eta \iota X ~, \qquad X,Y\in\alg{g} ~.
\end{equation}
Assuming that the decomposition \eqref{eq:dd} lifts to the group, that is the quotient $\widetilde{\grp{G}} \backslash\grp{D}$ can be identified with $\grp{G}$, we parametrise
\begin{equation}
\gdsl = \tilde{g} g ~, \qquad \tilde{g} \in \widetilde{\grp{G}} ~, \quad g \in \grp{G} ~,
\end{equation}
and use the gauge symmetry \eqref{eq:gsym} to fix $\tilde{g} = 1$ so that
\begin{equation}\label{eq:gaugechoice}
\gdsl = g \in \grp{G} ~.
\end{equation}

We now determine the action of the projectors $\mathcal{E}\mathcal{P}(\mathcal{E}\pm1)$, which are defined by their image and kernel.
In the current setup we have $\im \mathcal{E} \mathcal{P}(\mathcal{E}\pm 1) = \Ad_g^{-1} \tilde{\alg{g}}$ and $\ker \mathcal{E} \mathcal{P}(\mathcal{E}\pm 1) = \alg{e}_\mp$.
Observing that general elements of $\alg{e}_\mp$ take the form $X \pm \eta \iota X$, $X\in\alg{g}$, we consider the ansatz
\begin{equation}\label{eq:ansatzeta}
\mathcal{E}\mathcal{P}(\mathcal{E}\pm1)(X + \iota Y) = (\iota + \mathcal{R}_g)f_\pm(\mathcal{R}_g) (Y \mp \eta X) ~, \qquad X,Y\in\alg{g} ~,
\end{equation}
which ensures that $\alg{e}_\mp$ lies in the kernel.
Furthermore, recalling that $\iota + \mathcal{R} = \sigma : \alg{g} \to \tilde{\alg{g}}$ and noting that the commutation relations \eqref{eq:comiota} imply that $\Ad_g$ commutes with $\iota$, we also have that the image is contained in $\Ad_g \tilde{\alg{g}}$.
Demanding that $\mathcal{E}\mathcal{P}(\mathcal{E}\pm1)$ are projectors with the required image and kernel we find
\begin{equation}
f_\pm(\mathcal{R}_g) = \frac{1}{1\mp \eta \mathcal{R}_g} ~,
\end{equation}
and we arrive at the following expression for their action
\begin{equation}\label{eq:actprojepe}
\mathcal{E}\mathcal{P}(\mathcal{E}\pm1) (X + \iota Y) = (\iota + \mathcal{R}_g) \frac{1}{1\mp \eta \mathcal{R}_g} (Y\mp \eta X) ~,
\qquad X,Y \in \alg{g} ~.
\end{equation}

To conclude, we fix the gauge~\eqref{eq:gaugechoice} in the action~\eqref{eq:doubact1}.
Using the action of the projectors in eq.~\eqref{eq:actprojepe} and the invariant bilinear form~\eqref{eq:comiota} we find
\begin{equation}\begin{split}\label{eq:doubact2}
\Act_{\mathcal{E}_{\widetilde{\grp{G}}}} & = N \Big(\frac12 \int d^2x  \,
\big(
\lang g^{-1}\partial_+ g,
\iota \frac{-\eta}{1 -\eta \mathcal{R}_g} g^{-1}\partial_- g \rang
-
\lang g^{-1}\partial_- g,
\iota \frac{\eta}{1 +\eta \mathcal{R}_g} g^{-1}\partial_+ g \rang \big) \Big)
\\ & = N \Big(\frac12\int d^2 x \, \tr\big( g^{-1}\partial_+ g \frac{-\eta}{1-\eta \mathcal{R}_g} g^{-1}\partial_- g - g^{-1}\partial_-g \frac{\eta}{1+\eta \mathcal{R}_g} g^{-1}\partial_+ g\big) \Big)
\\ & = - \frac{\hay}{2} \int d^2 x \, \tr\big( g^{-1}\partial_+ g \frac{1}{1-\eta \mathcal{R}_g} g^{-1}\partial_- g \big)
\\ & = \Act_\indrm{YB} ~,
\end{split}\end{equation}
where we have set
\begin{equation}
N = \frac{\hay}{2\eta} ~,
\end{equation}
and used that $\alg{g}$ is a Lagrangian subalgebra with respect to the bilinear form $\lang \cdot,\cdot\rang$, hence
\begin{equation}
\lang g^{-1}\partial_i g,[[g^{-1}\partial_j g,g^{-1}\partial_k g]]\rang = 0 ~,
\end{equation}
along with the antisymmetry of the operator $\mathcal{R}$.
We have therefore recovered the action of the YB deformation of the PCM~\eqref{eq:ybdefgen} as claimed.

\subsection{Poisson-Lie T-duality}

The first-order formalism allows us to introduce the notion of Poisson-Lie T-duality.
For our purposes, Poisson-Lie T-dual models are found by integrating out the degrees of freedom associated to different Lagrangian subalgebras in the $\mathcal{E}$-model~\eqref{eq:doubact}.
For example, rather than integrating out the degrees of freedom associated to $\tilde{\alg{g}} \cong \alg{g}_{\mathcal{R}}$, we could instead choose to integrate out those associated to the Lagrangian subalgebra $\alg{g}$ to define a relativistic second-order model on $\grp{G}\backslash\grp{D}$.
Often there are many inequivalent Lagrangian subalgebras, leading to different Poisson-Lie T-dual models.
Poisson-Lie T-duality is a generalisation of T-duality and non-abelian T-duality to models without manifest global symmetries, but whose equations of motion can be written in the form of a non-commutative conservation law, such as in eq.~\eqref{eq:eomk}.
Moreover, all of these worldsheet dualities can be understood as canonical transformations on phase space, hence they preserve classical integrability.

As we have seen, starting from the $\mathcal{E}$-model~\eqref{eq:doubact} with $\mathcal{E}$ defined in~\eqref{eq:eop}, we can recover the YB deformation by integrating out the degrees of freedom associated to the Lagrangian subalgebra $\tilde{\alg{g}} \cong \alg{g}_{\mathcal{R}}$.
As a particular example of a Poisson-Lie T-dual model, let us instead integrate out the degrees of freedom associated to the Lagrangian subalgebra $\alg{g}$, that is we set
\begin{equation}
\grp{B} = \grp{G} ~, \qquad \alg{b} = \alg{g} ~.
\end{equation}
We focus here on the $c=1$ case, for which the Drinfel'd double is isomorphic to the real double $\alg{d} \cong \alg{g} \oplus \alg{g}$.
While the following construction goes through for any real form $\alg{g}$, if this real form admits a solution to the split mcYBe then the model we find is Poisson-Lie T-dual to the corresponding YB deformation.
If not, then the resulting model is simply not the Poisson-Lie T-dual of a YB deformation.
Analogous statements also hold for the $c=i$ case, for which the Drinfel'd double is isomorphic to the complex double $\alg{d} \cong \alg{g}^\Complex$.
The model that results from integrating out the degrees of freedom associated to the Lagrangian subalgebra $\alg{g}$ is an analytic continuation of the model we find starting from the real double.
When the real form $\alg{g}$ admits a solution to the non-split mcYBe this model is Poisson-Lie T-dual to the corresponding YB deformation.
If not, then again the resulting model is simply not the Poisson-Lie T-dual of a YB deformation.

We start by writing an element of the real double $\alg{d} = \alg{g} \oplus \alg{g}$ as
\begin{equation}\label{eq:realdoub}
\big(X,Y\big) \in \alg{d} ~, \qquad X,Y \in \alg{g} ~,
\end{equation}
where the Lagrangian subalgebra $\alg{g}$ is generated by the diagonal elements $\big(X,X\big)$ and the map $\iota:\alg{g} \to \alg{d}$ acts as $\iota\big(X,X\big) = \big(X,-X\big)$.
The Lie bracket and invariant bilinear form~\eqref{eq:comiota} are given by
\begin{equation}\begin{split}\label{eq:comiota2}
[[\big(X_1,Y_1\big),\big(X_2,Y_2\big)]] & = \big([X_1,X_2],[Y_1,Y_2]\big) ~,
\\
\lang \big(X_1,Y_1\big),\big(X_2,Y_2\big)\rang & = \frac{1}{2} \tr\big(X_1 X_2\big) - \frac{1}{2} \tr\big(Y_1Y_2\big) ~.
\end{split}\end{equation}
The action of the operator $\mathcal{E}$~\eqref{eq:eop} is then given by
\begin{equation}
\mathcal{E}\big(X,Y\big) = \big(- \frac{(\lambda^{-1}+\lambda)X - 2 Y}{\lambda^{-1}-\lambda} , \frac{(\lambda^{-1}+\lambda)Y -2 X}{\lambda^{-1}-\lambda}\big) ~, \qquad X,Y\in\alg{g}~,
\end{equation}
where we have introduced
\begin{equation}\label{eq:parammap}
\lambda = \frac{1-\eta}{1+\eta} ~.
\end{equation}
It will also be convenient to introduce the subalgebras $\alg{g}_+ \cong \alg{g}$ and $\alg{g}_- \cong \alg{g}$ generated by elements of the form $\big(X,0\big)$ and $\big(0,X\big)$ respectively, along with the associated Lie groups $\grp{G}_+ \cong \grp{G}$ and $\grp{G}_- \cong \grp{G}$.
We now parametrise
\begin{equation}
\gdsl = \big(g',g'\big)\big(g,1\big)~, \qquad \big(g',g'\big) \in \grp{G} ~, \quad \big(g,1\big) \in \grp{G}_+ ~,
\end{equation}
and use the gauge symmetry~\eqref{eq:gsym} to fix $\big(g',g'\big) = \big(1,1\big)$ so that
\begin{equation}\label{eq:gchoice}
\gdsl = \big(g,1\big) \in \grp{G}_+ ~.
\end{equation}
It remains to determine the action of the projectors $\mathcal{E} \mathcal{P} (\mathcal{E} \pm 1)$, which are defined by their image and kernel.
In the current setup we have $\im \mathcal{E} \mathcal{P} (\mathcal{E} \pm 1) = \Ad_{(g,1)}^{-1} \alg{g}$ and $\ker \mathcal{E} \mathcal{P} (\mathcal{E} \pm 1) = \alg{e}_\mp$.
Observing that general elements of $\alg{e}_\mp$ take the form $\big(X,\lambda^{\pm 1} X\big)$, $X\in\alg{g}$, we consider the ansatz
\begin{equation}
\mathcal{E} \mathcal{P} (\mathcal{E} \pm 1)\big(X,Y\big) = \big(\Ad_g^{-1}f_\pm(\Ad_g^{-1}) (Y - \lambda^{\pm 1} X) , f_\pm(\Ad_g^{-1}) (Y-\lambda^{\pm 1} X)\big) ~, \qquad X,Y \in\alg{g}~,
\end{equation}
which ensures that $\alg{e}_\mp$ lies in the kernel and the image is contained in $\Ad_{(g,1)}^{-1} \alg{g}$.
Demanding that $\mathcal{E}\mathcal{P}(\mathcal{E}\pm1)$ are projectors with the required image and kernel we find
\begin{equation}
f_\pm (\Ad_g^{-1} ) = \frac{1}{1-\lambda^{\pm 1}\Ad_g^{-1}} ~,
\end{equation}
and we arrive at the following expression for their action
\begin{equation}\begin{split}\label{eq:projlam}
& \mathcal{E}\mathcal{P}(\mathcal{E}\pm1)\big(X,Y\big)
= \big( \frac{\Ad_g^{-1}}{1-\lambda^{\pm 1}\Ad_g^{-1}} (Y-\lambda^{\pm 1} X) ,
\frac{1}{1-\lambda^{\pm 1}\Ad_g^{-1}} (Y-\lambda^{\pm 1} X) \big)
~, \qquad X,Y\in\alg{g} ~.
\end{split}\end{equation}

To conclude, we fix the gauge~\eqref{eq:gchoice} in the action~\eqref{eq:doubact1}.
Using the action of the projectors in eq.~\eqref{eq:projlam} and the invariant bilinear form~\eqref{eq:comiota2} we find
\begin{equation}\begin{split}\label{eq:doubact3}
\Act_{\mathcal{E}_{\grp{G}}} & = N \Big(\frac12\int d^2 x \,
\big(\lang \big(g^{-1}\partial_+ g,0\big) , \big(\frac{-\lambda\Ad_g^{-1}}{1 - \lambda\Ad_g^{-1}},
\frac{-\lambda}{1 - \lambda\Ad_g^{-1}}\big) \big(g^{-1}\partial_-g , g^{-1}\partial_- g \big) \rang
\\ & \qquad \qquad \qquad \quad
- \lang \big(g^{-1}\partial_- g , 0\big), \big(\frac{-\lambda^{-1}\Ad_g^{-1}}{1 - \lambda^{-1}\Ad_g^{-1}},\frac{-\lambda^{-1}}{1 - \lambda^{-1}\Ad_g^{-1}}\big) \big(g^{-1}\partial_+g , g^{-1}\partial_+ g \big) \rang \big)
\\ & \qquad \quad
+ \frac{1}{6} \int d^3 x \, \epsilon^{ijk} \lang \big(g^{-1}\partial_i g, 0\big),[[\big(g^{-1}\partial_j g, 0\big), \big(g^{-1}\partial_k g, 0\big)]] \rang\Big)
\\
& = N \Big(\frac14\int d^2 x \,
\tr \big(g^{-1}\partial_+ g \frac{-\lambda\Ad_g^{-1}}{1 - \lambda\Ad_g^{-1}} g^{-1}\partial_-g
- g^{-1}\partial_- g \frac{-\lambda^{-1}\Ad_g^{-1}}{1 - \lambda^{-1}\Ad_g^{-1}} g^{-1}\partial_+g \big)
\\ & \qquad \quad
+ \frac{1}{12} \int d^3 x \, \epsilon^{ijk} \tr\big( g^{-1}\partial_i g [\big(g^{-1}\partial_j g , g^{-1}\partial_k g] \big) \Big)
\\
& = -\frac{\kay}{2}\int d^2 x \,
\tr \big(g^{-1}\partial_+ g \frac{1+\lambda\Ad_g^{-1}}{1 - \lambda\Ad_g^{-1}} g^{-1}\partial_-g \big)
+ \frac{\kay}{6} \int d^3 x \, \epsilon^{ijk} \tr\big( g^{-1}\partial_i g [g^{-1}\partial_j g , g^{-1}\partial_k g] \big)
\\ & = \Act_{\indrm{\Lambda}} ~,
\end{split}\end{equation}
where we have set
\unskip\footnote{The parameters $(\kay,\lambda)$ are thus related to the parameters $(\hay,\eta)$ as
\begin{equation*}
\kay = \frac{\hay}{4\eta} ~, \qquad \lambda = \frac{1-\eta}{1+\eta} ~,
\end{equation*}}
\begin{equation}
N = 2\kay ~.
\end{equation}
Sending $\lambda \to 0$, or equivalently $\eta \to 1$, we see that this action becomes that of the WZW model, which is given by eq.~\eqref{eq:pcmwzterm} with $\hay = \kay$.
Moreover, expanding the action~\eqref{eq:doubact3} around $\lambda = 0$ we find
\begin{equation}\begin{split}\label{eq:lamex}
\Act_{\indrm{\Lambda}} &
= -\frac{\kay}{2}\int d^2 x \,
\tr \big(g^{-1}\partial_+ g g^{-1}\partial_-g \big)
+ \frac{\kay}{6} \int d^3 x \, \epsilon^{ijk} \tr\big( g^{-1}\partial_i g [g^{-1}\partial_j g , g^{-1}\partial_k g] \big)
\\ & \qquad
-\kay\lambda\int d^2 x \,
\tr \big(\partial_+ g g^{-1}  g^{-1}\partial_-g \big) + \Order(\lambda^2) ~.
\end{split}\end{equation}
Recalling that the equations of motion of the WZW model are
\begin{equation}
\partial_+ (g^{-1} \partial_- g)  = \partial_-(\partial_+ g g^{-1} ) = 0 ~,
\end{equation}
we see that the first subleading term in~\eqref{eq:lamex} can be understood as a current-current perturbation.

\section{Current-current deformations of the WZW model}\label{sec:wzwmodel}

We finish \iflectures these lecture notes \else this review \fi with a discussion of current-current deformations of the WZW model.
These are deformations that take the form
\begin{equation}\begin{split}
\Act_{\indrm{j \bar \jmath}} & = \Act_{\indrm{WZW}}
-\kay\lambda\int d^2 x \,
\tr \big(\partial_+ g g^{-1} \mathcal{O} g^{-1}\partial_-g \big) + \Order(\lambda^2) ~,
\end{split}\end{equation}
where $\mathcal{O}:\alg{g}\to\alg{g}$ is a constant linear operator.
We will focus on two examples of such deformations.
The first is the isotropic current-current deformation~\eqref{eq:lamex}, which has $\mathcal{O} = 1$, hence isotropic, and preserves integrability.
The second is the abelian current-current deformation, which preserves conformal invariance.
These are two examples of a wider class of such deformations whose integrability, scale invariance and conformality depend on the properties of the operator $\mathcal{O}$.

\subsection{Isotropic current-current deformation}

We start by considering the particular current-current deformation~\eqref{eq:doubact3} introduced at the end of \secref{sec:dd}.

\paragraph{Gauged WZW formulation and integrability.}
An alternative construction of the model~\eqref{eq:doubact3} starts from the $\grp{G}/\grp{G}$ gauged WZW model
\begin{equation}\begin{split}\label{eq:gwzw}
\Act_{\indrm{gWZW}} & = - \frac{\kay}{2} \int d^2x \, \tr \big(g^{-1}\partial_+ g g^{-1}\partial_- g\big)
+ \frac{\kay}{6} \int d^3 x \, \epsilon^{ijk} \tr\big( g^{-1}\partial_i g [g^{-1}\partial_j g , g^{-1}\partial_k g] \big)
\\
& \qquad + \kay \int d^2 x \, \tr\big(A_+ g^{-1} \partial_- g - A_- \partial_+gg^{-1} + A_+ g^{-1} A_- g - A_+ A_-\big) ~,
\end{split}\end{equation}
where the group-valued field $g \in \grp{G}$ and the gauge field $A_\pm \in \alg{g}$.
This action is invariant under the gauge transformations
\begin{equation}\label{eq:gaugegwzw}
g \to \bar g^{-1} g \bar g ~, \qquad A_\pm \to \bar g^{-1} A_\pm \bar g + \bar g^{-1} \partial_\pm \bar g ~, \qquad \bar g(t,x) \in \grp{G}~.
\end{equation}
Due to the $\grp{G}$ gauge symmetry there are no dynamical degrees of freedom.
Therefore, we consider the following modification of the action~\eqref{eq:gwzw}
\begin{equation}\begin{split}\label{eq:lambda}
\Act_{\indrm{\Lambda}} & = - \frac{\kay}{2} \int d^2x \, \tr \big(g^{-1}\partial_+ g g^{-1}\partial_- g\big)
+ \frac{\kay}{6} \int d^3 x \, \epsilon^{ijk} \tr\big( g^{-1}\partial_i g [g^{-1}\partial_j g , g^{-1}\partial_k g] \big)
\\
& \qquad + \kay \int d^2 x \, \tr\big(A_+ g^{-1} \partial_- g - A_- \partial_+gg^{-1} + A_+ g^{-1} A_- g - \lambda^{-1} A_+ A_-\big) ~.
\end{split}\end{equation}
Away from the point $\lambda = 1$, the action~\eqref{eq:lambda} is no longer invariant under the gauge symmetry~\eqref{eq:gaugegwzw} and there are $\dim \grp{G}$ dynamical degrees of freedom.
Moreover, in the limit $\lambda \to 0$ we recover the standard WZW model.
\unskip\footnote{To take this limit we first rescale, for example, $A_\pm \to \sqrt{\lambda} A_\pm$ such that after taking $\lambda \to 0$ the fields $A_\pm$ decouple and we are just left with the WZW model.}
It is natural to take the parameter $\lambda$ to lie in the range $[0,1]$, with $\lambda \to 1$ corresponding to the non-abelian T-dual limit, which, as we will see, is given by taking $\lambda \to 1$ and $g \to 1$ such that $\frac{g-1}{\lambda-1}$ remains finite.
The global part of the $\grp{G}$ gauge symmetry survives the deformation, hence the model has a $\grp{G}$ global symmetry acting as
\begin{equation}
g \to g_\ind{D}^{-1} g g_\ind{D} ~, \qquad
A_\pm \to g_\ind{D}^{-1} A_\pm g_\ind{D} ~, \qquad
g_\ind{D} \in \grp{G} ~.
\end{equation}

Since the fields $A_\pm$ enter the action~\eqref{eq:lambda} algebraically they can be easily integrated out.
Their equations of motion are
\begin{equation}\label{eq:apm}
(\lambda^{-1} - \Ad_g) A_+ = - \partial_+ g g^{-1} ~, \qquad
(\lambda^{-1} - \Ad_g^{-1}) A_- = g^{-1} \partial_- g ~.
\end{equation}
Substituting back into the action~\eqref{eq:lambda} we find
\begin{equation}\begin{split}\label{eq:lambda2}
\Act_{\indrm{\Lambda}} & = - \frac{\kay}{2} \int d^2x \, \tr \big(g^{-1}\partial_+ g g^{-1}\partial_- g\big)
+ \frac{\kay}{6} \int d^3 x \, \epsilon^{ijk} \tr\big( g^{-1}\partial_i g [g^{-1}\partial_j g , g^{-1}\partial_k g] \big)
\\
& \qquad - \kay \int d^2 x \, \tr\big(g^{-1} \partial_+ g \frac{\Ad_g^{-1}}{\lambda^{-1} - \Ad_g^{-1}} g^{-1} \partial_-g \big)
\\ & = - \frac{\kay}{2} \int d^2x \, \tr \big(g^{-1}\partial_+ g \frac{1+\lambda \Ad_g^{-1}}{1-\lambda \Ad_g^{-1}} g^{-1}\partial_- g\big)
+ \frac{\kay}{6} \int d^3 x \, \epsilon^{ijk} \tr\big( g^{-1}\partial_i g [g^{-1}\partial_j g , g^{-1}\partial_k g] \big) ~,
\end{split}\end{equation}
and we indeed recover the isotropic current-current deformation of the WZW model~\eqref{eq:doubact3}.

Writing the action in the form~\eqref{eq:lambda} has the advantage of making the integrability of the model more transparent.
In addition to the equations of motion for $A_\pm$~\eqref{eq:apm}, we also have the equation of motion for the group-valued field, which is
\begin{equation}
\partial_+(g^{-1}\partial_- g + g^{-1} A_- g) - \partial_- A_+ + [A_+,g^{-1}\partial_- g + g^{-1} A_- g] = 0 ~,
\end{equation}
or equivalently
\begin{equation}
\partial_-(-\partial_+ g g^{-1} + g A_+ g^{-1}) - \partial_+ A_- + [A_-,-\partial_+ g g^{-1} + g A_+ g^{-1}] = 0 ~.
\end{equation}
When $A_\pm$ are given by their on-shell expressions~\eqref{eq:apm} we can rewrite these two equations as
\begin{equation}
\lambda^{-1} \partial_+ A_- - \partial_- A_+ + \lambda^{-1}[A_+,A_-] = 0 ~,
\qquad
\lambda^{-1} \partial_- A_+ - \partial_+ A_- + \lambda^{-1} [A_-,A_+] = 0 ~.
\end{equation}
Taking their sum and difference we find
\begin{equation}
\partial_+ J_- + \partial_- J_+ = 0 ~,
\qquad
\partial_+ J_- - \partial_- J_+ + [J_+,J_-] = 0 ~,
\end{equation}
where
\begin{equation}
J_\pm = \frac{2}{1+\lambda} A_\pm ~.
\end{equation}
Therefore, $J_\pm$ is a conserved and flat current, which allows us to write down the following Lax connection of the isotropic current-current deformation of the WZW model
\begin{equation}
\Lax_\pm(z) = \frac{J_\pm}{1\mp z} ~.
\end{equation}

\paragraph{Non-abelian T-dual limit.}
Finally, let us return to the $\lambda \to 1$ limit of the action~\eqref{eq:lambda}.
According to the map between the parameters~\eqref{eq:parammap} this corresponds to the undeformed limit of the YB deformation~\eqref{eq:ybdefgen}, in which the $\grp{G}_\ind{L} \times \grp{G}_\ind{R}$ global symmetry is restored.
Therefore, we might expect the Poisson-Lie T-dual model to become the non-abelian T-dual of the PCM.
This is indeed the case, although the limit needs to be taken with care.
Specifically, we parametrise
\begin{equation}\label{eq:natdparam}
g = \exp\big(-\frac{\hay}{2\kay}v\big) ~, \qquad \lambda = \exp\big(-\frac{\hay}{2\kay}\big) ~, \qquad v \in \alg{g} ~,
\end{equation}
and take $\kay \to \infty$.
Taking this limit in the action~\eqref{eq:lambda} gives
\iflectures
\begin{equation}\label{eq:pcmint}
\Act_{\indrm{NATD-INT-PCM}} = -\frac{\hay}{2} \int d^2 x \, \tr\big( v (\partial_+ A_- - \partial_- A_+ + [A_+, A_-]) + A_+ A_- \big) ~.
\end{equation}
This is the first-order action that we find by gauging the left-acting $\grp{G}$ global symmetry of the PCM~\eqref{eq:pcmact}, $g \to g_\ind{L} g$, gauge-fixing $g = 1$, and adding a Lagrange multiplier $v \in \alg{g}$ enforcing that the gauge field is flat.
Note that integrating out the Lagrange multiplier field $v$ in the action~\eqref{eq:pcmint} we recover the PCM.
On the other hand, integrating out $A_\pm$ we find the non-abelian T-dual of the PCM
\begin{equation}\label{eq:pcmnatd}
\Act_{\indrm{NATD-PCM}} = - \frac{\hay}{2} \int d^2 x \, \tr \big(\partial_+ v \frac{1}{1-\ad_v} \partial_- v\big) ~.
\end{equation}

\question{Parametrising
\begin{equation*}
g = \exp\big(-\frac{\hay}{2\kay}v\big) ~, \qquad \lambda = \exp\big(-\frac{\hay}{2\kay}\big) ~, \qquad v \in \alg{g} ~,
\end{equation*}
in the actions~\eqref{eq:lambda} and~\eqref{eq:lambda2}, take the limit $\kay \to \infty$ to recover the actions~\eqref{eq:pcmint} and~\eqref{eq:pcmnatd}.
}

\answer{The parametrisation
\begin{equation*}
g = \exp\big(-\frac{\hay}{2\kay}v\big) ~, \qquad \lambda = \exp\big(-\frac{\hay}{2\kay}\big) ~, \qquad v \in \alg{g} ~,
\end{equation*}
implies that
\begin{equation*}
g^{-1}\partial_\pm g = -\frac{\hay}{2\kay} \partial_\pm v + \Order(\frac{1}{\kay^{2}}) ~,\qquad
\partial_\pm g g^{-1} = -\frac{\hay}{2\kay} \partial_\pm v + \Order(\frac{1}{\kay^{2}}) ~,\qquad
\lambda^{-1} = 1 + \frac{\hay}{2\kay} + \Order(\frac{1}{\kay^2}) ~.
\end{equation*}
Therefore, the two terms in the first line of the action~\eqref{eq:lambda} are of order $\frac{1}{\kay}$ and $\frac{1}{\kay^2}$, hence drop out in the limit $\kay \to \infty$.
Expanding the second line of the action~\eqref{eq:lambda} at large $\kay$ we find
\begin{equation*}\begin{split}
\Act_{\indrm{\Lambda}} & = \kay \int d^2 x \, \tr \big(- \frac{\hay}{2\kay} A_+ \partial_- v + \frac{\hay}{2\kay} A_- \partial_+ v + A_+ A_- + \frac{\hay}{2\kay} A_+ [v,A_-] - A_+ A_- - \frac{\hay}{2\kay} A_+ A_- + \Order(\frac{1}{\kay^2}) \big)
\\ & =
-\frac{\hay}{2} \int d^2 x \, \tr \big(A_+ \partial_- v - A_- \partial_+ v - A_+ [v,A_-] + A_+ A_- \big) + \Order(\frac{1}{\kay})
\\ & =
-\frac{\hay}{2} \int d^2 x \, \tr \big( v( \partial_+ A_- - \partial_- A_+ + [A_+,A_-]) + A_+ A_- \big) + \Order(\frac{1}{\kay}) ~,
\end{split}\end{equation*}
where we have integrated by parts.
Therefore, we indeed find the action~\eqref{eq:pcmint} in the limit $\kay \to \infty$.

Again using the parametrisation
\begin{equation*}
g = \exp\big(-\frac{\hay}{2\kay}v\big) ~, \qquad \lambda = \exp\big(-\frac{\hay}{2\kay}\big) ~, \qquad v \in \alg{g} ~,
\end{equation*}
we find
\begin{equation*}
\lambda \Ad_g^{-1} = 1- \frac{\hay}{2\kay} + \frac{\hay}{2\kay} \ad_v + \Order(\frac{1}{\kay^2}) ~,
\end{equation*}
hence
\begin{equation*}
\frac{1+\lambda \Ad_g^{-1}}{1-\lambda \Ad_g^{-1}} = \frac{4\kay}{\hay}\frac{1}{1-\ad_v} + \Order(1) ~.
\end{equation*}
Therefore, expanding the action~\eqref{eq:lambda2} at large $\kay$, we find
\begin{equation*}
\Act_{\indrm{\Lambda}} = - \frac{\hay}{2} \int d^2x \, \tr \big(\partial_+v \frac{1}{1-\ad_v}\partial_- v\big) + \Order(\frac{1}{\kay}) ~,
\end{equation*}
and we indeed recover the action~\eqref{eq:pcmnatd} in the limit $\kay \to \infty$.
}
\else
\unskip\footnote{We have $g^{-1}\partial_\pm g = -\frac{\hay}{2\kay} \partial_\pm v + \Order(\frac{1}{\kay^{2}})$, $\partial_\pm g g^{-1} = -\frac{\hay}{2\kay} \partial_\pm v + \Order(\frac{1}{\kay^{2}})$ and $\lambda^{-1} = 1 + \frac{\hay}{2\kay} + \Order(\frac{1}{\kay^2})$.
Therefore, the two terms in the first line of the action~\eqref{eq:lambda} are of order $\frac{1}{\kay}$ and $\frac{1}{\kay^2}$, hence drop out in the limit $\kay \to \infty$.
Expanding the second line of the action~\eqref{eq:lambda} at large $\kay$ and integrating by parts we find
\begin{equation*}\begin{split}
\Act_{\indrm{\Lambda}} & = \kay \int d^2 x \, \tr \big(- \frac{\hay}{2\kay} A_+ \partial_- v + \frac{\hay}{2\kay} A_- \partial_+ v + A_+ A_- + \frac{\hay}{2\kay} A_+ [v,A_-] - A_+ A_- - \frac{\hay}{2\kay} A_+ A_- + \Order(\frac{1}{\kay^2}) \big)
\\ & =
-\frac{\hay}{2} \int d^2 x \, \tr \big(A_+ \partial_- v - A_- \partial_+ v - A_+ [v,A_-] + A_+ A_- \big) + \Order(\frac{1}{\kay})
\\ & =
-\frac{\hay}{2} \int d^2 x \, \tr \big(v ( \partial_+ A_- - \partial_- A_+ + [A_+,A_-]) + A_+ A_- \big) + \Order(\frac{1}{\kay}) ~.
\end{split}\end{equation*}}
\begin{equation}\label{eq:pcmint}
\Act_{\indrm{NATD-INT-PCM}} = -\frac{\hay}{2} \int d^2 x \, \tr\big( v (\partial_+ A_- - \partial_- A_+ + [A_+, A_-]) + A_+ A_- \big) ~.
\end{equation}
This is the first-order action that we find by gauging the left-acting $\grp{G}$ global symmetry of the PCM~\eqref{eq:pcmact}, $g \to g_\ind{L} g$, gauge-fixing $g = 1$, and adding a Lagrange multiplier $v \in \alg{g}$ enforcing that the gauge field is flat.
Note that integrating out the Lagrange multiplier field $v$ in the action~\eqref{eq:pcmint} we recover the PCM.
On the other hand, integrating out $A_\pm$ we find the non-abelian T-dual of the PCM
\begin{equation}\label{eq:pcmnatd}
\Act_{\indrm{NATD-PCM}} = - \frac{\hay}{2} \int d^2 x \, \tr \big(\partial_+ v \frac{1}{1-\ad_v} \partial_- v\big) ~.
\end{equation}
We can also find the action~\eqref{eq:pcmnatd} directly by taking the $\kay \to \infty$ limit in the action~\eqref{eq:lambda2}.
\unskip\footnote{
Again using the parametrisation~\eqref{eq:natdparam} we have
$\lambda \Ad_g^{-1} = 1- \frac{\hay}{2\kay} + \frac{\hay}{2\kay} \ad_v + \Order(\frac{1}{\kay^2})$,
hence
\begin{equation*}
\frac{1+\lambda \Ad_g^{-1}}{1-\lambda \Ad_g^{-1}} = \frac{4\kay}{\hay}\frac{1}{1-\ad_v} + \Order(1) ~.
\end{equation*}
Therefore, expanding the action~\eqref{eq:lambda2} at large $\kay$, we find
\begin{equation*}
\Act_{\indrm{\Lambda}} = - \frac{\hay}{2} \int d^2x \, \tr \big(\partial_+v \frac{1}{1-\ad_v}\partial_- v\big) + \Order(\frac{1}{\kay}) ~.
\end{equation*}}
\fi

\paragraph{Examples based on $\grp{SL}(2)$.}
As for the YB deformation, it is instructive to look at explicit examples based on real forms of $\grp{G} = \grp{SL}(2,\Complex)$.
We start by considering the isotropic current-current deformation of the $\grp{SU}(2)$ WZW model.
Recalling the generators $T_1$, $T_2$ and $T_3$ of the compact real form defined in~\eqref{eq:su2gen}, we parametrise
\begin{equation}
g = \exp\big(\phi_1 (\cos \phi_2 T_3 + \sin\phi_2 (\cos\phi_3 T_1 + \sin \phi_3 T_2)) \big) ~.
\end{equation}
Substituting into the action~\eqref{eq:lambda2} with $\tr = \Tr$, we find that the deformed target-space metric and H-flux are given by
\begin{equation}\begin{split}
G & = 2\kay\big(\frac{1+\lambda}{1-\lambda} d\phi_1^2 + \frac{(1-\lambda^2)\sin^2\phi_1}{(1-\lambda)^2+ 4\lambda\sin^2\phi_1}(d\phi_2^2 + \sin^2\phi_2 d\phi_3^2) \big) ~,
\\
H & = -4\kay \frac{(1-\lambda^2)^2 + 2\lambda((1-\lambda)^2 + 4\lambda\sin^2\phi_1)}{((1-\lambda)^2 + 4\lambda\sin^2\phi_1)^2} \sin^2\phi_1\sin\phi_2 d\phi_1\wedge d\phi_2\wedge d\phi_3 ~.
\end{split}\end{equation}
The non-abelian T-dual limit is then given by setting $\lambda = \exp\big(-\frac{\hay}{2\kay}\big)$, rescaling $\phi_1 \to -\frac{\hay}{2\kay}\phi_1$ and taking $\kay \to \infty$.
Taking this limit in the deformed target-space metric and H-flux we find
\begin{equation}\begin{split}
G & = 2\hay\big(d\phi_1^2 + \frac{\phi_1^2}{1+4\phi_1^2}(d\phi_2^2 + \sin^2\phi_2 d\phi_3^2)\big) ~,
\\
H & = 4 \hay \frac{3+4\phi_1^2}{(1+4\phi_1^2)^2} \phi_1^2 \sin\phi_2 d\phi_1\wedge d\phi_2\wedge d\phi_3 ~.
\end{split}\end{equation}

We now consider the isotropic current-current deformation of the $\grp{SL}(2,\Real)$ WZW model.
Recall that for $\grp{G} = \grp{SL}(2,\Real)$ we reverse the overall sign of the action to ensure that the target-space metric has Lorentzian signature.
Taking the generators $S_0$, $S_+$ and $S_-$ of the split real form defined in~\eqref{eq:sl2gen}, we parametrise
\begin{equation}
h = \exp \big(\psi_1 (\cos \psi_2 S_0 + \sin\psi_2 (e^{\tau} S_+ + e^{-\tau} S_- ))\big) ~.
\end{equation}
Substituting into the action~\eqref{eq:lambda2} with $\tr = \Tr$ and the overall sign reversed, we find that the deformed target-space metric and H-flux are given by
\unskip\footnote{
Solving the $\AdS_3$ constraint equation $-Z_0^2 + Z_1^2 + Z_2^2 - Z_3^2 = - 1$ by setting
\begin{equation*}\begin{aligned}
Z_0 & = \cosh\psi_1 ~, \qquad &
Z_1 & = \sinh\psi_1\cos\psi_2 ~, \qquad &
Z_2 & = \sinh\psi_1\sin\psi_2\cosh\tau ~, \qquad &
Z_3 & = \sinh\psi_1\sin\psi_2\sinh\tau ~,
\end{aligned}\end{equation*}
and substituting into the metric on $\Real^{2,2}$, $G = -dZ_0^2 + dZ_1^2 + dZ_2^2 - dZ_3^2$, the induced metric on $\AdS_3$ is given by
\begin{equation*}
G = d\psi_1^2 + \sinh^2\psi_1(d\psi_2^2 - \sin^2\psi_2 d\tau^2) ~,
\end{equation*}
which agrees with the metric in eq.~\eqref{eq:sl2lambda} for $\lambda = 0$ and $\kay = \frac12$.
Furthermore, for $\lambda = 0$ the H-flux in eq.~\eqref{eq:sl2lambda} is proportional to the volume form of $\AdS_3$ as expected.}
\begin{equation}\begin{split}\label{eq:sl2lambda}
G & = 2\kay\big(\frac{1+\lambda}{1-\lambda} d\psi_1^2 + \frac{(1-\lambda^2)\sinh^2\psi_1}{(1-\lambda)^2 - 4\lambda\sinh^2\psi_1}(d\psi_2^2 - \sin^2\psi_2 d\tau^2) \big) ~,
\\
H & = - 4\kay \frac{(1-\lambda^2)^2 + 2\lambda((1-\lambda)^2 - 4\lambda\sinh^2\psi_1)}{((1-\lambda)^2 - 4\lambda\sinh^2\psi_1)^2} \sinh^2\psi_1\sin\psi_2 d\psi_1\wedge d\psi_2\wedge d\tau ~.
\end{split}\end{equation}
Note that this deformed background has a singularity at
\begin{equation}\label{eq:singularity}
\sinh^2\psi_1 = \frac{(1-\lambda)^2}{4\lambda} ~,
\end{equation}
which is related to the fact that the operator $1-\lambda \Ad_h^{-1}$ is not invertible at this point.
We will not address this singularity in detail, suffice to say that it defines for us two regions.
These are the region ``inside'' the singularity with $\sinh^2\psi_1 < \frac{(1-\lambda)^2}{4\lambda}$, which is connected to the undeformed limit $\lambda \to 0$, and the region ``outside'' the singularity with $\sinh^2\psi_1 > \frac{(1-\lambda)^2}{4\lambda}$.
Inside the singularity $\psi_1$ and $\psi_2$ are spacelike, while $\tau$ is timelike.
Outside the singularity $\psi_2$ becomes timelike, while $\tau$ becomes spacelike.
The non-abelian T-dual limit is again given by setting $\lambda = \exp\big(-\frac{\hay}{2\kay}\big)$, rescaling $\psi_1 \to -\frac{\hay}{2\kay}\psi_1$ and taking $\kay \to \infty$, resulting in the following target-space metric and H-flux
\begin{equation}\begin{split}
G & = 2\hay\big(d\psi_1^2 + \frac{\psi_1^2}{1-4\psi_1^2}(d\psi_2^2 - \sin^2\psi_2 d\tau^2)\big) ~,
\\
H & = 4 \hay \frac{3-4\psi_1^2}{(1-4\psi_1^2)^2} \psi_1^2 \sin\psi_2 d\psi_1\wedge d\psi_2\wedge d\tau ~.
\end{split}\end{equation}

A second interesting limit is given by setting $\psi_1 = \psi_1 + \gamma$ and taking $\gamma \to \infty$ in eq.~\eqref{eq:sl2lambda}.
This should be understood as a formal limit in the sense that it requires us to be ``outside'' the singularity~\eqref{eq:singularity}.
Doing so we find
\begin{equation}\begin{split}
G & = 2\kay\big(\frac{1+\lambda}{1-\lambda} d\psi_1^2 + \frac{1-\lambda^2}{4\lambda}(-d\psi_2^2 + \sin^2\psi_2 d\tau^2)\big) ~,
\\
H & = 2\kay \sin\psi_2 d\psi_1\wedge d\psi_2\wedge d\tau ~.
\end{split}\end{equation}
In this limit we find that the deformed model gains an additional symmetry corresponding to shifts of $\psi_1$.
Writing the deformed B-field as
\begin{equation}
B = 2 \kay \cos\psi_2 d\psi_1 \wedge d\tau ~,
\end{equation}
and T-dualising in $\psi_1$, the resulting background has vanishing B-field, while the T-dual deformed metric is given by
\begin{equation}
G = 2 \kay\big(\frac{1-\lambda}{1+\lambda}(d\tilde\psi_1 + \cos\psi_2 d\tau)^2 + \frac{1-\lambda^2}{4\lambda}(-d\psi_2^2 + \sin^2\psi_2 d\tau^2)\big) ~.
\end{equation}
Setting
\begin{equation}
\tilde\psi_1 = z_1 - z_2 ~, \qquad \tau = z_1+z_2 ~, \qquad \psi_2 = 2 z_0 ~, \qquad
\lambda = \frac{1-\eta}{1+\eta} ~, \qquad \kay = \frac{\hay}{4\eta} ~,
\end{equation}
this metric becomes
\begin{equation}
G= \frac{2\hay}{1-\eta^2}\big(-d z_0^2 + \cos^2 z_0(1-\eta^2 \cos^2z_0) dz_1^2 + \sin^2z_0(1-\eta^2\sin z_0^2) dz_2^2
+ 2\eta^2\sin^2 z_0 \cos^2 z_0 dz_1 dz_2 \big) ~.
\end{equation}
This is reminiscent of the deformed backgrounds of the inhomogeneous YB deformations constructed in \secref{sec:ybpcm}.
In fact, it is precisely the deformed metric of the inhomogeneous YB deformation of the $\grp{SL}(2,\Real)$ PCM based on the Drinfel'd-Jimbo solution to the split mcYBe~\eqref{eq:sl2rsplit}.
To see this explicitly we parametrise the group-valued field of the YB deformation as
\begin{equation}
g = \exp\big(\frac{z_1 - z_2}{2} S_0\big) \exp\big(z_0 (S_+ - S_-)\big) \exp\big(\frac{z_1 + z_2}{2} S_0\big) ~.
\end{equation}
Furthermore, the deformed B-field is closed.
Therefore, in this limit the isotropic current-current deformation of the $\grp{SL}(2,\Real)$ WZW model becomes the T-dual of the split inhomogeneous YB deformation of the $\grp{SL}(2,\Real)$ PCM up to a closed term.

This observation has an algebraic interpretation.
Recall that both the split inhomogeneous YB deformation of the $\grp{SL}(2,\Real)$ PCM and the isotropic current-current deformation of the $\grp{SL}(2,\Real)$ WZW model follow from the $\mathcal{E}$-model~\eqref{eq:doubact} on the real double $\alg{sl}(2,\Real) \oplus \alg{sl}(2,\Real)$.
They are found by integrating out the degrees of freedom associated to different Lagrangian subalgebras.
Taking
\begin{equation}
\{(H,0),(E,0),(F,0),(0,H),(0,E),(0,F)\} ~,
\end{equation}
as a basis for the real double, these Lagrangian subalgebras are $\tilde{\alg{g}} \cong \alg{g}_{\mathcal{R}}$ and $\alg{g}$ respectively, which are generated by
\unskip\footnote{Recall that $\alg{g}$ is generated by the diagonal elements $(X,X)$ and from eq.~\eqref{eq:ssol} we have $\mathcal{R}(H,H) = 0$, $\mathcal{R}(E,E) = (-E,-E)$ and $\mathcal{R}(F,F) = (F,F)$.
Elements of $\tilde{\alg{g}} \cong \alg{g}_{\mathcal{R}}$ take the form $\iota (X,X) + \mathcal{R}(X,X)$ where $\iota (X,X) = (X,-X)$.}
\begin{equation}
\tilde{\alg{g}} = \operatorname{span} \{(H,-H),(0,E),(F,0)\} ~, \qquad \alg{g} = \operatorname{span} \{(H,H),(E,E),(F,F)\} ~.
\end{equation}
The limit of interest is a limit of the isotropic current-current deformation and correspondingly can be related to the following contraction of the Lie algebra $\alg{g}$
\begin{equation}
\lim_{\gamma \to \infty} \Ad_{e^{\log\gamma (-H,H)}} \{(H,H),\gamma^{-2}(E,E),\gamma^{-2}(F,F)\}
= \{(H,H),(0,E),(F,0)\} ~.
\end{equation}
The only difference between this contracted Lie algebra and $\tilde{\alg{g}}$ is the linear combination of Cartan generators, $(H,H)$ versus $(H,-H)$, which can be shown to amount to a T-duality transformation.

\subsection{Abelian current-current deformations and TsT transformations}

As our final example of a deformed sigma model we consider the abelian current-current deformation of the WZW model.
In this case we deform by $\grp{U}(1)$ currents.
It turns out that this deformation can be understood as a TsT transformation of the WZW model, hence it preserves conformal invariance.

We can define the model in a similar way to the isotropic deformation by deforming the $\grp{G}/\grp{U}(1)$ gauged WZW model
\begin{equation}\begin{split}\label{eq:u1def}
\Act_{\indrm{\Lambda_{ab}}} & = - \frac{\kay}{2} \int d^2x \, \tr \big(g^{-1}\partial_+ g g^{-1}\partial_- g\big)
+ \frac{\kay}{6} \int d^3 x \, \epsilon^{ijk} \tr\big( g^{-1}\partial_i g [g^{-1}\partial_j g , g^{-1}\partial_k g] \big)
\\
& \qquad + \kay \int d^2 x \, \tr\big(A_+ g^{-1} \partial_- g - A_- \partial_+gg^{-1} + A_+ g^{-1} A_- g - \lambda^{-1} A_+ A_-\big) ~,
\end{split}\end{equation}
where, as before, the group-valued field $g \in \grp{G}$, but now the field $A_\pm \in \alg{u}(1) \subset \alg{g}$.
This model interpolates between the WZW model in the limit $\lambda \to 0$, the vectorially gauged WZW model in the limit $\lambda \to 1$ and the axially gauged WZW model in the limit $\lambda \to -1$.
The vectorially gauged WZW model is invariant under the gauge transformations
\begin{equation}
g \to \bar{h}^{-1} g \bar{h} ~, \qquad
A_\pm \to A_\pm + \bar{h}^{-1}\partial_\pm \bar{h} ~, \qquad \bar{h}(t,x) \in \grp{U}(1)~.
\end{equation}
Since $\grp{U}(1)$ is abelian we have $\bar{h}^{-1} A_\pm \bar{h} = A_\pm$ and $\bar{h}^{-1}\partial_\pm \bar{h} = \partial_\pm \bar{h} \bar{h}^{-1}$.
On the other hand, the gauge symmetry of the axially gauged WZW model is given by
\begin{equation}
g \to \bar{h} g \bar{h} ~, \qquad
A_\pm \to A_\pm \pm \bar{h}^{-1}\partial_\pm \bar{h} ~, \qquad \bar{h}(t,x) \in \grp{U}(1)~.
\end{equation}
Away from the points $\lambda = \pm1$ the action~\eqref{eq:u1def} is no longer invariant under the gauge symmetry and has $\dim \grp{G}$ dynamical degrees of freedom.
Again the global part of the gauge symmetry survives, hence the model has at least a $\grp{U}(1) \times \grp{U}(1)$ global symmetry acting as
\begin{equation}
g \to h_\ind{L} g h_\ind{R} ~, \qquad A_\pm \to A_\pm ~, \qquad h_\ind{L},h_\ind{R} \in \grp{U}(1) ~.
\end{equation}

Assuming that the $\alg{u}(1)$ subalgebra is not null, that is $\tr\big(T^2\big) \neq 0$ where $T$ is the $\alg{u}(1)$ generator, we can introduce the orthogonal projector $P:\alg{g}\to\alg{u}(1)$ and the equations of the motion for the fields $A_\pm$ are
\begin{equation}
(\lambda^{-1} - P \Ad_g) A_+ = -P \partial_+ g g^{-1} ~, \qquad
(\lambda^{-1} - P \Ad_g^{-1}) A_- = P g^{-1} \partial_- g ~.
\end{equation}
Substituting back into the action~\eqref{eq:u1def} we find
\begin{equation}\begin{split}\label{eq:u1def2}
\Act_{\indrm{\Lambda_{ab}}} & = - \frac{\kay}{2} \int d^2x \, \tr \big(g^{-1}\partial_+ g g^{-1}\partial_- g\big)
+ \frac{\kay}{6} \int d^3 x \, \epsilon^{ijk} \tr\big( g^{-1}\partial_i g [g^{-1}\partial_j g , g^{-1}\partial_k g] \big)
\\
& \qquad - \kay \int d^2 x \, \tr\big(\partial_+ g g^{-1} P \frac{1}{\lambda^{-1} - \Ad_g^{-1}P} g^{-1} \partial_- g \big) ~.
\end{split}\end{equation}
Expanding this action around $\lambda = 0$ gives
\begin{equation}\begin{split}\label{eq:u1def3}
\Act_{\indrm{\Lambda_{ab}}} & = - \frac{\kay}{2} \int d^2x \, \tr \big(g^{-1}\partial_+ g g^{-1}\partial_- g\big)
+ \frac{\kay}{6} \int d^3 x \, \epsilon^{ijk} \tr\big( g^{-1}\partial_i g [g^{-1}\partial_j g , g^{-1}\partial_k g] \big)
\\
& \qquad - \kay\lambda \int d^2 x \, \tr\big(\partial_+ g g^{-1} P g^{-1} \partial_- g \big) + \Order(\lambda^2) ~,
\end{split}\end{equation}
and we see that the first subleading term can be understood as an abelian current-current perturbation.

\iflectures
\question{Show that the abelian current-current deformation~\eqref{eq:u1def2} can also be found as a TsT transformation of the WZW model up to closed terms.
One approach is to parametrise
\begin{equation*}
g = e^{\Theta^1} \tilde{g} e^{\Theta^2} ~, \qquad \Theta^1,\Theta^2 \in \alg{u}(1) ~,
\end{equation*}
in the action of the WZW model and implement the TsT transformation in $\Theta^1$ and $\Theta^2$, that is first T-dualise $\Theta^1 \to \tilde \Theta_1$, then shift $\Theta^2 \to \Theta^2 - \gamma \tilde \Theta_1$ and finally T-dualise $\tilde\Theta_1 \to \Theta^1$.
To match with the abelian current-current deformation~\eqref{eq:u1def2} using the same parametrisation you will also need to redefine $\Theta^1 \to \Theta^1 + \gamma \Theta^2$ and $\Theta^2 \to \Theta^2 + \gamma \Theta^1$, add the closed term
\begin{equation*}
-\frac{\kay\gamma}{2} \int d^2x \,\tr\big(\partial_+\Theta^1\partial_-\Theta^2 - \partial_+ \Theta^2 \partial_- \Theta^1\big) ~,
\end{equation*}
and set
\begin{equation*}
\lambda = \frac{2\gamma}{1+\gamma^2} ~.
\end{equation*}
}

\answer{We start from the WZW model
\begin{equation*}
\Act_{\indrm{WZW}} = - \frac{\kay}{2} \int d^2x \, \tr \big(g^{-1}\partial_+ g g^{-1}\partial_- g\big)
+ \frac{\kay}{6} \int d^3 x \, \epsilon^{ijk} \tr\big( g^{-1}\partial_i g [g^{-1}\partial_j g , g^{-1}\partial_k g] \big) ~,
\end{equation*}
and set
\begin{equation*}
g = e^{\Theta^1} \tilde{g} e^{\Theta^2} ~,
\end{equation*}
which gives
\ifextra
\begin{equation*}\begin{split}
\Act_{\indrm{WZW}} & =
- \frac{\kay}{2} \int d^2x \, \tr\big(
\tilde{g}^{-1}\partial_+ \tilde{g} \tilde{g}^{-1}\partial_- \tilde{g}
+ \partial_+\Theta^1  \partial_- \Theta^1
+ \partial_+\Theta^2  \partial_- \Theta^2
\\ & \qquad \qquad \qquad \qquad
+ \partial_+ \Theta^1  \partial_- \tilde{g} \tilde{g}^{-1}
+ \partial_- \Theta^1  \partial_+ \tilde{g} \tilde{g}^{-1}
+ \partial_+ \Theta^2  \tilde{g}^{-1} \partial_- \tilde{g}
+ \partial_- \Theta^2  \tilde{g}^{-1} \partial_+ \tilde{g}
\\ & \qquad \qquad \qquad \qquad
+ \tilde{g}^{-1} \partial_+\Theta^1  \tilde{g} \partial_- \Theta^2
+ \tilde{g}^{-1} \partial_-\Theta^1  \tilde{g} \partial_+ \Theta^2  \big)
\\ & \quad
+ \frac{\kay}{6} \int d^3 x \, \epsilon^{ijk} \tr\big( \tilde{g}^{-1}\partial_i \tilde{g} [\tilde{g}^{-1}\partial_j \tilde{g} , \tilde{g}^{-1}\partial_k \tilde{g}]
+ 6 \tilde{g}^{-1}\partial_i \tilde{g} [ \tilde{g}^{-1} \partial_j \Theta^1  \tilde{g}, \partial_k \Theta^2 ]
\\ & \qquad \qquad \qquad \qquad \qquad
+3\partial_i\Theta^1  [\partial_j \tilde{g} \tilde{g}^{-1}, \partial_k \tilde{g} \tilde{g}^{-1}]
+3\partial_i\Theta^2  [\tilde{g}^{-1}\partial_j \tilde{g}, \tilde{g}^{-1}\partial_k \tilde{g}]
\big)
\\
& = - \frac{\kay}{2} \int d^2x \, \tr\big(
\tilde{g}^{-1}\partial_+ \tilde{g} \tilde{g}^{-1}\partial_- \tilde{g}
+ \partial_+\Theta^1  \partial_- \Theta^1
+ \partial_+\Theta^2  \partial_- \Theta^2
\\ & \qquad \qquad \qquad \qquad
+ \partial_+ \Theta^1  \partial_- \tilde{g} \tilde{g}^{-1}
+ \partial_- \Theta^1  \partial_+ \tilde{g} \tilde{g}^{-1}
+ \partial_+ \Theta^2  \tilde{g}^{-1} \partial_- \tilde{g}
+ \partial_- \Theta^2  \tilde{g}^{-1} \partial_+ \tilde{g}
\\ & \qquad \qquad \qquad \qquad
+ \tilde{g}^{-1} \partial_+\Theta^1  \tilde{g} \partial_- \Theta^2
+ \tilde{g}^{-1} \partial_-\Theta^1  \tilde{g} \partial_+ \Theta^2  \big)
\\ & \quad
+ \frac{\kay}{6} \int d^3 x \, \epsilon^{ijk} \tr\big( \tilde{g}^{-1}\partial_i \tilde{g} [\tilde{g}^{-1}\partial_j \tilde{g} , \tilde{g}^{-1}\partial_k \tilde{g}]
\\ & \qquad \qquad \qquad \qquad \qquad
+ 6 \partial_i (- \partial_j \Theta^1  \partial_k \tilde{g} \tilde{g}^{-1} + \partial_j \Theta^2  \tilde{g}^{-1} \partial_k \tilde{g} - \tilde{g}^{-1} \partial_j \Theta^1  \tilde{g} \partial_k \Theta^2  ) \big)
\\
& = - \frac{\kay}{2} \int d^2x \, \tr\big(
\tilde{g}^{-1}\partial_+ \tilde{g} \tilde{g}^{-1}\partial_- \tilde{g}
+ \partial_+\Theta^1  \partial_- \Theta^1
+ \partial_+\Theta^2  \partial_- \Theta^2
\\ & \qquad \qquad \qquad \qquad
+ \partial_+ \Theta^1  \partial_- \tilde{g} \tilde{g}^{-1}
+ \partial_- \Theta^1  \partial_+ \tilde{g} \tilde{g}^{-1}
+ \partial_+ \Theta^2  \tilde{g}^{-1} \partial_- \tilde{g}
+ \partial_- \Theta^2  \tilde{g}^{-1} \partial_+ \tilde{g}
\\ & \qquad \qquad \qquad \qquad
+ \tilde{g}^{-1} \partial_+\Theta^1  \tilde{g} \partial_- \Theta^2
+ \tilde{g}^{-1} \partial_-\Theta^1  \tilde{g} \partial_+ \Theta^2  \big)
\\ & \quad
+ \frac{\kay}{6} \int d^3 x \, \epsilon^{ijk} \tr\big( \tilde{g}^{-1}\partial_i \tilde{g} [\tilde{g}^{-1}\partial_j \tilde{g} , \tilde{g}^{-1}\partial_k \tilde{g}] \big)
\\ & \quad
+ \frac{\kay}{2} \int d^2 x \,
\tr \big(
\partial_+ \Theta^1  \partial_- \tilde{g} \tilde{g}^{-1}
- \partial_- \Theta^1  \partial_+ \tilde{g} \tilde{g}^{-1}
- \partial_+ \Theta^2  \tilde{g}^{-1} \partial_- \tilde{g}
+ \partial_- \Theta^2  \tilde{g}^{-1} \partial_+ \tilde{g}
\\ & \qquad \qquad \qquad \qquad
+ \tilde{g}^{-1} \partial_+ \Theta^1  \tilde{g} \partial_- \Theta^2
- \tilde{g}^{-1} \partial_- \Theta^1  \tilde{g} \partial_+ \Theta^2
\big)
\end{split}\end{equation*}
\fi
\begin{equation*}\begin{split}
\Act_{\indrm{WZW}} & =
- \frac{\kay}{2} \int d^2x \, \tr\big(
\tilde{g}^{-1}\partial_+ \tilde{g} \tilde{g}^{-1}\partial_- \tilde{g} \big)
+ \frac{\kay}{6} \int d^3 x \, \epsilon^{ijk} \tr\big( \tilde{g}^{-1}\partial_i \tilde{g} [\tilde{g}^{-1}\partial_j \tilde{g} , \tilde{g}^{-1}\partial_k \tilde{g}] \big)
\\ & \quad \,
- \kay \int d^2 x \, \tr\big(
\frac12 \partial_+\Theta^1  \partial_- \Theta^1 + \frac12 \partial_+\Theta^2  \partial_- \Theta^2
+ \partial_- \Theta^1  \partial_+ \tilde{g} \tilde{g}^{-1}
+ \partial_+ \Theta^2  \tilde{g}^{-1} \partial_- \tilde{g}
+ \tilde{g}^{-1} \partial_-\Theta^1  \tilde{g} \partial_+ \Theta^2  \big) ~,
\end{split}\end{equation*}
where we have used that $\alg{u}(1)$ is abelian.
T-dualising $\Theta^1 \to \tilde \Theta_1$ and then shifting $\Theta^2 \to \Theta^2 - \gamma \tilde \Theta_1$ the second line of this action becomes
\ifextra
\begin{equation*}\begin{split}
& - \kay \int d^2 x \, \tr\big(
\frac12 A_+ A_- + \frac12 \partial_+\Theta^2  \partial_- \Theta^2
+ A_-  \partial_+ \tilde{g} \tilde{g}^{-1}
+ \partial_+ \Theta^2  \tilde{g}^{-1} \partial_- \tilde{g}
+ \tilde{g}^{-1} A_-  \tilde{g} \partial_+ \Theta^2
\\ & \qquad \qquad \qquad \qquad
+ \frac12 \partial_+ \tilde \Theta_1 A_- - \frac12 \partial_- \tilde \Theta_1 A_+  \big)
\\
\to & - \kay \int d^2 x \, \tr\big(
\frac12 \partial_+ \tilde \Theta_1 \partial_- \tilde \Theta_1
+ \frac12 \partial_+\Theta^2  \partial_- \Theta^2
+ \partial_- \tilde \Theta_1 \partial_+ \tilde{g} \tilde{g}^{-1}
+ \partial_+ \Theta^2  \tilde{g}^{-1} \partial_- \tilde{g}
+ \tilde{g}^{-1} \partial_- \tilde \Theta_1  \tilde{g} \partial_+ \Theta^2
\big)
\end{split}\end{equation*}
\fi
\begin{equation*}\begin{split}
& - \kay \int d^2 x \, \tr\big(
\frac12 \partial_+ \tilde \Theta_1 \partial_- \tilde \Theta_1
+ \frac12 \partial_+\Theta^2  \partial_- \Theta^2
- \frac\gamma2 \partial_+\tilde\Theta_1  \partial_- \Theta^2
- \frac\gamma2 \partial_+\Theta^2  \partial_- \tilde\Theta_1
+ \frac{\gamma^2}2 \partial_+\tilde\Theta_1  \partial_- \tilde\Theta_1
\\ & \qquad \qquad \quad
+ \partial_- \tilde \Theta_1 \partial_+ \tilde{g} \tilde{g}^{-1}
+ \partial_+ \Theta^2  \tilde{g}^{-1} \partial_- \tilde{g}
- \gamma \partial_+ \tilde\Theta_1  \tilde{g}^{-1} \partial_- \tilde{g}
+ \tilde{g}^{-1} \partial_- \tilde \Theta_1  \tilde{g} \partial_+ \Theta^2
- \gamma \tilde{g}^{-1} \partial_- \tilde \Theta_1  \tilde{g} \partial_+ \tilde \Theta_1
\big) ~.
\end{split}\end{equation*}
Now T-dualising $\tilde \Theta_1 \to \Theta^1$ we find
\ifextra
\begin{equation*}\begin{split}
& - \kay \int d^2 x \, \tr\big(
A_+ \frac{1-2\gamma\Ad_{\tilde{g}}^{-1}+\gamma^2}2  A_-
+ \frac12 \partial_+\Theta^2  \partial_- \Theta^2
- \frac\gamma2 A_+  \partial_- \Theta^2
- \frac\gamma2 \partial_+\Theta^2 A_-
\\ & \qquad \qquad \quad
+ A_- \partial_+ \tilde{g} \tilde{g}^{-1}
+ \partial_+ \Theta^2  \tilde{g}^{-1} \partial_- \tilde{g}
- \gamma A_+  \tilde{g}^{-1} \partial_- \tilde{g}
+ \tilde{g}^{-1} A_-  \tilde{g} \partial_+ \Theta^2
+ \frac12 \partial_+ \Theta^1 A_- - \frac12 \partial_- \Theta^1 A_+
\big)
\end{split}\end{equation*}
\begin{equation*}
\frac{1+\gamma^2-2\gamma P \Ad_{\tilde{g}}^{-1}}2  A_- = \frac12 \partial_- \Theta^1 + \frac{\gamma}{2} \partial_- \Theta^2 + \gamma P \tilde{g}^{-1} \partial_- \tilde{g}
\end{equation*}
\begin{equation*}\begin{split}
& - \kay \int d^2 x \, \tr\big(
\frac12 \partial_+\Theta^2  \partial_- \Theta^2
- \frac\gamma2 \partial_+\Theta^2 A_-
+ A_- \partial_+ \tilde{g} \tilde{g}^{-1}
+ \partial_+ \Theta^2  \tilde{g}^{-1} \partial_- \tilde{g}
+ \tilde{g}^{-1} A_-  \tilde{g} \partial_+ \Theta^2
+ \frac12 \partial_+ \Theta^1 A_-
\big)
\\ & =
- \kay \int d^2 x \, \tr\big(
\frac12 \partial_+\Theta^2  \partial_- \Theta^2
+ \partial_+ \Theta^2  \tilde{g}^{-1} \partial_- \tilde{g}
+ (\frac12 \partial_+ \Theta^1
+ \partial_+ \tilde{g} \tilde{g}^{-1}
+ \Ad_{\tilde{g}}\partial_+ \Theta^2
-  \frac\gamma2 \partial_+\Theta^2
) A_-
\big)
\end{split}\end{equation*}
\fi
\begin{equation*}\begin{split}
& - \kay \int d^2 x \, \tr\big(
\frac12 \partial_+\Theta^2  \partial_- \Theta^2
+ \partial_+ \Theta^2  \tilde{g}^{-1} \partial_- \tilde{g}
\\ & \qquad \qquad \quad
+ (\frac12 \partial_+ (\Theta^1 -\gamma\Theta^2)
+ \partial_+ \tilde{g} \tilde{g}^{-1}
+ \Ad_{\tilde{g}}\partial_+ \Theta^2
)
P
\frac{\lambda\gamma^{-1}}{1-\lambda \Ad_{\tilde{g}}^{-1}P}
( \frac12 \partial_- (\Theta^1 + \gamma \Theta^2) + \gamma \tilde{g}^{-1} \partial_- \tilde{g}) ~,
\end{split}\end{equation*}
where we have introduced
\begin{equation*}
\lambda = \frac{2\gamma}{1+\gamma^2} ~.
\end{equation*}
Finally, redefining $\Theta^1 \to \Theta^1 + \gamma \Theta^2$ and $\Theta^2 \to \Theta^2 + \gamma \Theta^1$, the TsT transformed action takes the form
\begin{equation*}\begin{split}
& \Act_\indrm{TsT-WZW} =
\\ & \quad \,
- \frac{\kay}{2} \int d^2x \, \tr\big(
\tilde{g}^{-1}\partial_+ \tilde{g} \tilde{g}^{-1}\partial_- \tilde{g} \big)
+ \frac{\kay}{6} \int d^3 x \, \epsilon^{ijk} \tr\big( \tilde{g}^{-1}\partial_i \tilde{g} [\tilde{g}^{-1}\partial_j \tilde{g} , \tilde{g}^{-1}\partial_k \tilde{g}] \big)
\\ & \quad \,
- \kay \int d^2 x \, \tr\big(
\frac12 \partial_+\Theta^1  \partial_- \Theta^1 + \frac12 \partial_+\Theta^2  \partial_- \Theta^2
+ \partial_- \Theta^1  \partial_+ \tilde{g} \tilde{g}^{-1}
+ \partial_+ \Theta^2  \tilde{g}^{-1} \partial_- \tilde{g}
+ \tilde{g}^{-1} \partial_-\Theta^1  \tilde{g} \partial_+ \Theta^2  \big)
\\
& \quad \, - \kay \int d^2 x \, \tr\big(
(\partial_+ \Theta^1 + \partial_+ \tilde{g} \tilde{g}^{-1}
+ \Ad_{\tilde{g}} \partial_+ \Theta^2 )P \frac{1}{\lambda^{-1} - \Ad_{\tilde{g}}^{-1}P}
(\partial_-\Theta^2 + \tilde{g}^{-1} \partial_- \tilde{g} + \Ad_{\tilde{g}}^{-1} \partial_-\Theta^1  \big)
\\
& \quad \, + \frac{\kay \gamma}{2} \int d^2x \, \tr\big(\partial_+ \Theta^1 \partial_- \Theta^2 - \partial_+ \Theta^2 \partial_- \Theta^1 \big) ~.
\end{split}\end{equation*}
On the other hand, if we set
\begin{equation*}
g = e^{\Theta^1} \tilde{g} e^{\Theta^2} ~,
\end{equation*}
in the action of the abelian current-current perturbation of the WZW model~\eqref{eq:u1def2} we find
\begin{equation*}\begin{split}
\Act_{\indrm{\Lambda_{ab}}} & =
- \frac{\kay}{2} \int d^2x \, \tr\big(
\tilde{g}^{-1}\partial_+ \tilde{g} \tilde{g}^{-1}\partial_- \tilde{g} \big)
+ \frac{\kay}{6} \int d^3 x \, \epsilon^{ijk} \tr\big( \tilde{g}^{-1}\partial_i \tilde{g} [\tilde{g}^{-1}\partial_j \tilde{g} , \tilde{g}^{-1}\partial_k \tilde{g}] \big)
\\ & \quad \,
- \kay \int d^2 x \, \tr\big(
\frac12 \partial_+\Theta^1  \partial_- \Theta^1 + \frac12 \partial_+\Theta^2  \partial_- \Theta^2
+ \partial_- \Theta^1  \partial_+ \tilde{g} \tilde{g}^{-1}
+ \partial_+ \Theta^2  \tilde{g}^{-1} \partial_- \tilde{g}
+ \tilde{g}^{-1} \partial_-\Theta^1  \tilde{g} \partial_+ \Theta^2  \big)
\\
& \quad \, - \kay \int d^2 x \, \tr\big(
(\partial_+ \Theta^1 + \partial_+ \tilde{g} \tilde{g}^{-1}
+ \Ad_{\tilde{g}} \partial_+ \Theta^2 ) P \frac{1}{\lambda^{-1} - \Ad_{\tilde{g}}^{-1}P}
(\partial_-\Theta^2 + \tilde{g}^{-1} \partial_- \tilde{g} + \Ad_{\tilde{g}}^{-1} \partial_-\Theta^1  \big) ~,
\end{split}\end{equation*}
where we have used that $\Ad_{e^\Theta} P = P = P \Ad_{e^\Theta}$ for $\Theta \in \alg{u}(1)$.
This follows from the fact that $P$ is an orthogonal projector.
Therefore,
\begin{equation*}
\Act_{\indrm{\Lambda_{ab}}} = \Act_{\indrm{TsT-WZW}}
- \frac{\kay \gamma}{2} \int d^2x \, \tr\big(\partial_+ \Theta^1 \partial_- \Theta^2 - \partial_+ \Theta^2 \partial_- \Theta^1 \big) ~,
\end{equation*}
that is the two actions are equal up to closed terms.
}
\else

To show that the abelian current-current deformation~\eqref{eq:u1def2} can also be found as a TsT transformation of the WZW model up to closed terms we parametrise
\begin{equation}
g = e^{\Theta^1} \tilde{g} e^{\Theta^2} ~, \qquad \Theta^1,\Theta^2 \in \alg{u}(1) ~,
\end{equation}
T-dualise $\Theta^1 \to \tilde \Theta_1$, then shift $\Theta^2 \to \Theta^2 - \gamma \tilde \Theta_1$ and finally T-dualise $\tilde\Theta_1 \to \Theta^1$.
Explicitly, setting $g = e^{\Theta^1} \tilde{g} e^{\Theta^2}$ in the WZW model gives
\begin{equation}\begin{split}
\Act_{\indrm{WZW}} & = - \frac{\kay}{2} \int d^2x \, \tr \big(g^{-1}\partial_+ g g^{-1}\partial_- g\big)
+ \frac{\kay}{6} \int d^3 x \, \epsilon^{ijk} \tr\big( g^{-1}\partial_i g [g^{-1}\partial_j g , g^{-1}\partial_k g] \big)
\\
& =
- \frac{\kay}{2} \int d^2x \, \tr\big(
\tilde{g}^{-1}\partial_+ \tilde{g} \tilde{g}^{-1}\partial_- \tilde{g} \big)
+ \frac{\kay}{6} \int d^3 x \, \epsilon^{ijk} \tr\big( \tilde{g}^{-1}\partial_i \tilde{g} [\tilde{g}^{-1}\partial_j \tilde{g} , \tilde{g}^{-1}\partial_k \tilde{g}] \big)
\\ & \qquad
- \kay \int d^2 x \, \tr\big(
\frac12 \partial_+\Theta^1  \partial_- \Theta^1 + \frac12 \partial_+\Theta^2  \partial_- \Theta^2
\\ & \qquad \hphantom{- \kay \int d^2 x \, \tr\big( ~ }
+ \partial_- \Theta^1  \partial_+ \tilde{g} \tilde{g}^{-1}
+ \partial_+ \Theta^2  \tilde{g}^{-1} \partial_- \tilde{g}
+ \tilde{g}^{-1} \partial_-\Theta^1  \tilde{g} \partial_+ \Theta^2  \big) ~,
\end{split}\end{equation}
where we have used that $\alg{u}(1)$ is abelian.
T-dualising $\Theta^1 \to \tilde \Theta_1$, shifting $\Theta^2 \to \Theta^2 - \gamma \tilde \Theta_1$, T-dualising $\tilde \Theta_1 \to \Theta^1$, and finally  redefining $\Theta^1 \to \Theta^1 + \gamma \Theta^2$ and $\Theta^2 \to \Theta^2 + \gamma \Theta^1$, the TsT transformed action takes the form
\begin{equation}\begin{split}
& \Act_\indrm{TsT-WZW} =
\\
& \quad - \frac{\kay}{2} \int d^2x \, \tr\big(
\tilde{g}^{-1}\partial_+ \tilde{g} \tilde{g}^{-1}\partial_- \tilde{g} \big)
+ \frac{\kay}{6} \int d^3 x \, \epsilon^{ijk} \tr\big( \tilde{g}^{-1}\partial_i \tilde{g} [\tilde{g}^{-1}\partial_j \tilde{g} , \tilde{g}^{-1}\partial_k \tilde{g}] \big)
\\ & \qquad
- \kay \int d^2 x \, \tr\big(
\frac12 \partial_+\Theta^1  \partial_- \Theta^1 + \frac12 \partial_+\Theta^2  \partial_- \Theta^2
+ \partial_- \Theta^1  \partial_+ \tilde{g} \tilde{g}^{-1}
+ \partial_+ \Theta^2  \tilde{g}^{-1} \partial_- \tilde{g}
+ \tilde{g}^{-1} \partial_-\Theta^1  \tilde{g} \partial_+ \Theta^2  \big)
\\
& \qquad - \kay \int d^2 x \, \tr\big(
(\partial_+ \Theta^1 + \partial_+ \tilde{g} \tilde{g}^{-1}
+ \Ad_{\tilde{g}} \partial_+ \Theta^2 )P \frac{1}{\lambda^{-1} - \Ad_{\tilde{g}}^{-1}P}
(\partial_-\Theta^2 + \tilde{g}^{-1} \partial_- \tilde{g} + \Ad_{\tilde{g}}^{-1} \partial_-\Theta^1  \big)
\\
& \qquad + \frac{\kay \gamma}{2} \int d^2x \, \tr\big(\partial_+ \Theta^1 \partial_- \Theta^2 - \partial_+ \Theta^2 \partial_- \Theta^1 \big) ~,
\end{split}\end{equation}
where
\begin{equation}
\lambda = \frac{2\gamma}{1+\gamma^2} ~.
\end{equation}
On the other hand, if we set $g = e^{\Theta^1} \tilde{g} e^{\Theta^2}$ in the action of the abelian current-current perturbation of the WZW model~\eqref{eq:u1def2} we find
\begin{equation}
\Act_{\indrm{\Lambda_{ab}}} = \Act_{\indrm{TsT-WZW}}
- \frac{\kay \gamma}{2} \int d^2x \, \tr\big(\partial_+ \Theta^1 \partial_- \Theta^2 - \partial_+ \Theta^2 \partial_- \Theta^1 \big) ~,
\end{equation}
where we have used that $\Ad_{e^\Theta} P = P = P \Ad_{e^\Theta}$ for $\Theta \in \alg{u}(1)$, which follows from the fact that $P$ is an orthogonal projector.
Therefore, the two actions are equal up to closed terms.
\fi

\paragraph{Examples based on $\grp{SL}(2)$.}
We finish by briefly studying the abelian current-current deformation of the $\grp{SU}(2)$ WZW model.
Using the generators $T_1$, $T_2$ and $T_3$ of the compact real form~\eqref{eq:su2gen} we parametrise
\begin{equation}\label{eq:hparam}
g = \exp\big(\frac{\varphi+\phi}{2} T_3\big) \exp\big(\theta T_1\big) \exp\big(\frac{\varphi - \phi}{2} T_3\big) ~,
\end{equation}
and take the $\alg{u}(1)$ subalgebra to be generated by $T_3$.
Substituting into the action~\eqref{eq:u1def2} with $\tr = \Tr$, we find that the deformed target-space metric and H-flux are given by
\begin{equation}\begin{split}
G & = 2\kay \big(d\theta^2 + \frac{(1+\lambda)\cos^2\theta}{1-\lambda\cos2\theta} d\varphi^2 + \frac{(1-\lambda)\sin^2\theta}{1-\lambda\cos2\theta} d\phi^2 \big) ~,
\\
H & = -2\kay\frac{(1-\lambda^2)\sin2\theta}{(1-\lambda\cos 2\theta)^2} d\theta\wedge d\varphi \wedge d\phi ~,
\end{split}\end{equation}
such that for $\lambda = 0$ we recover the background of the $\grp{SU}(2)$ WZW model.
If we take the degenerate limits $\lambda \to \pm 1$ then the H-flux vanishes and the background metric becomes that of the $\grp{SU}(2)/\grp{U}(1)$ vectorially or axially gauged WZW models for $\lambda = 1$ and $\lambda = -1$ respectively.
Note that if we first rescale $\varphi \to \sqrt{\frac{2}{1+\lambda}} \varphi$ and $\phi \to \sqrt{\frac{2}{1-\lambda}}\phi$ then the deformed background becomes
\begin{equation}\begin{split}
G & = 2\kay \big(d\theta^2 + \frac{2\cos^2\theta}{1-\lambda\cos2\theta} d\varphi^2 + \frac{2\sin^2\theta}{1-\lambda\cos2\theta} d\phi^2 \big) ~,
\\
H & = -4\kay\frac{\sqrt{1-\lambda^2}\sin2\theta}{(1-\lambda\cos 2\theta)^2} d\theta\wedge d\varphi \wedge d\phi ~,
\end{split}\end{equation}
and the $\lambda \to \pm 1$ limits become non-degenerate.
The H-flux still vanishes, while for $\lambda \to 1$ the metric becomes
\begin{equation}
G = 2\kay\big (d\theta^2 + \cot^2 \theta d\varphi^2 + d\phi^2\big) ~,
\end{equation}
corresponding to the $\grp{SU}(2)/\grp{U}(1)$ vectorially gauged WZW model plus a free boson, and for $\lambda \to -1$ it becomes
\begin{equation}
G = 2\kay\big (d\theta^2 + \tan^2 \theta d\phi^2 + d\varphi^2\big) ~,
\end{equation}
corresponding to the $\grp{SU}(2)/\grp{U}(1)$ axially gauged WZW model plus a free boson.

\section{Bibliography and generalisations}\label{sec:generalisations}

In \iflectures these lecture notes \else this pedagogical review \fi we have focused on the PCM, the PCWZM and the WZW model for simple Lie group $\grp{G}$ and some of their integrable deformations and dualities.
This has allowed us to explore these models and the underlying algebraic structures in more detail, but it has also meant that we have only discussed a small corner of the space of integrable sigma models.
There is a large landscape of integrable sigma models and their classification continues to be an active area of study.

\subsection{Symmetric integrable sigma models}

In addition to the PCM~\cite{Zakharov:1973pp,Pohlmeyer:1975nb} and the PCWZM~\cite{Novikov:1982ei,Witten:1983ar}, another well-known integrable sigma model is the symmetric space sigma model (SSSM)~\cite{Eichenherr:1979ci}.
The PCM, PCWZM and SSSM have a long history and much is understood about their classical and quantum physics~\cite{Faddeev:1985qu,Faddeev:1987ph}.
A recent review can be found in~\cite{Zarembo:2017muf}.
The SSSM describes sigma models whose target space is a symmetric space.
This includes, for example, spheres and anti de Sitter spaces
\begin{equation}
\Sp^n \cong \frac{\grp{SO}(n+1)}{\grp{SO}(n)} ~, \qquad
\AdS_n \cong \frac{\grp{SO}(2,n-1)}{\grp{SO}(1,n-1)} ~.
\end{equation}
Symmetric spaces can be realised as $\Integer_2$ cosets
\begin{equation}\label{eq:z2coset}
\frac{\grp{G}}{\grp{H}} ~.
\end{equation}
Defining $\alg{g} = \Lie\grp{G}$ and $\alg{h} = \Lie\grp{H}$, the $\Integer_2$ grading implies that
\begin{equation}
\alg{g} = \alg{h} \dotplus \alg{p} ~, \qquad [\alg{h},\alg{h}] \subset \alg{h} ~, \quad
[\alg{h},\alg{p}] \subset \alg{p} ~, \quad [\alg{p},\alg{p}] \subset \alg{p} ~.
\end{equation}
Indeed, these commutation relations imply that the Lie algebra $\alg{g}$ admits the $\Integer_2$ automorphism $\sigma(\alg{h}) = \alg{h}$, $\sigma(\alg{p}) = - \alg{p}$, whose fixed points are elements of $\alg{h}$.
Moreover, they imply that the decomposition $\alg{g} = \alg{h} \dotplus \alg{p}$ is orthogonal with respect to the Killing form, that is $\tr\big(\alg{h}\alg{p}\big) = 0$.
The action of the SSSM is
\begin{equation}\label{eq:sssm}
\Act_{\indrm{SSSM}} = - \frac{\hay}{2} \int d^2x \, \tr \big(j_+ P_1 j_- \big) ~,
\end{equation}
where $j_\pm = g^{-1}\partial_\pm g$ is the pull-back of the left-invariant Maurer-Cartan form for the group-valued field $g \in \grp{G}$ and we denote the orthogonal projectors onto $\alg{h}$ and $\alg{p}$ by $P_0$ and $P_1$ respectively.
The Lax connection of the SSSM is given by
\begin{equation}
\Lax_\pm(z) = P_0 j_\pm + z^{\pm 1} P_1 j_\pm ~.
\end{equation}

The SSSM can be generalised to $\Integer_T$ cosets~\eqref{eq:z2coset}, for which
\begin{equation}\label{eq:ztdecomp}
\alg{g} = \alg{h} \dotplus \Big(\bigdotplus_{k=1}^{T-1} \alg{p}_k\Big) ~, \qquad
[\alg{p}_k , \alg{p}_l] \subset \alg{p}_{k+l \!\!\!\! \mod T} ~, \qquad
\tr\big(\alg{p}_k \alg{p}_l) = 0 \text{ if $k+l\neq 0 \!\!\!\!\mod T$}~,
\end{equation}
where we define $\alg{p}_0 = \alg{h}$.
In this case the complexified Lie algebra $\alg{g}^\Complex$ admits the $\Integer_T$ automorphism $\sigma(\alg{p}_k) = e^{\frac{2i\pi k}{T}} \alg{p}_k$, whose fixed points are again elements of $\alg{h}$.
Denoting the projectors onto $\alg{p}_k$ in the decomposition~\eqref{eq:ztdecomp} by $P_k$, the action
\begin{equation}\label{eq:z4cosets}
\Act_{\ind{\Integer_T}} = - \frac{\hay}{2} \int d^2x \, \tr \big(j_+ \mathcal{P}_- j_-
\big) ~, \qquad \mathcal{P}_\pm = \sum_{k=1}^{T-1} k P_{\frac{T}{2}\pm(\frac{T}{2}-k)} ~,
\end{equation}
on a $\Integer_T$ coset admits the Lax connection
\begin{equation}
\Lax_\pm(z) = P_0 j_\pm + \sum_{k=1}^{T-1} z^{\pm k} P_{\frac{T}{2}\pm(\frac{T}{2}-k)} j_\pm ~.
\end{equation}
This action and its Lax connection were constructed in \cite{Berkovits:2000fe,Berkovits:2000yr,Vallilo:2003nx} for $T=4$ and generalised to arbitrary $T$ in \cite{Young:2005jv}.
This is not the unique classically integrable sigma model on a $\Integer_T$ coset.
Indeed, an alternative is given by
\begin{equation}\label{eq:gs}
\Act'_{\ind{\Integer_T}} = - \frac{\hay}{2} \int d^2x \, \tr \big(j_+ \mathcal{Q}_- j_- \big) ~, \qquad \mathcal{Q}_\pm = \sum_{k=1}^{\lfloor\frac{T}{2}\rfloor} k  P_{\frac{T}{2}\pm(\frac{T}{2}-k)} - \sum_{k=\lfloor\frac{T}{2}\rfloor+1}^{T-1} (T-k)  P_{\frac{T}{2}\pm(\frac{T}{2}-k)} ~,
\end{equation}
which admits the Lax connection
\begin{equation}
\Lax_\pm(z) = P_0 j_\pm + \sum_{k=1}^{\lfloor\frac{T}{2}\rfloor} z^{\pm k} P_{\frac{T}{2}\pm(\frac{T}{2}-k)} j_\pm
+ \sum_{k=\lfloor\frac{T}{2}\rfloor+1}^{T-1} z^{\mp(T-k)}  P_{\frac{T}{2}\pm(\frac{T}{2}-k)} j_\pm ~.
\end{equation}
The action~\eqref{eq:gs} and its Lax connection were constructed in~\cite{Metsaev:1998it,Henneaux:1984mh,Berkovits:1999zq,Bena:2003wd} for $T=4$ and generalised to the case when $4$ divides $T$ in~\cite{Ke:2008zz}.
The most general case is, to the best of our knowledge, a new result.
For $T=2$, both $\Act_{\ind{\Integer_T}}$ and $\Act'_{\ind{\Integer_T}}$ coincide with the SSSM~\eqref{eq:sssm}, and for $T\geq3$ they have a non-vanishing B-field, which is needed to ensure their classical integrability.

By construction, the projectors $P_k$ commute with the adjoint action of $h \in \grp{H}$.
It follows that the SSSM~\eqref{eq:sssm} and the $\Integer_T$ coset models~\eqref{eq:z4cosets} and~\eqref{eq:gs} are invariant under the $\grp{H}$ gauge symmetry
\begin{equation}\label{eq:gaugeright}
g\to gh ~, \qquad h(t,x) \in \grp{H} ~.
\end{equation}
To better understand the structure of these models, let us parametrise the group-valued field $g = \exp(X)$ with $X \in \bigdotplus_{k=1}^{T-1}\alg{p}_k$, where we have fixed the $\grp{H}$ gauge symmetry by setting $P_0 X = 0$.
Expanding the actions~\eqref{eq:z4cosets} and~\eqref{eq:gs} to quadratic order in $X$ we find
\begin{equation}\begin{split}
\Act_{\ind{\Integer_T}} & = - \frac{\hay}{2} \int d^2x \, \tr \big(
\sum_{k=1}^{\lfloor\frac{T}{2}\rfloor} k \partial_+  X P_k \partial_- X
+ \sum_{k=1}^{\lfloor\frac{T-1}{2}\rfloor} (T-k) \partial_+  X P_k \partial_- X \big)
+ \Order(X^3) ~,
\\
\Act'_{\ind{\Integer_T}} & = - \frac{\hay}{2} \int d^2x \, \tr \big(
\sum_{k=1}^{\lfloor\frac{T}{2}\rfloor} k \partial_+  X P_k \partial_- X
-\sum_{k=1}^{\lfloor\frac{T-1}{2}\rfloor} k \partial_+  X P_k \partial_- X
\big) + \Order(X^3) ~.
\end{split}\end{equation}
For the action $\Act_{\ind{\Integer_T}}$ all $\dim \grp{G} - \dim \grp{H}$ components of $X$ have a canonical kinetic term and can be interpreted as physical degrees of freedom.
On the other hand, for $\Act'_{\ind{\Integer_T}}$ with even $T$, only those components in $P_{\frac{T}{2}} X$ have a canonical kinetic term, while for odd $T$ there are no components with a canonical kinetic term.
Moreover, for odd $T$, the target-space metric vanishes and the model is purely topological, while for even $T$ with $T\geq 3$, the target-space metric is degenerate.
\unskip\footnote{A related model is given by taking the WZ term~\eqref{eq:WZterm} as an action in its own right.
The equations of motion are given by $\partial_+ j_- - \partial_- j_+ = - [j_+,j_-] = 0$, which are encoded by the Lax connection $\Lax_\pm = z j_\pm$.}
This degenerate structure may be indicating that these models have additional local symmetries and their proper interpretation remains to be understood.

The SSSM~\eqref{eq:sssm} and the $\Integer_T$ coset models~\eqref{eq:z4cosets} and~\eqref{eq:gs} have a $\grp{G}$ global symmetry, which acts as $g \to g_\ind{L} g$.
Just as for the PCM and the PCWZM, the conserved current of the SSSM can be normalised such that it is also flat on-shell.
Therefore, it can be used to construct a Lax connection and an infinite number of local conserved charges in involution~\cite{Faddeev:1985qu,Faddeev:1987ph,Babelon:2003qtg,Evans:1999mj,Evans:2000hx,Evans:2000qx}, as required for classical integrability.
More generally, such charges can be constructed if the Poisson bracket of the Lax matrix, that is the spatial component of the Lax connection $\Lax_x$, with itself satisfies a Maillet bracket~\cite{Maillet:1985fn,Maillet:1985ek,Rajeev:1996kk} with twist function~\cite{Lacroix:2017isl,Lacroix:2018njs}.
This construction does not require the existence of a conserved and flat current, hence can be applied to the $\Integer_T$ coset models~\cite{Magro:2008dv,Ke:2011zzb,Ke:2011zzd,Lacroix:2018njs} and is also useful when studying integrable deformations that break the global symmetry.
As in the case of the PCM, there is evidence that the coset models, in particular the SSSM, also have a hidden symmetry that takes the form of a classical Yangian algebra.
An introduction to classical integrability is given in the accompanying \iflectures lecture notes \else review article \fi ``Introduction to classical and quantum integrability'' by A.~L.~Retore~\cite{Retore:2021wwh} and a more in depth exposition can be found in~\cite{Babelon:2003qtg}, while comprehensive reviews on Yangians include~\cite{Bernard:1992ya,MacKay:2004tc,Loebbert:2016cdm}.

Let us now briefly discuss some of the quantum properties of these models.
The one-loop renormalisability of the 2-d sigma model~\eqref{eq:gensigmamodel} with the factor of $\frac{1}{2\pi\alpha'}$ restored is well-known to be governed by the Ricci flow equation~\cite{Friedan:1980jm,Braaten:1985is}
\begin{equation}\label{eq:ricciflow}
\dot G_\ind{MN} + \dot B_\ind{MN} = \alpha' \hat R_\ind{MN} + \dots ~,
\end{equation}
where the derivative is with respect to the log of the renormalisation group (RG) mass scale, $\hat R^\ind{M}{}_\ind{NPQ}$ is the Riemannian curvature of the generalised connection $\hat \Gamma^\ind{M}{}_\ind{NP} = \Gamma^\ind{M}{}_\ind{NP} - \frac12 H^\ind{M}{}_\ind{NP}$ and the additional terms correspond to contributions from wavefunction renormalisation, $\alpha' \mathcal{L}_Z (G_\ind{MN} + B_\ind{MN})$ where $\mathcal{L}_Z$ denotes the Lie derivative with respect to an arbitrary vector $Z$, and boundary terms, $\alpha' (dY)_{\ind{MN}}$ where $Y$ is an arbitrary 1-form.
The running of the coupling $\hay$ at one loop in the PCM, PCWZM and SSSM is proportional to the dual Coxeter number $h^\vee$ of the group $\grp{G}$.
It is a well-established result that for the PCM~\eqref{eq:pcmact} and SSSM~\eqref{eq:sssm} we have
\begin{equation}\begin{aligned}
\dot\hay & = \frac{1}{2\pi}\frac{h^\vee}{2} ~, \qquad
\\
\dot\hay & = \frac{1}{2\pi} h^\vee ~, \qquad
\end{aligned}\end{equation}
while for the PCWZM~\eqref{eq:pcmwzterm}
\begin{equation}
\dot\hay = \frac{1}{2\pi}\frac{h^\vee}{2} (1-\frac{\kay^2}{\hay^2}) ~, \qquad \dot\kay = 0 ~.
\end{equation}
Moreover, based on their global symmetries we can argue that these models are renormalisable to all loops.
Less is known about the one-loop renormalisability of the $\Integer_T$ coset models~\eqref{eq:z4cosets} and~\eqref{eq:gs} since the global symmetries do not constrain the coefficients in the sums of projectors $\mathcal{P}_\pm$ and $\mathcal{Q}_\pm$.
Based on low-dimensional examples, a natural conjecture for the running of the coupling $\hay$ at one loop in the $\Integer_T$ coset model~\eqref{eq:z4cosets} is
\begin{equation}
\dot \hay = \frac{1}{2\pi}\frac{2h^\vee}{T} ~.
\end{equation}
For the $\Integer_T$ coset model~\eqref{eq:gs} the situation is more subtle since the degeneracy of the target-space metric means that the Ricci flow equation~\eqref{eq:ricciflow} cannot be used straightforwardly.
Instead, an alternative approach, such as the background field method~\cite{Adam:2007ws,Polyakov:2004br,Zarembo:2010sg,Appadu:2015nfa} can be used, see~\cite{Abbott:1981ke} for an introductory review.

Recalling that $\hay$ is the inverse coupling constant, these one-loop RG equations imply that the PCM, SSSM and PCWZM are all asymptotically free in the UV.
The PCM and SSSM become strongly coupled in the IR, while the PCWZM flows to the fixed point $\hay = \kay$, corresponding to the WZW model.
Note that if we reverse the sign of the action, as we did when considering $\grp{G} = \grp{SL}(2,\Real)$, then the sign of the $\beta$ function is also reversed and the UV and IR behaviour is modified accordingly.
For compact Lie groups the PCM and SSSM exhibit dynamical mass generation in the IR and the quantum theory is described by a massive S-matrix~\cite{Zamolodchikov:1978xm}.

For all the symmetric integrable models we have discussed we can also take the Lie groups $\grp{G}$ and $\grp{H}$ to be supergroups and the property of classical integrability still holds with the Lax connections taking the same form.
The analogue of taking the Lie group $\grp{G}$ to be simple is to take the Lie supergroup $\grp{G}$ to be basic.
As an example, we may consider sigma models whose target space is given by the supersphere
\begin{equation*}
\Sp^{n|m} \cong \frac{\grp{OSp}(n+1|2m)}{\grp{OSp}(n|2m)} ~,
\end{equation*}
which is a symmetric space.
For $T=4$ both the actions~\eqref{eq:z4cosets}~\cite{Berkovits:2000fe,Berkovits:2000yr,Vallilo:2003nx} and~\eqref{eq:gs}~\cite{Metsaev:1998it,Henneaux:1984mh,Berkovits:1999zq,Bena:2003wd} are of interest in string theory.
Whether a classical 2-d sigma model describes a worldsheet string theory depends on the quantum properties of the model.
If $\grp{G}$ is a supergroup with vanishing dual Coxeter number then $\dot\hay = 0$ and the models are scale-invariant at one loop, see~\cite{Kagan:2005wt,Zarembo:2010sg} and references therein.
\unskip\footnote{Note that for supergroups with vanishing dual Coxeter number the Killing form vanishes, hence $\tr$ should be taken to be a suitable invariant bilinear form, such as the matrix supertrace in the defining representation.}
Furthermore, if the $\Integer_4$ coset also has one-loop central charge less than or equal to 26, these models may describe sectors of string theories on semi-symmetric spaces in the pure-spinor and Green-Schwarz formalisms respectively.
This is a non-trivial result and additional considerations and properties, such as ghosts and BRST symmetry in the pure-spinor formalism and $\kappa$-symmetry in the Green-Schwarz formalism, need to be taken into account or verified to ensure unitarity.
In addition both scale invariance and the stronger property of Weyl invariance should hold exactly.

The most well-known example of such a $\Integer_4$ coset is $\frac{\grp{PSU}(2,2|4)}{\grp{Sp}(2,2) \times \grp{Sp}(4)}$, which describes the maximally supersymmetric $\AdS_5 \times \Sp^5$ type IIB superstring.
Other examples include $\frac{\grp{OSp}(6|4)}{\grp{U}(3) \times \grp{SL}(2,\Complex)}$, $\frac{\grp{PSU}(1,1|2)}{\grp{SO}(1,1) \times \grp{SO}(2)}$ and $\frac{\grp{D}(2,1;\alpha)}{\grp{SO}(1,1) \times \grp{SO}(2) \times \grp{SO}(2)}$, which describe sectors of string theories on $\AdS_4 \times \CP^3$, $\AdS_2 \times \Sp^2 \times \To^6$ and $\AdS_2 \times \Sp^2 \times \Sp^2 \times \To^4$ respectively.
For all these examples the Lie supergroup $\grp{G}$ is basic and a general classification can be found in~\cite{Zarembo:2010sg,Wulff:2014kja}.
These worldsheet string theories are also of interest in the context of the AdS/CFT correspondence, for example, the maximally supersymmetric $\AdS_5 \times \Sp^5$ type IIB superstring is the holographic dual of $\mathcal{N}=4$ super-Yang-Mills theory in 4 dimensions.
The integrability of these models in the free string, $g_s = 0$, or planar limit has proven to be a powerful tool, leading to a proposal for the exact spectrum of string energies.
For further references see the review articles~\cite{Arutyunov:2009ga,Beisert:2010jr,Mazzucato:2011jt,Levkovich-Maslyuk:2019awk}.

\subsection{Yang-Baxter deformations}

The trigonometric and elliptic deformations of the $\grp{SU}(2)$ PCM~\cite{Cherednik:1981df} were some of the earliest studied integrable deformations of sigma models.
Other early examples of integrable deformations of the $\grp{SU}(2)$ PCM and the $\grp{SU}(2)/\grp{U}(1)$ SSSM were constructed in~\cite{Fateev:1992tk,Fateev:1995ht,Fateev:1996ea} although their classical integrability was only established later in~\cite{Lukyanov:2012zt}.
In \secref{sec:su2pcm} we saw that if we add a boundary term to the trigonometric deformation of the $\grp{SU}(2)$ PCM, the conserved current associated to the right-acting $\grp{G}$ global symmetry is also flat on-shell~\cite{Kawaguchi:2010jg}.
This construction can be further generalised to include a non-vanishing WZ term~\cite{Kawaguchi:2011mz}.
The resulting models coincide with the YB deformations of the $\grp{SU}(2)$ PCM and PCWZM based on the Drinfel'd-Jimbo solution to the mcYBe.
The deformed models of~\cite{Fateev:1992tk,Fateev:1995ht,Fateev:1996ea} also turn out to be related to YB deformations in a similar way~\cite{Hoare:2014pna}.

It is interesting to note that, even though they only differ by a boundary term, the natural algebraic structures of the trigonometric and YB deformations of the $\grp{SU}(2)$ PCM differ.
For example, the Lax connection of the YB deformation is valued in the homogeneous gradation of the affinization of $\alg{su}(2)$, while for the trigonometric deformation it is valued in the principal gradation~\cite{Appadu:2017bnv}.
Furthermore, the conserved charges that come from expanding the monodromy matrix generate the classical affine q-deformed algebra in the homogeneous and principal gradations respectively~\cite{Appadu:2017bnv,Kawaguchi:2012ve,Kawaguchi:2013gma}.

The YB deformation of the PCM for general group $\grp{G}$~\eqref{eq:ybdefgen} was constructed in~\cite{Klimcik:2002zj} and the existence of a Lax connection encoding its equations of motion was shown in~\cite{Klimcik:2008eq}.
YB deformations are so called since they depend on a solution to the (m)cYBe.
While the full classification of such solutions is rather involved, antisymmetric solutions to the mcYBe have been classified for simple Lie algebras over $\Complex$~\cite{Belavin:1982,Belavin:1984}.
Classifications of antisymmetric solutions to the cYBe have also been carried out for low-rank simple Lie algebras over $\Complex$, including, for example, $\alg{sl}(2,\Complex)$ and $\alg{sl}(3,\Complex)$~\cite{Ogievetsky:1992ph,Stolin:1991a}.
YB deformations based on the Drinfel'd-Jimbo solution~\cite{Drinfeld:1985rx,Jimbo:1985zk} of the mcYBe~\eqref{eq:djcomplex} are often called $\eta$ deformations in the literature.

The YB deformation of the PCM can be generalised to include the WZ term to give the YB deformation of the PCWZM for general group $\grp{G}$.
First constructed in~\cite{Delduc:2014uaa} for the Drinfel'd-Jimbo solution to the mcYBe, in~\cite{Klimcik:2019kkf} the model was written in the compact form
\begin{equation}\label{eq:ybpcwzm}
\Act_\indrm{YB-PCWZM} = -\frac{\kay}{2} \int d^2x \, \tr\big(j_+ \frac{e^\chi + e^{\rho \mathcal{R}_g} }{e^\chi - e^{\rho \mathcal{R}_g}} j_-\big)
+ \frac{\kay}{6} \int d^3x  \, \epsilon^{ijk} \tr \big(j_i[j_j,j_k]\big) ~.
\end{equation}
It was shown in~\cite{Hoare:2020mpv} that this action defines an integrable deformation of the PCWZM for any antisymmetric operator $\mathcal{R}$ that satisfies the (m)cYBe if the Lie algebras $\alg{h}_\pm \equiv \im(\mathcal{R} \pm c)$ are solvable.
It turns out that not all antisymmetric solutions of the (m)cYBe can be used to deform the PCWZM and an additional cohomological condition needs to be satisfied.
While solvable $\alg{h}_\pm$ is sufficient for this to be the case, it has not been proven that it is also necessary.
The Lax connection of the YB deformation of the PCWZM~\eqref{eq:ybpcwzm} is
\begin{equation}
\Lax_\pm(z) = \frac{J_\pm}{1\mp z} ~, \qquad J_\pm = \pm \frac{2 (\cosh\chi - \cosh c\rho)}{\sinh\chi(e^{\pm \chi} e^{\rho R_g} - 1)} g^{-1}\partial_\pm g ~.
\end{equation}
The limit in which the WZ term vanishes and we recover the YB deformation of the PCM~\eqref{eq:ybdefgen} is given by setting
\begin{equation}\label{eq:paramchange}
\chi = \frac{2\kay}{\hay} ~,
\qquad \rho = \frac{2\eta\kay}{\hay} ~,
\end{equation}
and taking $\kay \to 0$.
To recover the PCWZM~\eqref{eq:pcmwzterm} we take the deformation parameter $\rho \to 0$ and set $\coth\frac{\chi}{2} = \frac{\hay}{\kay}$.

It is also possible to construct the YB deformation of the SSSM~\eqref{eq:sssm}~\cite{Delduc:2013fga,Matsumoto:2015jja}, the action of which is
\begin{equation}\label{eq:ybsssm}
\Act_{\indrm{YB-SSSM}} = - \frac{\hay}{2} \int d^2x \, \tr\big(j_+ P_1 \frac{1}{1-\eta \mathcal{R}_g P_1} j_-\big) ~.
\end{equation}
The Lax connection of the model is given by
\begin{equation}
\Lax_\pm(z) = P_0 J_\pm + z^{\pm 1}\sqrt{1-c^2\eta^2} P_1 J_\pm ~, \qquad J_\pm = \frac{1}{1\pm\eta \mathcal{R}_g P_1} j_\pm ~.
\end{equation}
Defining
\begin{equation}
\tilde \eta = \frac{1-c\eta}{1+c\eta} ~,
\end{equation}
the YB deformation of the action~\eqref{eq:z4cosets} on a $\Integer_T$ coset is
\begin{equation}\begin{split}\label{eq:ybzt}
\Act_{\indrm{YB-}\ind{\Integer_T}} & = - \frac{\hay}{2} \int d^2x \, \tr\big(j_+ \mathcal{P}^\eta_- \frac{1}{1-\eta \mathcal{R}_g \mathcal{P}^\eta_-} j_-\big) ~,
\\
\mathcal{P}^\eta_\pm & = \sum_{k=1}^{T-1} \frac{1+\tilde\eta}{1-\tilde \eta}\frac{1 - \tilde\eta^k}{1 + \tilde\eta^k} P_{\frac{T}{2} \pm (\frac{T}{2}-k)} ~,
\end{split}\end{equation}
with the Lax connection
\begin{equation}\begin{split}\label{eq:ybztlax}
\Lax_\pm(z) & = P_0 J_\pm + \sum_{k=1}^{T-1} \frac{2z^{\pm k}}{\tilde\eta^{\frac{k}{2}} + \tilde\eta^{-\frac{k}{2}} } P_{\frac{T}{2}\pm(\frac{T}{2}-k)} J_\pm ~,
\\
J_\pm & = \frac{1}{1 \pm \eta \mathcal{R}_g \mathcal{P}^\eta_\pm} j_\pm ~.
\end{split}\end{equation}
Similarly, the YB deformation of the action~\eqref{eq:gs} is
\begin{equation}\begin{split}\label{eq:ybgs}
\Act'_{\indrm{YB-}\ind{\Integer_T}} & = - \frac{\hay}{2} \int d^2x \, \tr \big(j_+ \mathcal{Q}^\eta_- \frac{1}{1-\eta \mathcal{R}_g \mathcal{Q}^\eta_-} j_-\big) ~,
\\ \mathcal{Q}^\eta_\pm & =  \sum_{k=1}^{\lfloor\frac{T}{2}\rfloor}
\frac{1+\tilde\eta}{1-\tilde \eta}
\frac{1 - \tilde\eta^k}{1 + \tilde\eta^k}
P_{\frac{T}{2} \pm (\frac{T}{2}-k)}
- \sum_{k=\lfloor\frac{T}{2}\rfloor+1}^{T-1}
\frac{1+\tilde\eta}{1-\tilde \eta}
\frac{1 - \tilde\eta^{T-k}}{1 + \tilde\eta^{T-k}}
P_{\frac{T}{2} \pm (\frac{T}{2}-k)} ~,
\end{split}\end{equation}
with the Lax connection
\begin{equation}\begin{split}
\Lax_\pm(z) & = P_0 J_\pm +  \sum_{k=1}^{\lfloor\frac{T}{2}\rfloor}
\frac{2z^{\pm k}}{\tilde\eta^{\frac{k}{2}} + \tilde\eta^{-\frac{k}{2}} }
P_{\frac{T}{2}\pm(\frac{T}{2}-k)} J_\pm
+  \sum_{k=\lfloor\frac{T}{2}\rfloor+1}^{T-1}
\frac{2z^{\mp (T-k)}}{\tilde\eta^{\frac{T-k}{2}} + \tilde\eta^{-\frac{T-k}{2}} }
P_{\frac{T}{2}\pm(\frac{T}{2}-k)} J_\pm
~, \qquad
\\
J_\pm & = \frac{1}{1\pm \eta \mathcal{R}_g \mathcal{Q}^\eta_\pm} j_\pm ~.
\end{split}\end{equation}
Both these YB deformations coincide with that of the SSSM~\eqref{eq:ybsssm} for $T=2$.
The deformed actions~\eqref{eq:ybsssm},~\eqref{eq:ybzt} and~\eqref{eq:ybgs} are still invariant under the $\grp{H}$ gauge symmetry~\eqref{eq:gaugeright}, ensuring that they have the same number of degrees of freedom as the undeformed models, while the left-acting $\grp{G}$ global symmetry is broken to a subgroup depending on $\mathcal{R}$.

That these deformations exist follows from the affine Gaudin model formalism~\cite{Vicedo:2017cge,Lacroix:2018njs}.
However, apart from certain special cases, including $\Act_{\indrm{YB-}\ind{\Integer_T}}$ with $T=4$, $c=0$~\cite{Benitez:2018xnh}, and $\Act'_{\indrm{YB-}\ind{\Integer_T}}$ with $T=4$, $c=i$~\cite{Delduc:2013qra}, $T=4$, $c=0$~\cite{Kawaguchi:2014qwa} and when 4 divides $T$ with $c=0$~\cite{Ke:2017wis}, for $T\geq 3$ these deformed actions and their Lax connections are, to the best of our knowledge, new results.
A complementary and related line of investigation is the study of sigma models on flag manifolds and their deformations~\cite{Bykov:2016ovg}, which has been recently reviewed in~\cite{Affleck:2021ypq}.

The construction of an infinite number of conserved charges in involution for YB deformations follows from the fact that the Poisson bracket of the Lax matrix $\Lax_x$ with itself satisfies a Maillet bracket with twist function~\cite{Lacroix:2017isl,Lacroix:2018njs}.
This ensures the classical integrability of these models.
There is extensive evidence that the YB deformations also have a hidden symmetry that takes the form of twists of the classical Yangian algebra for homogeneous deformations~\cite{Matsumoto:2015jja,Vicedo:2015pna} and the classical q-deformed algebra for inhomogeneous deformations~\cite{Delduc:2013fga,Delduc:2014kha,Delduc:2017brb}.
For split, $c=1$, and non-split, $c=i$, inhomogeneous YB deformations the q-deformation parameter is the exponential of a phase and real respectively.
More precisely, the classical q-deformation parameter is given by $\widehat q = q^{\hay}$ where
\begin{equation}\label{eq:qdefp}
q = \exp\big(\frac{ic\eta}{\hay}\big) ~.
\end{equation}

At this point it is worth emphasising that while the YB deformations describe a large class of integrable deformations of symmetric sigma models, they do not describe all such deformations.
For example, the elliptic deformation of the $\grp{SU}(2)$ PCM does not take the form of the YB deformation, and, strictly speaking, the same is also true of the trigonometric deformation.
Recent work exploring generalisations of the latter to higher-rank Lie groups includes~\cite{Fukushima:2020kta,Tian:2020ryu}.

Even though the YB deformations lead to non-trivial deformations of the background fields, the resulting 2-d sigma models remain renormalisable at one loop.
This has been established for the deformed PCM~\eqref{eq:ybdefgen}, SSSM~\eqref{eq:ybsssm} and PCWZM~\eqref{eq:ybpcwzm}, see, for example,~\cite{Squellari:2014jfa,Sfetsos:2015nya,Demulder:2017zhz,Klimcik:2019kkf}.
For the deformed PCM and SSSM we have
\begin{equation}\begin{aligned}\label{eq:rgyb}
\dot\hay & = \frac{1}{2\pi}\frac{h^\vee}{2} (1-c^2\eta^2)^2 ~, \qquad
&& (\eta\hay^{-1})\dot{} = 0 ~,
\\
\dot\hay & = \frac{1}{2\pi} h^\vee (1-c^2\eta^2) ~, \qquad
&& (\eta\hay^{-1})\dot{} = 0 ~,
\end{aligned}\end{equation}
while for the deformed PCWZM
\begin{equation}\begin{aligned}
\dot\chi & = - \frac{1}{2\pi}\frac{h^\vee (\cosh\chi - \cosh c\rho)^2}{\kay \sinh^2\chi} ~, \qquad
&& \dot \rho = 0 ~, \qquad \dot \kay = 0 ~.
\end{aligned}\end{equation}
Again, less is known about the one-loop renormalisability of the deformed $\Integer_T$ coset models~\eqref{eq:ybzt} and~\eqref{eq:ybgs}, however, based on low-dimensional examples, a natural conjecture for the model~\eqref{eq:ybzt} is
\begin{equation}\begin{aligned}
\dot\hay & = \frac{1}{2\pi} \frac{2h^\vee}{T} (1-c^2\eta^2) ~, \qquad
&& (\eta\hay^{-1})\dot{} = 0 ~.
\end{aligned}\end{equation}
As in the undeformed case, an alternative approach is needed to analyse the model~\eqref{eq:ybgs} due to the degeneracy of the target-space metric.

For $c=0$ the RG equations do not depend on the deformation parameters $\eta$ and $\rho$ and the RG behaviour is the same as for the undeformed models.
More generally, the deformation does not affect the IR behaviour, that is both the deformed PCM and SSSM become strongly coupled, while the deformed PCWZM flows to the WZW model.
On the other hand, the UV behaviour is modified.
If $c=1$ all three deformed models flow to the $\eta = 1$ or $\rho = \chi$ fixed point.
If $c=i$, the RG flow is cyclic and does not have a well-defined UV behaviour.
Consequently, it has been posited that these deformations may only exist as effective field theories~\cite{Appadu:2018ioy}.
Again, we note that if we reverse the sign of the action then the sign of the $\beta$ function is also reversed.

As discussed above, the case $T=4$ is of interest in string theory.
These YB deformations have been constructed in~\cite{Benitez:2018xnh} in the pure-spinor formalism for $c=0$, including analyses of ghosts and BRST symmetry, and in~\cite{Delduc:2013qra,Delduc:2014kha,Kawaguchi:2014qwa} in the Green-Schwarz formalism, including the demonstration of $\kappa$-symmetry.
For those models whose undeformed counterparts describe worldsheet string theories, this raises the important question of whether one-loop Weyl invariance is preserved by the deformation.
While scale invariance is in general preserved~\cite{Hoare:2015wia,Arutyunov:2015mqj,Borsato:2016ose}, whether Weyl invariance is preserved or not depends on the properties of $\mathcal{R}$.
A sufficient condition for Weyl invariance at one loop is that the trace of the structure constants of $\alg{g}_{\mathcal{R}}$, that is the Lie algebra whose Lie bracket is the R-bracket~\eqref{eq:rbracket}, vanishes~\cite{Borsato:2016ose}.
However, it should be noted that a full analysis turns out to be more subtle~\cite{Borsato:2018spz}.
For the $\AdS_5 \times \Sp^5$ superstring, homogeneous YB deformations with this property were explored in~\cite{Borsato:2016ose,vanTongeren:2019dlq}, while inhomogeneous YB deformations have been investigated in~\cite{Hoare:2018ngg}.
For recent reviews discussing these models in more detail see~\cite{Orlando:2019his,Seibold:2020ouf}.

\subsection{Dualities and current-current deformations}

The relation between deformations and dualities dates back to early work on TsT transformations~\cite{Horne:1991gn,Giveon:1991jj}.
In a model with two commuting isometries these deformations are constructed by first T-dualising~\cite{Buscher:1987qj,Cvetic:1999zs,Kulik:2000nr} $\Theta^1 \to \tilde \Theta_1$, then shifting $\Theta^2 \to \Theta^2 - \eta\tilde\Theta_1$ and finally T-dualising back $\tilde \Theta_1 \to \Theta^1$.
Abelian homogeneous YB deformations are equivalent to TsT transformations~\cite{Matsumoto:2014nra,Osten:2016dvf}, while homogeneous YB deformations more generally are closely related to non-abelian T-dual models~\cite{Hoare:2016wsk,Borsato:2016pas,Borsato:2017qsx}.
Non-abelian T-duality~\cite{delaOssa:1992vci} generalises T-duality to models with non-abelian global symmetry groups.
Both T-duality and non-abelian T-duality preserve one-loop renormalisability.
However, non-abelian T-duality may introduce a Weyl anomaly if the structure constants of the algebra with respect to which we are T-dualising have non-vanishing trace~\cite{Alvarez:1994np,Elitzur:1994ri}.

Poisson-Lie T-duality~\cite{Klimcik:1995ux,Klimcik:1995jn} further generalises non-abelian T-duality to models without manifest global symmetries, but whose equations of motion can be written in the form of a non-commutative conservation law
\begin{equation}
\partial_+ K_- + \partial_- K_+ + [K_+,K_-]' = 0 ~.
\end{equation}
Here $K_\pm$ are valued in the Lie algebra $\alg{g}$ and $[\cdot,\cdot]'$ is the Lie bracket of a dual Lie algebra $\alg{g}'$.
The two Lie brackets $[\cdot,\cdot]$ and $[\cdot,\cdot]'$ satisfy a compatibility condition such that the vector space direct sum $\alg{d} = \alg{g} \dotplus \alg{g}'$ has the structure of a Drinfel'd double.
We say that the model has a $\grp{G}$ global Poisson-Lie symmetry with respect to $\grp{G}'$.
This construction includes models with $\grp{G}$ global symmetry, in which case $\grp{G}'$ is abelian.
YB deformations are examples of Poisson-Lie symmetric models.
The dual Lie algebra is given by $\alg{g}_\mathcal{R}$, the Lie algebra whose Lie bracket is the R-bracket~\eqref{eq:rbracket}.
Poisson-Lie T-duality is a canonical transformation, which implies that it preserves the property of classical integrability.
Furthermore, Poisson-Lie symmetric models are renormalisable at one loop, a property that is also preserved by Poisson-Lie T-duality~\cite{Valent:2009nv,Sfetsos:2009dj,Sfetsos:2009vt,Klimcik:2018vhl,Severa:2018pag,Klimcik:2019kkf}.
For recent reviews on T-duality and its generalisations see~\cite{Thompson:2019ipl,Klimcik:2021bjy}.

Current-current deformations of the WZW model preserving scale and conformal invariance, and their relation to TsT transformations and homogeneous YB deformations have been extensively studied, see, for example,~\cite{Borsato:2018spz} and references therein.
The isotropic current-current deformation of the WZW model~\eqref{eq:lambda}, which in general does not preserve scale or conformal invariance, is related to the non-abelian Thirring model~\cite{Dashen:1974hp,Kutasov:1989aw}.
The all-order in deformation parameter action was first constructed in~\cite{Sfetsos:2013wia}.
This model and its generalisations are often called $\lambda$ deformations in the literature.

It was later shown that the isotropic current-current deformation of the WZW model is related by Poisson-Lie T-duality to split inhomogeneous YB deformations for suitable real forms $\alg{g}$~\cite{Vicedo:2015pna}.
It can also be related to non-split inhomogeneous YB deformations by combining Poisson-Lie T-duality with analytic continuation~\cite{Hoare:2015gda,Sfetsos:2015nya,Klimcik:2015gba}.
There also exist analogous deformations of gauged WZW models that are Poisson-Lie T-dual to split inhomogeneous YB deformations of the SSSM~\eqref{eq:ybsssm} and the $\Integer_T$ coset models~\eqref{eq:ybzt} and~\eqref{eq:ybgs}.
As for the PCM, the resulting deformations can be considered for any real form $\alg{g}$, including the compact real form.
The actions of the deformed models are
\begin{equation}\begin{split}\label{eq:lambdasssm}
\Act_{\indrm{\Lambda-SSSM}} & = - \frac{\kay}{2} \int d^2x \, \tr \big(g^{-1}\partial_+ g g^{-1}\partial_- g\big)
+ \frac{\kay}{6} \int d^3 x \, \epsilon^{ijk} \tr\big( g^{-1}\partial_i g [g^{-1}\partial_j g , g^{-1}\partial_k g] \big)
\\
& \qquad + \kay \int d^2 x \, \tr\big(A_+ g^{-1} \partial_- g - A_- \partial_+gg^{-1} + A_+ g^{-1} A_- g - A_+ (P_0 + \lambda P_1)^{-1} A_-\big) ~,
\end{split}\end{equation}
for the SSSM~\cite{Sfetsos:2013wia,Hollowood:2014rla},
\begin{equation}\begin{split}\label{eq:lambdazt}
\Act_{\indrm{\Lambda-}\ind{\Integer_T}} & = - \frac{\kay}{2} \int d^2x \, \tr \big(g^{-1}\partial_+ g g^{-1}\partial_- g\big)
+ \frac{\kay}{6} \int d^3 x \, \epsilon^{ijk} \tr\big( g^{-1}\partial_i g [g^{-1}\partial_j g , g^{-1}\partial_k g] \big)
\\
& \qquad + \kay \int d^2 x \, \tr\big(A_+ g^{-1} \partial_- g - A_- \partial_+gg^{-1} + A_+ g^{-1} A_- g - A_+ (P_0 + \mathcal{P}^\lambda_-)^{-1} A_- \big) ~, \qquad
\\ \mathcal{P}^\lambda_\pm & = \sum_{k=1}^{T-1} \lambda^{k} P_{\frac{T}{2} \pm (\frac{T}{2}-k)} ~,
\end{split}\end{equation}
for the $\Integer_T$ coset model~\eqref{eq:z4cosets}, and
\begin{equation}\begin{split}\label{eq:lambdags}
\Act'_{\indrm{\Lambda-}\ind{\Integer_T}} & = - \frac{\kay}{2} \int d^2x \, \tr \big(g^{-1}\partial_+ g g^{-1}\partial_- g\big)
+ \frac{\kay}{6} \int d^3 x \, \epsilon^{ijk} \tr\big( g^{-1}\partial_i g [g^{-1}\partial_j g , g^{-1}\partial_k g] \big)
\\
& \qquad + \kay \int d^2 x \, \tr\big(A_+ g^{-1} \partial_- g - A_- \partial_+gg^{-1} + A_+ g^{-1} A_- g - A_+ (P_0 + \mathcal{Q}^\lambda_-)^{-1} A_- \big) ~, \qquad
\\ \mathcal{Q}^\lambda_\pm & = \sum_{k=1}^{\lfloor\frac{T}{2}\rfloor} \lambda^{k} P_{\frac{T}{2} \pm (\frac{T}{2}-k)} + \sum_{k=\lfloor\frac{T}{2}\rfloor+1}^{T-1} \lambda^{k-T} P_{\frac{T}{2} \pm (\frac{T}{2}-k)} ~,
\end{split}\end{equation}
for the $\Integer_T$ coset model~\eqref{eq:gs}, where we recall that the group-valued field $g \in \grp{G}$ and $A_\pm \in \alg{g}$.
Again the two actions~\eqref{eq:lambdazt} and~\eqref{eq:lambdags} both coincide with the action~\eqref{eq:lambdasssm} for $T = 2$.
All of these models have a $\grp{H}$ gauge symmetry acting as
\begin{equation}\label{eq:gaugelambda}
g \to h^{-1} g h ~, \qquad
A_\pm \to h^{-1} A_\pm h + h^{-1} \partial_\pm h ~, \qquad
h(t,x) \in \grp{H}~,
\end{equation}
hence they have the same number of degrees of freedom as their Poisson-Lie T-dual counterparts.
Typically they have no global symmetries.
The Lax connections for these models are
\begin{equation}
\Lax_\pm(z) = P_0 A_\pm + z^{\pm 1} \lambda^{-\frac12} P_1 A_\pm ~,
\end{equation}
\begin{equation}
\Lax_\pm(z) = P_0 A_\pm + \sum_{k=1}^{T-1} z^{\pm k} \lambda^{-\frac{k}{2}} P_{\frac{T}{2} \pm (\frac{T}{2} - k)} A_\pm ~,
\end{equation}
and
\begin{equation}
\Lax_\pm(z) = P_0 A_\pm
+ \sum_{k=1}^{\lfloor\frac{T}{2}\rfloor} z^{\pm k} \lambda^{-\frac{k}{2}} P_{\frac{T}{2} \pm (\frac{T}{2} - k)} A_\pm
+ \sum_{k=\lfloor\frac{T}{2}\rfloor+1}^{T-1} z^{\pm (k-T)} \lambda^{\frac{T-k}{2}} P_{\frac{T}{2} \pm (\frac{T}{2} - k)} A_\pm ~,
\end{equation}
respectively, where $A_\pm$ are given in terms of $g$ by their on-shell expressions.
For $T\geq 3$, and apart from the case $T=4$~\cite{Hollowood:2014qma}, these deformations and their Lax connections are, to the best of our knowledge, new results.

The fields $A_\pm$ enter the actions~\eqref{eq:lambdasssm},~\eqref{eq:lambdazt} and~\eqref{eq:lambdags} algebraically, hence can be easily integrated out to give
\begin{equation}\begin{split}\label{eq:l1}
\Act_{\indrm{\Lambda-SSSM}} & = - \frac{\kay}{2} \int d^2x \, \tr \big(g^{-1}\partial_+ g \frac{1+\Ad_g^{-1}(P_0 + \lambda P_1)}{1-\Ad_g^{-1}(P_0+\lambda P_1)} g^{-1}\partial_- g\big)
\\ & \qquad
+ \frac{\kay}{6} \int d^3 x \, \epsilon^{ijk} \tr\big( g^{-1}\partial_i g [g^{-1}\partial_j g , g^{-1}\partial_k g] \big) ~,
\end{split}\end{equation}
\begin{equation}\begin{split}\label{eq:l2}
\Act_{\indrm{\Lambda-}\ind{\Integer_T}} & = - \frac{\kay}{2} \int d^2x \, \tr \big(g^{-1}\partial_+ g \frac{1+\Ad_g^{-1}(P_0+\mathcal{P}^\lambda_-)}{1-\Ad_g^{-1}(P_0+\mathcal{P}^\lambda_-)} g^{-1}\partial_- g\big)
\\ & \qquad
+ \frac{\kay}{6} \int d^3 x \, \epsilon^{ijk} \tr\big( g^{-1}\partial_i g [g^{-1}\partial_j g , g^{-1}\partial_k g] \big) ~,
\end{split}\end{equation}
and
\begin{equation}\begin{split}\label{eq:l3}
\Act'_{\indrm{\Lambda-}\ind{\Integer_T}} & = - \frac{\kay}{2} \int d^2x \, \tr \big(g^{-1}\partial_+ g \frac{1+\Ad_g^{-1}(P_0+\mathcal{Q}^\lambda_-)}{1-\Ad_g^{-1}(P_0+\mathcal{Q}^\lambda_-)} g^{-1}\partial_- g\big)
\\ & \qquad
+ \frac{\kay}{6} \int d^3 x \, \epsilon^{ijk} \tr\big( g^{-1}\partial_i g [g^{-1}\partial_j g , g^{-1}\partial_k g] \big) ~.
\end{split}\end{equation}
These are the analogues of the action~\eqref{eq:lambda2} of the isotropic current-current deformation of the WZW model.
Observing that all these actions, including eq.~\eqref{eq:lambda2}, take the general form
\begin{equation}\begin{split}
- \frac{\kay}{2} \int d^2x \, \tr \big(g^{-1}\partial_+ g \frac{1+\Ad_g^{-1}\mathcal{O}(\lambda)}{1-\Ad_g^{-1}\mathcal{O}(\lambda)} g^{-1}\partial_- g\big)
+ \frac{\kay}{6} \int d^3 x \, \epsilon^{ijk} \tr\big( g^{-1}\partial_i g [g^{-1}\partial_j g , g^{-1}\partial_k g] \big) ~,
\end{split}\end{equation}
where $\mathcal{O}(\lambda^{-1}) = \mathcal{O}(\lambda)^{-1}$, it follows that they are invariant under the following $\Integer_2$ transformation~\cite{Itsios:2014lca,Hoare:2015gda,Kutasov:1989aw,Tseytlin:1993hm}.
\begin{equation}
g \to g^{-1} ~, \qquad \lambda \to \lambda^{-1} ~, \qquad \kay \to - \kay ~.
\end{equation}
This invariance has been particularly useful in the study of anomalous dimensions and correlation functions exact in the deformation parameter $\lambda$ at large $\kay$~\cite{Georgiou:2015nka,Georgiou:2016iom}.

Simply taking $\lambda \to 1$, the three models~\eqref{eq:lambdasssm},~\eqref{eq:lambdazt} and~\eqref{eq:lambdags} all limit to the $\grp{G}/\grp{G}$ gauged WZW model, the same as for the isotropic current-current deformation of the WZW model~\eqref{eq:lambda}.
Taking $\lambda \to 0$ in the actions~\eqref{eq:lambdasssm} and~\eqref{eq:lambdazt} we end up with the $\grp{G}/\grp{H}$ gauged WZW model.
In particular, it has been argued that $\Act_{\indrm{\Lambda-SSSM}}$ can be understood as a parafermion-parafermion deformation of this gauged WZW model~\cite{Sfetsos:2013wia}.
On the other hand, taking $\lambda \to 0$ in the action~\eqref{eq:lambdags} is more subtle, at least for $T \geq 3$.
We find that $P_k A_- = P_{T-k} A_+ = 0$ for $k=1,\dots,\lfloor\frac{T}{2}\rfloor$ and the action becomes
\begin{equation}\begin{split}\label{eq:lambdagslim}
\lim_{\lambda\to 0}\Act'_{\indrm{\Lambda-}\ind{\Integer_T}} & = - \frac{\kay}{2} \int d^2x \, \tr \big(g^{-1}\partial_+ g g^{-1}\partial_- g\big)
+ \frac{\kay}{6} \int d^3 x \, \epsilon^{ijk} \tr\big( g^{-1}\partial_i g [g^{-1}\partial_j g , g^{-1}\partial_k g] \big)
\\
& \qquad + \kay \int d^2 x \, \tr\big(A_+ g^{-1} \partial_- g - A_- \partial_+gg^{-1} + A_+ g^{-1} A_- g - A_+ P_0 A_- \big) ~,
\end{split}\end{equation}
where $A_+ \in \alg{h} \dotplus \big(\bigdotplus_{k=1}^{\lfloor \frac{T-1}{2}\rfloor} \alg{p}_k\big)$
and $A_- \in \alg{h} \dotplus \big(\bigdotplus_{k=\lfloor \frac{T}{2}\rfloor + 1}^{T-1} \alg{p}_k\big)$.

The non-abelian T-dual limit is given by parametrising the group-valued field $g$ and $\lambda$ as in eq.~\eqref{eq:natdparam} and taking $\kay \to \infty$.
Integrating out the Lagrange multiplier field $v$ we recover the SSSM~\eqref{eq:sssm} and the $\Integer_T$ coset models~\eqref{eq:z4cosets} and \eqref{eq:gs}, while integrating out $A_\pm$ we find their non-abelian T-duals, whose actions are given by
\begin{equation}
\Act_{\indrm{NATD-SSSM}} = - \frac{\hay}{2} \int d^2x \, \tr \big(\partial_+ v \frac{1}{P_1-\ad_v} \partial_- v\big) ~,
\end{equation}
\begin{equation}
\Act_{\indrm{NATD-}{\ind{\Integer_T}}} = - \frac{\hay}{2} \int d^2x \, \tr \big(\partial_+ v \frac{1}{\mathcal{P}_- - \ad_v} \partial_- v\big) ~,
\end{equation}
and
\begin{equation}
\Act'_{\indrm{NATD-}{\ind{\Integer_T}}} = - \frac{\hay}{2} \int d^2x \, \tr \big(\partial_+ v \frac{1}{\mathcal{Q}_- - \ad_v} \partial_- v\big) ~,
\end{equation}
with $\mathcal{P}_-$ and $\mathcal{Q}_-$ defined in eq.~\eqref{eq:z4cosets} and eq.~\eqref{eq:gs} respectively.

In models with global symmetries we can often non-abelian T-dualise with respect to subgroups of the global symmetry group, treating the remaining fields as spectators.
It is also possible to Poisson-Lie T-dualise YB deformations with respect to ``subgroups'' of the global Poisson-Lie symmetry.
However, not every non-abelian T-duality transformation of the undeformed model can be extended a Poisson-Lie T-duality transformation of the YB deformation~\cite{Hoare:2017ukq,Lust:2018jsx,Hoare:2018ebg}.
When it is possible to Poisson-Lie T-dualise, this leads to new examples of integrable sigma models.

The classical integrability of the isotropic current-current deformation of the WZW model~\eqref{eq:lambda} and its generalisations to $\Integer_2$ and $\Integer_T$ cosets~\eqref{eq:lambdasssm},~\eqref{eq:lambdazt} and~\eqref{eq:lambdags}, including the existence of an infinite number of conserved charges in involution, follows from the fact that the Poisson bracket of the Lax matrix $\Lax_x$ with itself again satisfies a Maillet bracket with twist function~\cite{Lacroix:2017isl,Lacroix:2018njs}.
This can also be understood as a consequence of their relation to split inhomogeneous YB deformations by Poisson-Lie T-duality for suitable real forms $\alg{g}$.
Similarly, there is evidence that these deformed models have a hidden symmetry that takes the form of a classical q-deformed algebra where the q-deformation parameter is the exponential of a phase~\cite{Hollowood:2015dpa}.
Using the relation between the parameters $(\kay,\lambda)$ and $(\hay,\eta)$
\begin{equation}\label{eq:paramrel}
\kay = \frac{\hay}{4\eta} ~, \qquad \lambda = \frac{1-\eta}{1+\eta} ~,
\end{equation}
and setting $c=1$, the q-deformation parameter~\eqref{eq:qdefp} is
\begin{equation}
q = \exp\big(\frac{i}{4\kay}\big) = \exp\big(\frac{i \pi}{k}\big) ~,
\end{equation}
where $k$ is the level of the WZ term.
When $k$ is integer-valued we see that $q$ is a root of unity.
Analytic continuations of the models~\eqref{eq:lambda2},~\eqref{eq:l1},~\eqref{eq:l2} and~\eqref{eq:l3} also exist, which coincide with Poisson-Lie T-duals of non-split YB deformations for suitable real forms $\alg{g}$.
They are also classically integrable and have a hidden symmetry that takes the form of a classical q-deformed algebra with real q-deformation parameter.

The deformed models~\eqref{eq:lambda} and~\eqref{eq:lambdasssm} are renormalisable at one loop~\cite{Itsios:2014lca,Sfetsos:2014jfa}.
Furthermore, the RG equations are the same as for the YB deformations~\eqref{eq:rgyb} with $c=1$ and the parameters related as in~\eqref{eq:paramrel}.
This again follows from the fact that the models are related by Poisson-Lie T-duality, hence the same should also be true of the deformed models~\eqref{eq:lambdazt} and~\eqref{eq:lambdags}.
As expected, we find that $\dot\kay =0$, hence the level of the WZ term does not run.
Note that the UV fixed point, $\eta = 1$, corresponds to $\lambda = 0$, that is the WZW and gauged WZW models.
We can again ask if the deformed models~\eqref{eq:lambdazt} and~\eqref{eq:lambdags} with $T=4$ can be used to describe worldsheet string theories in the pure spinor and Green-Schwarz formalisms respectively.
In the Green-Schwarz case this is indeed possible with invariance under $\kappa$-symmetry shown in~\cite{Hollowood:2014qma}, where this action was first constructed, and scale and Weyl invariance at one loop demonstrated in~\cite{Appadu:2015nfa} and~\cite{Borsato:2016ose} respectively.

\subsection{\texorpdfstring{$\mathcal{E}$}{E}-models}

In \secref{sec:dd} we saw that the YB deformation of the PCM~\eqref{eq:ybdefgen} and the isotropic current-current deformation of the WZW model~\eqref{eq:lambda2} can both be found from a first-order model on a Drinfel'd double~\eqref{eq:doubact}.
In general, this first-order model leads to different Poisson-Lie T-dual models upon integrating out the degrees of freedom associated to different Lagrangian subalgebras.
Such first-order models were first introduced in~\cite{Klimcik:1995dy,Klimcik:1996nq}, generalising earlier models on doubled space-times found in the context of T-duality~\cite{Tseytlin:1990va}, and in the modern literature are often termed $\mathcal{E}$-models.
One appealing feature of $\mathcal{E}$-models is that the one-loop renormalisability of the different Poisson-Lie T-dual models following from the same $\mathcal{E}$-model is simply characterised by a flow equation for the constant linear operator $\mathcal{E}$~\cite{Sfetsos:2009vt,Klimcik:2018vhl,Severa:2018pag,Klimcik:2019kkf}.

For the YB deformation of the PCM the Drinfel'd double $\alg{d}$ is isomorphic to the real double $\alg{g}\oplus \alg{g}$, the complex double $\alg{g}^\Complex$, or the semi-abelian double $\alg{g} \ltimes \alg{g}^{\text{ab}}$, for $c=1$, $c=i$ and $c=0$ respectively.
The invariant bilinear form on $\alg{d}$ and the linear operator $\mathcal{E}$ are defined in eq.~\eqref{eq:comiota} and eq.~\eqref{eq:eop} respectively.
To recover the YB deformation of the PCM~\eqref{eq:ybdefgen} from the $\mathcal{E}$-model~\eqref{eq:doubact} we integrate out the degrees of freedom associated to the Lagrangian subalgebra $\tilde{\alg{g}} \cong \alg{g}_{\mathcal{R}}$
\begin{equation}
\tilde{\alg{g}} = \{ (\iota + \mathcal{R})X ~:~ X\in\alg{g}\}~.
\end{equation}
The isotropic current-current deformation of the WZW model~\eqref{eq:lambda2} follows from the same $\mathcal{E}$-model on the real double upon integrating out the degrees of freedom associated to the Lagrangian subalgebra $\alg{g}$ for any real form $\alg{g}$.
Therefore, for those real forms $\alg{g}$ that admit a solution to the split mcYBe the model is Poisson-Lie T-dual to the corresponding YB deformation.
It is also possible to integrate out the degrees of freedom associated to the Lagrangian subalgebra $\alg{g}$ starting from the $\mathcal{E}$-model on the complex double.
The resulting model is an analytic continuation of the isotropic current-current deformation of the WZW model and when the real form $\alg{g}$ admits a solution to the non-split mcYBe it is Poisson-Lie T-dual to the corresponding YB deformation.

The YB deformation of the PCWZM~\eqref{eq:ybpcwzm} can also be recovered from an $\mathcal{E}$-model.
The Drinfel'd double is still isomorphic to the real double $\alg{g}\oplus \alg{g}$, the complex double $\alg{g}^\Complex$, or the semi-abelian double $\alg{g} \ltimes \alg{g}^{\text{ab}}$, for $c=1$, $c=i$ and $c=0$ respectively.
To introduce the WZ term we deform the invariant bilinear form on $\alg{d}$ and modify the operator $\mathcal{E}$ accordingly~\cite{Klimcik:2017ken,Hoare:2020mpv}.
The deformed invariant bilinear form is given by
\begin{equation}\label{eq:bldef}
\lang X_1 + \iota Y_1, X_2 + \iota Y_2\rang_\rho = \cosh c\rho \big(\tr\big(X_1 Y_2\big) + \tr\big(Y_1 X_2\big)\big)
+\frac{\sinh c\rho}{c}\big(\tr\big(X_1 X_2\big) + c^2\tr\big(Y_1 Y_2\big)\big) ~,
\end{equation}
while the operator $\mathcal{E}$ is modified to $\mathcal{E}_\rho$, which acts as
\begin{equation}\begin{split}\label{eq:erhodef}
\mathcal{E}_\rho (X+\iota Y) = \big(\frac{\cosh c\rho - \cosh\chi}{\sinh \chi} X & + \frac{c(\cosh c\rho - e^\chi)(\cosh c\rho - e^{-\chi})}{\sinh c\rho\sinh\chi} Y\big)
\\ & - \iota\big(\frac{\sinh c\rho}{c\sinh \chi} X + \frac{\cosh c\rho - \cosh\chi}{\sinh \chi} Y \big) ~, \qquad
X,Y\in\alg{g} ~.
\end{split}\end{equation}
This operator still squares to the identity, $\mathcal{E}_\rho^2 = 1$, and is symmetric with respect to the deformed invariant bilinear form
\begin{equation}
\lang \mathcal{E}_\rho \mathscr{X},\mathscr{Y} \rang_\rho
=
\lang \mathscr{X},\mathcal{E}_\rho \mathscr{Y} \rang_\rho ~, \qquad \mathscr{X},\mathscr{Y}\in\alg{d} ~.
\end{equation}
If we parametrise $\chi$ and $\rho$ in terms of $\hay$, $\eta$ and $\kay$ as in eq.~\eqref{eq:paramchange} and take $\kay \to 0$, then $\lang\cdot,\cdot\rang_\rho \to \lang\cdot,\cdot\rang$ as defined in eq.~\eqref{eq:comiota}, and $\mathcal{E}_\rho \to \mathcal{E}$ as defined in eq.~\eqref{eq:eop}.

For an antisymmetric operator $\mathcal{R}$ that satisfies the (m)cYBe with solvable $\alg{h}_\pm \equiv \im(\mathcal{R} \pm c)$, the operator $\widehat{\mathcal{R}} : \alg{g} \to \alg{g}$ defined as
\begin{equation}\label{eq:hatr}
\widehat{\mathcal{R}} = \frac{c}{\sinh c\rho} (e^{\rho \mathcal{R}} -\cosh c\rho) = \mathcal{R} + \Order(\rho) ~,
\end{equation}
also solves the (m)cYBe, although it is not antisymmetric.
It then follows that the subspace
\begin{equation}
\tilde{\alg{g}}_\rho = \{(\iota + \widehat{\mathcal{R}})X ~:~ X\in\alg{g} \} ~,
\end{equation}
is a Lagrangian subalgebra of $\alg{d}$ with respect to the deformed invariant bilinear form~\eqref{eq:bldef}.
Moreover, as $\kay \to 0$ we have that $\tilde{\alg{g}}_\rho \to \tilde{\alg{g}} \cong \alg{g}_{\mathcal{R}}$.

Starting from the $\mathcal{E}$-model~\eqref{eq:doubact} with the deformed invariant bilinear form~\eqref{eq:bldef} and modified operator $\mathcal{E}_\rho$~\eqref{eq:erhodef}, we can integrate out the degrees of freedom associated to the Lagrangian subalgebra $\tilde{\alg{g}}_\rho$.
Doing so, we arrive at the relativistic second-order model on $\grp{B} \backslash \grp{D}$~\eqref{eq:doubact1} with $\grp{B} = \widetilde{\grp{G}}_\rho$ and $\alg{b} = \tilde{\alg{g}}_\rho$, still with the deformed invariant bilinear form~\eqref{eq:bldef} and modified operator $\mathcal{E}_\rho$~\eqref{eq:erhodef}.
Assuming that the decomposition $\alg{d} = \alg{g}\dotplus\tilde{\alg{g}}_\rho$ lifts to the group, we parametrise $\gdsl = \tilde{g}_\rho g$, $\tilde{g}_\rho \in \widetilde{\grp{G}}_\rho$, $g \in \grp{G}$, and use the gauge symmetry~\eqref{eq:gsym} to fix $\tilde{g}_\rho = 1$ so that $\gdsl = g \in \grp{G}$.
The action of the projectors $\mathcal{E}_\rho \mathcal{P}(\mathcal{E}_\rho \pm 1)$ is then given by
\begin{equation}\label{eq:actprojdef}
\mathcal{E}_\rho\mathcal{P}(\mathcal{E}_\rho \pm 1)(X+\iota Y) = (\iota + \widehat{\mathcal{R}}_g)
\frac{1}{1 + \frac{\sinh c\rho}{c(\cosh c\rho-e^{\pm \chi})} \widehat{\mathcal{R}}_g } (Y + \frac{\sinh c\rho}{c(\cosh c\rho-e^{\pm \chi})} X ) ~, \qquad X,Y\in \alg{g} ~,
\end{equation}
where $\widehat{\mathcal{R}}_g = \Ad_g^{-1} \widehat{\mathcal{R}} \Ad_g$.
Finally, we fix the gauge $\gdsl = g \in \grp{G}$ in the action~\eqref{eq:doubact1}.
Using the action of the projectors in eq.~\eqref{eq:actprojdef}, the deformed invariant bilinear form~\eqref{eq:bldef}, and the definition of $\widehat{\mathcal{R}}$~\eqref{eq:hatr}, we recover the YB deformation of the PCWZM~\eqref{eq:ybpcwzm} when we set
\begin{equation}
N = \frac{\kay c}{\sinh c\rho} ~.
\end{equation}

In order to describe the coset models using $\mathcal{E}$-models we need to account for the gauge symmetry.
One approach to doing this is to gauge the symmetry directly in the action~\eqref{eq:doubact} following~\cite{Klimcik:2019kkf}.
We take a subgroup $\grp{H}$ of $\grp{D}$ where $\Lie\grp{H} = \alg{h}$ such that
\begin{equation}\label{eq:cond1}
\mathcal{E} \Ad_h^{-1} \mathscr{X} = \Ad_h^{-1} \mathcal{E}\mathscr{X} ~, \qquad h \in \grp{H} ~, \qquad \mathscr{X} \in \alg{d} ~,
\end{equation}
and
\begin{equation}\label{eq:cond2}
\lang X,Y\rang = 0 ~, \qquad X,Y \in \alg{h} ~.
\end{equation}
The first condition~\eqref{eq:cond1} ensures that $\gdsl \to \gdsl h$ is a global symmetry that can be gauged, while the second condition~\eqref{eq:cond2} ensures that this gauging is not anomalous~\cite{Klimcik:1994wp}.
After gauging the symmetry we find the action
\begin{equation}\begin{split}\label{eq:doubactgauge}
\Act_{{\mathcal{E}}} & = N \Big(\int d^2 x \, \lang \gdsl^{-1} \partial_t \gdsl, \gdsl^{-1}\partial_x \gdsl \rang
+ \frac{1}{6} \int d^3 x \, \epsilon^{ijk} \lang \gdsl^{-1}\partial_i\gdsl,[[\gdsl^{-1}\partial_j\gdsl, \gdsl^{-1} \partial_k\gdsl]] \rang
\\ & \qquad \quad
- 2 \int d^2 x\, \lang A_t , \gdsl^{-1}\partial_x \gdsl \rang
- \int d^2 x\; \lang (\gdsl^{-1} \partial_x \gdsl -A_x), \mathcal{E} (\gdsl^{-1} \partial_x \gdsl - A_x) \rang \Big) ~,
\end{split}\end{equation}
where $\gdsl$ is still valued in the Drinfel'd double $\grp{D}$ and $A_\mu$ is a gauge field valued in the Lie algebra $\alg{h}$.
This action is invariant under the following gauge transformations
\begin{equation}
\gdsl \to \gdsl h ~, \qquad A_{\mu} \to h^{-1} A_{\mu} h + h^{-1} \partial_{\mu} h ~, \qquad h(t,x) \in \grp{H} ~,
\end{equation}
hence defines a first-order model on the coset $\grp{D}/\grp{H}$.
The operator $\mathcal{E}$ is still required to satisfy $\mathcal{E}^2 = 1$ and $\mathcal{E}^t = \mathcal{E}$.
Assuming that the bilinear form $\lang \cdot,\mathcal{E}\cdot\rang$ is non-degenerate on $\alg{h}$, the degenerate $\mathcal{E}$-models of~\cite{Klimcik:1996np,Sfetsos:1999zm,Squellari:2011dg} are recovered upon integrating out the gauge field $A_{\mu}\in\alg{h}$.
Here, we will continue to work with the action~\eqref{eq:doubactgauge} and only integrate out the gauge field at the final stage.

As in the ungauged construction, we can take any Lagrangian subalgebra $\alg{b}$ of $\alg{d}$ and redefine $\gdsl \to b \gdsl$, $b\in\grp{B}$.
The action~\eqref{eq:doubactgauge} then only depends on $b$ through $b^{-1}\partial_x b \in \alg{b}$.
If $\Ad_\gdsl^{-1} \alg{b}$ and $\mathcal{E}\Ad_\gdsl^{-1} \alg{b}$ have trivial intersection, we can integrate out the degrees of freedom in $b$ to obtain the action
\begin{equation}\begin{split}\label{eq:doubact1gauge}
\Act_{\mathcal{E}_{\grp{B}}} & = N \Big(\frac12\int d^2 x \, \big(\lang (\gdsl^{-1} \partial_+ \gdsl - A_+) , \mathcal{E}\mathcal{P}(\mathcal{E}+1) (\gdsl^{-1}\partial_- \gdsl - A_-) \rang
\\ & \qquad \qquad \qquad \quad
- \lang (\gdsl^{-1} \partial_- \gdsl - A_-), \mathcal{E}\mathcal{P}(\mathcal{E}-1) (\gdsl^{-1}\partial_+ \gdsl - A_+) \rang\big)
\\ & \qquad \quad
+ \int d^2x \, \epsilon^{\mu\nu} \lang \gdsl^{-1}\partial_\mu \gdsl, A_\nu\rang
+ \frac{1}{6} \int d^3 x \, \epsilon^{ijk} \lang \gdsl^{-1}\partial_i \gdsl,[[\gdsl^{-1}\partial_j \gdsl, \gdsl^{-1} \partial_k \gdsl]] \rang\Big) ~,
\end{split}\end{equation}
where $A_\pm = A_t \pm A_x$ and, as before, $\mathcal{P}$ is the projector with $\im \mathcal{P} = \mathcal{E} \Ad_\gdsl^{-1} \alg{b}$ and $\ker \mathcal{P} = \Ad_\gdsl^{-1} \alg{b}$.
This means that the operators $\mathcal{E}\mathcal{P}(\mathcal{E}\pm1)$ are projectors with $\im \mathcal{E}\mathcal{P}(\mathcal{E}\pm1) = \Ad_\gdsl^{-1} \alg{b}$ and $\ker \mathcal{E}\mathcal{P}(\mathcal{E}\pm1) = \alg{e}_\mp$ where $\alg{e}_\pm$ are the eigenspaces of $\mathcal{E}$ with eigenvalues $\pm 1$.
The full gauge symmetry of the action~\eqref{eq:doubact1gauge} is
\begin{equation}\label{eq:gsymgauge}
\gdsl \to b \gdsl h ~, \qquad A_\pm \to h^{-1} A_\pm h + h^{-1}\partial_\pm h ~, \qquad b(t,x) \in \grp{B} ~, \quad h(t,x) \in \grp{H} ~,
\end{equation}
hence $\Act_{\mathcal{E}_{\grp{B}}}$ describes a relativistic second-order model on $\grp{B} \backslash \grp{D} / \grp{H}$.

To show that the YB deformations of the SSSM~\eqref{eq:ybsssm} and $\Integer_T$ coset models~\eqref{eq:ybzt} and \eqref{eq:ybgs} can be written as $\mathcal{E}$-models we take the Drinfel'd double to be the real double $\alg{g}\oplus \alg{g}$, the complex double $\alg{g}^\Complex$, or the semi-abelian double $\alg{g} \ltimes \alg{g}^{\text{ab}}$, for $c=1$, $c=i$ and $c=0$ respectively.
The invariant bilinear form is still given by~\eqref{eq:comiota} and $\alg{h}$ is taken to be the subalgebra of the Lagrangian subalgebra $\alg{g}$ spanned by the fixed points of the $\Integer_2$ or $\Integer_T$ automorphism.
This ensures that the condition~\eqref{eq:cond2} is satisfied.
We take the operator $\mathcal{E}$ to act on an element of the Drinfel'd double as
\begin{equation}\label{eq:ee1}
\begin{aligned}
& \mathcal{E} (X+ \iota Y) = - \big((\mathcal{P}^\eta_\ind{G})^{-1} Y - (\mathcal{P}_\ind{G}^\eta)^{-1} \mathcal{P}_\ind{B}^\eta X\big) 
- \iota \big((\mathcal{P}^\eta_\ind{G} - \mathcal{P}_\ind{B}^\eta(\mathcal{P}^\eta_\ind{G})^{-1}\mathcal{P}_\ind{B}^\eta) X + \mathcal{P}_\ind{B}^\eta (\mathcal{P}^\eta_\ind{G})^{-1} Y \big) ~, \qquad X,Y\in\alg{g} ~,
\\ & \mathcal{P}^\eta_\ind{G} = \eta P_0 + \frac\eta2 (\mathcal{P}^\eta_- + \mathcal{P}^\eta_+) ~, \qquad
\mathcal{P}^\eta_\ind{B} = \frac\eta2 (\mathcal{P}^\eta_- - \mathcal{P}^\eta_+) ~.
\end{aligned}\end{equation}
for the coset model~\eqref{eq:ybzt}, and the same with $\mathcal{P}^\eta_\pm$ replaced by $\mathcal{Q}^\eta_\pm$ for the coset model~\eqref{eq:ybgs}.
Recall that these two models both coincide with the SSSM for $T=2$ and $\mathcal{P}_\pm^\eta$ and $\mathcal{Q}_\pm^\eta$ are defined in eq.~\eqref{eq:ybzt} and eq.~\eqref{eq:ybgs} respectively.
The condition~\eqref{eq:cond1} is also satisfied since $\mathcal{E}$ is built from the projectors $P_k$ and $\iota$, which commute with $\Ad_h^{-1}$, $h\in\grp{H}$, by construction.
Taking $\grp{B} = \widetilde{\grp{G}}$, $\alg{b} = \tilde{\alg{g}} \cong \alg{g}_{\mathcal{R}}$, we parametrise $\gdsl = \tilde{g} g$, $\tilde{g} \in \widetilde{\grp{G}}$, $g \in \grp{G}$, and use the gauge symmetry~\eqref{eq:gsymgauge} to fix $\tilde{g} = 1$ so that $\gdsl = g \in \grp{G}$.
The residual $\grp{H}$ gauge symmetry preserving this gauge choice acts as
\begin{equation}
\gdsl \to \gdsl h ~, \qquad g \to g h ~, \qquad h(t,x) \in \grp{H} ~,
\end{equation}
which coincides with the gauge transformations~\eqref{eq:gaugeright}.
The action of the projectors $\mathcal{E} \mathcal{P} (\mathcal{E} \pm 1)$ is then given by
\begin{equation}\label{eq:ztproj}
\mathcal{E} \mathcal{P} (\mathcal{E} \pm 1)(X+\iota Y) = (\iota + \mathcal{R}_g)\frac{1}{1 \mp \eta (P_0 + \mathcal{P}^{\eta}_\mp) \mathcal{R}_g}(Y \mp  \eta (P_0 + \mathcal{P}^{\eta}_\mp) X) ~, \qquad X,Y \in \alg{g} ~.
\end{equation}
Fixing the gauge $\gdsl = g \in \grp{G}$ in the action~\eqref{eq:doubact1gauge}, and using the action of the projectors in eq.~\eqref{eq:ztproj} together with the invariant bilinear form~\eqref{eq:comiota}, we find
\begin{equation}
\Act_{\mathcal{E}_{\widetilde{\grp{G}}}} = - N \eta \int d^2x \, \tr\big((g^{-1}\partial_+ g - A_+) (P_0 + \mathcal{P}^{\eta}_-) \frac{1}{1-\eta \mathcal{R}_g (P_0 + \mathcal{P}^{\eta}_-)}(g^{-1}\partial_- g - A_-)\big) ~.
\end{equation}
After integrating out the gauge field $A_\pm \in \alg{h}$ and setting $N = \frac{\hay}{2\eta}$, we recover the $\Integer_T$ coset model~\eqref{eq:ybzt}.
Similarly, the $\Integer_T$ coset model~\eqref{eq:ybgs} follows from the same derivation with $\mathcal{P}^\eta_\pm$ replaced by $\mathcal{Q}^\eta_\pm$.

Starting from the same $\mathcal{E}$-models on the real double, we can also find the coset models~\eqref{eq:l1},~\eqref{eq:l2} and~\eqref{eq:l3}.
Writing an element of the real double as $\big(X,Y\big)\in\alg{d}$, $X,Y\in\alg{g}$, the action of the operator $\mathcal{E}$~\eqref{eq:ee1} is given by
\begin{equation}\begin{split}\label{eq:ee3}
& \mathcal{E}\big(X,Y\big) =
\big( -
\big((\widetilde{\mathcal{P}}_+^{\lambda})^{-1} - \widetilde{\mathcal{P}}_-^\lambda \big)^{-1}
\big(
\big( (\widetilde{\mathcal{P}}_+^{\lambda})^{-1} + \widetilde{\mathcal{P}}_-^\lambda \big)X - 2 Y\big) ,
\\ & \qquad \qquad \qquad \qquad
\big( (\widetilde{\mathcal{P}}_-^\lambda)^{-1} - \widetilde{\mathcal{P}}_+^{\lambda}\big)^{-1}
\big(  
\big( (\widetilde{\mathcal{P}}_-^\lambda)^{-1} + \widetilde{\mathcal{P}}_+^{\lambda}\big)Y - 2X \big) ~, \qquad
X,Y\in\alg{g} ~,
\\
& \widetilde{\mathcal{P}}_\pm^\lambda = \lambda P_0 + \mathcal{P}_\pm^\lambda ~.
\end{split}\end{equation}
Similarly, the operator $\mathcal{E}$~\eqref{eq:ee1} with $\mathcal{P}^\eta_\pm$ replaced by $\mathcal{Q}^\eta_\pm$ is given by the expression~\eqref{eq:ee3} with $\mathcal{P}_\pm^\lambda$ replaced by $\mathcal{Q}_\pm^\lambda$.
Here $\mathcal{P}_\pm^\lambda$ and $\mathcal{Q}_\pm^\lambda$ are defined in~\eqref{eq:lambdazt} and~\eqref{eq:lambdags} and we recall that
\begin{equation}
\lambda = \frac{1-\eta}{1+\eta} ~.
\end{equation}
Taking $\grp{B} = \grp{G}$, $\alg{b} = \alg{g}$, we parametrise $\gdsl = \big(g',g'\big)\big(g,1\big)$, $\big(g',g'\big) \in \grp{G}$, $\big(g,1\big) \in \grp{G}_+$, and use the gauge symmetry~\eqref{eq:gsymgauge} to fix $\big(g',g'\big) = \big(1,1\big)$ so that $\gdsl = \big(g,1\big) \in \grp{G}_+$.
The residual $\grp{H}$ gauge symmetry that preserves this gauge choice acts as
\begin{equation}
\gdsl \to \big(h^{-1},h^{-1}\big)\gdsl\big(h,h\big) ~, \qquad g \to h^{-1} g h ~, \qquad h(t,x) \in \grp{H} ~,
\end{equation}
which coincides with the gauge transformations~\eqref{eq:gaugelambda}.
The action of the projectors $\mathcal{E} \mathcal{P} (\mathcal{E} \pm 1)$ is then given by
\begin{equation}\begin{split}\label{eq:lamproja}
\mathcal{E} \mathcal{P} (\mathcal{E} \pm 1)\big(X,Y\big) & = 
\big(\Ad_g^{-1}\frac{1}{1-(\lambda P_0 + \mathcal{P}_\mp^\lambda)^{\pm 1} \Ad_g^{-1}}, \frac{1}{1-(\lambda P_0 + \mathcal{P}_\mp^\lambda)^{\pm 1} \Ad_g^{-1}}\big)
\\ & \qquad \qquad \qquad \qquad \quad
\big(Y-(\lambda P_0 + \mathcal{P}_\mp^\lambda)^{\pm 1} X, Y - (\lambda P_0 + \mathcal{P}_\mp^\lambda)^{\pm 1} X \big) ~, \qquad X,Y\in\alg{g} ~.
\end{split}\end{equation}
Fixing the gauge $\gdsl = \big(g,1) \in \grp{G}_+$ in the action~\eqref{eq:doubact1gauge}, and using the action of the projectors in eq.~\eqref{eq:lamproja} together with the invariant bilinear form~\eqref{eq:comiota2}, we find
\begin{equation}\begin{split}
\Act_{\mathcal{E}_{\grp{G}}} & = - \frac{N}{4} \int d^2x \, \big( (g^{-1}\partial_+ g - A_+ + g^{-1} A_+ g) \frac{1+\Ad_g^{-1} (\lambda P_0 + \mathcal{P}_-^\lambda)}{1-\Ad_g^{-1} (\lambda P_0 + \mathcal{P}_-^\lambda)} (g^{-1}\partial_- g - A_- + g^{-1} A_- g)\big)
\\ & \qquad + \frac{N}{4} \int d^2x \, \tr\big(A_+ (g^{-1}\partial_- g + \partial_- g g^{-1}) - A_- (\partial_+ g g^{-1} + g^{-1}\partial_+ g) + A_+ g^{-1} A_- g - A_+ g A_- g^{-1}\big)
\\ & \qquad
+ \frac{N}{12} \int d^3 x \, \epsilon^{ijk} \tr\big( g^{-1}\partial_i g [g^{-1}\partial_j g , g^{-1}\partial_k g] \big) ~.
\end{split}\end{equation}
After integrating out the gauge field $A_\pm \in \alg{h}$ and setting $N = 2\kay$, we recover the $\Integer_T$ coset model~\eqref{eq:l2}.
Similarly, the $\Integer_T$ coset model~\eqref{eq:l3} follows from the same derivation with $\mathcal{P}^\lambda_\pm$ replaced by $\mathcal{Q}^\lambda_\pm$.
Recall that the actions of these two models both coincide with the action~\eqref{eq:l1} for $T=2$.
Finally, we note that analytic continuations of the deformed models~\eqref{eq:lambda2},~\eqref{eq:l1},~\eqref{eq:l2} and \eqref{eq:l3}, which are Poisson-Lie T-duals of non-split YB deformations for suitable real forms $\alg{g}$, are found by starting from the complex double and integrating out the degrees of freedom associated to the Lagrangian subalgebra $\alg{g}$.

\subsection{More general deformations}

Before we conclude, we would like to emphasise that there are many more integrable deformations of sigma models than those that we have discussed up to this point.
We have largely focused on deformations that deform a simple Lie group (or basic Lie supergroup) $\grp{G}$.
However, it is also possible to consider deformations of semi-simple or non-semi-simple symmetry groups.
The YB deformation of the PCM discussed in \secref{sec:ybpcm} preserves the right-acting $\grp{G}$ global symmetry by construction.
It can be generalised such that both the left- and right-acting symmetries are independently deformed.
This model is known as the bi-YB deformation~\cite{Klimcik:2008eq,Klimcik:2014bta} and its action is given by
\begin{equation*}
\Act_{\indrm{biYB}} = -\frac{\hay}{2} \int d^2 x \, \tr \big(j_+ \frac{1}{1-\eta_\ind{L} \mathcal{R}_g - \eta_\ind{R} \tilde{\mathcal{R}}} j_- \big) ~,
\end{equation*}
where both $\mathcal{R}$ and $\tilde{\mathcal{R}}$ are antisymmetric solutions of the (m)cYBe, and $\eta_{\ind{L}}$ and $\eta_{\ind{R}}$ control the breaking of the left-acting and right-acting symmetries respectively.
It is also possible to construct Poisson-Lie T-duals of the bi-YB deformation.
For example, if $\mathcal{R}$ solves the split mcYBe, we can Poisson-Lie T-dualise with respect to the left-acting $\grp{G}$ global Poisson-Lie symmetry to give~\cite{Sfetsos:2015nya}
\begin{equation}\begin{split}
\Act_{\indrm{\Lambda YB}} & = - \frac{\kay}{2} \int d^2x \, \tr \big(g^{-1}\partial_+ g g^{-1}\partial_- g\big)
+ \frac{\kay}{6} \int d^3 x \, \epsilon^{ijk} \tr\big( g^{-1}\partial_i g [g^{-1}\partial_j g , g^{-1}\partial_k g] \big)
\\
& \qquad + \kay \int d^2 x \, \tr\big(A_+ g^{-1} \partial_- g - A_- \partial_+gg^{-1} + A_+ g^{-1} A_- g - \lambda_\ind{L}^{-1} A_+ \frac{1}{1-\eta_\ind{R}\tilde{\mathcal{R}}} A_- \big) ~.
\end{split}\end{equation}
This action can then be considered for any real form $\alg{g}$ with $\widetilde{\mathcal{R}}$ an antisymmetric solution of the (m)cYBe on $\alg{g}$.

The bi-YB deformation of the PCWZM with both operators $\mathcal{R}$ given by the same Drinfel'd-Jimbo solution to the mcYBe was initially constructed in~\cite{Delduc:2017fib}.
In~\cite{Klimcik:2019kkf} this model was rewritten in the compact form
\begin{equation}\label{eq:biybpcwzm}
\Act_\indrm{biYB-PCWZM} = -\frac{\kay}{2} \int d^2x \, \tr\big(j_+ \frac{e^\chi + e^{\rho_\ind{L} \mathcal{R}_g} e^{\rho_\ind{R} \tilde{\mathcal{R}}}}{e^\chi - e^{\rho_\ind{L} \mathcal{R}_g} e^{\rho_\ind{R} \tilde{\mathcal{R}}}} j_-\big)
+ \frac{\kay}{6} \int d^3x  \, \epsilon^{ijk} \tr \big(j_i[j_j,j_k]\big) ~,
\end{equation}
with $\tilde{\mathcal{R}} = \mathcal{R}$.
The limit in which the WZ term vanishes is given by setting
\begin{equation}
\chi = \frac{2\kay}{\hay} ~,
\qquad \rho_\ind{L} = \frac{2\eta_\ind{L}\kay}{\hay} ~,
\qquad \rho_\ind{R} = \frac{2\eta_\ind{R}\kay}{\hay} ~,
\end{equation}
and taking $\kay \to 0$.
It is natural to expect that, as for the YB deformation of the PCWZM~\eqref{eq:ybpcwzm}, the action~\eqref{eq:biybpcwzm} is classically integrable for independent antisymmetric solutions $\mathcal{R}$ and $\tilde{\mathcal{R}}$ of the (m)cYBe, so long as the Lie algebras $\alg{h}_\pm \equiv \im(\mathcal{R} \pm c)$ and $\tilde{\alg{h}}_\pm \equiv \im(\tilde{\mathcal{R}} \pm \tilde{c})$ are solvable.

The PCM and PCWZM can also be deformed in a way that mixes the left- and right-acting symmetries.
A simple example of this would be to implement a TsT transformation in two coordinates whose shifts are associated to the action of generators of $\grp{G}_\ind{L}$ and $\grp{G}_\ind{R}$~\cite{Delduc:2017fib}.
More generally, the Lie group $\grp{G}$ can be understood as a $\Integer_2$ coset $\frac{\grp{G} \times \grp{G}}{\grp{G}}$ where the gauge group is the diagonal subgroup.
We can then consider deformations of the corresponding SSSM.
Doing so puts these models in a form that allows us to Poisson-Lie T-dualise with respect to the $\grp{G}_\ind{L} \times \grp{G}_\ind{R}$ global Poisson-Lie symmetry.
It is worth noting that in this case the direct product structure of the symmetry group and the existence of a second invariant bilinear form allows a WZ term to be added to the SSSM, such that after gauge fixing the PCWZM is recovered.

In the context of worldsheet string theories, the $\Integer_4$ cosets $\frac{\grp{PSU}(1,1|2) \times \grp{PSU}(1,1|2)}{\grp{SU}(1,1) \times \grp{SU}(2)}$ and $\frac{\grp{D}(2,1;\alpha) \times \grp{D}(2,1;\alpha)}{\grp{SU}(1,1) \times \grp{SU}(2) \times \grp{SU}(2)}$ describe sectors of string theories on $\AdS_3 \times \Sp^3 \times \To^4$ and $\AdS_3 \times \Sp^3 \times \Sp^3 \times \Sp^1$ respectively.
The direct product structure of the symmetry group and the existence of a second invariant bilinear form again means that a WZ term, which cannot be written as a globally well-defined and invariant integral over the 2-d worldsheet, can be added to the $\Integer_4$ coset actions~\eqref{eq:z4cosets} and~\eqref{eq:gs} while preserving classical integrability~\cite{Cagnazzo:2012se}.
In the context of string theory this corresponds to supporting the $\AdS_3 \times \Sp^3 \times \To^4$ and $\AdS_3 \times \Sp^3 \times \Sp^3 \times \Sp^1$ backgrounds with a mix of Ramond-Ramond and Neveu-Schwarz-Neveu-Schwarz fluxes.
The bi-YB deformation of this model with both operators $\mathcal{R}$ given by the same Drinfel'd-Jimbo solution to the mcYBe has also been constructed in~\cite{Hoare:2014oua,Delduc:2018xug} and the Weyl invariance of these models has been investigated in~\cite{Seibold:2019dvf}.

More generally, if we have a $\Integer_T$ coset $\frac{\grp{G}}{\grp{H}}$ then we can consider models on
\begin{equation}
\frac{\grp{G}^{\times N}}{\grp{H}} ~,
\end{equation}
where $\grp{H}$ is again diagonally embedded and we formally include $\grp{H}= \grp{G}$ as the case $T=1$.
In general, the structure of these cosets and their automorphisms is more involved than those based on a simple Lie group (or basic Lie supergroup) $\grp{G}$.
This additional structure means that these models admit a richer variety of integrable couplings and deformations, see, for example,~\cite{Georgiou:2018gpe,Delduc:2019bcl,Bassi:2019aaf,Georgiou:2020wwg,Arutyunov:2020sdo}.

\subsection{Outlook}

In \iflectures these lecture notes \else this pedagogical review \fi we have explored integrable deformations of sigma models that preserve the sigma model form, that is the couplings of the model are all classically marginal.
While beyond the scope of \iflectures these notes, \else this review, \fi it is important to note that there are other classes of interesting integrable deformations of sigma models found by perturbing by relevant or irrelevant operators.
For example, massive perturbations of conformal field theories~\cite{Zamolodchikov:1989hfa} or $T\bar{T}$ deformations and their generalisations~\cite{Smirnov:2016lqw,Cavaglia:2016oda}.

In this final section we have outlined some of the many generalisations of the PCM, the PCWZM and the WZW model, their deformations and their duals.
A full classification of all integrable sigma models remains a open problem, with more modern formalisms such as those based on affine Gaudin models and higher-dimensional Chern-Simons theories~\cite{Vicedo:2017cge,Costello:2019tri,Vicedo:2019dej,Bittleston:2020hfv} providing new insights and promising avenues to explore, as discussed in the accompanying \iflectures lecture notes \else review article \fi ``4-dimensional Chern-Simons theory and integrable field theories'' by S.~Lacroix~\cite{Lacroix:2021iit}.
Such new formalisms also have the potential to help investigate the quantum physics of these models, about which much remains to be understood.

Possibly the most successful approach to quantizing integrable sigma models has been exact S-matrix theory~\cite{Zamolodchikov:1978xm,Ogievetsky:1987vv}.
This employs the bootstrap principle to conjecture exact results and uses a toolkit of associated methods, including the Bethe ansatz and its generalisations, to compute observables.
Reviews on exact S-matrices include~\cite{Mussardo:1992uc,Dorey:1996gd,Zarembo:2017muf}.
In spite of this success, due to non-ultralocal terms in the Maillet bracket, a first-principles canonical quantization of integrable sigma models remains a challenging goal.
For certain models it is possible to find a formal gauge transformation of the Lax connection that removes the non-ultralocal terms in the Maillet bracket, a direction that has been explored in~\cite{Brodbeck:1999ib,Bykov:2016rdv,Bazhanov:2017nzh,Delduc:2019lpe}.

An alternative approach to quantization is the construction of dual integrable massive models~\cite{Fateev:1992tk,Fateev:1995ht,Fateev:1996ea,Fateev:2018yos,Litvinov:2018bou}.
This is based on first deforming the integrable sigma model introducing a UV fixed point.
The dual model is then constructed by perturbing the UV conformal field theory by relevant operators that preserve the same symmetry as the original deformation.
This approach has the advantage that for classically integrable sigma models that are anomalous~\cite{Abdalla:1982yd}, quantum integrability can be restored by introducing additional degrees of freedom that decouple in the classical limit~\cite{Litvinov:2019rlv,Fateev:2019xuq}.
Corresponding developments for sigma models on flag manifolds have been reviewed in~\cite{Affleck:2021ypq}.

A related question is the precise relationship between integrability and renormalisability.
It has long been known that there is a connection between classical integrability and one-loop renormalisability.
In particular, for all known examples, the space of integrable models is closed under one-loop RG flow.
Understanding the deeper origins of this, for example, if Ward identities for hidden symmetries constrain the RG flow, remains an interesting open problem.
Recent work has shown that the renormalisability of these models persists to higher loops~\cite{Hoare:2019mcc,Borsato:2019oip,Georgiou:2019nbz,Hassler:2020xyj} subject to the addition of $\hbar$ corrections, which are related to $\hbar$ corrections to Poisson-Lie T-duality transformations~\cite{Hassler:2020tvz,Borsato:2020wwk,Codina:2020yma}.
A systematic and universal explanation of these $\hbar$ corrections remains to be found.

Finally, the search for new instances of solvable string theories has been one of the main driving forces behind many of the recent developments in the field.
As we pursue the goal of classifying integrable sigma models it is likely that we will come across new candidates for such theories.
While the application of integrability methods to the free-string limit of the $\AdS_5 \times \Sp^5$ superstring has undoubtedly been one of the great success stories in high-energy theoretical physics over the last 25 years, the generalisation to other integrable worldsheet string theories, their deformations and their duals has highlighted that much remains to be understood about how integrability works in these models.
Moreover, the status of any holographic interpretation for many of these worldsheet string theories remains to be determined, see, for example,~\cite{vanTongeren:2015uha,Araujo:2017jap} and references therein.
A truly universal application of integrability methods to worldsheet string theories, together with a proof of quantum integrability, remains an open challenge.

\iflectures
\medskip

\noindent\textbf{Note from the author.}
\textit{In this final section we have discussed generalisations of the results explored in these lecture notes.
We have also attempted to give an overview of the literature.
The literature is vast and only certain aspects of integrable deformations of sigma models have been covered.
Nevertheless, if there is a reference whose addition would be useful for the reader, please contact the author who will endeavour to include it in a future version.}
\fi

\section*{Acknowledgements}
We would like to thank S.~Demulder, S.~Driezen, S.~Lacroix, J.~L.~Miramontes and A.~L.~Retore for helpful discussions on the contents of \iflectures these lecture notes \else this pedagogical review \fi and S.~Demulder, S.~Driezen, S.~Lacroix, N.~Levine and A.~A.~Tseytlin for comments on draft versions.
\iflectures These notes were \else This review is based on notes \fi written for lectures delivered at the school ``Integrability, Dualities and Deformations,'' which ran from 23 to 27 August 2021 in Santiago de Compostela and virtually.
The website for the school can be found at \href{https://indico.cern.ch/e/IDD2021}{\texttt{https://indico.cern.ch/e/IDD2021}}.
We would also like to thank the organisers of the school for the opportunity to present these lectures and the participants of the school for many interesting discussions.
This work was supported by a UKRI Future Leaders Fellowship (grant number MR/T018909/1).

\begin{bibtex}[\jobname]

@book{Faddeev:1987ph,
author = "Faddeev, L. D. and Takhtajan, L. A.",
title = "{Hamiltonian Methods in the Theory of Solitons}",
year = "1987",
publisher = "Springer"
}

@article{Zakharov:1973pp,
author = "Zakharov, V. E. and Mikhailov, A. V.",
title = "{Relativistically Invariant Two-Dimensional Models in Field Theory Integrable by the Inverse Problem Technique}",
journal = "Sov. Phys. JETP",
volume = "47",
pages = "1017--1027",
year = "1978"
}

@article{Novikov:1982ei,
author = "Novikov, S. P.",
title = "{The Hamiltonian formalism and a many valued analog of Morse theory}",
doi = "10.1070/RM1982v037n05ABEH004020",
journal = "Usp. Mat. Nauk",
volume = "37N5",
number = "5",
pages = "3--49",
year = "1982"
}

@article{Witten:1983ar,
author = "Witten, Edward",
editor = "Stone, M.",
title = "{Non-abelian Bosonization in Two Dimensions}",
reportNumber = "PRINT-83-0934 (PRINCETON)",
doi = "10.1007/BF01215276",
journal = "Commun. Math. Phys.",
volume = "92",
pages = "455--472",
year = "1984"
}

@article{Eichenherr:1979ci,
author = "Eichenherr, H. and Forger, M.",
title = "{On the Dual Symmetry of the Nonlinear Sigma Models}",
reportNumber = "FREIBURG-THEP 79/2a",
doi = "10.1016/0550-3213(79)90276-1",
journal = "Nucl. Phys. B",
volume = "155",
pages = "381--393",
year = "1979"
}

@article{Young:2005jv,
author = "Young, Charles A. S.",
title = "{Non-local charges, $\Integer_m$ gradings and coset space actions}",
eprint = "hep-th/0503008",
archivePrefix = "arXiv",
doi = "10.1016/j.physletb.2005.10.090",
journal = "Phys. Lett. B",
volume = "632",
pages = "559--565",
year = "2006"
}

@article{Henneaux:1984mh,
author = "Henneaux, Marc and Mezincescu, Luca",
title = "{A Sigma Model Interpretation of Green-Schwarz Covariant Superstring Action}",
reportNumber = "UTTG-26-84",
doi = "10.1016/0370-2693(85)90507-6",
journal = "Phys. Lett. B",
volume = "152",
pages = "340--342",
year = "1985"
}

@article{Berkovits:1999zq,
author = "Berkovits, N. and Bershadsky, M. and Hauer, T. and Zhukov, S. and Zwiebach, B.",
title = "{Superstring theory on $AdS_2 \times S^2$ as a coset supermanifold}",
eprint = "hep-th/9907200",
archivePrefix = "arXiv",
reportNumber = "IFT-P-060-99, HUTP-99-A044, MIT-CTP-2878, CTP-MIT-2878",
doi = "10.1016/S0550-3213(99)00683-5",
journal = "Nucl. Phys. B",
volume = "567",
pages = "61--86",
year = "2000"
}

@article{Metsaev:1998it,
author = "Metsaev, R. R. and Tseytlin, Arkady A.",
title = "{Type IIB superstring action in $AdS_5 \times S^5$ background}",
eprint = "hep-th/9805028",
archivePrefix = "arXiv",
reportNumber = "FIAN-TD-98-21, IMPERIAL-TP-97-98-44, NSF-ITP-98-055",
doi = "10.1016/S0550-3213(98)00570-7",
journal = "Nucl. Phys. B",
volume = "533",
pages = "109--126",
year = "1998"
}

@article{Bena:2003wd,
author = "Bena, Iosif and Polchinski, Joseph and Roiban, Radu",
title = "{Hidden symmetries of the $AdS_5 \times S^5$ superstring}",
eprint = "hep-th/0305116",
archivePrefix = "arXiv",
reportNumber = "NSF-KITP-03-34, UCLA-03-TEP-14",
doi = "10.1103/PhysRevD.69.046002",
journal = "Phys. Rev. D",
volume = "69",
pages = "046002",
year = "2004"
}

@article{Zarembo:2017muf,
author = "Zarembo, K.",
title = "{Integrability in Sigma-Models}",
booktitle = "{Les Houches Summer School: Integrability: From Statistical Systems to Gauge Theory, Les Houches, France, 6th June - 1st July 2016}",
eprint = "1712.07725",
archivePrefix = "arXiv",
primaryClass = "hep-th",
reportNumber = "NORDITA-2017-137, UUITP-04-17",
doi = "10.1093/oso/9780198828150.003.0005",
month = "12",
year = "2017"
}

@article{Cherednik:1981df,
author = "Cherednik, I. V.",
title = "{Relativistically Invariant Quasiclassical Limits of Integrable Two-dimensional Quantum Models}",
doi = "10.1007/BF01086395",
journal = "Theor. Math. Phys.",
volume = "47",
pages = "422--425",
year = "1981"
}

@article{Appadu:2017bnv,
author = "Appadu, Calan and Hollowood, Timothy J. and Price, Dafydd and Thompson, Daniel C.",
title = "{Yang Baxter and Anisotropic Sigma and Lambda Models, Cyclic RG and Exact S-Matrices}",
eprint = "1706.05322",
archivePrefix = "arXiv",
primaryClass = "hep-th",
doi = "10.1007/JHEP09(2017)035",
journal = "JHEP",
volume = "09",
pages = "035",
year = "2017"
}

@article{Klimcik:2002zj,
author = "Klimcik, Ctirad",
title = "{Yang-Baxter sigma models and dS/AdS T duality}",
eprint = "hep-th/0210095",
archivePrefix = "arXiv",
reportNumber = "IML-02-XY",
doi = "10.1088/1126-6708/2002/12/051",
journal = "JHEP",
volume = "12",
pages = "051",
year = "2002"
}

@article{Klimcik:2008eq,
author = "Klimcik, Ctirad",
title = "{On integrability of the Yang-Baxter sigma-model}",
eprint = "0802.3518",
archivePrefix = "arXiv",
primaryClass = "hep-th",
doi = "10.1063/1.3116242",
journal = "J. Math. Phys.",
volume = "50",
pages = "043508",
year = "2009"
}

@article{Klimcik:2014bta,
author = "Klimcik, Ctirad",
title = "{Integrability of the bi-Yang-Baxter sigma-model}",
eprint = "1402.2105",
archivePrefix = "arXiv",
primaryClass = "math-ph",
doi = "10.1007/s11005-014-0709-y",
journal = "Lett. Math. Phys.",
volume = "104",
pages = "1095--1106",
year = "2014"
}

@article{Kawaguchi:2011mz,
author = "Kawaguchi, Io and Orlando, Domenico and Yoshida, Kentaroh",
title = "{Yangian symmetry in deformed WZNW models on squashed spheres}",
eprint = "1104.0738",
archivePrefix = "arXiv",
primaryClass = "hep-th",
reportNumber = "KUNS-2328, IPMU11-0054",
doi = "10.1016/j.physletb.2011.06.007",
journal = "Phys. Lett. B",
volume = "701",
pages = "475--480",
year = "2011"
}

@article{Delduc:2014uaa,
author = "Delduc, Francois and Magro, Marc and Vicedo, Benoit",
title = "{Integrable double deformation of the principal chiral model}",
eprint = "1410.8066",
archivePrefix = "arXiv",
primaryClass = "hep-th",
doi = "10.1016/j.nuclphysb.2014.12.018",
journal = "Nucl. Phys. B",
volume = "891",
pages = "312--321",
year = "2015"
}

@article{Klimcik:2019kkf,
author = "Klimcik, Ctirad",
title = "{Dressing cosets and multi-parametric integrable deformations}",
eprint = "1903.00439",
archivePrefix = "arXiv",
primaryClass = "hep-th",
doi = "10.1007/JHEP07(2019)176",
journal = "JHEP",
volume = "07",
pages = "176",
year = "2019"
}

@article{Hoare:2020mpv,
author = "Hoare, B. and Lacroix, S.",
title = "{Yang-Baxter deformations of the principal chiral model plus Wess-Zumino term}",
eprint = "2009.00341",
archivePrefix = "arXiv",
primaryClass = "hep-th",
reportNumber = "ZMP-HH/20-17",
doi = "10.1088/1751-8121/abc43d",
journal = "J. Phys. A",
volume = "53",
number = "50",
pages = "505401",
year = "2020"
}

@article{Delduc:2013fga,
author = "Delduc, Francois and Magro, Marc and Vicedo, Benoit",
title = "{On classical $q$-deformations of integrable sigma-models}",
eprint = "1308.3581",
archivePrefix = "arXiv",
primaryClass = "hep-th",
doi = "10.1007/JHEP11(2013)192",
journal = "JHEP",
volume = "11",
pages = "192",
year = "2013"
}

@article{Delduc:2013qra,
author = "Delduc, Francois and Magro, Marc and Vicedo, Benoit",
title = "{An integrable deformation of the $AdS_5 \times S^5$ superstring action}",
eprint = "1309.5850",
archivePrefix = "arXiv",
primaryClass = "hep-th",
doi = "10.1103/PhysRevLett.112.051601",
journal = "Phys. Rev. Lett.",
volume = "112",
number = "5",
pages = "051601",
year = "2014"
}

@article{Delduc:2014kha,
author = "Delduc, Francois and Magro, Marc and Vicedo, Benoit",
title = "{Derivation of the action and symmetries of the $q$-deformed $AdS_{5} \times S^{5}$ superstring}",
eprint = "1406.6286",
archivePrefix = "arXiv",
primaryClass = "hep-th",
doi = "10.1007/JHEP10(2014)132",
journal = "JHEP",
volume = "10",
pages = "132",
year = "2014"
}

@article{Vallilo:2003nx,
author = "Vallilo, Brenno Carlini",
title = "{Flat currents in the classical $AdS_5 \times S^5$ pure spinor superstring}",
eprint = "hep-th/0307018",
archivePrefix = "arXiv",
reportNumber = "IFT-P-029-2003",
doi = "10.1088/1126-6708/2004/03/037",
journal = "JHEP",
volume = "03",
pages = "037",
year = "2004"
}

@article{Berkovits:2000fe,
author = "Berkovits, Nathan",
title = "{Super Poincare covariant quantization of the superstring}",
eprint = "hep-th/0001035",
archivePrefix = "arXiv",
reportNumber = "IFT-P-005-2000",
doi = "10.1088/1126-6708/2000/04/018",
journal = "JHEP",
volume = "04",
pages = "018",
year = "2000"
}

@article{Berkovits:2000yr,
author = "Berkovits, Nathan and Chandia, Osvaldo",
title = "{Superstring vertex operators in an $AdS_5 \times S^5$ background}",
eprint = "hep-th/0009168",
archivePrefix = "arXiv",
reportNumber = "IFT-P-079-2000, UCCHEP-14-00",
doi = "10.1016/S0550-3213(00)00697-0",
journal = "Nucl. Phys. B",
volume = "596",
pages = "185--196",
year = "2001"
}

@article{Evans:1999mj,
author = "Evans, J. M. and Hassan, M. and MacKay, N. J. and Mountain, A. J.",
title = "{Local conserved charges in principal chiral models}",
eprint = "hep-th/9902008",
archivePrefix = "arXiv",
reportNumber = "DAMTP-98-171, IMPERIAL-TP-98-99-21",
doi = "10.1016/S0550-3213(99)00489-7",
journal = "Nucl. Phys. B",
volume = "561",
pages = "385--412",
year = "1999"
}

@article{Faddeev:1985qu,
author = "Faddeev, L. D. and Reshetikhin, N. Yu.",
title = "{Integrability of the Principal Chiral Field Model in (1+1) Dimension}",
reportNumber = "LOMI-E-2-85",
doi = "10.1016/0003-4916(86)90201-0",
journal = "Annals Phys.",
volume = "167",
pages = "227",
year = "1986"
}

@article{Pohlmeyer:1975nb,
author = "Pohlmeyer, K.",
title = "{Integrable Hamiltonian Systems and Interactions Through Quadratic Constraints}",
reportNumber = "DESY-75-28",
doi = "10.1007/BF01609119",
journal = "Commun. Math. Phys.",
volume = "46",
pages = "207--221",
year = "1976"
}

@article{Friedan:1980jm,
author = "Friedan, Daniel Harry",
title = "{Nonlinear Models in Two + Epsilon Dimensions}",
reportNumber = "LBL-11517, UMI-81-13038",
doi = "10.1016/0003-4916(85)90384-7",
journal = "Annals Phys.",
volume = "163",
pages = "318",
year = "1985"
}

@article{Zarembo:2010sg,
author = "Zarembo, K.",
title = "{Strings on Semisymmetric Superspaces}",
eprint = "1003.0465",
archivePrefix = "arXiv",
primaryClass = "hep-th",
reportNumber = "ITEP-TH-12-10, LPTENS-10-12, UUITP-05-10",
doi = "10.1007/JHEP05(2010)002",
journal = "JHEP",
volume = "05",
pages = "002",
year = "2010"
}

@article{Wulff:2014kja,
author = "Wulff, Linus",
title = "{Superisometries and integrability of superstrings}",
eprint = "1402.3122",
archivePrefix = "arXiv",
primaryClass = "hep-th",
reportNumber = "IMPERIAL-TP-LW-2014-01",
doi = "10.1007/JHEP05(2014)115",
journal = "JHEP",
volume = "05",
pages = "115",
year = "2014"
}

@article{Benitez:2018xnh,
author = "Benittez, Hector A. and Rivelles, Victor O.",
title = "{Yang-Baxter deformations of the $AdS_{5}\times S^{5}$ pure spinor superstring}",
eprint = "1807.10432",
archivePrefix = "arXiv",
primaryClass = "hep-th",
doi = "10.1007/JHEP02(2019)056",
journal = "JHEP",
volume = "02",
pages = "056",
year = "2019"
}

@article{Kawaguchi:2014qwa,
author = "Kawaguchi, Io and Matsumoto, Takuya and Yoshida, Kentaroh",
title = "{Jordanian deformations of the $AdS_5 x S^5$ superstring}",
eprint = "1401.4855",
archivePrefix = "arXiv",
primaryClass = "hep-th",
reportNumber = "KUNS-2477, ITP-UU-14-05, SPIN-14-05",
doi = "10.1007/JHEP04(2014)153",
journal = "JHEP",
volume = "04",
pages = "153",
year = "2014"
}

@article{Valent:2009nv,
author = "Valent, Galliano and Klimcik, Ctirad and Squellari, Romain",
title = "{One loop renormalizability of the Poisson-Lie sigma models}",
eprint = "0902.1459",
archivePrefix = "arXiv",
primaryClass = "hep-th",
doi = "10.1016/j.physletb.2009.06.001",
journal = "Phys. Lett. B",
volume = "678",
pages = "143--148",
year = "2009"
}

@article{Sfetsos:2009dj,
author = "Sfetsos, Konstadinos and Siampos, Konstadinos",
title = "{Quantum equivalence in Poisson-Lie T-duality}",
eprint = "0904.4248",
archivePrefix = "arXiv",
primaryClass = "hep-th",
doi = "10.1088/1126-6708/2009/06/082",
journal = "JHEP",
volume = "06",
pages = "082",
year = "2009"
}

@article{Squellari:2014jfa,
author = "Squellari, Romain",
title = "{Yang-Baxter $\sigma$ model: Quantum aspects}",
eprint = "1401.3197",
archivePrefix = "arXiv",
primaryClass = "hep-th",
doi = "10.1016/j.nuclphysb.2014.02.009",
journal = "Nucl. Phys. B",
volume = "881",
pages = "502--513",
year = "2014"
}

@article{Demulder:2017zhz,
author = "Demulder, Saskia and Driezen, Sibylle and Sevrin, Alexander and Thompson, Daniel C.",
title = "{Classical and Quantum Aspects of Yang-Baxter Wess-Zumino Models}",
eprint = "1711.00084",
archivePrefix = "arXiv",
primaryClass = "hep-th",
doi = "10.1007/JHEP03(2018)041",
journal = "JHEP",
volume = "03",
pages = "041",
year = "2018"
}

@article{Borsato:2016ose,
author = "Borsato, Riccardo and Wulff, Linus",
title = "{Target space supergeometry of $\eta$ and $\lambda$-deformed strings}",
eprint = "1608.03570",
archivePrefix = "arXiv",
primaryClass = "hep-th",
reportNumber = "IMPERIAL-TP-LW-2016-03",
doi = "10.1007/JHEP10(2016)045",
journal = "JHEP",
volume = "10",
pages = "045",
year = "2016"
}

@article{Hoare:2018ngg,
author = "Hoare, Ben and Seibold, Fiona K.",
title = "{Supergravity backgrounds of the $\eta$-deformed AdS$_2 \times S^2 \times T^6 $ and AdS$_5 \times S^5$ superstrings}",
eprint = "1811.07841",
archivePrefix = "arXiv",
primaryClass = "hep-th",
doi = "10.1007/JHEP01(2019)125",
journal = "JHEP",
volume = "01",
pages = "125",
year = "2019"
}

@article{vanTongeren:2019dlq,
author = "van Tongeren, Stijn J.",
title = "{Unimodular jordanian deformations of integrable superstrings}",
eprint = "1904.08892",
archivePrefix = "arXiv",
primaryClass = "hep-th",
doi = "10.21468/SciPostPhys.7.1.011",
journal = "SciPost Phys.",
volume = "7",
pages = "011",
year = "2019"
}

@article{Delduc:2017fib,
author = "Delduc, Francois and Hoare, Ben and Kameyama, Takashi and Magro, Marc",
title = "{Combining the bi-Yang-Baxter deformation, the Wess-Zumino term and TsT transformations in one integrable $\sigma$-model}",
eprint = "1707.08371",
archivePrefix = "arXiv",
primaryClass = "hep-th",
doi = "10.1007/JHEP10(2017)212",
journal = "JHEP",
volume = "10",
pages = "212",
year = "2017"
}

@article{Horne:1991gn,
author = "Horne, James H. and Horowitz, Gary T.",
title = "{Exact black string solutions in three dimensions}",
eprint = "hep-th/9108001",
archivePrefix = "arXiv",
reportNumber = "UCSBTH-91-39",
doi = "10.1016/0550-3213(92)90536-K",
journal = "Nucl. Phys. B",
volume = "368",
pages = "444--462",
year = "1992"
}

@article{Giveon:1991jj,
author = "Giveon, Amit and Rocek, Martin",
title = "{Generalized duality in curved string backgrounds}",
eprint = "hep-th/9112070",
archivePrefix = "arXiv",
reportNumber = "IASSNS-HEP-91-84, ITP-SB-91-67",
doi = "10.1016/0550-3213(92)90518-G",
journal = "Nucl. Phys. B",
volume = "380",
pages = "128--146",
year = "1992"
}

@article{Hollowood:2014qma,
author = "Hollowood, Timothy J. and Miramontes, J. Luis and Schmidtt, David M.",
title = "{An Integrable Deformation of the $AdS_5 \times S^5$ Superstring}",
eprint = "1409.1538",
archivePrefix = "arXiv",
primaryClass = "hep-th",
doi = "10.1088/1751-8113/47/49/495402",
journal = "J. Phys. A",
volume = "47",
number = "49",
pages = "495402",
year = "2014"
}

@article{Hollowood:2014rla,
author = "Hollowood, Timothy J. and Miramontes, J. Luis and Schmidtt, David M.",
title = "{Integrable Deformations of Strings on Symmetric Spaces}",
eprint = "1407.2840",
archivePrefix = "arXiv",
primaryClass = "hep-th",
doi = "10.1007/JHEP11(2014)009",
journal = "JHEP",
volume = "11",
pages = "009",
year = "2014"
}

@article{Vicedo:2015pna,
author = "Vicedo, Benoit",
title = "{Deformed integrable $\sigma$-models, classical R-matrices and classical exchange algebra on Drinfel'd doubles}",
eprint = "1504.06303",
archivePrefix = "arXiv",
primaryClass = "hep-th",
doi = "10.1088/1751-8113/48/35/355203",
journal = "J. Phys. A",
volume = "48",
number = "35",
pages = "355203",
year = "2015"
}

@article{Hoare:2016wsk,
author = "Hoare, B. and Tseytlin, A. A.",
title = "{Homogeneous Yang-Baxter deformations as non-abelian duals of the $AdS_5$ $\sigma$-model}",
eprint = "1609.02550",
archivePrefix = "arXiv",
primaryClass = "hep-th",
reportNumber = "IMPERIAL-TP-AT-2016-03",
doi = "10.1088/1751-8113/49/49/494001",
journal = "J. Phys. A",
volume = "49",
number = "49",
pages = "494001",
year = "2016"
}

@article{Hoare:2015gda,
author = "Hoare, B. and Tseytlin, A. A.",
title = "{On integrable deformations of superstring sigma models related to $AdS_n \times S^n$ supercosets}",
eprint = "1504.07213",
archivePrefix = "arXiv",
primaryClass = "hep-th",
reportNumber = "IMPERIAL-TP-AT-2015-02, HU-EP-15-21",
doi = "10.1016/j.nuclphysb.2015.06.001",
journal = "Nucl. Phys. B",
volume = "897",
pages = "448--478",
year = "2015"
}

@article{Klimcik:2015gba,
author = "Klimcik, Ctirad",
title = "{$\eta$ and $\lambda$ deformations as $\mathcal{E}$-models}",
eprint = "1508.05832",
archivePrefix = "arXiv",
primaryClass = "hep-th",
doi = "10.1016/j.nuclphysb.2015.09.011",
journal = "Nucl. Phys. B",
volume = "900",
pages = "259--272",
year = "2015"
}

@article{Sfetsos:2015nya,
author = "Sfetsos, Konstantinos and Siampos, Konstantinos and Thompson, Daniel C.",
title = "{Generalised integrable $\lambda$- and $\eta$-deformations and their relation}",
eprint = "1506.05784",
archivePrefix = "arXiv",
primaryClass = "hep-th",
doi = "10.1016/j.nuclphysb.2015.08.015",
journal = "Nucl. Phys. B",
volume = "899",
pages = "489--512",
year = "2015"
}

@article{Sfetsos:2013wia,
author = "Sfetsos, Konstadinos",
title = "{Integrable interpolations: From exact CFTs to non-abelian T-duals}",
eprint = "1312.4560",
archivePrefix = "arXiv",
primaryClass = "hep-th",
reportNumber = "DMUS-MP-13-23, DMUS--MP--13-23",
doi = "10.1016/j.nuclphysb.2014.01.004",
journal = "Nucl. Phys. B",
volume = "880",
pages = "225--246",
year = "2014"
}

@article{Affleck:2021ypq,
author = "Affleck, Ian and Bykov, Dmitri and Wamer, Kyle",
title = "{Flag manifold sigma models: spin chains and integrable theories}",
eprint = "2101.11638",
archivePrefix = "arXiv",
primaryClass = "hep-th",
month = "1",
year = "2021"
}

@article{Tseytlin:1990va,
author = "Tseytlin, Arkady A.",
title = "{Duality symmetric closed string theory and interacting chiral scalars}",
reportNumber = "KCL-TP-1990-3",
doi = "10.1016/0550-3213(91)90266-Z",
journal = "Nucl. Phys. B",
volume = "350",
pages = "395--440",
year = "1991"
}

@article{Buscher:1987qj,
author = "Buscher, T. H.",
title = "{Path Integral Derivation of Quantum Duality in Nonlinear Sigma Models}",
reportNumber = "ITP-SB-87-61",
doi = "10.1016/0370-2693(88)90602-8",
journal = "Phys. Lett. B",
volume = "201",
pages = "466--472",
year = "1988"
}

@article{Cvetic:1999zs,
author = "Cvetic, Mirjam and Lu, Hong and Pope, C. N. and Stelle, K. S.",
title = "{T duality in the Green-Schwarz formalism, and the massless/massive IIA duality map}",
eprint = "hep-th/9907202",
archivePrefix = "arXiv",
reportNumber = "UPR-0852-T, CTP-TAMU-31-99, SISSA-88-99-EP, IMPERIAL-TP-98-99-63, NSF-ITP-99-086",
doi = "10.1016/S0550-3213(99)00740-3",
journal = "Nucl. Phys. B",
volume = "573",
pages = "149--176",
year = "2000"
}

@article{Kulik:2000nr,
author = "Kulik, Bogdan and Roiban, Radu",
title = "{T duality of the Green-Schwarz superstring}",
eprint = "hep-th/0012010",
archivePrefix = "arXiv",
reportNumber = "ITP-SB-00-80",
doi = "10.1088/1126-6708/2002/09/007",
journal = "JHEP",
volume = "09",
pages = "007",
year = "2002"
}

@article{Cagnazzo:2012se,
author = "Cagnazzo, A. and Zarembo, K.",
title = "{B-field in $AdS_3$/$CFT_2$ Correspondence and Integrability}",
eprint = "1209.4049",
archivePrefix = "arXiv",
primaryClass = "hep-th",
reportNumber = "NORDITA-2012-67, UUITP-24-12",
doi = "10.1007/JHEP11(2012)133",
journal = "JHEP",
volume = "11",
pages = "133",
year = "2012",
note = "[Erratum: JHEP 04, 003 (2013)]"
}

@article{Klimcik:1995ux,
author = "Klimcik, C. and Severa, P.",
title = "{Dual non-abelian duality and the Drinfel'd double}",
eprint = "hep-th/9502122",
archivePrefix = "arXiv",
reportNumber = "CERN-TH-95-39, CERN-TH-95-039",
doi = "10.1016/0370-2693(95)00451-P",
journal = "Phys. Lett. B",
volume = "351",
pages = "455--462",
year = "1995"
}

@article{Klimcik:1995jn,
author = "Klimcik, C.",
editor = "Gava, E. and Narain, K. S. and Vafa, C.",
title = "{Poisson-Lie T-duality}",
eprint = "hep-th/9509095",
archivePrefix = "arXiv",
reportNumber = "CERN-TH-95-248",
doi = "10.1016/0920-5632(96)00013-8",
journal = "Nucl. Phys. B Proc. Suppl.",
volume = "46",
pages = "116--121",
year = "1996"
}

@article{Matsumoto:2014nra,
author = "Matsumoto, Takuya and Yoshida, Kentaroh",
title = "{Lunin-Maldacena backgrounds from the classical Yang-Baxter equation - towards the gravity/CYBE correspondence}",
eprint = "1404.1838",
archivePrefix = "arXiv",
primaryClass = "hep-th",
reportNumber = "KUNS-2488, ITP-UU-14-12, SPIN-14-12",
doi = "10.1007/JHEP06(2014)135",
journal = "JHEP",
volume = "06",
pages = "135",
year = "2014"
}

@article{Osten:2016dvf,
author = "Osten, David and van Tongeren, Stijn J.",
title = "{Abelian Yang-Baxter deformations and TsT transformations}",
eprint = "1608.08504",
archivePrefix = "arXiv",
primaryClass = "hep-th",
reportNumber = "HU-EP-16-29",
doi = "10.1016/j.nuclphysb.2016.12.007",
journal = "Nucl. Phys. B",
volume = "915",
pages = "184--205",
year = "2017"
}

@article{Borsato:2016pas,
author = "Borsato, Riccardo and Wulff, Linus",
title = "{Integrable Deformations of T-Dual $\sigma$-Models}",
eprint = "1609.09834",
archivePrefix = "arXiv",
primaryClass = "hep-th",
reportNumber = "IMPERIAL-TP-RB-2016-06",
doi = "10.1103/PhysRevLett.117.251602",
journal = "Phys. Rev. Lett.",
volume = "117",
number = "25",
pages = "251602",
year = "2016"
}

@article{Hoare:2014oua,
author = "Hoare, Ben",
title = "{Towards a two-parameter $q$-deformation of $AdS_3 \times S^3 \times M^4$ superstrings}",
eprint = "1411.1266",
archivePrefix = "arXiv",
primaryClass = "hep-th",
reportNumber = "HU-EP-14-44",
doi = "10.1016/j.nuclphysb.2014.12.012",
journal = "Nucl. Phys. B",
volume = "891",
pages = "259--295",
year = "2015"
}

@article{Delduc:2018xug,
author = "Delduc, F. and Hoare, B. and Kameyama, T. and Lacroix, S. and Magro, M.",
title = "{Three-parameter integrable deformation of $\mathbb{Z}_4$ permutation supercosets}",
eprint = "1811.00453",
archivePrefix = "arXiv",
primaryClass = "hep-th",
reportNumber = "ZMP-HH/18-22",
doi = "10.1007/JHEP01(2019)109",
journal = "JHEP",
volume = "01",
pages = "109",
year = "2019"
}

@article{Borsato:2017qsx,
author = "Borsato, Riccardo and Wulff, Linus",
title = "{On non-abelian T-duality and deformations of supercoset string sigma-models}",
eprint = "1706.10169",
archivePrefix = "arXiv",
primaryClass = "hep-th",
reportNumber = "NORDITA-2017-065",
doi = "10.1007/JHEP10(2017)024",
journal = "JHEP",
volume = "10",
pages = "024",
year = "2017"
}

@article{delaOssa:1992vci,
author = "de la Ossa, Xenia C. and Quevedo, Fernando",
title = "{Duality symmetries from non-abelian isometries in string theory}",
eprint = "hep-th/9210021",
archivePrefix = "arXiv",
reportNumber = "NEIP-92-004",
doi = "10.1016/0550-3213(93)90041-M",
journal = "Nucl. Phys. B",
volume = "403",
pages = "377--394",
year = "1993"
}

@article{Klimcik:1995dy,
author = "Klimcik, C. and Severa, P.",
title = "{Poisson-Lie T-duality and loop groups of Drinfel'd doubles}",
eprint = "hep-th/9512040",
archivePrefix = "arXiv",
reportNumber = "CERN-TH-95-330",
doi = "10.1016/0370-2693(96)00025-1",
journal = "Phys. Lett. B",
volume = "372",
pages = "65--71",
year = "1996"
}

@article{Klimcik:1996nq,
author = "Klimcik, C. and Severa, P.",
title = "{Non-abelian momentum winding exchange}",
eprint = "hep-th/9605212",
archivePrefix = "arXiv",
reportNumber = "CERN-TH-96-142",
doi = "10.1016/0370-2693(96)00755-1",
journal = "Phys. Lett. B",
volume = "383",
pages = "281--286",
year = "1996"
}

@article{Borsato:2018spz,
author = "Borsato, Riccardo and Wulff, Linus",
title = "{Marginal deformations of WZW models and the classical Yang-Baxter equation}",
eprint = "1812.07287",
archivePrefix = "arXiv",
primaryClass = "hep-th",
reportNumber = "NORDITA 2018-122",
doi = "10.1088/1751-8121/ab1b9c",
journal = "J. Phys. A",
volume = "52",
number = "22",
pages = "225401",
year = "2019"
}

@article{Kagan:2005wt,
author = "Kagan, David and Young, Charles A. S.",
title = "{Conformal sigma-models on supercoset targets}",
eprint = "hep-th/0512250",
archivePrefix = "arXiv",
reportNumber = "DAMTP-2005-130",
doi = "10.1016/j.nuclphysb.2006.02.027",
journal = "Nucl. Phys. B",
volume = "745",
pages = "109--122",
year = "2006"
}

@article{Fateev:1992tk,
author = "Fateev, V. A. and Onofri, E. and Zamolodchikov, Alexei B.",
title = "{Integrable deformations of the $O(3)$ sigma model. The sausage model}",
reportNumber = "PAR-LPTHE-92-46, LPTHE-92-46",
doi = "10.1016/0550-3213(93)90001-6",
journal = "Nucl. Phys. B",
volume = "406",
pages = "521--565",
year = "1993"
}

@article{Fateev:1996ea,
author = "Fateev, V. A.",
title = "{The sigma model (dual) representation for a two-parameter family of integrable quantum field theories}",
doi = "10.1016/0550-3213(96)00256-8",
journal = "Nucl. Phys. B",
volume = "473",
pages = "509--538",
year = "1996"
}

@article{Hoare:2014pna,
author = "Hoare, B. and Roiban, R. and Tseytlin, A. A.",
title = "{On deformations of $AdS_n \times S^n$ supercosets}",
eprint = "1403.5517",
archivePrefix = "arXiv",
primaryClass = "hep-th",
reportNumber = "IMPERIAL-TP-AT-2014-02, HU-EP-14-10",
doi = "10.1007/JHEP06(2014)002",
journal = "JHEP",
volume = "06",
pages = "002",
year = "2014"
}

@article{Lukyanov:2012zt,
author = "Lukyanov, Sergei L.",
title = "{The integrable harmonic map problem versus Ricci flow}",
eprint = "1205.3201",
archivePrefix = "arXiv",
primaryClass = "hep-th",
reportNumber = "RUNHETC-2012-10",
doi = "10.1016/j.nuclphysb.2012.08.002",
journal = "Nucl. Phys. B",
volume = "865",
pages = "308--329",
year = "2012"
}

@article{Braaten:1985is,
author = "Braaten, Eric and Curtright, Thomas L. and Zachos, Cosmas K.",
title = "{Torsion and Geometrostasis in Nonlinear Sigma Models}",
reportNumber = "UFTP-85-01, ANL-HEP-PR-85-03",
doi = "10.1016/0550-3213(86)90196-3",
journal = "Nucl. Phys. B",
volume = "260",
pages = "630",
year = "1985",
note = "[Erratum: Nucl.Phys.B 266, 748--748 (1986)]"
}

@article{Appadu:2018ioy,
author = "Appadu, Calan and Hollowood, Timothy J. and Price, Dafydd and Thompson, Daniel C.",
title = "{Quantum Anisotropic Sigma and Lambda Models as Spin Chains}",
eprint = "1802.06016",
archivePrefix = "arXiv",
primaryClass = "hep-th",
doi = "10.1088/1751-8121/aadc6d",
journal = "J. Phys. A",
volume = "51",
number = "40",
pages = "405401",
year = "2018"
}

@article{Georgiou:2018gpe,
author = "Georgiou, George and Sfetsos, Konstantinos",
title = "{The most general $\lambda$-deformation of CFTs and integrability}",
eprint = "1812.04033",
archivePrefix = "arXiv",
primaryClass = "hep-th",
doi = "10.1007/JHEP03(2019)094",
journal = "JHEP",
volume = "03",
pages = "094",
year = "2019"
}

@article{Delduc:2019bcl,
author = "Delduc, F. and Lacroix, S. and Magro, M. and Vicedo, B.",
title = "{Assembling integrable $\sigma$-models as affine Gaudin models}",
eprint = "1903.00368",
archivePrefix = "arXiv",
primaryClass = "hep-th",
reportNumber = "ZMP-HH/19-4",
doi = "10.1007/JHEP06(2019)017",
journal = "JHEP",
volume = "06",
pages = "017",
year = "2019"
}

@article{Elitzur:1994ri,
author = "Elitzur, S. and Giveon, A. and Rabinovici, E. and Schwimmer, A. and Veneziano, G.",
title = "{Remarks on non-abelian duality}",
eprint = "hep-th/9409011",
archivePrefix = "arXiv",
reportNumber = "CERN-TH-7414-94, RI-9-94, WIS-7-94",
doi = "10.1016/0550-3213(94)00426-F",
journal = "Nucl. Phys. B",
volume = "435",
pages = "147--171",
year = "1995"
}

@article{Alvarez:1994np,
author = "Alvarez, Enrique and Alvarez-Gaume, Luis and Lozano, Yolanda",
title = "{On non-abelian duality}",
eprint = "hep-th/9403155",
archivePrefix = "arXiv",
reportNumber = "CERN-TH-7204-94",
doi = "10.1016/0550-3213(94)90093-0",
journal = "Nucl. Phys. B",
volume = "424",
pages = "155--183",
year = "1994"
}

@article{Itsios:2014lca,
author = "Itsios, Georgios and Sfetsos, Konstadinos and Siampos, Konstantinos",
title = "{The all-loop non-abelian Thirring model and its RG flow}",
eprint = "1404.3748",
archivePrefix = "arXiv",
primaryClass = "hep-th",
doi = "10.1016/j.physletb.2014.04.061",
journal = "Phys. Lett. B",
volume = "733",
pages = "265--269",
year = "2014"
}

@article{Sfetsos:2014jfa,
author = "Sfetsos, Konstadinos and Siampos, Konstadinos",
title = "{Gauged WZW-type theories and the all-loop anisotropic non-abelian Thirring model}",
eprint = "1405.7803",
archivePrefix = "arXiv",
primaryClass = "hep-th",
doi = "10.1016/j.nuclphysb.2014.06.012",
journal = "Nucl. Phys. B",
volume = "885",
pages = "583--599",
year = "2014"
}

@article{Orlando:2019his,
author = "Orlando, Domenico and Reffert, Susanne and Sakamoto, Jun-ichi and Sekiguchi, Yuta and Yoshida, Kentaroh",
title = "{Yang-Baxter deformations and generalized supergravity -- a short summary}",
eprint = "1912.02553",
archivePrefix = "arXiv",
primaryClass = "hep-th",
doi = "10.1088/1751-8121/abb510",
journal = "J. Phys. A",
volume = "53",
number = "44",
pages = "443001",
year = "2020"
}

@article{Evans:2000hx,
author = "Evans, J. M. and Hassan, M. and MacKay, N. J. and Mountain, A. J.",
title = "{Conserved charges and supersymmetry in principal chiral and WZW models}",
eprint = "hep-th/0001222",
archivePrefix = "arXiv",
reportNumber = "PUPT-1908, DAMTP-99-178, IMPERIAL-TP-99-00-15",
doi = "10.1016/S0550-3213(00)00257-1",
journal = "Nucl. Phys. B",
volume = "580",
pages = "605--646",
year = "2000"
}

@article{Lacroix:2017isl,
author = "Lacroix, Sylvain and Magro, Marc and Vicedo, Benoit",
title = "{Local charges in involution and hierarchies in integrable sigma-models}",
eprint = "1703.01951",
archivePrefix = "arXiv",
primaryClass = "hep-th",
doi = "10.1007/JHEP09(2017)117",
journal = "JHEP",
volume = "09",
pages = "117",
year = "2017"
}

@article{Evans:2000qx,
author = "Evans, J. M. and Mountain, A. J.",
title = "{Commuting charges and symmetric spaces}",
eprint = "hep-th/0003264",
archivePrefix = "arXiv",
reportNumber = "PUPT-1922, DAMTP-2000-28, IMPERIAL-TP-99-00-20",
doi = "10.1016/S0370-2693(00)00566-9",
journal = "Phys. Lett. B",
volume = "483",
pages = "290--298",
year = "2000"
}

@article{Loebbert:2016cdm,
author = "Loebbert, Florian",
title = "{Lectures on Yangian Symmetry}",
eprint = "1606.02947",
archivePrefix = "arXiv",
primaryClass = "hep-th",
reportNumber = "HU-EP-16-12",
doi = "10.1088/1751-8113/49/32/323002",
journal = "J. Phys. A",
volume = "49",
number = "32",
pages = "323002",
year = "2016"
}

@article{Bernard:1992ya,
author = "Bernard, Denis",
title = "{An Introduction to Yangian Symmetries}",
eprint = "hep-th/9211133",
archivePrefix = "arXiv",
reportNumber = "SACLAY-SPH-T-92-134",
doi = "10.1142/S0217979293003371",
journal = "Int. J. Mod. Phys. B",
volume = "7",
pages = "3517--3530",
year = "1993"
}

@article{MacKay:2004tc,
author = "MacKay, N. J.",
title = "{Introduction to Yangian symmetry in integrable field theory}",
eprint = "hep-th/0409183",
archivePrefix = "arXiv",
reportNumber = "ESI-1514",
doi = "10.1142/S0217751X05022317",
journal = "Int. J. Mod. Phys. A",
volume = "20",
pages = "7189--7218",
year = "2005"
}

@article{Kawaguchi:2010jg,
author = "Kawaguchi, Io and Yoshida, Kentaroh",
title = "{Hidden Yangian symmetry in sigma model on squashed sphere}",
eprint = "1008.0776",
archivePrefix = "arXiv",
primaryClass = "hep-th",
reportNumber = "KUNS-2286",
doi = "10.1007/JHEP11(2010)032",
journal = "JHEP",
volume = "11",
pages = "032",
year = "2010"
}

@article{Arutyunov:2009ga,
author = "Arutyunov, Gleb and Frolov, Sergey",
title = "{Foundations of the $AdS_{5} \times S^{5}$ Superstring. Part I}",
eprint = "0901.4937",
archivePrefix = "arXiv",
primaryClass = "hep-th",
reportNumber = "ITP-UU-09-05, SPIN-09-05, TCD-MATH-09-06, HMI-09-03",
doi = "10.1088/1751-8113/42/25/254003",
journal = "J. Phys. A",
volume = "42",
pages = "254003",
year = "2009"
}

@article{Beisert:2010jr,
author = "Beisert, Niklas and others",
title = "{Review of AdS/CFT Integrability: An Overview}",
eprint = "1012.3982",
archivePrefix = "arXiv",
primaryClass = "hep-th",
reportNumber = "AEI-2010-175, CERN-PH-TH-2010-306, HU-EP-10-87, HU-MATH-2010-22, KCL-MTH-10-10, UMTG-270, UUITP-41-10",
doi = "10.1007/s11005-011-0529-2",
journal = "Lett. Math. Phys.",
volume = "99",
pages = "3--32",
year = "2012"
}

@article{Levkovich-Maslyuk:2019awk,
author = "Levkovich-Maslyuk, Fedor",
title = "{A review of the AdS/CFT Quantum Spectral Curve}",
eprint = "1911.13065",
archivePrefix = "arXiv",
primaryClass = "hep-th",
doi = "10.1088/1751-8121/ab7137",
journal = "J. Phys. A",
volume = "53",
number = "28",
pages = "283004",
year = "2020"
}

@article{Thompson:2019ipl,
author = "Thompson, Daniel C.",
editor = "Anagnostopoulos, Konstantinos and others",
title = "{An Introduction to Generalised Dualities and their Applications to Holography and Integrability}",
eprint = "1904.11561",
archivePrefix = "arXiv",
primaryClass = "hep-th",
doi = "10.22323/1.347.0099",
journal = "PoS",
volume = "CORFU2018",
pages = "099",
year = "2019"
}

@article{Ogievetsky:1992ph,
author = "Ogievetsky, O.",
title = "{Hopf structures on the Borel subalgebra of sl(2)}",
editor = "Bure\v{s}, J. and Sou\v{c}ek, J.",
booktitle = "{Proceedings of the 13th Winter School `Geometry and Physics', Zd\'{i}kov, Czech Republic, 1993}",
publisher = "Circolo Matematico di Palermo",
address = "Palermo",
journal = "Suppl. Rend. Circ. Mat. Palermo, II. Ser.",
volume = "37",
year = "1994",
pages = "185",
}

@article{Stolin:1991a,
author = "Stolin, A",
title = "{Constant solutions of Yang-Baxter equation for sl(2) and sl(3)}",
doi = "10.7146/math.scand.a-12370",
journal = "Math. Scand.",
volume = "69",
pages = "81",
year = "1991",
}

@article{Drinfeld:1985rx,
author = "Drinfeld, V.G.",
title = "{Hopf algebras and the quantum Yang-Baxter equation}",
journal = "Sov. Math. Dokl.",
volume = "32",
pages = "254--258",
year = "1985"
}

@article{Jimbo:1985zk,
author = "Jimbo, Michio",
title = "{A q-Difference Analogue of U(g) and the Yang-Baxter Equation}",
doi = "10.1007/BF00704588",
journal = "Lett. Math. Phys.",
volume = "10",
pages = "63--69",
year = "1985"
}

@article{Belavin:1982,
author = "Belavin, A. A. and Drinfel'd, V. G.",
title = "{Solutions of the classical Yang-Baxter equation for simple Lie algebras}",
doi = "10.1007/BF01081585",
journal = "Funct. Anal. Appl.",
volume = "16",
pages = "159--180",
year = "1982",
}

@article{Belavin:1984,
author = "Belavin, A. A. and Drinfel'd, V. G.",
title = "{Triangle equations and simple Lie algebras}",
journal = "Sov. Sci. Rev.",
volume = "C4",
pages = "93",
year = "1984",
}

@article{Fukushima:2020kta,
author = "Fukushima, Osamu and Sakamoto, Jun-ichi and Yoshida, Kentaroh",
title = "{Comments on $\eta$-deformed principal chiral model from 4D Chern-Simons theory}",
eprint = "2003.07309",
archivePrefix = "arXiv",
primaryClass = "hep-th",
reportNumber = "KUNS-2802",
doi = "10.1016/j.nuclphysb.2020.115080",
journal = "Nucl. Phys. B",
volume = "957",
pages = "115080",
year = "2020"
}

@article{Tian:2020ryu,
author = "Tian, Jia",
title = "{Comments on $\lambda$--deformed models from 4D Chern-Simons theory}",
eprint = "2005.14554",
archivePrefix = "arXiv",
primaryClass = "hep-th",
month = "5",
year = "2020"
}

@article{Hoare:2015wia,
author = "Hoare, B. and Tseytlin, A. A.",
title = "{Type IIB supergravity solution for the T-dual of the $\eta$-deformed $AdS_{5} \times S^{5}$ superstring}",
eprint = "1508.01150",
archivePrefix = "arXiv",
primaryClass = "hep-th",
reportNumber = "HU-EP-15-34, IMPERIAL-TP-AT-2015-05",
doi = "10.1007/JHEP10(2015)060",
journal = "JHEP",
volume = "10",
pages = "060",
year = "2015"
}

@article{Arutyunov:2015mqj,
author = "Arutyunov, G. and Frolov, S. and Hoare, B. and Roiban, R. and Tseytlin, A. A.",
title = "{Scale invariance of the $\eta$-deformed $AdS_5 \times S^5$ superstring, T-duality and modified type II equations}",
eprint = "1511.05795",
archivePrefix = "arXiv",
primaryClass = "hep-th",
reportNumber = "ZMP-HH-15-27, TCDMATH-15-12, IMPERIAL-TP-AT-2015-08",
doi = "10.1016/j.nuclphysb.2015.12.012",
journal = "Nucl. Phys. B",
volume = "903",
pages = "262--303",
year = "2016"
}

@article{Appadu:2015nfa,
author = "Appadu, Calan and Hollowood, Timothy J.",
title = "{Beta function of k deformed $AdS_{5} \times S^{5}$ string theory}",
eprint = "1507.05420",
archivePrefix = "arXiv",
primaryClass = "hep-th",
doi = "10.1007/JHEP11(2015)095",
journal = "JHEP",
volume = "11",
pages = "095",
year = "2015"
}

@article{Bassi:2019aaf,
author = "Bassi, Cristian and Lacroix, Sylvain",
title = "{Integrable deformations of coupled $\sigma$-models}",
eprint = "1912.06157",
archivePrefix = "arXiv",
primaryClass = "hep-th",
reportNumber = "ZMP-HH/19-26",
doi = "10.1007/JHEP05(2020)059",
journal = "JHEP",
volume = "05",
pages = "059",
year = "2020"
}

@article{Arutyunov:2020sdo,
author = "Arutyunov, Gleb and Bassi, Cristian and Lacroix, Sylvain",
title = "{New integrable coset sigma models}",
eprint = "2010.05573",
archivePrefix = "arXiv",
primaryClass = "hep-th",
reportNumber = "ZMP-HH/20-19",
doi = "10.1007/JHEP03(2021)062",
journal = "JHEP",
volume = "03",
pages = "062",
year = "2021"
}

@article{Georgiou:2020wwg,
author = "Georgiou, George",
title = "{Webs of integrable theories}",
eprint = "2006.12525",
archivePrefix = "arXiv",
primaryClass = "hep-th",
doi = "10.1016/j.nuclphysb.2021.115340",
journal = "Nucl. Phys. B",
volume = "965",
pages = "115340",
year = "2021"
}

@article{Seibold:2019dvf,
author = "Seibold, Fiona K.",
title = "{Two-parameter integrable deformations of the $AdS_3 \times S^3 \times T^4$ superstring}",
eprint = "1907.05430",
archivePrefix = "arXiv",
primaryClass = "hep-th",
doi = "10.1007/JHEP10(2019)049",
journal = "JHEP",
volume = "10",
pages = "049",
year = "2019"
}

@article{Vicedo:2017cge,
author = "Vicedo, Benoit",
title = "{On integrable field theories as dihedral affine Gaudin models}",
eprint = "1701.04856",
archivePrefix = "arXiv",
primaryClass = "hep-th",
doi = "10.1093/imrn/rny128",
journal = "Int. Math. Res. Not.",
volume = "2020",
number = "15",
pages = "4513--4601",
year = "2020"
}

@article{Costello:2019tri,
author = "Costello, Kevin and Yamazaki, Masahito",
title = "{Gauge Theory And Integrability, III}",
eprint = "1908.02289",
archivePrefix = "arXiv",
primaryClass = "hep-th",
reportNumber = "IPMU19-0110",
month = "8",
year = "2019"
}

@article{Vicedo:2019dej,
author = "Vicedo, Benoit",
title = "{Holomorphic Chern-Simons theory and affine Gaudin models}",
eprint = "1908.07511",
archivePrefix = "arXiv",
primaryClass = "hep-th",
month = "8",
year = "2019"
}

@article{Bittleston:2020hfv,
author = "Bittleston, Roland and Skinner, David",
title = "{Twistors, the ASD Yang-Mills equations, and 4d Chern-Simons theory}",
eprint = "2011.04638",
archivePrefix = "arXiv",
primaryClass = "hep-th",
month = "11",
year = "2020"
}

@article{Kawaguchi:2012ve,
author = "Kawaguchi, Io and Matsumoto, Takuya and Yoshida, Kentaroh",
title = "{The classical origin of quantum affine algebra in squashed sigma models}",
eprint = "1201.3058",
archivePrefix = "arXiv",
primaryClass = "hep-th",
reportNumber = "KUNS-2379",
doi = "10.1007/JHEP04(2012)115",
journal = "JHEP",
volume = "04",
pages = "115",
year = "2012"
}

@article{Kawaguchi:2013gma,
author = "Kawaguchi, Io and Yoshida, Kentaroh",
title = "{A deformation of quantum affine algebra in squashed Wess-Zumino-Novikov-Witten models}",
eprint = "1311.4696",
archivePrefix = "arXiv",
primaryClass = "hep-th",
reportNumber = "KUNS-2467",
doi = "10.1063/1.4880341",
journal = "J. Math. Phys.",
volume = "55",
pages = "062302",
year = "2014"
}

@article{Matsumoto:2015jja,
author = "Matsumoto, Takuya and Yoshida, Kentaroh",
title = "{Yang-Baxter sigma models based on the CYBE}",
eprint = "1501.03665",
archivePrefix = "arXiv",
primaryClass = "hep-th",
reportNumber = "KUNS-2534",
doi = "10.1016/j.nuclphysb.2015.02.009",
journal = "Nucl. Phys. B",
volume = "893",
pages = "287--304",
year = "2015"
}

@article{Delduc:2017brb,
author = "Delduc, Francois and Kameyama, Takashi and Magro, Marc and Vicedo, Benoit",
title = "{Affine $q$-deformed symmetry and the classical Yang-Baxter  $\sigma$-model}",
eprint = "1701.03691",
archivePrefix = "arXiv",
primaryClass = "hep-th",
doi = "10.1007/JHEP03(2017)126",
journal = "JHEP",
volume = "03",
pages = "126",
year = "2017"
}

@phdthesis{Lacroix:2018njs,
author = "Lacroix, Sylvain",
title = "{Integrable models with twist function and affine Gaudin models}",
eprint = "1809.06811",
archivePrefix = "arXiv",
primaryClass = "hep-th",
reportNumber = "tel-01900498, 2018LYSEN014",
journal = "Doctoral",
volume = "Thesis",
pages = "ENS de Lyon",
school = "Lyon, Ecole Normale Superieure",
year = "2018"
}

@phdthesis{Seibold:2020ouf,
author = "Seibold, Fiona K.",
title = "{Integrable deformations of sigma models and superstrings}",
doi = "10.3929/ethz-b-000440825",
journal = "Doctoral",
volume = "Thesis",
pages = "ETH Zurich",
school = "Zurich, ETH",
year = "2020"
}

@article{Maillet:1985fn,
author = "Maillet, Jean Michel",
title = "{Kac-moody Algebra and Extended {Yang-Baxter} Relations in the O($N$) Nonlinear $\sigma$ Model}",
reportNumber = "PAR LPTHE 85-22",
doi = "10.1016/0370-2693(85)91075-5",
journal = "Phys. Lett. B",
volume = "162",
pages = "137--142",
year = "1985"
}

@article{Maillet:1985ek,
author = "Maillet, Jean Michel",
title = "{New Integrable Canonical Structures in Two-dimensional Models}",
reportNumber = "PAR/LPTHE-85-32",
doi = "10.1016/0550-3213(86)90365-2",
journal = "Nucl. Phys. B",
volume = "269",
pages = "54--76",
year = "1986"
}

@article{Hollowood:2015dpa,
author = "Hollowood, Timothy J. and Miramontes, J. Luis and Schmidtt, David M.",
title = "{S-Matrices and Quantum Group Symmetry of k-Deformed Sigma Models}",
eprint = "1506.06601",
archivePrefix = "arXiv",
primaryClass = "hep-th",
doi = "10.1088/1751-8113/49/46/465201",
journal = "J. Phys. A",
volume = "49",
number = "46",
pages = "465201",
year = "2016"
}

@article{Hoare:2017ukq,
author = "Hoare, Ben and Seibold, Fiona K.",
title = "{Poisson-Lie duals of the $\eta$ deformed symmetric space sigma model}",
eprint = "1709.01448",
archivePrefix = "arXiv",
primaryClass = "hep-th",
doi = "10.1007/JHEP11(2017)014",
journal = "JHEP",
volume = "11",
pages = "014",
year = "2017"
}

@article{Lust:2018jsx,
author = {L\"ust, Dieter and Osten, David},
title = "{Generalised fluxes, Yang-Baxter deformations and the O(d,d) structure of non-abelian T-duality}",
eprint = "1803.03971",
archivePrefix = "arXiv",
primaryClass = "hep-th",
reportNumber = "LMU-ASC 11/18, MPP-2018-35, LMU-ASC-11-18",
doi = "10.1007/JHEP05(2018)165",
journal = "JHEP",
volume = "05",
pages = "165",
year = "2018"
}

@article{Hoare:2018ebg,
author = "Hoare, Ben and Seibold, Fiona K.",
title = "{Poisson-Lie duals of the $\eta$-deformed $AdS_2 \times S^2 \times T^6$ superstring}",
eprint = "1807.04608",
archivePrefix = "arXiv",
primaryClass = "hep-th",
doi = "10.1007/JHEP08(2018)107",
journal = "JHEP",
volume = "08",
pages = "107",
year = "2018"
}

@article{Klimcik:1996np,
author = "Klimcik, C. and Severa, P.",
title = "{Dressing cosets}",
eprint = "hep-th/9602162",
archivePrefix = "arXiv",
reportNumber = "CERN-TH-96-43",
doi = "10.1016/0370-2693(96)00669-7",
journal = "Phys. Lett. B",
volume = "381",
pages = "56--61",
year = "1996"
}

@article{Squellari:2011dg,
author = "Squellari, R.",
title = "{Dressing cosets revisited}",
eprint = "1105.0162",
archivePrefix = "arXiv",
primaryClass = "hep-th",
doi = "10.1016/j.nuclphysb.2011.07.025",
journal = "Nucl. Phys. B",
volume = "853",
pages = "379--403",
year = "2011"
}

@article{Klimcik:2017ken,
author = "Klimcik, Ctirad",
title = "{Yang-Baxter $\sigma$-model with WZNW term as ${ \mathcal E}$-model}",
eprint = "1706.08912",
archivePrefix = "arXiv",
primaryClass = "hep-th",
doi = "10.1016/j.physletb.2017.07.051",
journal = "Phys. Lett. B",
volume = "772",
pages = "725--730",
year = "2017"
}

@article{Bazhanov:2017nzh,
author = "Bazhanov, Vladimir V. and Kotousov, Gleb A. and Lukyanov, Sergei L.",
title = "{Quantum transfer-matrices for the sausage model}",
eprint = "1706.09941",
archivePrefix = "arXiv",
primaryClass = "hep-th",
doi = "10.1007/JHEP01(2018)021",
journal = "JHEP",
volume = "01",
pages = "021",
year = "2018"
}

@article{Brodbeck:1999ib,
author = "Brodbeck, Othmar and Zagermann, Marco",
title = "{Dimensionally reduced gravity, Hermitian symmetric spaces and the Ashtekar variables}",
eprint = "gr-qc/9911118",
archivePrefix = "arXiv",
reportNumber = "PSU-TH-223",
doi = "10.1088/0264-9381/17/14/310",
journal = "Class. Quant. Grav.",
volume = "17",
pages = "2749--2764",
year = "2000"
}

@article{Bykov:2016rdv,
author = "Bykov, Dmitri",
title = "{Complex structures and zero-curvature equations for $\sigma$-models}",
eprint = "1605.01093",
archivePrefix = "arXiv",
primaryClass = "hep-th",
doi = "10.1016/j.physletb.2016.06.071",
journal = "Phys. Lett. B",
volume = "760",
pages = "341--344",
year = "2016"
}

@article{Delduc:2019lpe,
author = "Delduc, Francois and Kameyama, Takashi and Lacroix, Sylvain and Magro, Marc and Vicedo, Benoit",
title = "{Ultralocal Lax connection for para-complex $\mathbb{Z}_T$-cosets}",
eprint = "1909.00742",
archivePrefix = "arXiv",
primaryClass = "hep-th",
doi = "10.1016/j.nuclphysb.2019.114821",
journal = "Nucl. Phys. B",
volume = "949",
pages = "114821",
year = "2019"
}

@article{Fateev:1995ht,
author = "Fateev, V. A.",
title = "{The Duality between two-dimensional integrable field theories and sigma models}",
doi = "10.1016/0370-2693(95)00883-M",
journal = "Phys. Lett. B",
volume = "357",
pages = "397--403",
year = "1995"
}

@article{Fateev:2018yos,
author = "Fateev, V. A. and Litvinov, A. V.",
title = "{Integrability, Duality and Sigma Models}",
eprint = "1804.03399",
archivePrefix = "arXiv",
primaryClass = "hep-th",
doi = "10.1007/JHEP11(2018)204",
journal = "JHEP",
volume = "11",
pages = "204",
year = "2018"
}

@article{Litvinov:2018bou,
author = "Litvinov, A. V. and Spodyneiko, L. A.",
title = "{On dual description of the deformed $O(N)$ sigma model}",
eprint = "1804.07084",
archivePrefix = "arXiv",
primaryClass = "hep-th",
doi = "10.1007/JHEP11(2018)139",
journal = "JHEP",
volume = "11",
pages = "139",
year = "2018"
}

@article{Litvinov:2019rlv,
author = "Litvinov, A. V.",
title = "{Integrable $\mathfrak{gl}(n|n)$ Toda field theory and its sigma-model dual}",
eprint = "1901.04799",
archivePrefix = "arXiv",
primaryClass = "hep-th",
doi = "10.1134/S0021364019230048",
journal = "Pisma Zh. Eksp. Teor. Fiz.",
volume = "110",
number = "11",
pages = "723--726",
year = "2019"
}

@article{Fateev:2019xuq,
author = "Fateev, Vladimir",
title = "{Classical and Quantum Integrable Sigma Models. Ricci Flow, \textquotedblleft{}Nice Duality\textquotedblright{} and Perturbed Rational Conformal Field Theories}",
eprint = "1902.02811",
archivePrefix = "arXiv",
primaryClass = "hep-th",
doi = "10.1134/S1063776119100042",
journal = "J. Exp. Theor. Phys.",
volume = "129",
number = "4",
pages = "566--590",
year = "2019"
}

@article{Abdalla:1982yd,
author = "Abdalla, E. and Forger, M. and Gomes, M.",
title = "{On the Origin of Anomalies in the Quantum Nonlocal Charge for the Generalized Nonlinear $\sigma$ Models}",
reportNumber = "IFUSP/P-329",
doi = "10.1016/0550-3213(82)90238-3",
journal = "Nucl. Phys. B",
volume = "210",
pages = "181--192",
year = "1982"
}

@article{Hoare:2019mcc,
author = "Hoare, Ben and Levine, Nat and Tseytlin, Arkady A.",
title = "{Integrable sigma models and 2-loop RG flow}",
eprint = "1910.00397",
archivePrefix = "arXiv",
primaryClass = "hep-th",
reportNumber = "Imperial-TP-AT-2019-06",
doi = "10.1007/JHEP12(2019)146",
journal = "JHEP",
volume = "12",
pages = "146",
year = "2019"
}

@article{Borsato:2019oip,
author = "Borsato, Riccardo and Wulff, Linus",
title = "{Two-loop conformal invariance for Yang-Baxter deformed strings}",
eprint = "1910.02011",
archivePrefix = "arXiv",
primaryClass = "hep-th",
doi = "10.1007/JHEP03(2020)126",
journal = "JHEP",
volume = "03",
pages = "126",
year = "2020"
}

@article{Georgiou:2019nbz,
author = "Georgiou, George and Sagkrioti, Eftychia and Sfetsos, Konstantinos and Siampos, Konstantinos",
title = "{An exact symmetry in $\lambda$-deformed CFTs}",
eprint = "1911.02027",
archivePrefix = "arXiv",
primaryClass = "hep-th",
reportNumber = "CERN-TH-2019-053",
doi = "10.1007/JHEP01(2020)083",
journal = "JHEP",
volume = "01",
pages = "083",
year = "2020"
}

@article{Hassler:2020xyj,
author = "Hassler, Falk",
title = "{RG flow of integrable $\mathcal{E}$-models}",
eprint = "2012.10451",
archivePrefix = "arXiv",
primaryClass = "hep-th",
doi = "10.1016/j.physletb.2021.136367",
journal = "Phys. Lett. B",
volume = "818",
pages = "136367",
year = "2021"
}

@article{Hassler:2020tvz,
author = "Hassler, Falk and Rochais, Thomas",
title = "{$\alpha'$-Corrected Poisson-Lie T-Duality}",
eprint = "2007.07897",
archivePrefix = "arXiv",
primaryClass = "hep-th",
doi = "10.1002/prop.202000063",
journal = "Fortsch. Phys.",
volume = "68",
number = "9",
pages = "2000063",
year = "2020"
}

@article{Borsato:2020wwk,
author = "Borsato, Riccardo and Wulff, Linus",
title = "{Quantum Correction to Generalized $T$ Dualities}",
eprint = "2007.07902",
archivePrefix = "arXiv",
primaryClass = "hep-th",
doi = "10.1103/PhysRevLett.125.201603",
journal = "Phys. Rev. Lett.",
volume = "125",
number = "20",
pages = "201603",
year = "2020"
}

@article{Codina:2020yma,
author = "Codina, Tomas and Marques, Diego",
title = "{Generalized Dualities and Higher Derivatives}",
eprint = "2007.09494",
archivePrefix = "arXiv",
primaryClass = "hep-th",
doi = "10.1007/JHEP10(2020)002",
journal = "JHEP",
volume = "10",
pages = "002",
year = "2020"
}

@article{vanTongeren:2015uha,
author = "van Tongeren, Stijn J.",
title = "{Yang-Baxter deformations, AdS/CFT, and twist-noncommutative gauge theory}",
eprint = "1506.01023",
archivePrefix = "arXiv",
primaryClass = "hep-th",
reportNumber = "HU-EP-15-27, HU-MATH-15-08",
doi = "10.1016/j.nuclphysb.2016.01.012",
journal = "Nucl. Phys. B",
volume = "904",
pages = "148--175",
year = "2016"
}

@article{Araujo:2017jap,
author = "Araujo, Thiago and Bakhmatov, Ilya and O~Colgain, Eoin and Sakamoto, Jun-ichi and Sheikh-Jabbari, Mohammad M. and Yoshida, Kentaroh",
title = "{Conformal twists, Yang-Baxter $\sigma$-models \& holographic noncommutativity}",
eprint = "1705.02063",
archivePrefix = "arXiv",
primaryClass = "hep-th",
reportNumber = "IPM-P-2017-016, KUNS-2676, APCTP-PRE2017-004",
doi = "10.1088/1751-8121/aac195",
journal = "J. Phys. A",
volume = "51",
number = "23",
pages = "235401",
year = "2018"
}

@article{Mazzucato:2011jt,
author = "Mazzucato, Luca",
title = "{Superstrings in AdS}",
eprint = "1104.2604",
archivePrefix = "arXiv",
primaryClass = "hep-th",
doi = "10.1016/j.physrep.2012.08.001",
journal = "Phys. Rept.",
volume = "521",
pages = "1--68",
year = "2012"
}

@article{Ke:2017wis,
author = "Ke, San-Min and Yang, Wen-Li and Jang, Ke-Xia and Wang, Chun and Shuai, Xue-Min and Wang, Zhan-Yun and Shi, Gang",
title = "{Yang-Baxter deformations of supercoset sigma models with \ensuremath{\mathbb{Z}}(4)m( ) grading}",
doi = "10.1088/1674-1137/41/11/113101",
journal = "Chin. Phys. C",
volume = "41",
number = "11",
pages = "113101",
year = "2017"
}

@article{Magro:2008dv,
author = "Magro, Marc",
title = "{The Classical Exchange Algebra of $AdS_5 \times S^5$}",
eprint = "0810.4136",
archivePrefix = "arXiv",
primaryClass = "hep-th",
reportNumber = "AEI-2008-085",
doi = "10.1088/1126-6708/2009/01/021",
journal = "JHEP",
volume = "01",
pages = "021",
year = "2009"
}

@article{Ke:2011zzb,
author = "Ke, San-Min and Li, Xin-Ying and Wang, Chun and Yue, Rui-Hong",
title = "{Classical exchange algebra of the nonlinear sigma model on a supercoset target with $\mathbb{Z}_{2n}$ grading}",
doi = "10.1088/0256-307X/28/10/101101",
journal = "Chin. Phys. Lett.",
volume = "28",
pages = "101101",
year = "2011"
}

@article{Ke:2011zzd,
author = "Ke, San-Min and Yang, Wen-Li and Wang, Chun and Jiang, Ke-Xia and Shi, Kang-Jie",
title = "{The classical exchange algebra of a Green-Schwarz sigma model on supercoset target space with $\mathbb{Z}_{4m}$ grading}",
doi = "10.1063/1.3626193",
journal = "J. Math. Phys.",
volume = "52",
pages = "083511",
year = "2011"
}

@article{Ke:2008zz,
author = "Ke, San-Min and Shi, Kang-Jie and Wang, Chun",
title = "{Flat currents of a Green-Schwarz sigma model on supercoset targets with $\mathbb{Z}_{4m}$ grading}",
doi = "10.1142/S0217751X08040378",
journal = "Int. J. Mod. Phys. A",
volume = "23",
pages = "4219--4243",
year = "2008"
}

@article{Bykov:2016ovg,
author = "Bykov, D. V.",
title = "{Cyclic gradings of Lie algebras and Lax pairs for $\sigma$-models}",
doi = "10.1134/S0040577916120060",
journal = "Theor. Math. Phys.",
volume = "189",
number = "3",
pages = "1734--1741",
year = "2016"
}

@book{Babelon:2003qtg,
author = "Babelon, Olivier and Bernard, Denis and Talon, Michel",
title = "{Introduction to Classical Integrable Systems}",
doi = "10.1017/CBO9780511535024",
isbn = "978-0-521-03670-2, 978-0-511-53502-4",
publisher = "Cambridge University Press",
series = "Cambridge Monographs on Mathematical Physics",
year = "2003"
}

@article{Zamolodchikov:1978xm,
author = "Zamolodchikov, Alexander B. and Zamolodchikov, Alexei B.",
editor = "Khalatnikov, I. M. and Mineev, V. P.",
title = "{Factorized S-Matrices in Two-Dimensions as the Exact Solutions of Certain Relativistic Quantum Field Models}",
reportNumber = "ITEP-35-1978",
doi = "10.1016/0003-4916(79)90391-9",
journal = "Annals Phys.",
volume = "120",
pages = "253--291",
year = "1979"
}

@article{Klimcik:2021bjy,
author = "Klimcik, Ctirad",
title = "{Brief lectures on duality, integrability and deformations}",
eprint = "2101.05230",
archivePrefix = "arXiv",
primaryClass = "hep-th",
doi = "10.1142/S0129055X21300041",
journal = "Rev. Math. Phys.",
volume = "33",
number = "06",
pages = "2130004",
year = "2021"
}

@article{Zamolodchikov:1989hfa,
author = "Zamolodchikov, A. B.",
editor = "Jimbo, M. and Miwa, T. and Tsuchiya, A.",
title = "{Integrable field theory from conformal field theory}",
journal = "Adv. Stud. Pure Math.",
volume = "19",
pages = "641--674",
year = "1989"
}

@article{Smirnov:2016lqw,
author = "Smirnov, F. A. and Zamolodchikov, A. B.",
title = "{On space of integrable quantum field theories}",
eprint = "1608.05499",
archivePrefix = "arXiv",
primaryClass = "hep-th",
doi = "10.1016/j.nuclphysb.2016.12.014",
journal = "Nucl. Phys. B",
volume = "915",
pages = "363--383",
year = "2017"
}

@article{Cavaglia:2016oda,
author = "Cavaglia, Andrea and Negro, Stefano and Szecsenyi, Istvan M. and Tateo, Roberto",
title = "{$T \bar{T}$-deformed 2D Quantum Field Theories}",
eprint = "1608.05534",
archivePrefix = "arXiv",
primaryClass = "hep-th",
doi = "10.1007/JHEP10(2016)112",
journal = "JHEP",
volume = "10",
pages = "112",
year = "2016"
}

@article{Retore:2021wwh,
author = "Retore, Ana L.",
title = "{Introduction to classical and quantum integrability}",
eprint = "2109.14280",
archivePrefix = "arXiv",
primaryClass = "hep-th",
month = "9",
year = "2021"
}

@article{Lacroix:2021iit,
author = "Lacroix, Sylvain",
title = "{4-dimensional Chern-Simons theory and integrable field theories}",
eprint = "2109.14278",
archivePrefix = "arXiv",
primaryClass = "hep-th",
month = "9",
year = "2021"
}

@article{Dashen:1974hp,
author = "Dashen, Roger F. and Frishman, Y.",
title = "{Four Fermion Interactions and Scale Invariance}",
reportNumber = "Print-75-0205 (IAS,PRINCETON)",
doi = "10.1103/PhysRevD.11.2781",
journal = "Phys. Rev. D",
volume = "11",
pages = "2781",
year = "1975"
}

@article{Kutasov:1989aw,
author = "Kutasov, D.",
title = "{Duality Off the Critical Point in Two-dimensional Systems With Nonabelian Symmetries}",
doi = "10.1016/0370-2693(89)91325-7",
journal = "Phys. Lett. B",
volume = "233",
pages = "369--373",
year = "1989"
}

@article{Georgiou:2015nka,
author = "Georgiou, George and Sfetsos, Konstantinos and Siampos, Konstantinos",
title = "{All-loop anomalous dimensions in integrable \ensuremath{\lambda}-deformed \ensuremath{\sigma}-models}",
eprint = "1509.02946",
archivePrefix = "arXiv",
primaryClass = "hep-th",
doi = "10.1016/j.nuclphysb.2015.10.007",
journal = "Nucl. Phys. B",
volume = "901",
pages = "40--58",
year = "2015"
}

@article{Georgiou:2016iom,
author = "Georgiou, George and Sfetsos, Konstantinos and Siampos, Konstantinos",
title = "{All-loop correlators of integrable \ensuremath{\lambda}-deformed \ensuremath{\sigma}-models}",
eprint = "1604.08212",
archivePrefix = "arXiv",
primaryClass = "hep-th",
doi = "10.1016/j.nuclphysb.2016.05.018",
journal = "Nucl. Phys. B",
volume = "909",
pages = "360--393",
year = "2016"
}

@article{Tseytlin:1993hm,
author = "Tseytlin, Arkady A.",
title = "{On A 'Universal' class of WZW type conformal models}",
eprint = "hep-th/9311062",
archivePrefix = "arXiv",
reportNumber = "CERN-TH-7068-93",
doi = "10.1016/0550-3213(94)90243-7",
journal = "Nucl. Phys. B",
volume = "418",
pages = "173--194",
year = "1994"
}

@article{Sfetsos:1999zm,
author = "Sfetsos, Konstadinos",
title = "{Duality invariant class of two-dimensional field theories}",
eprint = "hep-th/9904188",
archivePrefix = "arXiv",
reportNumber = "CERN-TH-99-112",
doi = "10.1016/S0550-3213(99)00485-X",
journal = "Nucl. Phys. B",
volume = "561",
pages = "316--340",
year = "1999"
}

@article{Klimcik:1994wp,
author = "Klimcik, C. and Tseytlin, Arkady A.",
title = "{Exact four-dimensional string solutions and Toda like sigma models from 'null gauged' WZNW theories}",
eprint = "hep-th/9402120",
archivePrefix = "arXiv",
reportNumber = "IMPERIAL-TP-93-94-17, PRA-HEP-94-1",
doi = "10.1016/0550-3213(94)90089-2",
journal = "Nucl. Phys. B",
volume = "424",
pages = "71--96",
year = "1994"
}

@article{Sfetsos:2009vt,
author = "Sfetsos, K. and Siampos, K. and Thompson, Daniel C.",
title = "{Renormalization of Lorentz non-invariant actions and manifest T-duality}",
eprint = "0910.1345",
archivePrefix = "arXiv",
primaryClass = "hep-th",
reportNumber = "QMUL-PH-09-22",
doi = "10.1016/j.nuclphysb.2009.11.001",
journal = "Nucl. Phys. B",
volume = "827",
pages = "545--564",
year = "2010"
}

@article{Severa:2018pag,
author = "Severa, Pavol and Valach, Fridrich",
title = "{Courant Algebroids, Poisson\textendash{}Lie T-Duality, and Type II Supergravities}",
eprint = "1810.07763",
archivePrefix = "arXiv",
primaryClass = "math.DG",
doi = "10.1007/s00220-020-03736-x",
journal = "Commun. Math. Phys.",
volume = "375",
number = "1",
pages = "307--344",
year = "2020"
}

@article{Polyakov:2004br,
author = "Polyakov, A. M.",
editor = "Shifman, M. and Vainshtein, A. and Wheater, J.",
title = "{Conformal fixed points of unidentified gauge theories}",
eprint = "hep-th/0405106",
archivePrefix = "arXiv",
reportNumber = "PUPT-2118",
doi = "10.1142/S0217732304015129",
journal = "Mod. Phys. Lett. A",
volume = "19",
pages = "1649--1660",
year = "2004"
}

@article{Adam:2007ws,
author = "Adam, Ido and Dekel, Amit and Mazzucato, Luca and Oz, Yaron",
title = "{Integrability of Type II Superstrings on Ramond-Ramond Backgrounds in Various Dimensions}",
eprint = "hep-th/0702083",
archivePrefix = "arXiv",
reportNumber = "TAUP-2849-07",
doi = "10.1088/1126-6708/2007/06/085",
journal = "JHEP",
volume = "06",
pages = "085",
year = "2007"
}

@article{Abbott:1981ke,
author = "Abbott, L. F.",
title = "{Introduction to the Background Field Method}",
reportNumber = "CERN-TH-3113",
journal = "Acta Phys. Polon. B",
volume = "13",
pages = "33",
year = "1982"
}

@article{Klimcik:2018vhl,
author = "Klimcik, Ctirad",
title = "{Affine Poisson and affine quasi-Poisson T-duality}",
eprint = "1809.01614",
archivePrefix = "arXiv",
primaryClass = "hep-th",
doi = "10.1016/j.nuclphysb.2018.12.008",
journal = "Nucl. Phys. B",
volume = "939",
pages = "191--232",
year = "2019"
}

@article{Dorey:1996gd,
author = "Dorey, P.",
title = "{Exact S matrices}",
booktitle = "{Eotvos Summer School in Physics: Conformal Field Theories and Integrable Models}",
eprint = "hep-th/9810026",
archivePrefix = "arXiv",
reportNumber = "DTP-98-69",
month = "8",
year = "1996"
}

@article{Ogievetsky:1987vv,
author = "Ogievetsky, E. and Wiegmann, P. and Reshetikhin, N.",
title = "{The Principal Chiral Field in Two-Dimensions on Classical Lie Algebras: The Bethe Ansatz Solution and Factorized Theory of Scattering}",
doi = "10.1016/0550-3213(87)90138-6",
journal = "Nucl. Phys. B",
volume = "280",
pages = "45--96",
year = "1987"
}

@article{Rajeev:1996kk,
author = "Rajeev, S. G. and Stern, A. and Vitale, P.",
title = "{Integrability of the Wess-Zumino-Witten model as a nonultralocal theory}",
eprint = "hep-th/9602149",
archivePrefix = "arXiv",
reportNumber = "UAHEP-9603, DSFNA-T-9606, ESI-310",
doi = "10.1016/S0370-2693(96)01224-5",
journal = "Phys. Lett. B",
volume = "388",
pages = "769--775",
year = "1996"
}

@article{Mussardo:1992uc,
author = "Mussardo, Giuseppe",
title = "{Off critical statistical models: Factorized scattering theories and bootstrap program}",
reportNumber = "SISSA-37-92-EP",
doi = "10.1016/0370-1573(92)90047-4",
journal = "Phys. Rept.",
volume = "218",
pages = "215--379",
year = "1992"
}

\end{bibtex}

\bibliographystyle{nb}
\bibliography{\jobname}

\end{document}